\documentclass[a4paper,11pt,twoside]{report}

\pdfoutput=1

\usepackage[UKenglish]{babel}   

\usepackage[T1]{fontenc}        
\usepackage{mathptmx}           
\usepackage{helvet}             
\usepackage{courier}            

\usepackage{amsmath,amssymb}  
\usepackage{microtype}        
\usepackage{pbox}             
\usepackage{textcomp}
\usepackage{bm}                 

\usepackage{float}
\usepackage{subcaption}       
\usepackage{graphicx}

\usepackage{fancyhdr}         
\usepackage{titlesec}         
\titleformat{\chapter}[hang]{\LARGE\bfseries}{\thechapter\hspace{26pt}}{0pt}{\LARGE\bfseries}
\titlespacing*{\chapter}{0pt}{3.5\baselineskip}{1.8\baselineskip}

\usepackage[font=small,labelfont=bf,labelsep=period]{caption}  

\usepackage[top=30mm, bottom=30mm, inner=30mm, outer=30mm, bindingoffset=0mm, marginparsep=3mm, marginparwidth=20mm, heightrounded]{geometry}
\newlength{\logobox}
\usepackage{comment}

\usepackage{color}
\usepackage{soul}
\usepackage{multirow}

\usepackage{siunitx} 
\usepackage{booktabs,dcolumn} 
\newcolumntype{,}{D{,}{,}{-1}} 
\usepackage{ragged2e}  
\usepackage{tabularx}  
\newcolumntype{Y}{>{\RaggedRight\arraybackslash}X}
\usepackage{threeparttable}		

\usepackage{nicefrac}

\usepackage{braket}

\usepackage[dvipsnames]{xcolor}

\usepackage[version=4]{mhchem}	

\usepackage{miller}

\usepackage{natbib}

\usepackage[colorlinks=true,allcolors=blue,bookmarks=true,bookmarksnumbered=true]{hyperref}

\hypersetup{                    
pdfauthor = {The OC18 consortium},
pdftitle = {Guidelines for developing optical clocks with 1E-18 fractional frequency uncertainty},
pdfsubject = {Optical clocks},
pdfkeywords = {optical clock, ion clock, lattice clock},
pdfstartview = {FitV},
}

\newcommand{\authorlist}[1]{#1 \vspace{0.5\baselineskip}}

\newcommand{\affil}[2]{\noindent {\footnotesize $^{#1}$ \emph{#2} \par}}
\newcommand{\corr}[1]{\noindent {\footnotesize $^\dag$ e-mail: \href{mailto:#1}{#1} \par}}
\newcommand{\presaddr}[1]{\noindent {\footnotesize $^\ddag$ Present address: #1 \par}}

\newcommand{\chapstart}{\vspace{1.5\baselineskip} \noindent}

\newcommand{\ptb}{1}
\newcommand{\umk}{2}
\newcommand{\roa}{3}
\newcommand{\cmi}{4}
\newcommand{\npl}{5}
\newcommand{\tubitak}{{6}}
\newcommand{\op}{7}
\newcommand{\vtt}{8}
\newcommand{\luh}{9}
\newcommand{\inrim}{{10}}
\newcommand{\bu}{{11}}
\newcommand{\nbi}{{12}}

\newcommand{\PTBaff}{Physikalisch-Technische Bundesanstalt, Bundesallee 100, 38116 Braunschweig, Germany}
\newcommand{\LUHaff}{Leibniz Universit\"at Hannover, Welfengarten 1, 30167 Hannover, Germany}
\newcommand{\VTTaff}{VTT Technical Research Centre of Finland Ltd, National Metrology Institute VTT MIKES, P.O.\ Box 1000, FI-02044 VTT, Finland}
\newcommand{\NPLaff}{National Physical Laboratory, Hampton Road, Teddington, TW11 0LW, United Kingdom}
\newcommand{\NBIaff}{Copenhagen University, Niels Bohr Institute, Blegdamsvej 17, 2100 Copenhagen, Denmark}
\newcommand{\OPaff}{LNE-SYRTE, Observatoire de Paris, Universit\'e PSL, CNRS, Sorbonne Universit\'e, 61 Avenue de l'Observatoire, 75014 Paris, France}
\newcommand{\ROAaff}{Secci\'{o}n de Hora, Real Instituto y Observatorio de la Armada, San Fernando, Spain}
\newcommand{\TUBITAKaff}{T{\"U}B\.{I}TAK National Metrology Institute (UME), 41470, Kocaeli, Turkey}
\newcommand{\BUaff}{Department of Physics, Bo\u{g}azi\c{c}i University, 34342, Istanbul, Turkey}
\newcommand{\INRIMaff}{Istituto Nazionale di Ricerca Metrologica, Strada delle Cacce 91, 10135 Torino, Italy}
\newcommand{\CMIaff}{Czech Metrology Institute, V Botanice 4, 150\;72 Prague, Czech Republic}
\newcommand{\UMKaff}{Institute of Physics, Faculty of Physics, Astronomy and Informatics, Nicolaus Copernicus University, Grudzi\c{a}dzka 5, PL-87-100 Toru\'n, Poland}

\newcommand*\loadev{\sigma_\mathrm{L}}
\newcommand*\clkadev{\sigma_\mathrm{C}}
\newcommand*\wnom{\omega}

\newcommand*\Tnought{Z}

\newcommand{\state}[3]{{}^{#1}\mathrm{#2}_{#3}}
\newcommand{\isotope}[2]{{}^{#1}\mathrm{#2}}
\newcommand{\Sr}{\isotope{87}{Sr}}
\newcommand{\Erec}{E_\mathrm{r}} 						
\newcommand{\Ngnd}{N_\mathrm{g}}						
\newcommand{\Nexc}{N_\mathrm{e}}						
\newcommand{\thold}{t_\mathrm{h}}						
\newcommand{\Upeak}{U_0}								
\newcommand{\traplossGS}{\Gamma_{\mathrm{bg}}}			
\newcommand{\traplossES}{\Gamma_{\mathrm{bg}}^\prime}	
\newcommand{\traplossDIFF}{\Delta\Gamma_{\mathrm{bg}}}	
\newcommand{\decayratecoeffLAT}{\gamma_{\mathrm{L}}}	
\newcommand{\decayrateOTHER}{\Gamma_{\mathrm{0}}}		
\newcommand{\Utherm}{U}									
\newcommand{\matrixelement}[3]{\braket{#1|#2|#3}}

\newcommand{\dd}{\mathrm{d}}
\newcommand{\ped}[1]{_\mathrm{#1}}

\newcommand{\kB}{k_\mathrm{B}}

\newcommand{\aEO}{\alpha_\text{E1}}

\begin{document}

\graphicspath{{Title/}}

\enlargethispage{1\baselineskip}
\thispagestyle{empty}
\vspace*{-2\baselineskip}
\pbox{0.12\textwidth}{\vspace{-5mm}\hspace{-5mm}\href{http://www.oc18.eu}{\includegraphics[width=0.12\textwidth]{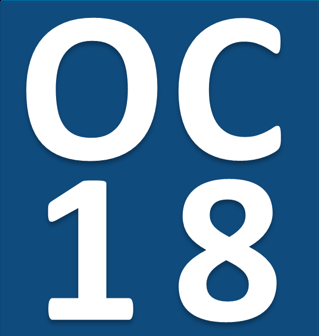}}}
\hfill
\pbox{0.65\textwidth}{\vspace{-6mm}\hspace{1mm}\href{http://www.empir-online.eu/}{\includegraphics[width=0.65\textwidth]{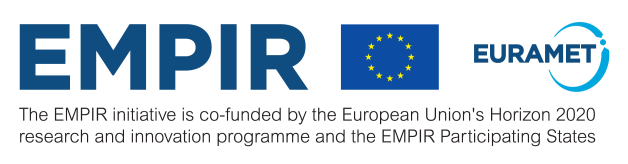}}}

\vspace*{3.5\baselineskip}

\begin{flushleft}
{\huge \textbf{Guidelines for developing optical clocks with $\mathbf{10^{-18}}$ fractional frequency uncertainty}}\\  
\pdfbookmark[0]{Guidelines for developing optical clocks with 1E-18 fractional frequency uncertainty}{title}
\vspace*{1.5\baselineskip}

\noindent
Moustafa Abdel-Hafiz$^\ptb$,
Piotr Ablewski$^\umk$,
Ali Al-Masoudi$^\ptb$,
H\'{e}ctor \'{A}lvarez Mart\'{i}nez$^\roa$,
Petr~Balling$^\cmi$,
Geoffrey Barwood$^\npl$,
Erik Benkler$^\ptb$,
Marcin Bober$^\umk$,
Mateusz~Borkowski$^\umk$,
William~Bowden$^\npl$,
Roman Ciury\l{}o$^\umk$,
Hubert Cybulski$^\umk$,
Alexandre Didier$^\ptb$,
Miroslav Dole\v{z}al$^\cmi$,
S\"oren D\"orscher$^\ptb$,
Stephan Falke$^{\ptb}$,
Rachel~M.~Godun$^\npl$,
Ramiz Hamid$^\tubitak$,
Ian~R.~Hill$^\npl$,
Richard~Hobson$^\npl$,
Nils~Huntemann$^\ptb$,
Yann Le Coq$^\op$,
Rodolphe Le Targat$^\op$,
Thomas Legero$^\ptb$,
Thomas Lindvall$^\vtt$,
Christian Lisdat$^\ptb$,
J\'er\^ome~Lodewyck$^\op$,
Helen~S.~Margolis$^\npl$,
Tanja~E.~Mehlst\"aubler$^{\ptb}$,
Ekkehard Peik$^\ptb$,
Lennart Pelzer$^{\ptb ,\luh}$,
Marco Pizzocaro$^\inrim$,
Benjamin Rauf$^\inrim$,
Antoine~Rolland$^\npl$,
Nils~Scharnhorst$^{\ptb ,\luh}$,
Marco Schioppo$^\npl$,
Piet O.~Schmidt$^{\ptb ,\luh}$,
Roman Schwarz$^\ptb$,
\c{C}a\u{g}r{\i}~\c{S}enel$^{\tubitak ,\bu}$,
Nicolas~Spethmann$^\ptb$,
Uwe~Sterr$^\ptb$,
Christian~Tamm$^\ptb$,
Jan W.~Thomsen$^\nbi$,
Alvise~Vianello$^\npl$
and Micha\l{}~Zawada$^\umk$ \\

\vspace*{0.7\baselineskip}

\noindent
{Edited by Thomas~Lindvall$^\vtt$}\\

\vspace*{0.7\baselineskip}

\affil{\ptb}{\PTBaff}
\affil{\umk}{\UMKaff}
\affil{\roa}{\ROAaff}
\affil{\cmi}{\CMIaff}
\affil{\npl}{\NPLaff}
\affil{\tubitak}{\TUBITAKaff}
\affil{\op}{\OPaff}
\affil{\vtt}{\VTTaff}
\affil{\luh}{\LUHaff}
\affil{\inrim}{\INRIMaff}
\affil{\bu}{\BUaff}
\affil{\nbi}{\NBIaff}


\end{flushleft}

\vfill

\noindent
\begin{tabular}{c @{\hspace{0.065\textwidth}} c @{\hspace{0.045\textwidth}} c  @{\hspace{0.045\textwidth}} c}
\centering
\vspace{5mm}
\pbox{\logobox}{\href{http://www.npl.co.uk/}{\includegraphics[width=0.2\textwidth]{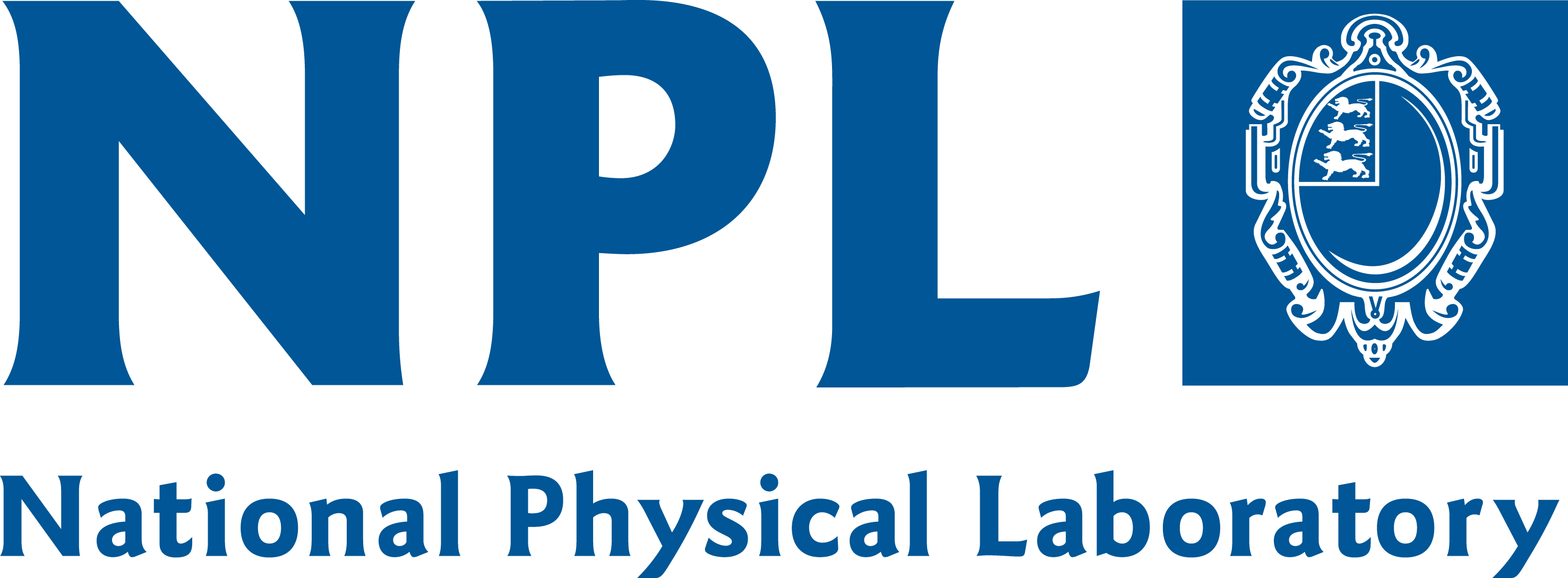}}} &
\pbox{\logobox}{\vspace{-2mm}\href{https://www.cmi.cz/?language=en}{\includegraphics[width=0.2\textwidth]{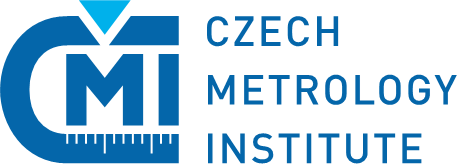}}} &
\pbox{\logobox}{\vspace{-2mm}\href{https://www.inrim.eu/}{\includegraphics[width=0.2\textwidth]{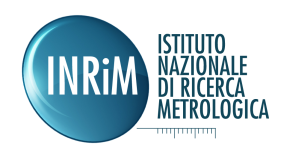}}} &
\pbox{\logobox}{\vspace{-2mm}\href{https://www.lne.fr/en}{\includegraphics[width=0.2\textwidth]{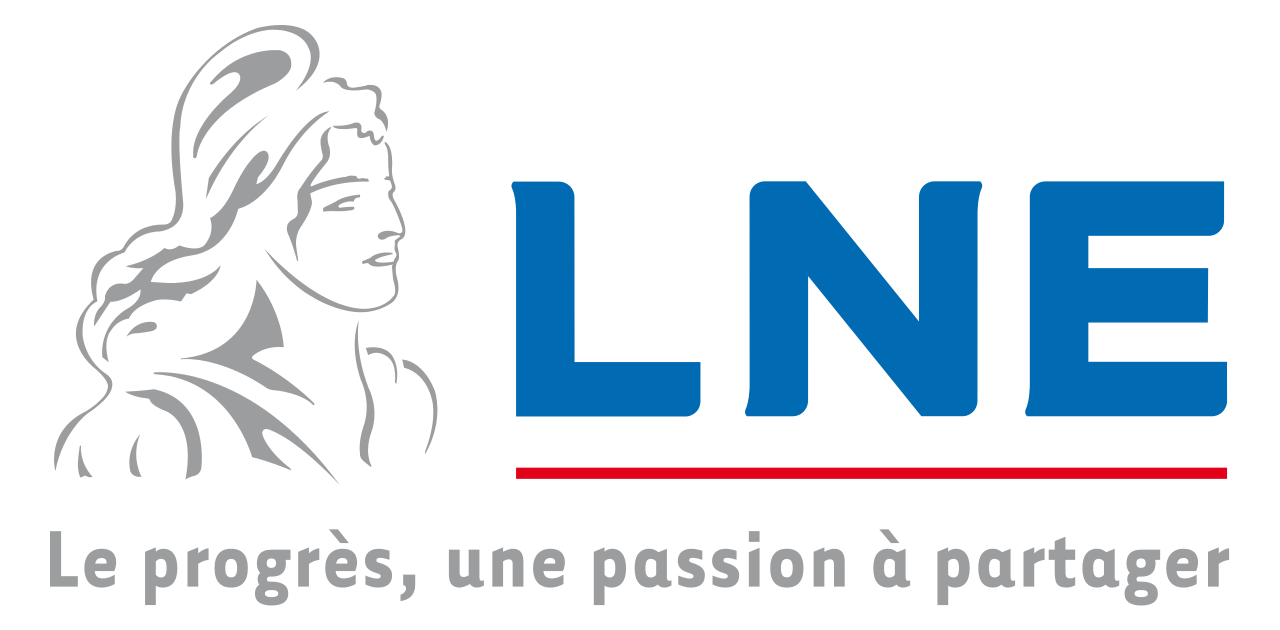}}} \\
\vspace{3mm}
\pbox{\logobox}{\href{https://www.obspm.fr/?lang=en}{\includegraphics[width=0.22\textwidth]{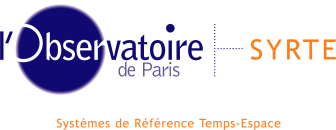}}} &
\pbox{\logobox}{\href{http://www.ptb.de/cms/en.html}{\includegraphics[width=0.2\textwidth]{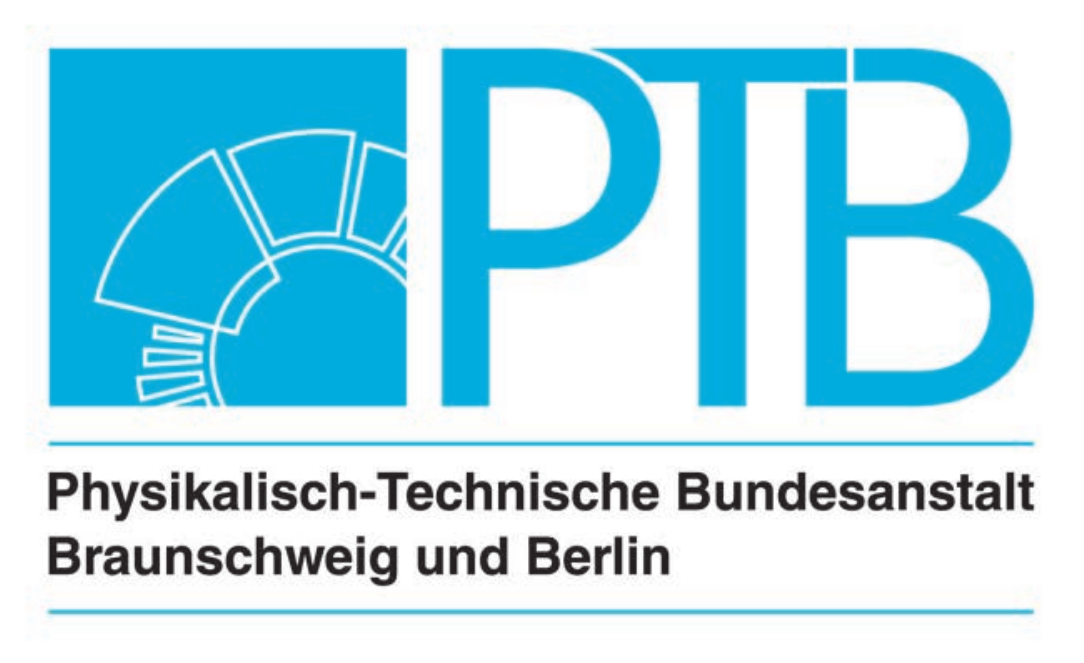}}} &
\pbox{\logobox}{\href{http://www.tubitak.gov.tr/en}{\includegraphics[width=0.11\textwidth]{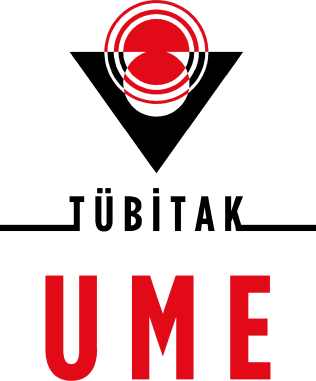}}} &
\pbox{\logobox}{\href{http://www.mikes.fi/en}{\includegraphics[width=0.20\textwidth]{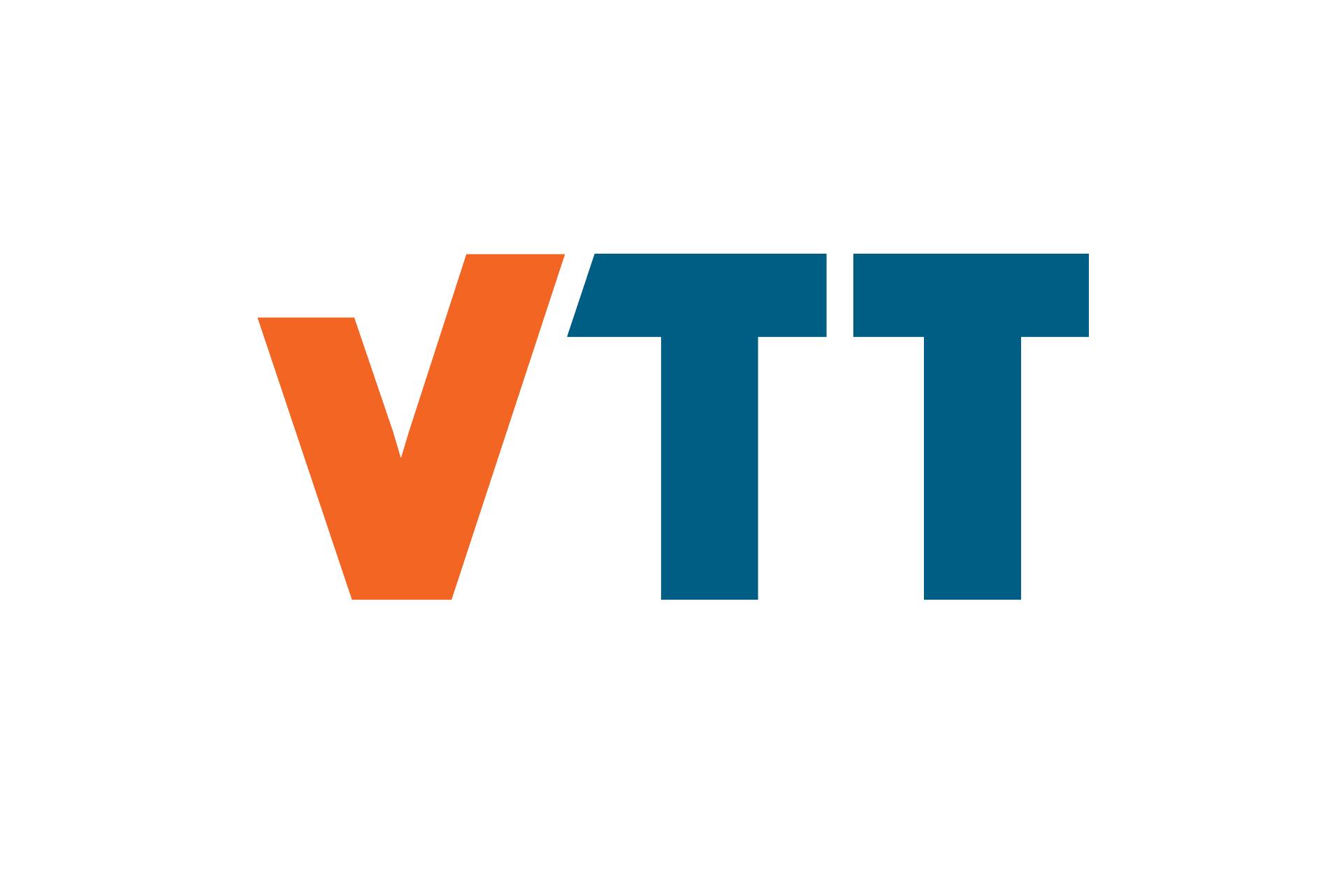}}} \\
\pbox{\logobox}{\href{http://www.ku.dk/english/}{\includegraphics[width=0.12\textwidth]{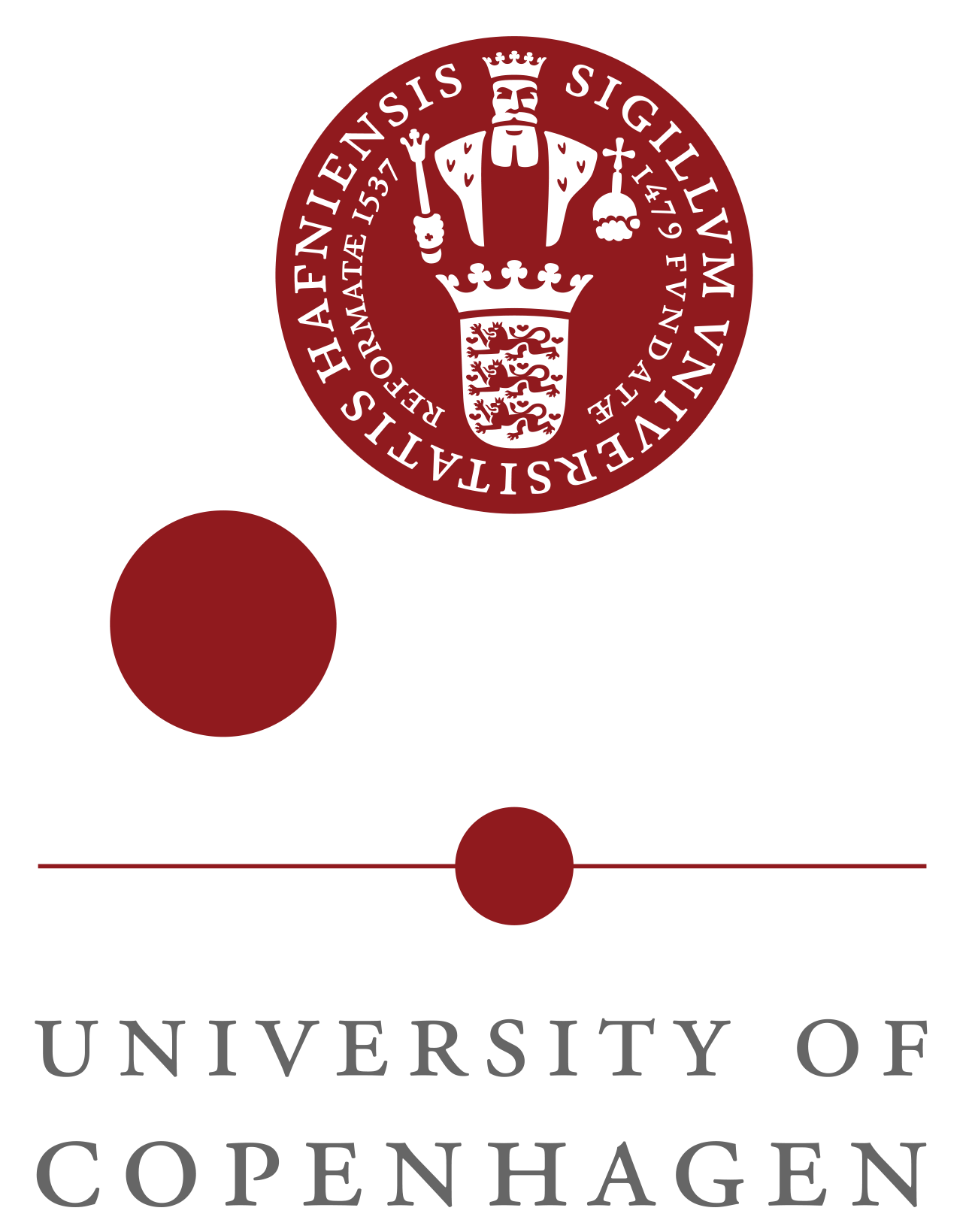}}} &
\pbox{\logobox}{\href{http://www.uni-hannover.de/en/}{\includegraphics[width=0.20\textwidth]{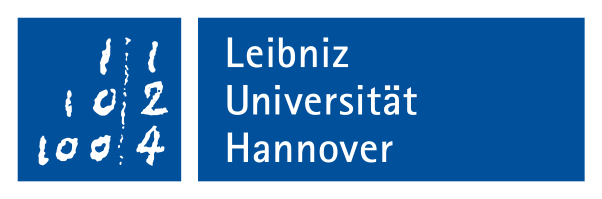}}} &
\pbox{\logobox}{\href{https://www.umk.pl/en/}{\includegraphics[width=0.22\textwidth]{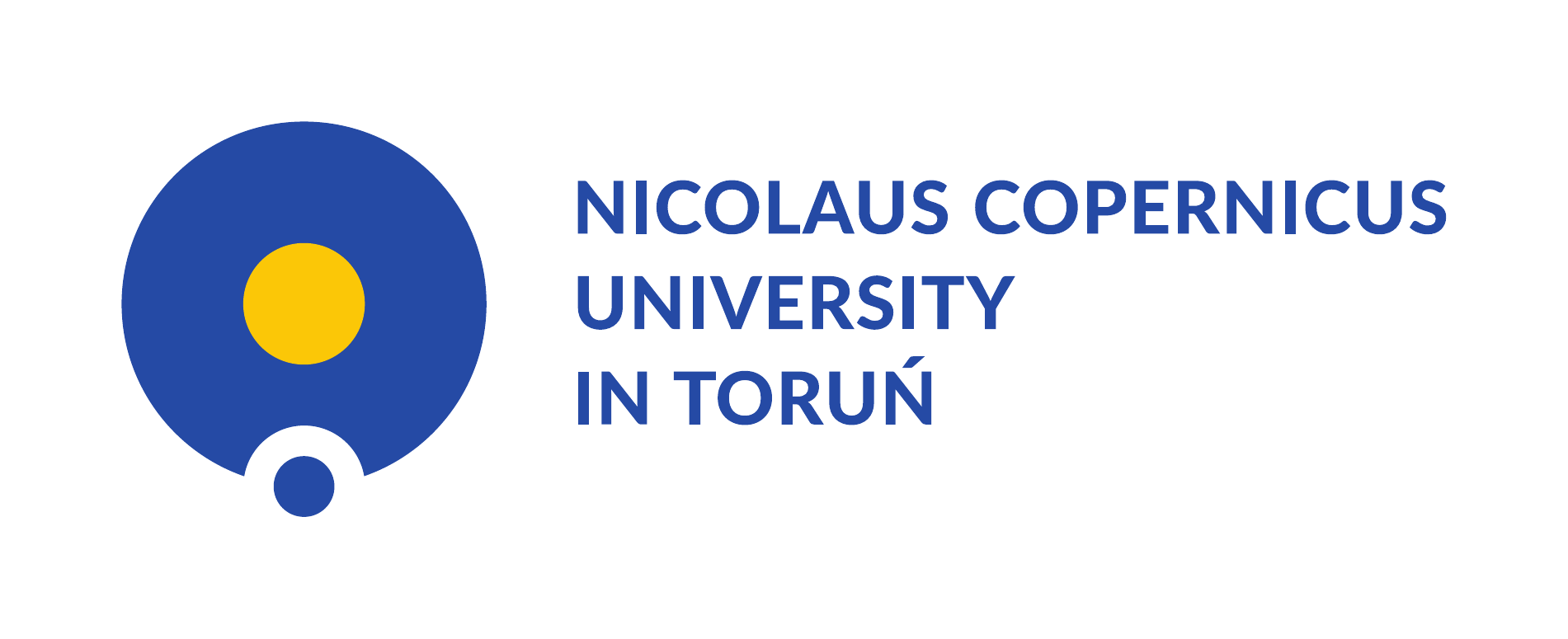}}} &
\pbox{\logobox}{\href{http://www.cnrs.fr/index.php/en}{\includegraphics[width=0.12\textwidth]{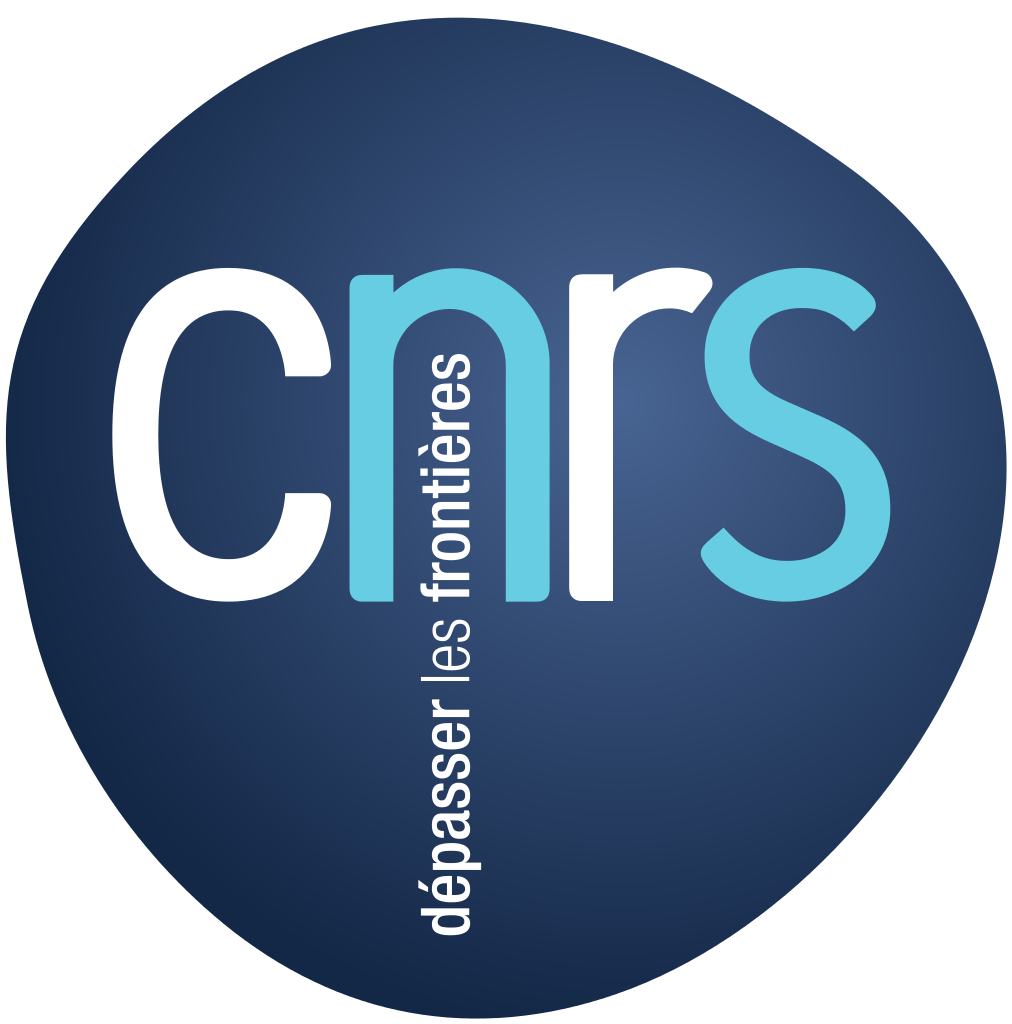}}} \\

\end{tabular}

\clearpage

\thispagestyle{plain}
\vspace*{5\baselineskip}

Guidelines for developing optical clocks with $10^{-18}$ fractional frequency uncertainty \par
{\href{http://www.oc18.eu}{http://www.oc18.eu}}\par
2019\par
\vspace{0.5\baselineskip}
Edited by Thomas~Lindvall\par
e-mail: \href{mailto:thomas.lindvall@vtt.fi}{thomas.lindvall@vtt.fi}\par

\vfill

\noindent
This work was carried out within the EMPIR project 15SIB03 OC18. This project has received funding from the EMPIR programme co-financed by the Participating States and from the European Union's Horizon 2020 research and innovation programme. 
\clearpage

\cleardoublepage
\phantomsection
\addcontentsline{toc}{chapter}{Contents}
\tableofcontents

\chapter*{Preface}
\addcontentsline{toc}{chapter}{Preface}

There has been tremendous progress in the performance of optical frequency standards since the early proposals to carry out precision spectroscopy on trapped, single ions \citep{Dehmelt1973,Dehmelt1982}.
The estimated fractional frequency uncertainty of today's leading optical standards is currently in the $10^{-18}$ range, approximately two orders of magnitude better than that of the best caesium-fountain microwave primary frequency standards. Due to this progress, the International Committee for Weights and Measures (CIPM) recommended four optical clock transitions as \emph{secondary representations of the second} in 2006 \citep{CIPM2006}. Currently, there are eight optical secondary representations \citep{Riehle2018,BIPM2019} and, ultimately, this development is expected to lead to a \emph{redefinition of the second} based on some optical transition(s) \citep{Gill2011,Riehle2015,Riehle2016}.

The exceptional accuracy and stability offered by optical frequency standards is resulting in a growing number of research groups developing optical clocks. While good review papers covering the topic already exist (see for example \citet{Ludlow2015} and references therein), more practical guidelines are needed as a complement. The purpose of this document is therefore to provide technical guidance for researchers starting in the field of optical clocks. The target audience includes national metrology institutes (NMIs) wanting to set up optical clocks (or subsystems thereof) and PhD students and postdocs entering the field. Another potential audience is academic groups with experience in atomic physics and atom or ion trapping, but with less experience of time and frequency metrology and optical clock requirements.

These guidelines have arisen from the scope of the EMPIR project ``Optical clocks with \num{1e-18} uncertainty'' (\href{http://www.oc18.eu}{OC18}). Therefore, the examples are from European laboratories even though similar work is carried out all over the world.

The goal of OC18 was to push the development of optical clocks by improving each of the necessary subsystems:  ultrastable lasers, transfer of frequency stability from the laser to the atoms, atom/ion traps and interrogation techniques. This document shares the knowledge acquired by the OC18 project consortium and gives practical guidance on each of these aspects in the following chapters.

\vspace{2\baselineskip}
\noindent
June 2019\\


\noindent\begin{minipage}[t]{0.25\linewidth}
Rachel M. Godun\\
OC18 coordinator
\end{minipage}
\begin{minipage}[t]{0.4\linewidth}
Thomas Lindvall\\
OC18 impact work package leader
\end{minipage}

\clearpage

\graphicspath{{D2_ultrastable/}}

\chapter{Ultrastable lasers for optical clocks \label{chap:ultrastable}}

\authorlist{%
Thomas Legero$^{1,\dag}$,
Alexandre Didier$^{1}$,
Yann Le Coq$^2$,
Thomas Lindvall$^3$,
Tanja E.\ Mehlst\"aubler$^{1}$,
Marco Schioppo$^4$ and
Jan W.\ Thomsen$^5$
}

\affil{1}{\PTBaff}
\affil{2}{\OPaff}
\affil{3}{\VTTaff}
\affil{4}{\NPLaff}
\affil{5}{\NBIaff}

\corr{Thomas.Legero@ptb.de}

\chapstart
Ultrastable lasers are a key component in optical atomic clocks. They serve as the interrogation laser for atomic clock transitions and largely determine the stability of the clocks. However, the coherence time of today's most stable lasers is on the order of a few seconds, which is still an order of magnitude lower than the coherence time of optical clock transitions in neutral atoms or ions. Thus, the frequency fluctuations of the lasers still limit the resolved linewidth of the clock transitions and, due to the discontinuous interrogation, also affect the clock stability via the Dick effect.

Due to these limitations, great effort has been put into the realisation of laser sources with low frequency noise. The most common method used so far is the stabilisation of the laser frequency to a passive high-finesse Fabry--Perot cavity. With well-designed servo electronics, the fractional frequency instability of the laser is then identical to the fractional optical-path-length instability of the cavity. Accordingly, any environmental perturbation, such as temperature fluctuations or vibrations, which affects the optical path length must be largely suppressed.
In Section~\ref{sec:room_temp_glass_cavities} we discuss the performance and limitations of this approach with room-temperature glass cavities and show how to reach instabilities of $10^{-16}$ or below after \SI{1}{s}.

The fundamental limit of the frequency instability of the stabilised laser is given by length fluctuations due to inevitable Brownian thermal noise of the cavity components. Placing the cavity in a cryogenic environment can largely decrease the amount of thermal noise. Section~\ref{sec:cryogenic_silicon_cavity} illustrates the performance of ultrastable lasers employing cavities made of single-crystal silicon at cryogenic temperatures, reaching state-of-the-art instabilities of $4 \times 10^{-17}$.

In parallel, novel laser-stabilisation techniques that do not require such extreme control of the length of a Fabry--Perot cavity are also being explored. One of these new schemes for generating ultrastable optical frequencies uses cavity-QED systems, while another one uses spectral hole burning in cryogenically cooled crystals. We will discuss these approaches in Sections~\ref{sec:active_resonators} and \ref{sec:spectral_hole_burning}, respectively.

\section{Room temperature glass cavities \label{sec:room_temp_glass_cavities}}

Among the different systems presented in this chapter, ultrastable lasers obtained by stabilising the frequency of a CW laser to a high-finesse Fabry--Perot cavity made from ultralow thermal expansion materials, such as ULE$\textsuperscript{\textregistered}$, are broadly used and already commercialised~\citep{StableLaser,Menlo}.
The important feature of these cavities is the direct relation between the fractional frequency instability of a longitudinal cavity mode and its fractional length instability, expressed in Equation~\eqref{eq:Fabry_Perot} by the corresponding power spectral densities
\begin{equation}
\label{eq:Fabry_Perot}
	\frac{S_\nu}{\nu^{2}}=\frac{S_{L}}{L^{2}}.
\end{equation}
Here $\nu$ is the mode frequency and $L$ the length between the cavity mirrors. For a perfect control loop, the laser frequency instability is thus due to the cavity-length instability alone. Hence, important work has been performed over the years to develop Fabry--Perot cavities of extremely stable length.
This section addresses the major contributions to the instability of lasers stabilised to room-temperature glass cavities and discusses the techniques required to reach instabilities at the level of $10^{-16}$ or below. In Figure~\ref{fig:instability_summary} we will then see an example of how the different contributions add up to the final fractional frequency instability of a laser \citep{Keller2014}.

\subsection{Thermal noise \label{subsec:Thermal_Noise}}

The fundamental limit of the length stability of a cavity is imposed by the Brownian motion of its components. One can link the volume fluctuations of a material at equilibrium with its response to applied mechanical forces and how the energy is dissipated within this system~\citep{Callen1952}, thus obtaining a relation between the thermal noise of a material and its mechanical losses. The important equations used for estimating this thermal noise and its related power spectral density are described in~\citet{Numata2004,Kessler2012} and summarised below.

The power spectral density of the length fluctuations of a cavity can be approximated by the sum of the contributions from its components: one spacer, two mirror substrates (``sub'') and their coatings (``coat''),
\begin{equation}
\label{eq:Thermal_Noise}
S_{x}(f)=S_{x}^{\mathrm{spacer}}(f)+ 2 S_{x}^{\mathrm{sub}}(f) + 2 S_{x}^{\mathrm{coat}}(f).
\end{equation}
The thermal noise of the mirror substrate is given by
\begin{equation}
\label{eq:Thermal_Noise_Substrate}
S_{x}^{\mathrm{sub}}(f) =
\frac{4\kB T}{\pi f}
\frac{1-\sigma_{\mathrm{sub}}^{2}}{2\sqrt{\pi}E_{\mathrm{sub}}w}\phi_{\mathrm{sub}},
\end{equation}
 where $\kB$ is the Boltzmann constant, $T$ the temperature, and $w$ the beam waist on the mirrors. For all the equations, $\sigma_i$ denotes the Poisson coefficient of the material $i$, $E_i$ its Young's modulus and $\phi_i$ its mechanical losses.

In this section we consider only the thermal noise of a dielectric coating, given by
\begin{equation}
\label{eq:Thermal_Noise_Coating}
S_{x}^{\mathrm{coat}}(f) =
S_{x}^{\mathrm{sub}}(f)
\frac{2}{\sqrt{\pi}}
\frac{1-2\sigma_{\mathrm{sub}}}{1-\sigma_{\mathrm{sub}}}
\frac{\phi_{\mathrm{coat}}}{\phi_{\mathrm{sub}}}
\frac{d_{\mathrm{coat}}}{w},
\end{equation}
where $d_{\mathrm{coat}}$ is the coating thickness.
Finally, the thermal noise of a cylindrical spacer of length $L$, radius $R$ and central bore radius $r$ is given by
\begin{equation}
\label{eq:Thermal_Noise_Spacer}
S_{x}^{\mathrm{spacer}}(f) =
\frac{4\kB T}{\pi f}
\frac{L}{2\pi E_{\mathrm{spacer}}(R^{2}-r^{2})}\phi_{\mathrm{spacer}}.
\end{equation}
The thermal noise therefore has a flicker-frequency behaviour, which converts to a floor in the Allan deviation~\citep{Allan1966} and is given by
\begin{equation}
\sigma_{y}^{\mathrm{floor}} = \sqrt{2 \ln(2) f S_{y}(f)} =
\sqrt{2 \ln(2) f \frac{S_{x}(f)}{L^{2}}}.
\end{equation}

From these equations one can understand the different approaches that have been used to reduce the thermal noise and that recently led to instabilities $\leq 1\times 10^{-16}$. These include the use of mixed-material cavities with mirror substrates of lower loss factor, such as fused-silica (FS) mirror substrates on ULE spacers~\citep{Millo2009,Dawkins2010}, longer cavities \citep{Jiang2011,Nicholson2012,Keller2014,Haefner2015}, larger beams on the cavity mirrors~\citep{DAmbrosio2003,Mours2006,Amairi2012}, higher-performance crystalline coating materials~\citep{Cole2016} and materials with very low mechanical loss factor, such as single-crystal silicon operated at cryogenic temperatures~\citep{Kessler2012,Matei2017,Zhang2017} (see Section~\ref{sec:cryogenic_silicon_cavity}).

\begin{figure}[tb]
	\centering
    \includegraphics[width=1\textwidth]{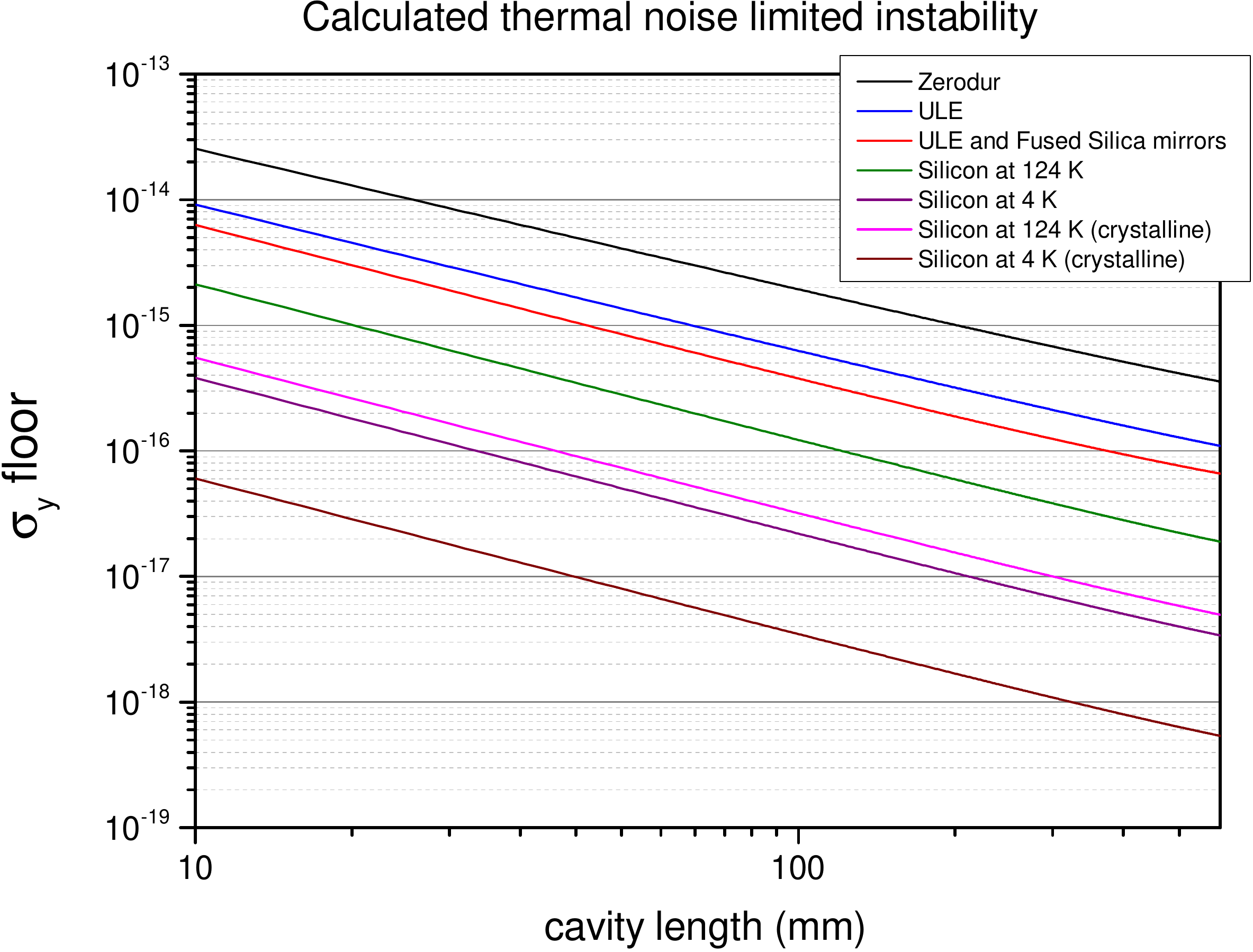}
	\caption{Calculated flicker-frequency-floor instability due to thermal noise, as a function of the cavity length. The curves are for dielectric high-reflective coatings (SiO$_{2}$/Ta$_{2}$O$_{5}$) centred at \SI{1542}{nm} and at a temperature of \SI{300}{K}, except for silicon which is plotted for two different cryogenic temperatures. In addition we plot curves for crystalline (Al$_{x}$Ga$_{1-x}$As) high-reflective coatings attached to a silicon cavity.
	\label{fig:Instab_Thermal_Noise}}
\end{figure}

As an example, we plot in Figure~\ref{fig:Instab_Thermal_Noise} the calculated thermal noise for a cylindrical cavity with 50-mm diameter and a central bore of \SI{11}{mm} as a function of the cavity length for different materials at \SI{300}{K} and for silicon at cryogenic temperatures. For simplicity we have neglected the noise contribution of the spacer arising from local deformations at the surface of the mirror contact \citep{Kessler2012a}. This contribution becomes important, for instance, in very short cavities where the spacer's thermal noise can dominate the substrate and coating contributions. The mirror coatings are dielectric (SiO$_{2}$/Ta$_{2}$O$_{5}$) unless specified as crystalline (Al$_{x}$Ga$_{1-x}$As) and are centred at \SI{1542}{nm}. The radius of curvature of the curved mirror is kept constant at \SI{1}{m}. Such a radius could be used for long or short cavities. One can note that even a radius of \SI{10}{m} has been successfully used for a very compact cavity~\citep{Davila-Rodriguez2017}.

\subsection{Finesse \label{sec:Finesse}}

Typical Fabry--Perot cavities used for ultrastable lasers requiring fractional frequency instabilities of $10^{-15}$ or below usually exhibit high finesse on the order of a few \num{100000}.
The finesse $\mathcal{F}$ of a Fabry--Perot resonator is given by the reflectivity $r$ of its mirrors as $\mathcal{F}\approx\pi/(1-r^{2})$.
In the case of the Pound--Drever--Hall (PDH) technique for locking laser frequencies to cavity modes, the error signal can near resonance be expressed by~\citep{Black2001}
\begin{equation}
\label{eq:Finesse_PDH}
\epsilon = D \, \delta f, \space ~\mathrm{with}~ \space D = \frac{-8\sqrt{P_\mathrm{c} P_\mathrm{s}}}{\delta \nu_\mathrm{c}} = \frac{-16 \mathcal{F} \sqrt{P_\mathrm{c} P_\mathrm{s}}\, L}{c}.
\end{equation}
Here $\delta f$ corresponds to the frequency fluctuations around resonance, $P_\mathrm{c}$ and $P_\mathrm{s}$ are, respectively, the powers on the carrier and sideband of the phase-modulated input beam, $\delta \nu_\mathrm{c}$ is the full width at half maximum (FWHM) of the cavity resonance, $\mathcal{F}$ is its finesse, $L$ its length, and $c$ the speed of light. Hence, high-finesse resonators help to achieve high gains in the control loop.
In addition, perturbations added to the error signal, like residual amplitude modulation (Section~\ref{sec:RAM}), will have a smaller effect on the control loop if the finesse is high.

Using Equation~(\ref{eq:Finesse_PDH}), one can express the relation between the power spectral density of the frequency fluctuations $S_{\delta f}$ and that of the electric noise $S_{\epsilon}$ as $S_{\delta f}=S_{\epsilon}/D^{2}$. Therefore, near resonance the frequency noise due to the electric noise of the error signal will scale with $1/\mathcal{F}^{2}$. Hence, using high-finesse cavities will relax the requirements on the noise of the electronics involved.
One drawback is that high finesses lead to high intra-cavity powers, which can lead to thermal processes because of the mirror losses and thus to a sensitivity to fluctuations in the optical power. Therefore, active intensity stabilisation can be required, as described in Section~\ref{sec:Optical_Intensity}.

\subsection{Vacuum level \label{subsec:vacuum_level}}

A vacuum pressure fluctuation changes the index of refraction $n$ of the residual gas and thus the effective length $L$ and the resonance frequency $\nu$ of the cavity as
\begin{equation}
    \frac{\Delta n}{n} = \frac{\Delta L}{L} = - \frac{\Delta \nu}{\nu}.
\end{equation}
For a gas that can be approximated as an ideal gas, such as nitrogen, the fractional change in the index of refraction for a given pressure change $\Delta P$ in mbar at \SI{30}{\degreeCelsius} can be given by \citep{Egan2015}
\begin{equation}
  \frac{\Delta n}{n} \approx 2.65\times10^{-7}\Delta P\,[\text{mbar}].
\end{equation}
This proportionality is valid at low pressures, where $n-1 \ll 1$ .
For example, for a target instability of $1\times 10^{-16}$ at room temperature, the pressure instability would have to be ${\lesssim} \SI{4e-10}{\milli\bar}$.
A pressure sensitivity of ${\sim} 5\times10^{-7}\;\mathrm{mbar}^{-1}$ has been experimentally measured for a cryogenic silicon cavity~\citep{Matei2016} by switching off the vacuum pump and observing the related frequency drift.

In practice, the pressure is not measured inside the cavity spacer but at the location of the ion pump and is estimated from the pump current, which can be sensitive to leakage currents and electrical noise within the pump, its controller and their connections.
A common approach is to reach low enough absolute pressures to relax the requirements on the pressure fluctuations. At an absolute pressure of~${\sim} 10^{-8}$\;mbar, which is usually aimed for, fractional pressure fluctuations of 4\% are compatible with fractional frequency instabilities below $10^{-16}$.
This level of vacuum does not require advanced vacuum techniques; it can be achieved with a standard ion pump and commercial vacuum chambers or custom aluminium chambers with indium or lead sealing.

\subsection{Temperature instability \label{Temperature Instability}}

For a cavity near room temperature and in vacuum, the main mechanism of heat transfer is through thermal radiation. At room temperature the flux through conduction is at least one order of magnitude smaller using standard interface materials such as FKM fluoroelastomers (for instance Viton\texttrademark) and PEEK polymers. The most effective strategy to isolate the cavity from the environmental thermal radiation is to use low-emissivity thermal shields (typically polished or gold-coated aluminium).

\begin{figure}[tb]
	\centering
    \includegraphics[width=0.88\textwidth]{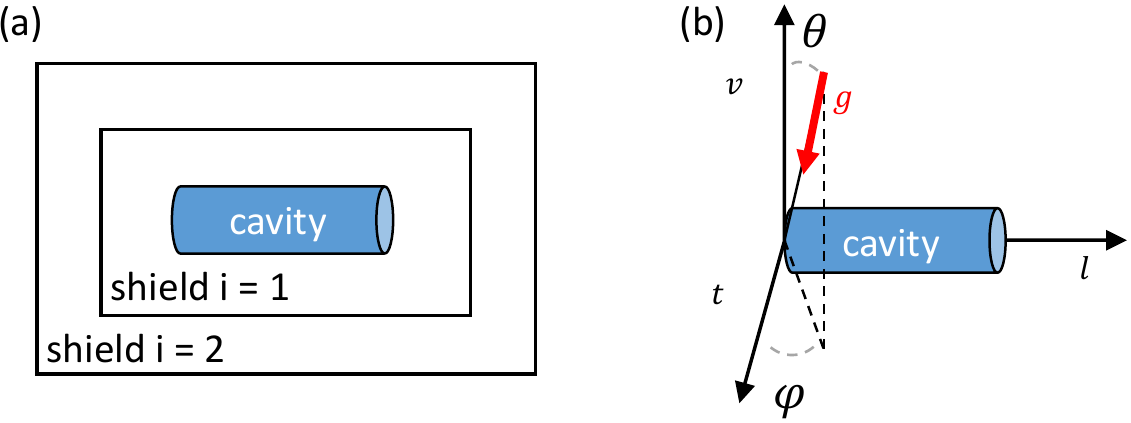}
	\caption{ (a) Simplified diagram to describe the configuration of the thermal shields.  (b) Relevant axes of the optical cavity with respect to the gravitational acceleration $g$.
	\label{fig:shield_tilt}}
\end{figure}

In the case of small temperature changes around room temperature, it is possible to derive an approximate analytic expression for the characteristic thermal response time  $\tau$ of a cavity surrounded by $N$ shields,
\begin{equation}
    \tau \approx \sum_{k=0}^{N} \left( c_k V_k \sum_{i=k}^{N} R_i \right).
\end{equation}
Here $c_k$ and $V_k$ are, respectively, the specific thermal capacitance and the volume of shield $k$, whereas $R_i$ is the  emissive thermal resistance  between the external surface of shield $i$ and the internal surface of shield $i+1$,  given by
\begin{equation}
    R_i = \frac{1}{4\sigma T^3}\left(\frac{1-\varepsilon_{i+1}}{\varepsilon_{i+1} S_{i+1, \text{int}}} + \frac{1}{S_{i, \text{ext}}} + \frac{1-\varepsilon_{i}}{\varepsilon_{i} S_{i, \text{ext}}}\right).
\end{equation}
Here $\sigma$ is the Stefan-Boltzmann constant, $T$ is the temperature, $S_{i, \text{int}}$ ($S_{i, \text{ext}}$) is the internal (external) area of shield $i$ and $\varepsilon_{i}$ is its emissivity. The index~$0$ corresponds to the cavity, while the index $N+1$ corresponds to the vacuum chamber, see Figure~\ref{fig:shield_tilt}(a).

In addition, external temperature fluctuations could also be strongly transmitted to the cavity by thermal conduction through the mountings of the shields and of the cavity.
One can filter the temperature fluctuations by using a combination of supporting points with very high thermal resistance and thermal shields of high thermal capacity to increase the global thermal time constant to the cavity. In the case of the cavity presented in~\cite{Keller2014}, the measured thermal time constant between the temperature-controlled thermal shield outside the vacuum chamber and the cavity is on the order of 50~hours. The temperature sensitivity of the cavity at the temperature set-point was measured by applying small temperature steps to the temperature-regulated shield and monitoring the frequency response of the cavity. With these two values, one can estimate the contribution of the temperature fluctuations on the fractional frequency instability as depicted in Figure~\ref{fig:instability_summary} for this cavity.

\subsection{Vibrations, acceleration sensitivities and effect of a tilt \label{Vibrations}}

\begin{figure}[t]
	\centering
    \includegraphics[width=0.8\textwidth]{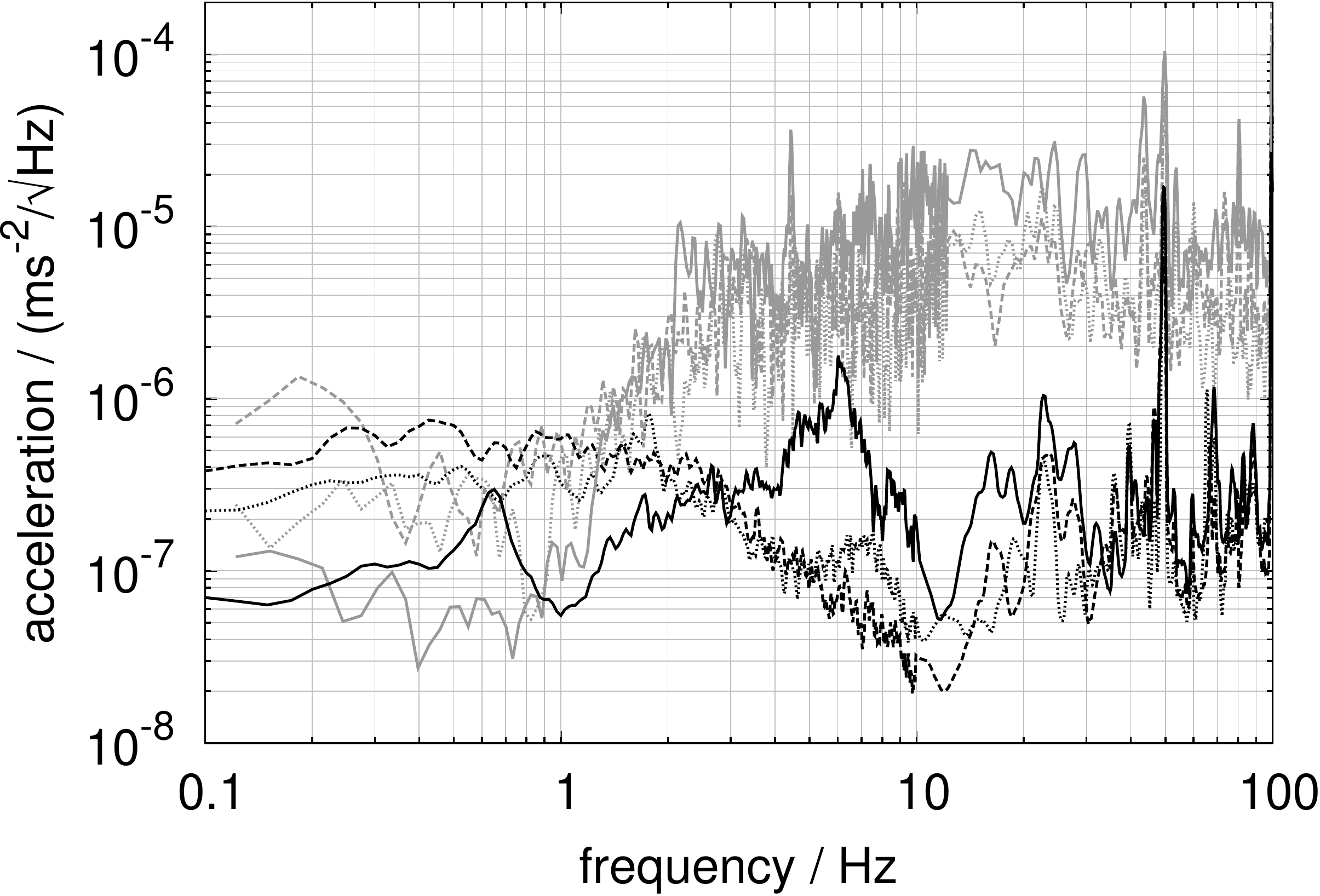}
	\caption{Acceleration spectra due to environmental vibrations. The grey curves were recorded on the laboratory floor, whereas the black curves show the suppressed vibrations on the passive isolation platform supporting the cavity. Solid lines correspond to the vertical, dashed and dotted lines to the horizontal directions. The peak at 50\;Hz is due to electronic noise. Figure and caption reprinted from \citet{Keller2014}. }
	\label{fig:vibrations}
\end{figure}

The deformations due to environmental vibrations are one of the dominant instability contributions for long cavities~\citep{Haefner2015} and a major concern for ultrastable lasers aiming to reach instabilities below $10^{-15}$. They are affected by the choice of geometry and materials. If we take the example of a simple cylinder of ULE glass supported by four Viton rings placed approximately at the Airy points of the spacer, as described in~\cite{Keller2014}, we can estimate a vertical sensitivity due to the Poisson effect of $2.8\times10^{-10}\;(\mathrm{m}/\mathrm{s}^{2})^{-1}$. The isolation offered by a passive vibration-isolation table and an acoustic isolation box are usually sufficient to obtain short-term fractional frequency instabilities slightly below $10^{-15}$ with such a system. To reach instabilities in the low $10^{-16}$, an optimised spacer and support geometry that exploits symmetries~\citep{Notcutt2005,Nazarova2006,Ludlow2007,Webster2008,Millo2009,Keller2014} has to be chosen.
To give an idea of the typical vibration environment in a laboratory, Figure~\ref{fig:vibrations} presents the level of vibrations that has been measured at PTB in Germany and the isolation that can be achieved with a commercial vibration-isolation platform (Minus-K Technology 150BM-1).

Given the sensitivity of the cavity fractional frequency to vibrations along the vertical, transverse and longitudinal directions, $k_v$, $k_t$ and $k_l$, respectively, the sensitivity to a spurious small tilt $\theta$ with respect to the gravitational axis can be written as
\begin{equation}
    \frac{\Delta \nu}{\nu} \approx g \left[k_v + \theta \left(k_t \cos{\varphi} + k_l \sin{\varphi}\right) - \frac{\theta^2}{2} k_v\right],
\end{equation}
where $g$ is the gravitational acceleration and  $\varphi$ is the projection of the angular tilt on the $t$\nobreakdash--$l$~plane, see Figure~\ref{fig:shield_tilt}(b). To lowest order in $\theta$, the tilt sensitivity is
\begin{equation}
    \frac{d}{d\theta}\frac{\Delta \nu}{\nu} \approx g \left(k_t \cos{\varphi} + k_l \sin{\varphi}\right).
\end{equation}
As an example, for a sensitivity value measured in an existing cavity, $k_l = 1.7 \times 10^{-10}/g$~\citep{Haefner2015}, the angular tilt $\theta$ has to be stable within $0.1$\;arcsec ($3\times 10^{-5}$~degrees) to have a fractional stability of $1\times 10^{-16}$.

\subsection{Optical intracavity power}
\label{sec:Optical_Intensity}

The intracavity power of a Fabry--Perot cavity increases with the finesse of its mirrors. As a consequence, the instabilities of the optical power of the laser beam coupled into a high-finesse cavity are amplified within the cavity.
Due to the mirror losses, the intracavity power is converted to a local heating of the cavity, thus changing its length through thermal expansion of the mirrors, which results in a shift of the resonance frequency.

\begin{figure}[bth]
	\centering
    \includegraphics[width=.7\textwidth]{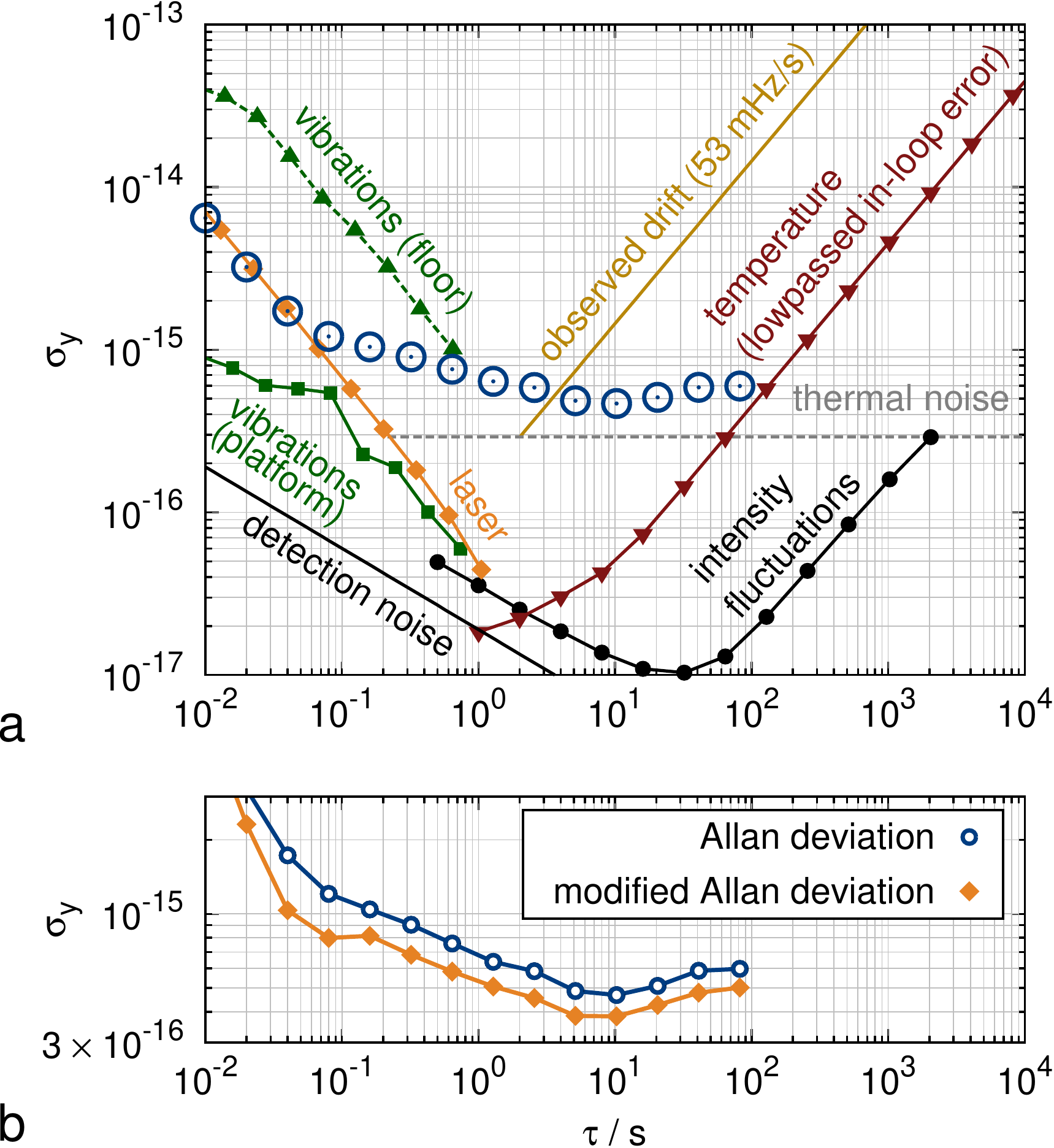}
	\caption{(a) Summary of instability contributions of an extended-cavity diode laser (ECDL) stabilised to a $12$-cm-long ULE cavity. Instabilities are given as fractional-frequency Allan deviations. The blue open circles show the resulting instability of the laser (after removing the linear drift) determined in a three-cornered-hat measurement with two other stable lasers \citep{Kessler2012,Haefner2015}. (b) The modified Allan deviation reveals a peak at $0.2$~s due to additional vibrations transmitted by a cable. Figure and caption reprinted from~\cite{Keller2014}. }
	\label{fig:instability_summary}
\end{figure}

We take the example of a 12-cm-long ULE cavity with FS mirrors assembled at PTB~\citep{Keller2014}. The cavity has a measured finesse of around 280\,000 and the mirror transmission at 822\;nm is $1.2\times10^{-6}$.
The sensitivity to the intracavity power has been determined by applying a step of $\pm \SI{5}{\micro\watt}$ to the input power and monitoring the resulting frequency shifts. This gives a sensitivity of $\SI{2e-13}{W^{-1}}$ to the intracavity power.
For a typical input power around \SI{50}{\micro\watt}, this sets a limit of around $1\times10^{-4}$ on the fractional instability of the intracavity power to reach fractional frequency instabilities in the low $10^{-16}$. Power stabilisation can be realised by using the optical power transmitted  through the cavity on resonance to provide an error signal, and then actuating on an acousto-optic modulator (AOM) placed before the cavity.
One can estimate the contribution of the intracavity power on the frequency stability by monitoring the cavity-transmitted power and knowing the optical power coupled into the cavity and the sensitivity to the intracavity power, as depicted for the 12-cm-long ULE cavity in Figure~\ref{fig:instability_summary}.

\subsection{Residual amplitude modulation (RAM)}
\label{sec:RAM}

The frequency stabilisation of the majority of cavity-stabilised lasers is performed via the Pound--Drever--Hall (PDH) locking technique, which requires the input beam to be phase modulated.
An additional amplitude modulation at the same frequency can be caused by etalon effects or polarisation fluctuations if an electro-optic modulator (EOM) and a polarisation-selective optical component is used~\citep{Whittaker1985,Zhang2014}, which is very often the case.
This RAM creates an unstable offset on the PDH error signal which transfers directly to a frequency instability of the locked laser.
The voltage fluctuations of the error signal away from resonance can be converted to a contribution to the laser's fractional frequency instability as
\begin{equation}
\label{eq:RAM}
\sigma_y(\tau) = \frac{\sigma_{\delta V}  D}{\nu},
\end{equation}
where $\sigma_{\delta V}$ is the Allan deviation of the voltage fluctuations,  $D$ is the discriminant of the error signal around resonance, defined in Section~\ref{sec:Finesse}, and $\nu$ is the laser frequency.

Etalon effects inside the optical setup can be reduced by slightly tilting optical components. The RAM produced by the EOM can be actively compensated~\citep{Zhang2014} to a level of $1\times 10^{-6}$. Another alternative is the use of Brewster-cut EOMs~\citep{Dooley2012,Tai2016} that separate the ordinary and extra-ordinary polarisation components of the light, thus avoiding interference in polarisation-selective elements. With these EOM geometries, etalon effects inside the crystal itself can be avoided, and intrinsic RAM as low as $2\times 10^{-6}$ can be obtained. At these levels, the RAM contribution to the fractional frequency instability is in the $10^{-17}$ range.

\subsection{Fibre-length stabilisation}

For most of the ultrastable lasers, the laser itself and the resonator are separated by optical fibres of a few metres length. The stabilised light can also be sent to other buildings via longer fibres ($100$\;m or more).
Shocks, vibrations and temperature variations produce local disturbances of the refractive index of the fibres, thus varying the optical path of the laser. This results in a variation of the laser's frequency at the end of the fibre. Fortunately, the frequency fluctuations induced by the travel through the fibre can be easily corrected by schemes involving a Michelson interferometer and an AOM placed before or after the fibre~\citep{Lopez2012}.

\begin{figure}[bt]
	\centering
    \includegraphics[width=\textwidth]{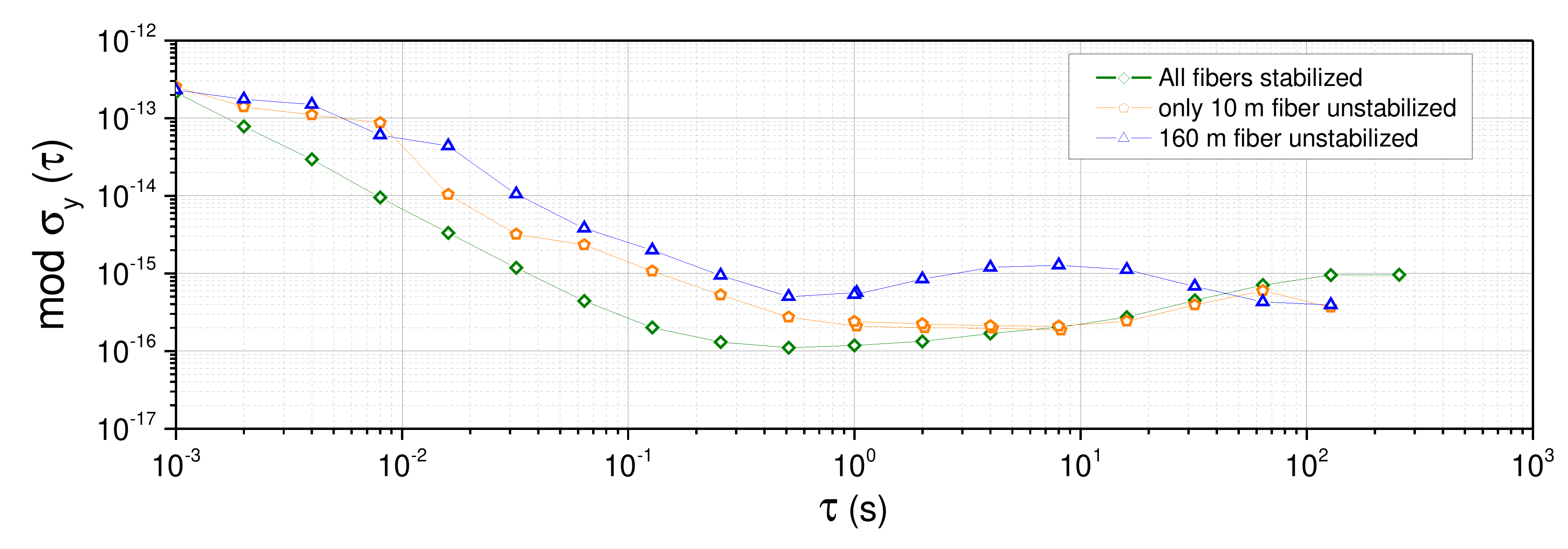}
	\caption{Fractional frequency instability of a 946-nm laser stabilised to a 30-cm ULE Fabry--Perot cavity, measured by a comparison with a 1542-nm laser stabilised to a silicon cavity. The measurement has been performed with and without fibre stabilisation of fibres of the 946-nm setup. Figure and caption reprinted from \citet{Didier2019}. }
	\label{fig:fiber_length}
\end{figure}

The realisation of such a scheme is useful if short-term fractional frequency instabilities below $10^{-14}$ are expected. As an example, we show in Figure~\ref{fig:fiber_length} the instability of a 946-nm laser stabilised to a 30-cm-long ULE cavity~\citep{Didier2019}, which has been measured by comparison with a 1542-nm laser stabilised to a cryogenic silicon cavity with a fractional frequency instability below $10^{-16}$ from 50\;ms to 500\;s. The instability of the beat note is therefore dominated by the 946-nm laser. The laser and the resonator are separated by a 10-m fibre inside the same room. The laser is sent to another building through a 160-m fibre, where it is measured via a femtosecond laser against the 1542-nm laser.
From this measurement, one can get an idea of the typical instabilities achievable without fibre-length stabilisation. The fibres have a non-negligible contribution when instabilities below $10^{-15}$ are expected, especially for short integration times. For an instability of $10^{-16}$ or below, one should consider stabilising shorter fibres as well.

For more discussion on optical-path-length stabilisation, see Section~\ref{sec:delivering}, which covers transfer of frequency stability from the laser source to the atoms.

\section{Cryogenic silicon cavities
} \label{sec:cryogenic_silicon_cavity}

As discussed above, the fundamentally limiting thermal noise in Fabry--Perot cavities is proportional to the cavity temperature $T$ and the ratio $\varphi/E$ of the mechanical loss factor $\varphi$ and the Young's modulus $E$ of the cavity constituents. Therefore, we search for a cavity material suitable for cryogenic temperatures in combination with a very small mechanical loss factor and a large Young's modulus. Monocrystalline silicon has proven to be an excellent choice, since it has a zero crossing of its coefficient of thermal expansion (CTE) around 124\;K, a mechanical loss factor of only $10^{-8}$ at low temperatures (three orders of magnitude smaller than that of ULE glass) and a large Young's modulus of up to $187.5$\;GPa. This allows for record-low thermal-noise-limited fractional frequency instabilities in the $10^{-17}$ range.

\citet{Kessler2012} first demonstrated a 212-mm-long cryogenic single-crystal silicon cavity used for laser stabilisation at 1542\;nm. This system realised a fractional frequency instability of around $10^{-16}$, still limited by technical noise. We finally demonstrated thermal-noise-limited operation of two identical cavity-stabilised laser systems (in the following called Si2 and Si3) with a fractional frequency instability of $4 \times 10^{-17}$ \citep{Matei2017}. Going to even lower temperatures around 4\;K, at which the CTE of silicon asymptotically approaches zero, \citet{Zhang2017} demonstrated the operation of a 60-mm-long silicon cavity in a closed-cycle cryostat with a fractional frequency instability of $1 \times 10^{-16}$. In the following we sum up the technical requirements and specifications needed to realise ultrastable operation of cryogenic silicon cavities.

\subsection{Cavity design and experimental setup}

All systems cited above employ vertically mounted cavities made of single-crystal silicon with its crystallographic \hkl<111> direction oriented along the cavity axis. The crystal orientation was chosen to maximise Young's modulus along this direction. In addition, the crystal shows a three-fold symmetry with respect to the cavity axis, which perfectly fits to an unambiguously plane-defining three-point support and allows for additional tuning and minimisation of the vertical acceleration sensitivity just by varying the azimuthal angle between the crystal orientation and the support points \citep{Matei2016}.

It is advantageous to machine the spacer and mirror substrates from the same low-impurity silicon rod with a resistivity of ${\sim} \SI{30}{\kilo\ohm.\centi\meter}$ or above in order to prevent aging-related length drifts of the final cavity. The cavity mirrors consist of low-loss dielectric Ta$_2$O$_5$/SiO$_2$ multilayers, IBS (ion beam sputtering) coated on the silicon substrates, allowing for a cavity finesse around 500\,000. The crystal orientation of the spacer and the optically contacted silicon mirrors were aligned to each other in order to maintain the single-crystal behaviour of the entire cavity.

\begin{figure}[tb]
	\centering
    \includegraphics[width=\textwidth]{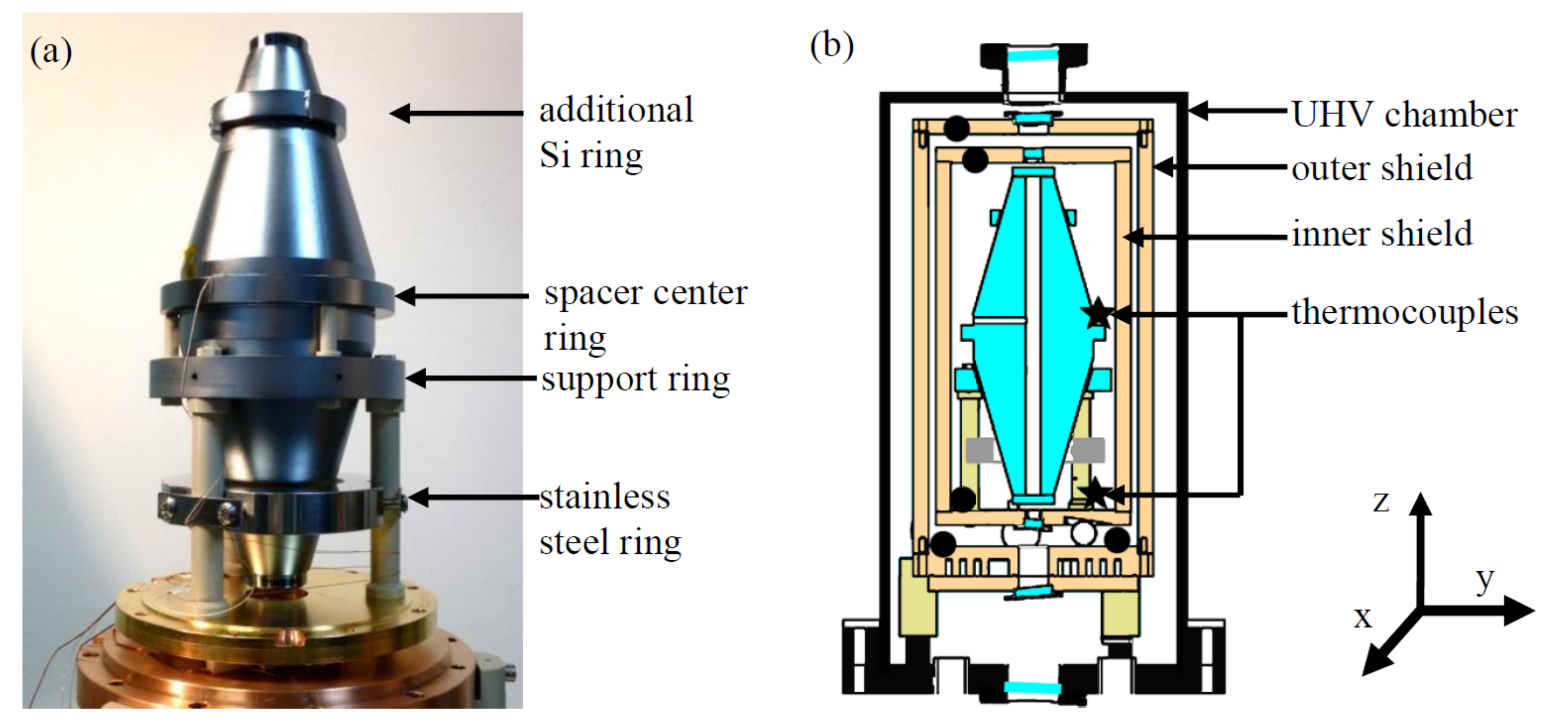}
	\caption{Details of the 212-mm-long cavity setup. The photo (a) shows the vertically oriented silicon cavity supported near its middle plane by a tripod made of PEEK. The support ring is made of silicon to minimise thermal stress during cooling. A stainless-steel ring clamps the PEEK tripod in order to maximise the frequency of mechanical resonances of the support. The sketch (b) shows the thermal concept with the actively cooled ``outer shield'' and a passive ``inner shield'', both made of gold-plated copper. The temperature of the shields is measured at different spots by means of Pt100 temperature sensors (black filled circles), while the cavity temperature is monitored with a pair of thermocouples (stars). The whole setup is placed in a vacuum chamber with a residual pressure of $10^{-10}$ mbar. Figure reprinted from \cite{Matei2016} under \href{https://creativecommons.org/licenses/by/3.0/}{CC BY license}.}
	\label{fig:Silicon-Cavity-Setup}
\end{figure}

For minimal vibration sensitivity, see Section~\ref{Vibrations}, the optimal cavity shape and the exact vertical position of the cavity support need to be identified by means of finite-element-modelling (FEM) simulations. The vertically oriented silicon cavities used in \citet{Matei2017} and \citet{Zhang2017} are designed as a ``double cone'' supported near its middle plane (see Figure~\ref{fig:Silicon-Cavity-Setup}). For the 212-mm-long cavities an additional silicon ring placed around the upper part of the cavity was used to optimise the vertical vibration sensitivity. The tripod support is made from PEEK and is designed for high stiffness in order to shift mechanical resonances to higher frequencies around~70\;Hz.

The 212-mm-long cavities rest on a coldplate, which is temperature controlled to 124\;K by means of cold nitrogen gas from a liquid-nitrogen Dewar. An additional gold-plated, low-emissivity thermal heat shield between the actively cooled surrounding and the cavity is used to suppress residual temperature fluctuations of the temperature servo (compare Section~\ref{Temperature Instability}).

Reaching even lower temperatures below 10\;K requires more elaborate cooling devices.  As discussed in \cite{Zhang2017}, a Gifford--McMahon closed-cycle cryocooler is employed to temperature stabilise a 60-mm-long silicon cavity to 4\;K. Residual temperature fluctuations on the order of 20\;mK at the cryocooler's cold head were damped by means of a multistage heat shield system. A cylindrical plate made of holmium copper (HoCu$_2$) placed between the cryocooler's cold finger and the bottom of the active shield allows for additional thermal damping due to its high specific heat at low temperatures.

The closed-cycle cryosystem generates significant vibrations. On top of the cryocooler, accelerations of about $\SI{1}{\meter\per\second^2}$  peak-to-peak are measured. Therefore, the cryocooler is fixed to the concrete floor of the lab and separated from the cavity chamber by bellows. The thermal link between the cold finger and the 4-K heat shield as well as the link between the first cooling stage and the 30-K shield are obtained by flexible copper braids. Furthermore, the vibration sensitivity of the cavity must be extremely small. As the sensitivity scales with the cavity length, we have designed a short cavity of 60-mm length.

We frequency stabilise commercial erbium-doped distributed-feedback (DFB) fibre lasers with a wavelength of 1542\;nm to the cavities using the Pound--Drever--Hall (PDH) method. A servo with a locking bandwidth of around 200\;kHz is achieved using fibre-coupled acousto-optic modulators (AOMs) driven by voltage-controlled oscillators as fast actuators. Residual amplitude modulation as discussed in Section~\ref{sec:RAM} is actively compensated to keep the corresponding fractional frequency fluctuations below the thermal noise level of the systems.

\subsection{Technical specifications}

To realise thermal-noise-limited operation of a cavity-stabilised laser system, one has to control and minimise several sources of technical noise as discussed in Section~\ref{sec:room_temp_glass_cavities}. The following sections will sum up the impact of residual noise on the silicon systems and might be useful as a guideline for designing next-generation silicon cavities.

\subsubsection{Pressure fluctuations}

As discussed in Section~\ref{subsec:vacuum_level}, fluctuations in the residual vacuum pressure affect the index of refraction and thus the optical path length of the cavity. To ensure a fractional length instability of the silicon cavities below the thermal noise limit, one needs a base pressure well below $10^{-9}$\;mbar. Even with the large surface of all the components in the vacuum chamber, a standard ion pump with approximately $\SI{30}{l/s}$ pumping speed is sufficient to realise this level because of the cryopumping provided by the cold environment. Pressure variations affect the long-term stability of the system. From the observed pressure fluctuations we estimate a frequency instability below $4 \times 10^{-17}$ for averaging times up to 2000~seconds \citep{Matei2016}.

\subsubsection{Temperature stability}

To minimise the impact of temperature fluctuations, the operating temperature of the cavity must be set as close as possible to the zero-crossing point of the coefficient of thermal expansion (CTE) of the silicon cavity. Close to the zero crossing, the CTE can be approximated as linear with a slope of around $1.7 \times 10^{-8}\;\mathrm{K}^{-2}$. Thus, a temperature offset of 50\;mK leads to a CTE of around $8.5 \times 10^{-10}\;\mathrm{K}^{-1}$. To ensure a length stability at the low $10^{-17}$ level, a temperature stability of about 10\;nK is required.

For the Si2 and Si3 systems, the temperature fluctuations of the actively cooled heat shield were on the order of 1\;mK. Therefore, we need a thermal low-pass filter with a time constant on the order of a few days to reduce residual temperature fluctuations of the temperature-controlled environment. As shown in Figure~\ref{fig:Silicon-Cavity-Setup}, we use one additional passive heat shield surrounding the cavity. The time constants for the heat flow between cavity and passive shield and passive shield and active shield are 1.3~days and 6.5~days, respectively. The temperature fluctuations of the cavity are below 1\;nK for averaging times of a few seconds and affect the length stability only for times of thousands of seconds and longer.

The long-term stability is also affected by the blackbody radiation from the environment reaching the cavity through the windows of the vacuum chamber and heat shields. This becomes more important in cryogenic systems, as the heat transfer between the environment at room-temperature $T_\mathrm{env}$ and the cavity at the cryogenic temperature $T_\mathrm{cav}$ scales with $T_\mathrm{env}^4 - T_\mathrm{cav}^4$. Due to the good thermal insulation and the large time constants, any thermal radiation from outside leads to a significant temperature offset of the cavity. Therefore, a careful blocking of blackbody radiation by using appropriate glass windows is mandatory. Additionally limiting the solid angle by small tubes at the windows of the passive heat shield will further reduce the radiation flux.

\subsubsection{Vibration noise}

As shown in Figure~\ref{fig:vibrations}, the level of environmental vibrations increases for Fourier frequencies above~1\;Hz. Therefore, vibrations usually affect the laser performance most at short averaging times. Using passive or active vibration-isolation platforms reduces the power spectral density (PSD) of vibrations by about one to two orders of magnitude for frequencies above 0.5\;Hz.

Assuming uncorrelated vibrations, the PSD of the frequency noise of the laser system, $S_{\nu}$, is given by the sum of the vibration-noise PSDs in all three directions, $G_x$, $G_y$ and $G_z$, weighted by the corresponding vibration sensitivities $k_x$, $k_y$ and $k_z$ of the cavity setup
\begin{equation}
    \label{eq:seismic_noise}
S_{\nu} = G_x  k_x^2 + G_y k_y^2 + G_z k_z^2.
\end{equation}
The sensitivities also include mechanical resonances of the whole setup. Therefore, the cavity support must be designed for large stiffness, pushing resonances to higher frequencies. Besides any resonances, the seismic spectrum above 1\;Hz approximately leads to white frequency noise of the laser system. Hence, the short-term instability of most cavity-stabilised laser systems shows a typical $1/\sqrt{\tau}$ behaviour. Depending on the thermal noise level of the cavity, the vibration-noise-limited short-term stability completely determines the linewidth and thus the coherence time of the laser (see Section~\ref{subsc:frequency_instability}).

\begin{table}[b]
\centering
\caption{Measured horizontal ($x$ and $y$) and vertical ($z$) vibration sensitivities of the 212-mm-long silicon cavities Si1 \citep{Kessler2012} and Si2/3 \citep{Matei2017}.}
\label{tab:vibration_sens}
\begin{tabular}{lSSS}
\toprule
$10^{-12}\;(\mathrm{m/s}^2)^{-1}$  &  {Si1}  & {Si2} & {Si3} \\
\midrule
horizontal k$_x$  & 8.6 & 2.5 & 8.6  \\
horizontal k$_y$  & 6.9 & 0.7 & 4.0 \\
vertical k$_z$    & 5.6 & 0.4 & 0.8 \\
\bottomrule
\end{tabular}
\end{table}

The vibration sensitivities for the 212-mm-long silicon cavities are listed in Table~\ref{tab:vibration_sens}. The vertical vibration sensitivities of Si2 and Si3 were experimentally minimised to below $10^{-12}\;(\mathrm{m/s}^2)^{-1}$ by adjusting the azimuthal angle between the crystal orientation and the support points. The horizontal sensitivities can be caused by, for example, mechanical tolerances, unbalanced forces on the support points and the impact of mirror tilt due to a geometrical offset between cavity mode and cavity axis. The impact of mirror tilt increases with increasing cavity length. FEM simulations show that a 1-mm offset in the 212-mm-long silicon cavities leads to a sensitivity of about $30 \times 10^{-12}\;(\mathrm{m/s}^2)^{-1}$. Correspondingly, for the shorter cavity in \citet{Zhang2017}, we only expect about $4 \times 10^{-12}\;(\mathrm{m/s}^2)^{-1}$ per mm offset.

For cryogenic systems, the employed cooling devices potentially lead to additional vibrations. The 124-K temperature of the 212-mm-long cavities is simply achieved by cold nitrogen gas flowing through a cooling base plate. The slow, laminar gas flow does not significantly increase the vibration level. However, pulsed cryocoolers used for 4-K environments add significant noise to the system. The  two-stage Gifford--McMahon closed-cycle cryocooler (Montana Instruments) used for the 60-mm-long silicon cavity \citep{Zhang2017} leads to residual vibrations on top of the vibration-isolation platform of around $3 \times 10^{-5}\;(\mathrm{m/s}^2)/\sqrt{\mathrm{Hz}}$ between 10\;Hz and 400\;Hz. This is about two orders of magnitude more noise than on a vibration-isolation table without additional cooling (compare with Figure \ref{fig:vibrations}). Thus, the cryocooler must be mechanically decoupled from the cavity setup. The cavity system is connected to the cryosystem only by a vacuum bellow and flexible copper braids for thermally linking the cold finger to the cavity heat shield and the first cooling stage to the 30-K shield. The cavity setup rests on an active vibration-isolation platform. With this setup we could reduce the cryocooler vibrations by up to five  orders of magnitude between 1\;Hz and 10\;Hz and by three orders of magnitude between 10\;Hz and 400\;Hz.

\subsection{Frequency instability and coherence time}
\label{subsc:frequency_instability}

\begin{figure}[tb]
	\centering
    \includegraphics[width=0.8\textwidth]{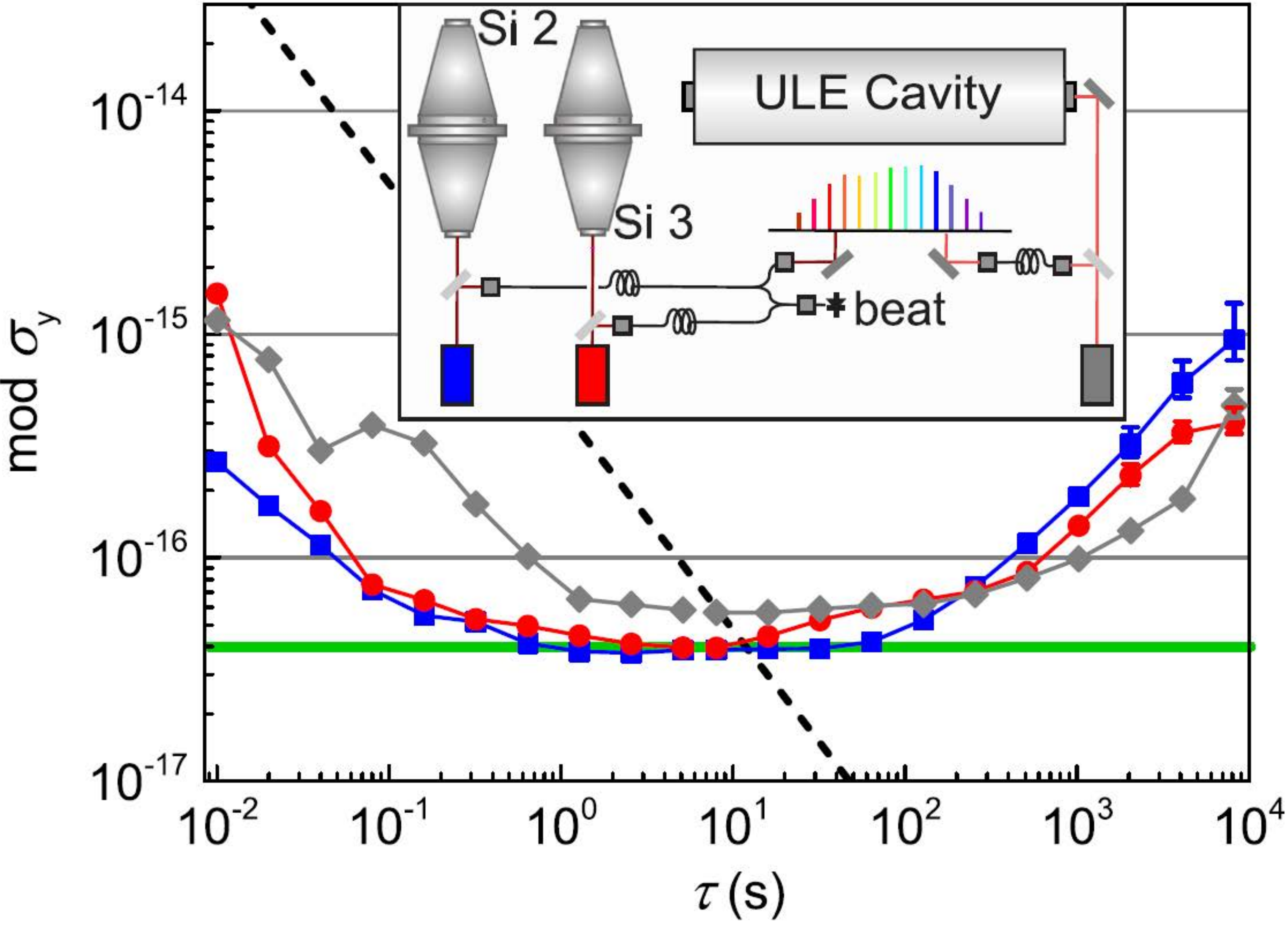}
	\caption{Fractional frequency instability of the two 212-mm-long silicon cavities measured within a three-cornered-hat comparison with the 698-nm clock laser for PTB's strontium lattice clock \citep{Haefner2015}. The dashed line illustrates the instability where the RMS phase fluctuations are 1\;rad for a given $\tau$. The intersections with the instability curves of the Si lasers result in coherence times of around 11\;s. Linear frequency drifts in each data set were subtracted. Figure reprinted with permission from \cite{Matei2017}. Copyright (2017) by the American Physical Society.}
	\label{fig:Silicon-modADEV}
\end{figure}

Figure~\ref{fig:Silicon-modADEV} shows the result of the frequency instability measurement of the two 212-mm-long silicon cavities Si2 and Si3. Both systems reach their thermal-noise-limited flicker floor in the modified Allan deviation for averaging times between 0.8\;s and a few tens of seconds. For short averaging times, the systems are partly limited by the impact of residual vibration noise. The long-term stability is affected by residual temperature and pressure fluctuations at the cavity and parasitic etalons within the optics.

The coherence time of the lasers can be estimated from the instability measurement.
Apart from a factor of $\sqrt{2}$, the following deduction is commonly used in radio astronomy to derive the coherence time from measurements of the frequency instability \citep{Rogers1981,Thompson2017}.

We consider two-pulse Ramsey interrogation of atoms. In this scheme, one estimates the average frequency of the laser from the preceding interval and uses that value for the current interrogation in order to minimise the phase excursions between the pulses. The Allan deviation is actually defined as the variance between successive frequencies ${\nu}_{i}$, which can also be expressed in terms of  successive phase differences $\Delta \phi_i$
\begin{equation}
    \label{eq:ADEV_Freq}
\sigma^2_y(T_0) = \frac{1}{\nu_0^2} \left\langle \frac{1}{2} \left(\bar{\nu}_{i-1} - \bar{\nu}_i\right) \right\rangle = \frac{1}{4 \pi^2 T_0^2 \nu_0^2} \left\langle \frac{1}{2} \left(\Delta \phi_{i-1} - \Delta \phi_i \right)^2 \right\rangle.
\end{equation}
The corrections between the pulses mean that $\Delta \phi_{i-1} = 0$, so we get
\begin{equation}
    \label{eq:ADEV_Freq_1}
\left\langle \Delta \phi_i^2 \right\rangle = 8 \pi^2 \nu_0^2 T_0^2 \sigma^2_y(T_0),
\end{equation}
which is actually the square of the RMS value of the phase deviations, $\Delta \phi_\mathrm{rms}^2$, and allows estimating the coherence time from the measured frequency stability as shown in Figure~\ref{fig:Silicon-modADEV}. A phase deviation of $\Delta \phi_\mathrm{rms} = 1$\;rad corresponds to an Allan deviation of
\begin{equation}
    \label{eq:coherence_time}
\sigma_y(T_0) = \frac{1}{2 \sqrt{2} \pi \nu_0 T_0},
\end{equation}
which is shown in Figure~\ref{fig:Silicon-modADEV} as the dashed line.

The coherence time is defined by the intersection of $\sigma_y(T_0)$ with the individual frequency stabilities of the lasers, which for the silicon systems Si2 and Si3 in Figure~\ref{fig:Silicon-modADEV} gives a value of about 11\;s. Note that this value is not limited by the short-term stability of the lasers, but only by their thermal-noise flicker floor. There is still some margin for increased vibration-induced white frequency noise before this contribution will limit the coherence properties of the laser. For the next-generation optical coatings (see Section~\ref{sec:nextgen_coat}) with even smaller thermal-noise flicker floor, the long-term stability of the cavity becomes the limiting factor for the coherence time of the laser system.

\subsection{Next-generation optical coatings \label{sec:nextgen_coat}}

\begin{table}[b]
\centering
\caption{Thermal displacement noise of spacer, substrate and coatings according to Section~\ref{subsec:Thermal_Noise} for a 212-mm-long single-crystal silicon cavity at 124\;K. Using crystalline mirror coatings reduces the flicker floor of the ADEV, $\sigma_y^{\mathrm{floor}}$, by a factor of five.}
\label{tab:thermal_noise}
\begin{tabular}{p{25mm}
S[table-format=3.3e+2, table-align-exponent = false]
S[table-format=1.3e+2, table-align-exponent = false]}
\toprule
                 &  {Ta$_2$O$_5$/SiO$_2$}  & {Al$_{x}$Ga$_{1-x}$As} \\
           &  {dielectric coating}  & {crystalline coating} \\
\midrule
S$_{x}^{\mathrm{spacer}}$ \hfill  $[\mathrm{m^2/Hz}]$  & \multicolumn{2}{c}{\tablenum[table-format=1.3e+2]{0.005e-36}}  \\
S$_{x}^{\mathrm{sub}}$    \hfill  $[\mathrm{m^2/Hz}]$  & \multicolumn{2}{c}{\tablenum[table-format=1.3e+2]{0.146e-36}} \\
S$_{x}^{\mathrm{coat}}$   \hfill  $[\mathrm{m^2/Hz}]$  & 116.933e-36 & 4.872e-36 \\
\midrule
S$_{x}$                   \hfill  $[\mathrm{m^2/Hz}]$  & 117.084e-36 & 5.023e-36  \\
\midrule
ADEV $\sigma_y^{\mathrm{floor}}$   & 5e-17 & 1e-17 \\
\bottomrule
\end{tabular}
\end{table}

The loss factor $\varphi$ of silicon of around $10^{-8}$ is so much smaller than the loss factor of the dielectric mirror coatings of about $4 \times 10^{-4}$ that the few-\si{\micro\meter}-thick coatings can dominate the thermal noise of the whole cavity. The difference between both contributions is very large. As shown in Table~\ref{tab:thermal_noise}, the thermal noise of the silicon spacer and the substrates is completely negligible compared to that of the coatings. Thus, there is an urgent need for low-loss coatings.

Crystalline Al$_{x}$Ga$_{1-x}$As multilayers show a much smaller loss factor of only $17 \times 10^{-6}$. At room temperature, these coatings have already demonstrated a large reduction in Brownian thermal noise \citep{Cole2013}. As shown in Figure~\ref{fig:Instab_Thermal_Noise}, combining these coatings with a 212-mm-long silicon cavity at 124 K allows for a thermal-noise-limited frequency instability of about $1 \times 10^{-17}$. At temperatures of 4\;K or below, the thermal noise limit can even be pushed to about $2 \times 10^{-18}$ and below.

Thus, crystalline mirror coatings are a promising candidate for cryogenic cavities. However, the coatings are not directly grown on the substrates, but are pressed in contact for a spontaneous van der Waals bond. The thermally driven strain due to the CTE mismatch between the silicon substrates and the Al$_{x}$Ga$_{1-x}$As multilayers can potentially exceed the bonding strength. Therefore, we have tested a pair of silicon mirrors employing crystalline mirror coatings at temperatures down to 100\;K. The 1-inch, plano-concave mirrors with a radius of curvature of 2\;m were optically contacted to a 100-mm-long silicon test spacer. Passing several cooling cycles, no degradation of the mirrors was observed.

Due to mismatch strain between the high- and low-index components of the multilayer, the coatings show a strong birefringence. For the 100-mm-long test cavity and a wavelength of 1542\;nm, the birefringence resulted in a line splitting of the fundamental TEM$_{00}$ mode of around 420\;kHz. The internal coating strain seems much stronger than a potential thermal-driven mismatch strain between coating and silicon substrate as we see a constant line splitting down to 100\;K.

The coating transmission is optimised for cryogenic temperatures, thus the measured finesse increased from around 300\,000 at room temperature to almost 400\,000 at 100\;K. The new mirrors will be used to replace the current dielectric mirrors on one of our silicon cavities in order to further improve the frequency stability of our cavity-stabilised laser system.

\section{Active resonators
} \label{sec:active_resonators}

We now turn our attention to an entirely different scheme for generating stable-frequency light.  Active resonators are based on cold atoms with narrow optical transitions placed inside an optical cavity.
The atomic sample is assumed to have a temperature in the millikelvin to microkelvin range, provided by a magneto-optical trap (MOT), and is not necessarily placed in an optical lattice. This cavity-atom system takes advantage of the strong coupling between atoms and cavity, where the narrow atomic transition dominates the frequency response of the system. In the strong-coupling regime, collective effects will dominate the dynamics. This has been suggested in a number of publications as a new scheme for improving the spectral purity of light in connection with optical clocks.

\begin{figure}[tb]
	\centering
    \includegraphics[width=0.85\textwidth]{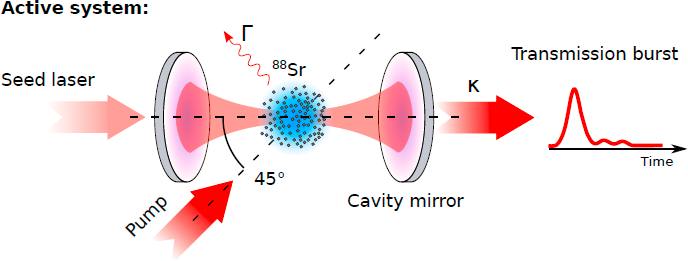}
	\caption{Proposed active system where an ensemble of cold atoms provided by a MOT is excited by a $\pi$-pulse.}
	\label{fig:active_system}
\end{figure}

Below we list a number of important relations related to cavity-assisted collective phenomena \citep{Kuppens1994,Wang2000,Bohnet2012}. We imagine cold atoms possessing a narrow-transition line, such as group-two elements, placed in an optical cavity with finesse $\mathcal{F}$, see Figure~\ref{fig:active_system}. The cavity linewidth is denoted by $\kappa$ while the atomic transition linewidth is given by~$\Gamma$.

\subsection{Bad-cavity regime}
\label{subsec:bad cavity regime}

In most cases one considers the so-called bad-cavity regime. Here the cavity linewidth is significantly larger than the atomic transition linewidth, $\kappa \gg \Gamma$. Generally the system of cavity plus atoms will oscillate at the geometric mean of the atomic resonance frequency and the cavity resonance frequency, leading to a cavity jitter suppression factor
\begin{equation}
\label{eq:aplha}
\alpha = \frac{2 \Gamma}{2 \Gamma + \kappa} \simeq \frac{2 \Gamma}{\kappa},
\end{equation}
which for narrow transitions may be significantly smaller than unity, often of the order $10^{-3}$ to~$10^{-5}$.

\subsection{Collective phenomena}
\label{subsec:collective phenomena}

In Figure~\ref{fig:active_system} we depict a typical scenario of atoms placed inside a cavity.
For the atom-cavity system the most important parameter is the single-atom cooperativity
\begin{equation}
\label{eq:cooperativity}
C_0 = \frac{4 g_0^2}{\kappa \Gamma} = \mathcal{F} \frac{6}{\pi^3} \frac{\lambda^2}{w_0^2},
\end{equation}
where $g_0$ is the single-atom coupling strength, $\lambda$ the transition wavelength and $w_0$ the cavity waist where the atoms are placed. Typical cooperativity values for a cavity with $\mathcal{F} = 1000$ are $C_0= 10^{-4}$ to $10^{-5}$.

One way to interpret the cooperativity is simply as the ratio of the optical cross section to the cavity waist area scaled by the finesse. Another way is to see the physics as diffraction-limited coherent emission into the cavity mode, which in one dimension would scale as $\lambda/w_0$ and in two dimensions as $(\lambda/w_0)^2$, again scaled with the finesse.

For the number of atoms $N$ in the cavity volume, the collective optical depth is given by $C = N C_0$. For systems involving MOT samples, $N$ may be as high as $10^6\ldots 10^7$, making the collective cooperativity $C$ as high as 1000 or more.

\subsection{Collective emission into the cavity}
\label{subsec:collective emission into the cavity}

The amount of photons emitted into the cavity mode by $N$ atoms is a key parameter for evaluating possible collective and even superradiance-like behaviour. This amount may be estimated by $N  C_0 \Gamma$.
For the case of a MOT with $10^6\ldots 10^7$ atoms, a typical number would be $10^6\ldots 10^7\;\mathrm{s}^{-1}$.
This number is to be compared with the typical decoherence rate $\Gamma_\mathrm{decoh}$ in the system, for example, spontaneous emission out of the cavity mode (at the rate $\Gamma$) or Doppler decoherence, which in the case of a MOT sample is the dominant part. For the MOT sample this is usually a few MHz. Another decoherence parameter one may consider is the transit time of the atoms moving out of the cavity waist volume.
If one can establish
\begin{equation}
\label{eq:decoherence}
N  C_0 \Gamma \gg \Gamma_\mathrm{decoh},
\end{equation}
collective effects are prone to appear and even dominate the physics of the system.

\subsection{Cavity collective-emission linewidth}
\label{subsec:cavity collective emission line width}

For the situation where the atoms in the cavity are excited by a $\pi$-pulse (in practice more than 50\% of the entire atom sample should be excited) and meet the above criteria, the atoms will emit a burst of light into the cavity mode.  In this case one may calculate the linewidth \citep{Chen2009,meiser2009,Martin2011,Bohnet2012,Norcia2016a,Norcia2016b}.

One may show that the linewidth of superradiant light emitted from the cavity mode in atom-cavity collective operation is given by
\begin{equation}
\label{eq:supperradiant line width}
\Delta \nu = \frac{C_0 \Gamma}{\pi},
\end{equation}
provided that one meets the criterion given by Equation~(\ref{eq:decoherence}).
Thus, decreasing the single-atom cooperativity will make the linewidth smaller. This is done by having a cavity with lower finesse (worse mirrors), which at first seems counterintuitive. However, if the finesse is decreased, the number of atoms must be correspondingly increased, which is not always an easy task. For typical values one may have a linewidth of 1--10\;Hz or lower.

\begin{figure}[tb]
	\centering
    \includegraphics[width=0.59\textwidth]{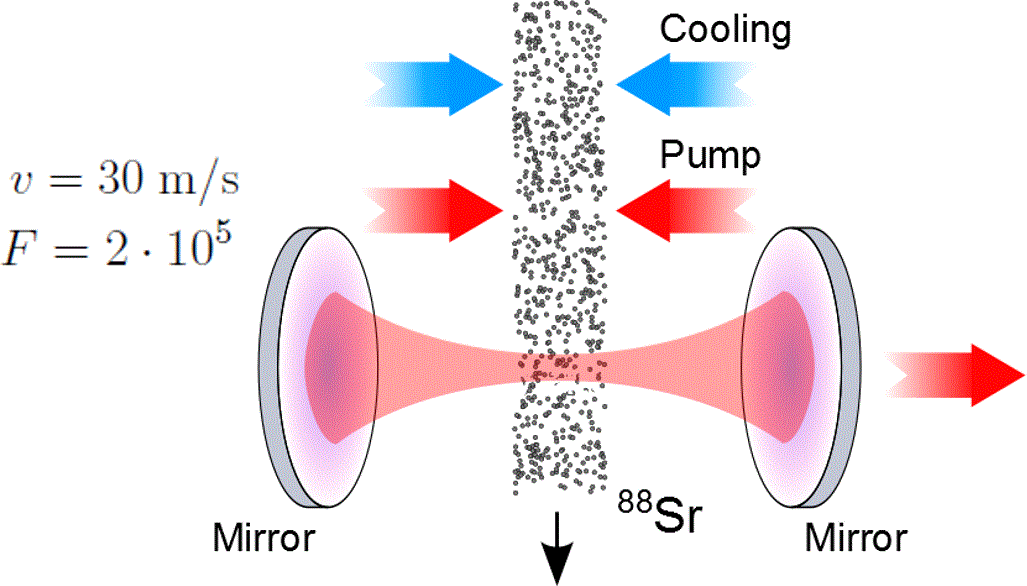}
	\caption{Schematic view of a beam of slow atoms traversing an optical cavity. The purpose is to create a condition for continuous superradiant operation --- an optical maser. To reduce decoherence as introduced by Doppler effects, several cooling stages of the atomic beam must be introduced.}
	\label{fig:continuous_superradiant_scheme}
\end{figure}

\subsection{Continuous superradiant mode}
\label{subsec:continuous superradiant mode}

One of the great challenges is to find a way to operate the atom-cavity collective phenomena continuously, rather than in a burst mode. Several authors have worked out numerical examples based on the basic physics above \citep{Yu2008,Chen2009,Kazakov2013,Kazakov2015}. In Figure~\ref{fig:continuous_superradiant_scheme} we show schematically a case where cold pre-cooled atoms traverse an optical cavity. In addition to longitudinal cooling of the atomic beam, transverse cooling stages are also important to reduce possible decoherence effects. Before the atoms enter the cavity region they are optically prepared in the excited state.

Following \cite{Kazakov2015}, we may determine the atomic flux threshold (number of atoms entering the cavity per second) for sustained superradiant operation
\begin{equation}
\label{eq:atomic number threshold}
R_\mathrm{th}= \frac{\pi^3 v^2}{3 \lambda^2 \Gamma \mathcal{F}},
\end{equation}
where $v$ is the atom velocity. With a beam velocity of $\SI{30}{m.s^{-1}}$ and a cavity finesse of 200\,000 we find $R_\mathrm{th} = 2 \times 10^{6}\;\mathrm{s}^{-1}$; a number that should be realistic from an experimental point of view.

This technique for generating stable laser light is very much in its infancy compared with the more established techniques using high-finesse optical cavities.  Nevertheless, the science is progressing well and active resonators remain a promising approach to pursue into the future.

\section{Spectral hole burning
} \label{sec:spectral_hole_burning}

Another novel approach for stabilising laser frequencies is that of spectral hole burning, which relies on ultra-narrow spectral features in rare-earth-doped crystals at cryogenic temperatures (typically below 20\;K). A hyperfine splitting of the ground state in some rare-earth elements leads to the appearance of metastable substates with lifetimes of several hours at temperatures of 4\;K and below.  Perturbations from the crystalline matrix in which these atoms are held lead to inhomogeneous broadening of the linewidth up to a few GHz. By applying a narrow-linewidth (\SI{<100}{Hz}) laser to the inhomogeneously broadened absorption spectrum, resonant rare-earth atoms will be optically pumped to these metastable states.  This generates a narrow spectral hole in the inhomogeneously broadened spectrum, which persists for several hours after extinction of the ``burning'' laser.
Subsequent laser probing of this narrow spectral feature allows an error signal to be generated, which can be used to frequency lock the ``probing'' laser to the spectral-hole frequency~\citep{Julsgaard2007}. In this application the burning and probing lasers are at the same frequency, but their optical powers are orders of magnitude apart, so as to keep the residual burning effect of the probing laser as small as possible.

As a case study, we present the details of our spectral-hole-burning setup at LNE-SYRTE.
We have utilised $0.1\%$ doped Eu$^{3+}$:Y$_2$SiO$_5$ crystals, which exhibit an inhomogeneously broadened absorption spectrum near 580-nm vacuum wavelength on the $^7\mathrm{F}_0 \rightarrow ^5\mathrm{D}_0$ transition. The $5 \times 8 \times 8$\;mm$^3$ crystal is maintained in a pulse-tube-based closed-cycle cryostat, specially designed for low vibrations of the crystal. The burning and probing beams propagate along the $b$ crystalline axis and are polarised along the $D1$ axis, a configuration which provides the largest absorption~\citep{Ferrier2016}. In this setup, we have realised \SI{<2}{kHz} narrow spectral features at 3.5\;K (although typical operation with \SI{\approx 3.5}{kHz} linewidth is currently preferred for maximal contrast-to-linewidth ratio). Frequency locking the probing laser based on the spectral hole and measuring the fractional frequency stability (overlapping Allan variance, 1-kHz measurement bandwidth) with an optical frequency comb, phase locked to a 1.5-\si{\micro\meter} Fabry--Perot-stabilised laser, leads to a combined instability of $1.8\times10^{-15}$ at 1\;s. From independent assessment of the 1.5-\si{\micro\meter} CW laser, this implies an intrinsic instability of the spectral-hole-burning-stabilised laser near $1.7\times10^{-15}$ at 1\;s.

We have measured the effect of temperature fluctuations on the frequency shift of spectral holes and have confirmed previous measurements by NIST ($15\;\mathrm{kHz/K}$ sensitivity for site 1). In the current development at LNE-SYRTE, this effect is identified as the main contribution to the currently measured instability at the low $10^{-15}$ level. Additional work on active temperature stabilisation of the cryostat and passive isolation of the crystal temperature from the environment should lead to a substantial improvement in the near future. Furthermore, cancellation of the thermal sensitivity to first order using He background gas following the method used at NIST \citep{Thorpe2011} should lead to several orders-of-magnitude improvement of this contribution in the future.

We have measured the effect of uniaxial strain applied to the crystal (ranging from 6 to $120\;\mathrm{Hz/Pa}$ frequency shift for the different directions and crystallographic sites), which provides a quantitative assessment of the effect of residual vibrations (accelerations) on the crystal (estimated with a piezoelectric accelerometer at room temperature with the pulse tube on). Preliminary measurements imply that this effect is compatible with $\num{< e-16}$ instability at 1\;s. Specially designed mounting of the crystal, similar to what is commonly done with Fabry--Perot-stabilised lasers \citep{Millo2009}, should lead to the impact of residual vibrations on instability being at $\num{< e-17}$ level at 1\;s.

The detection noise of the measurement of the mismatch between the probing-laser frequency and the spectral-hole central frequency was estimated to be compatible with a $5\times10^{-16}$ instability at 1\;s, approximately a factor of five larger than the fundamental limit imposed by photon shot noise under our probing conditions and compatible with thermal-noise and photo-diode-efficiency calculations. Further improvement of the design of the measurement chain should allow reaching the shot-noise limit in the near future. Increasing the effective probing power, thereby lowering the shot- and thermal-noise-induced limit would allow the detection-noise contribution to be below the $10^{-16}$ level, although at the cost of increasing the residual burning effect of the probe laser, which already currently limits the time of optimal operation to about 1\;h (after which the holes are broadened by an ``over-burning'' effect by a factor of two or more and need to be re-generated). Utilising several spectral holes in parallel so as to increase the effective probing optical power without increasing the power \emph{per hole} will be utilised in the future to further decrease detection noise. Furthermore, decreasing the linewidth of the spectral holes and increasing their contrast, which may be achieved at lower temperature and in the presence of strong polarising magnetic fields \citep{Equall1994}, would also permit a further decrease in the detection noise. Thus, a total contribution of a few $10^{-18}$ seems achievable.

Further work will also be required to investigate, quantify and minimise the impact of electric- and magnetic-field fluctuations from the environment \citep{Thorpe2013}. When these technical challenges have been addressed, we will be able to explore the fundamental limits of the stability of spectral-hole-burning-stabilised lasers.

\section{Summary}
\label{sec:summary}

The tremendous progress in ultrastable lasers during the last few years has allowed essential improvements in the stability and accuracy of optical atomic clocks. Nowadays laser systems stabilised to room-temperature glass cavities realise frequency instabilities in the low $10^{-16}$ range for averaging times of 1\;s. This was only possible by carefully addressing and minimising the impact of all technical noise sources affecting the length stability of the Fabry--Perot cavities. We have given an overview of the potential noise sources and showed how to reduce their impact.
It can be shown that the frequency stability and the coherence properties of the lasers is then fundamentally limited by the cavities' Brownian thermal noise. This noise is determined by the cavity dimensions, material parameters and the operating temperature. With less noisy materials, such as crystalline silicon placed in a cryogenic environment, we have pushed the thermal-noise-limited frequency instability down to a record-low level of $4 \times 10^{-17}$. With even lower temperatures and with novel, low-noise crystalline mirror coatings, laser systems with frequency instabilities below $10^{-18}$ are feasible.

However, the technical challenge becomes increasingly demanding. Therefore, alternative approaches for generating ultrastable optical frequencies are being explored. The ``active resonator'' scheme employs cold atoms placed inside optical cavities. The strong coupling between the cavity mode and the chosen narrow optical transition of the atoms then allows for collective phenomena leading to the emission of narrow-linewidth superradiant light. In a second approach we have investigated spectral hole burning in rare-earth-doped crystals at cryogenic temperatures. With spectral-hole-burning-stabilised lasers, frequency instabilities at the low $10^{-15}$ level have already been realised. Potentially, this technique will also allow ultrastable lasers with $10^{-18}$ frequency instabilities.

\clearpage

\graphicspath{{D1_transfer/}}

\chapter{Transferring frequency stability from the laser source to the atoms \label{chap:transfer}} 

\authorlist{%
Rodolphe Le Targat$^{1,\dag}$,
H\'{e}ctor \'{A}lvarez Mart\'{i}nez$^2$,
Erik Benkler$^3$,
Ramiz Hamid$^4$,
Helen S.\ Margolis$^5$,
Lennart Pelzer$^{3,6}$,
Benjamin Rauf$^7$,
Antoine Rolland$^5$,
Piet O.\ Schmidt$^{3,6}$,
\c{C}a\u{g}r{\i} \c{S}enel$^{4,8}$,
Nicolas Spethmann$^3$ and
Uwe Sterr$^3$
}

\affil{1}{\OPaff}
\affil{2}{\ROAaff}
\affil{3}{\PTBaff}
\affil{4}{\TUBITAKaff}
\affil{5}{\NPLaff}
\affil{6}{\LUHaff}
\affil{7}{\INRIMaff}
\affil{8}{\BUaff}
\corr{rodolphe.letargat@obspm.fr}

\chapstart
This chapter explains how to transfer the frequency stability from an ultrastable laser source to the beam probing the atoms in an optical clock, whilst adding no more than $1 \times 10^{-17}$ to the laser noise after 1\;s. In Section~\ref{sec:transfer_spectrum}, we describe how to transfer the frequency stability from one wavelength to another, using a femtosecond optical frequency comb.  This is necessary for cases where the laser source is stabilised to a reference, such as an optical cavity at a particular wavelength, but the atoms must be probed at a different wavelength.  In Section~\ref{sec:delivering}, the challenges of setting up the beam path to deliver the laser light to the atoms with minimum propagation noise are described.

\section{Transfer of spectral purity}\label{sec:transfer_spectrum}

\subsection{Outline of the problem}\label{sec:transfer_Description}

In optical atomic clocks, the residual frequency noise of the ultrastable laser source probing the narrow atomic resonance is a key element of the stability of the clock. For instance, it is the sampling of this frequency noise, the Dick effect ~\citep{Santarelli1998}, that limits the stability of even the most advanced lattice clocks to a few $10^{-16}/\sqrt{\tau\,[\mathrm{s}]}$, while the quantum projection noise would allow levels below $10^{-17}/\sqrt{\tau\,[\mathrm{s}]}$ for $10^4$ atoms probed simultaneously. Therefore, the potential of optical clocks has not yet been fully explored, and further progress in laser stability, such as that pursued in the OC18 project, is required.

It is often convenient to build ultrastable lasers at a wavelength where components are easily available and where narrow-laser technologies already exist in order to reduce the requirements on the electronic system necessary to reduce the linewidth. Typically,  1064\;nm (Nd:YAG laser) and the 1530--1565-nm band (C-band, used for optical infrared telecommunications) are very favourable domains. The most traditional method to achieve ultrastable frequencies is to lock a laser to a narrow-linewidth linear optical resonator, composed of two mirrors contacted to a ultra-low-expansion (ULE) glass spacer (see Section~\ref{sec:room_temp_glass_cavities}). This system exhibits a thermal noise floor due to the Brownian motion of the elements composing the cavity, and it is often the contribution of the mirror coatings that dominates. For a given geometry (spacer length, radii of curvature of the mirrors), the size of the laser beams on the mirrors are larger for infrared light; therefore, the thermal noise is better averaged. Hence, it is favourable to build a laser in the infrared domain and to transfer its frequency stability to the clock-interrogation light field with high fidelity, i.e., adding as little as possible excess noise from the transfer setup.

\begin{figure}[t]
    \centering
    \includegraphics[trim={3mm 3mm 2mm 5mm},clip,width=1\columnwidth]{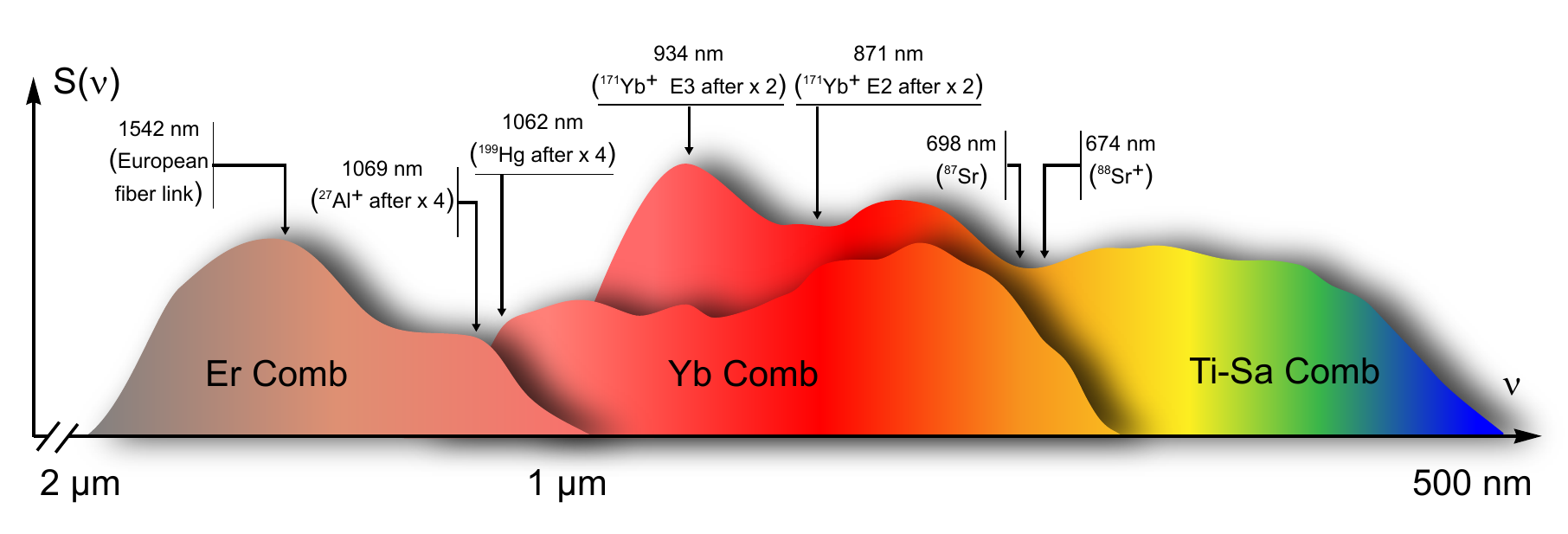}
    \caption{Most common technologies of frequency combs and examples of clock frequencies. Frequency combs are based on mode-locked femtosecond lasers: titanium-sapphire lasers (range covered after spectral broadening: 500--1000\;nm), ytterbium-doped fibre lasers (650--1450\;nm), or erbium-doped fibre lasers (1000--2000\;nm). The first two types offer a lot of power in a single-output configuration, while the last one is highly reliable but needs specific outputs after frequency conversion to reach the metrological wavelengths necessary for optical clocks.}
    \label{fig:Combs_Transitions}
\end{figure}

The transfer process is based on the transfer-oscillator method~\citep{Telle2002} applied to a frequency comb with teeth of frequencies $\nu_N=N f_{\text{rep}}+f_0$, where $f_{\text{rep}}$ is the repetition rate of the comb and $f_0$ the carrier-envelope-offset (CEO) frequency. This method allows the spectral properties of a master (M) laser (frequency $\nu_\mathrm{M}$) to be transferred to a slave (S) laser (frequency $\nu_\mathrm{S}$) in the Fourier frequency range covered by the bandwidth of the feedback loop. The choice of frequency-comb technology is motivated by the frequencies $\nu_\mathrm{M}$ and $\nu_\mathrm{S}$ involved: they must be part of the spectral range covered by the comb, either directly or after frequency conversion (Figure~\ref{fig:Combs_Transitions}). In what follows, we present the practical details of the implementation and the pros and cons of the alternative approaches.

\subsection{The transfer oscillator in practice} \label{sec:transfer_principle}

\begin{figure}[t]
    \centering
    \includegraphics[width=1\columnwidth]{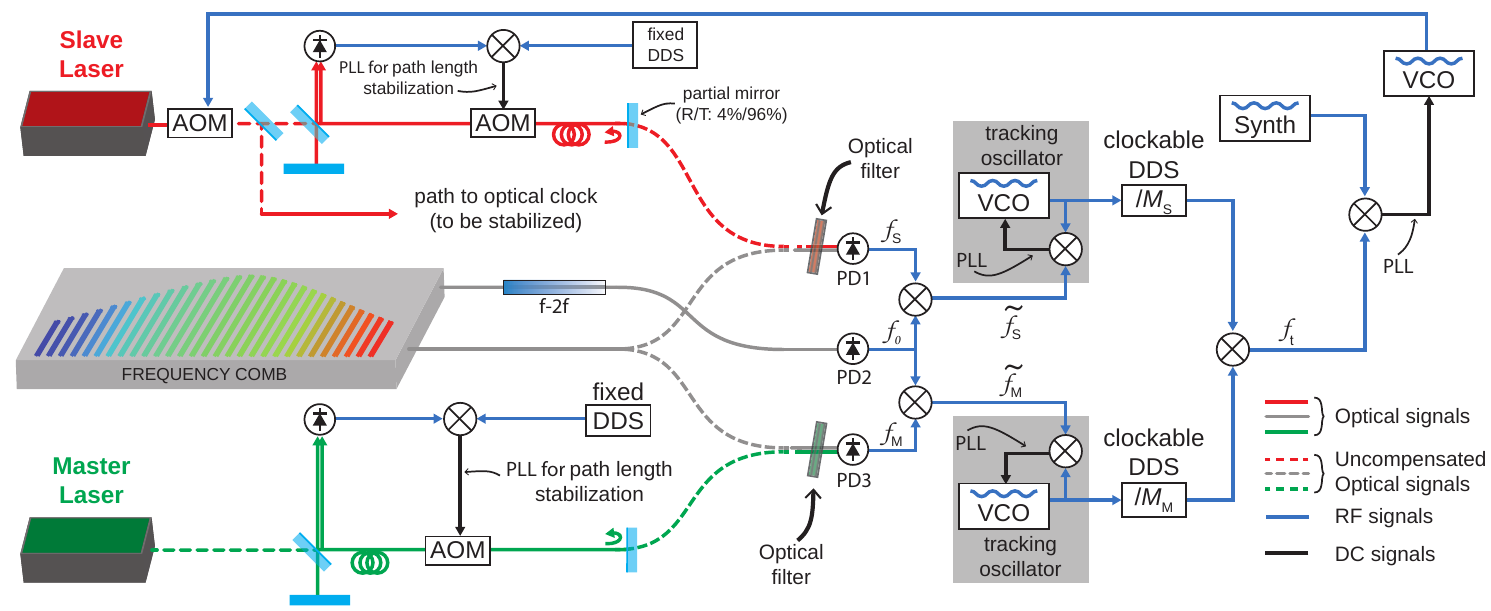}
    \caption{Principle of the transfer of spectral purity. The optical beat notes (comb vs slave and comb vs master) are detected by the photodetectors PD1 and PD3, respectively. The carrier-envelope-offset frequency $f_0$ of the comb is detected by PD2 and mixed out of the optical beat notes. These $f_0$-free beats are first regenerated by tracking oscillators before re-scaling by a division factor $M$ and mixing to form the transfer beat $f_\mathrm{t}$. This carries all the information on the relative fluctuations of the master and slave lasers in the bandwidth chosen by the trackings. Finally, the transfer beat is locked to, e.g., a synthesiser, and the feedback to the AOM will force the slave laser to follow the residual fluctuations of the master laser. The propagation noise of some of the optical paths is not compensated (dashed lines). The strategy to tackle this issue is described in Section~\ref{sec:Fiber_Stab}. VCO: voltage-controlled oscillator, PLL: phase-locked loop, AOM: acousto-optic modulator, PD: photodiode.}
    \label{fig:Principle_TSP}
\end{figure}

The transfer-oscillator technique is based on the photodetection of the beat notes between the frequency comb and the master on the one hand ($f_\mathrm{M}=\nu_\mathrm{M}-N_\mathrm{M}  f_{\text{rep}}-f_0$) and the slave on the other hand ($f_\mathrm{S}=\nu_\mathrm{S}-N_\mathrm{S} f_{\text{rep}}-f_0$), see Figure~\ref{fig:Principle_TSP}. The technical parameters of the comb, $f_{\text{rep}}$ and $f_0$, are not \emph{a priori} ultrastable quantities; therefore, it is necessary to eliminate them.

The classical technique is to detect $f_0$ with an $f$--$2f$ interferometer and to mix $f_0$ out of $f_\mathrm{M}$ and $f_\mathrm{S}$ in order to form $\widetilde{f_\mathrm{M}}=\nu_\mathrm{M}-N_\mathrm{M} f_{\text{rep}}$ and $\widetilde{f_\mathrm{S}}=\nu_\mathrm{S}-N_\mathrm{S} f_{\text{rep}}$. Recently, CEO-free combs have appeared. These are based on difference-frequency generation between the two far ends of an initial comb with nonzero CEO frequency. In this case, the first step to remove the CEO frequency is not necessary \citep{Puppe2016}.

Subsequently, $\widetilde{f_\mathrm{M}}$ and $\widetilde{f_\mathrm{S}}$ are divided by float numbers $M_\mathrm{M}$ and $M_\mathrm{S}$, respectively, such that
\begin{equation}\label{eq:Mnumbers}
    \frac{N_\mathrm{M}}{M_\mathrm{M}}=\frac{N_\mathrm{S}}{M_\mathrm{S}}.
\end{equation}
These rescaled beat notes are mixed together to form the transfer frequency $f_\mathrm{t}$, which leads to the elimination of $f_{\text{rep}}$,
\begin{equation}
    f_\mathrm{t} = \frac{\widetilde{f_\mathrm{S}}}{M_\mathrm{S}}-\frac{\widetilde{f_\mathrm{M}}}{M_\mathrm{M}} = \frac{\nu_\mathrm{S}}{M_\mathrm{S}}-\frac{\nu_\mathrm{M}}{M_\mathrm{M}}.
\end{equation}
This leads to a direct relation between $\nu_\mathrm{M}$ and $\nu_\mathrm{S}$, thus justifying that the comb is only used as an intermediate oscillator. It must be noted that the sign of the quantities $f_\mathrm{M}$ and $f_\mathrm{S}$ is not known \emph{a priori}; the relevant component at the output of the mixer forming $f_\mathrm{t}$ is the one immune to a change of $f_{\text{rep}}$.

The rescaling itself is best performed with a direct digital synthesiser (DDS), with a number of bits that is typically $K=32$ or $48$. The beat notes $\widetilde{f_\mathrm{M}}$ and $\widetilde{f_\mathrm{S}}$ can be filtered by RF tracking oscillators in order to clean up the signal and to regenerate the signal-to-noise ratio (SNR). The bandwidth of the tracking must be adjusted to avoid copying the electronic noise limiting the SNR of $\widetilde{f_\mathrm{M}}$ and $\widetilde{f_\mathrm{S}}$. Furthermore, the tracking bandwidths of the two beats should be matched in order to eliminate the transfer oscillator in the same range of Fourier frequencies. The DDSs are then clocked by these filtered signals and act as frequency dividers: the possible DDS output frequencies are $f_{\text{out}}={f_{\text{in}}}/{M}$, with $M=2^K/n$ and $n$ an integer in the range $[0,2^{K-1}]$. In practice, it is best to keep the signals at high frequencies and to use $M$ numbers that are as small as possible. The simplest alternative is to divide the beat note corresponding to the laser with the smaller frequency by $M=2$ and to adapt the other division factor in order to satisfy Equation~\eqref{eq:Mnumbers}, resulting in a second $M$ number between 2 and 4 if the lasers' frequencies are within the octave covered by the comb.

Finally, the transfer frequency is phase locked to a stable enough reference, e.g., a synthesiser or a hydrogen maser. The actuator to act on the frequency of the slave laser can be, for instance, an AOM. In this way, any pre-stabilisation of the slave laser can be decoupled from the correction from the master laser. The final relation between the two lasers reads
\begin{equation}\label{Eq:Transfer}
    \frac{\nu_\mathrm{S}}{M_\mathrm{S}}-\frac{\nu_\mathrm{M}}{M_\mathrm{M}}=f_\text{synth}.
\end{equation}
\noindent
The quantity $f_\text{synth}$, typically a few tens of MHz, is seven orders of magnitude smaller than $\nu_\mathrm{M}$, so its absolute noise can be neglected even if the fractional stability of the synthesiser is not better than a few times~$10^{-12}$. Noise analysis of Equation~\eqref{Eq:Transfer} leads to
\begin{equation}
    \sigma\left( \nu_\mathrm{S} \right)=M_\mathrm{S}\sqrt{\frac{ \sigma^2\left( \nu_\mathrm{M} \right)}{M_\mathrm{M}^2}+\sigma^2\left( f_\text{synth}\right)}\simeq \frac{M_\mathrm{S}}{M_\mathrm{M}} \sigma\left( \nu_\mathrm{M} \right)\simeq\frac{\nu_\mathrm{S}}{\nu_\mathrm{M}} \sigma\left( \nu_\mathrm{M} \right).
\end{equation}
Considering that photodiodes detect phase signals and that the feedback is implemented via a phase-locked loop,  Equation~\eqref{Eq:Transfer} can be integrated and gives the same deterministic relation between the phase of the two lasers and of the synthesiser.

Since the division factors $M$ are at least $2$, if $N_\mathrm{M}$ and $N_\mathrm{S}$ differ by less than a factor two, it is necessary to rescale both $\widetilde{f_\mathrm{M}}$ and $\widetilde{f_\mathrm{S}}$. If, on the other hand, $N_\mathrm{M}$ and $N_\mathrm{S}$ are more than a factor two apart, rescaling the beat note of highest $N$ by a division factor $\max{\left({N_\mathrm{M},N_\mathrm{S}}\right)}/\min{\left({N_\mathrm{M},N_\mathrm{S}}\right)}$ is sufficient, and the other beat note can be left unchanged.

\subsection{Limitations of the method}

This technique nevertheless has limitations that must be carefully considered. These are due either to an incomplete elimination of the comb parameters or to technical noise added by the successive loops necessary for the transfer.

The $M$ numbers are critical: as the choice of $n$ (``tuning word'' of the DDS) is discrete, there can be a residual contribution of $f_{\text{rep}}$
\begin{equation}
\nu_\mathrm{S}={M_\mathrm{S}} \left( \frac{\nu_\mathrm{M}}{M_\mathrm{M}} + f_\text{synth} +  \epsilon f_{\text{rep}} \right), \quad \text{with}\; \epsilon=\frac{N_\mathrm{M}}{M_\mathrm{M}}-\frac{N_\mathrm{S}}{M_\mathrm{S}}.
\end{equation}
It can easily be shown that $\epsilon$ is at most $\max{\left(N_\mathrm{M},N_\mathrm{S}\right)}/2^K$, i.e., on the order of $10^{-8}$ for a $K=48$~bit DDS. In order to ensure that the noise contribution of $\epsilon f_{\text{rep}}$ is small compared to that from $\nu_\mathrm{M}/M_\mathrm{M}$ at all timescales, the usual technique is to stabilise $f_{\text{rep}}$ to an external reference (optical or microwave), or to choose a combination of $N$ and $M$ numbers leading to a further rejection of $f_{\text{rep}}$.

Often it is not possible to form all the beat notes with the same output of the comb, because the master or the slave laser is not directly within reach of the spectrum of the comb. Additional dedicated optical amplifiers can then be used in order to amplify a specific spectral region of the comb output before frequency conversion towards a target metrological wavelength. This is the case, for instance, with the erbium comb and the strontium lattice clock: the atomic resonance (698 nm) is not accessible with this type of comb (1000--2000\;nm), but it is possible to add an erbium-doped fibre amplifier (EDFA) followed by spectral shifting or broadening  to 1396\;nm, typically in a highly nonlinear fibre, and by a doubling crystal that finally converts it to the relevant window of the spectrum around 698\;nm. However, adding specific amplifiers comes at the expense of losing common-mode noise rejection: when the beat notes $f_\mathrm{M}$ and $f_\mathrm{S}$ are formed with independent outputs of the comb, the approach is said to be multi-branch, in contrast to the single-branch technique, where a unique comb output is sufficient to form beat notes with the master and the slave, thus suppressing the influence of path-length fluctuations.

The CEO beat can be generated using an $f$--$2f$ interferometer in a separate comb branch, because it is not sensitive to path-length fluctuations, but only to dispersion fluctuations. Thus, elimination of $f_0$ is easier, as it is a small quantity offsetting a beat note in the optical domain, while the noise on $f_{\text{rep}}$ is multiplied by a large lever arm (the $N$ number).

\subsection{Single-branch transfer} \label{sec:transfer_single}

In the single-branch configuration, excess noise mainly results from differential path-length fluctuations between the light fields at the two optical frequencies involved and from beat-detection noise. The OC18 team has developed setups for superposition of the CW light fields with the comb light, efficiently eliminating differential path-length fluctuations along all critical paths.
This enables unsurpassed residual instability of the transfer setup consisting of the fibres starting at the optical frequency-comb generator and at the reference and clock light sources.

In the context of OC18, the single-branch method has been used at PTB to transfer the stability of a 1542-nm Si-cavity-stabilised laser (frequency $\nu_\mathrm{Si}$) to the Sr-lattice-clock ($\nu_\mathrm{Sr}$) or Yb$^+$-ion-clock interrogation light field and at LNE-SYRTE to transfer the stability of a 1542-nm laser to the Sr (698\;nm) and Hg (1062\;nm before quadrupling) lattice clocks.

The single-branch principle is illustrated in Figure~\ref{fig:Principle_TSP}. In what follows, we study two examples of transfer from 1542\;nm towards metrological wavelengths (698\;nm and 1062\;nm) with an erbium comb.

\subsubsection{1542\;nm towards 698\;nm}

The transfer from the infrared domain towards a metrological wavelength in the visible range, such as 1542\;nm towards the Sr resonance at 698\;nm is complicated, since 698\;nm is not directly accessible by EDFAs.
When a single branch is used, the comb light fields near both frequencies $(\nu_\mathrm{Si}$ and $\nu_\mathrm{Sr}$) propagate on a common path until the superposition beam combiner, i.e., they stem from a single branch of the comb setup. For the Si--Sr stability transfer, this branch consists of a single EDFA, a nonlinear fibre for spectral broadening, and a PPLN (periodically-poled lithium niobate) crystal frequency doubler in series. Besides the efficiently generated comb light near $\nu_\mathrm{Sr}$, the fundamental comb light near $\nu_\mathrm{Si}$ is present behind the frequency doubler, and the light fields propagate entirely on a common path. Hence, path-length fluctuations are differentially cancelled out due to common-mode rejection.

\begin{figure}[t]
    \centering
    \includegraphics[width=1\columnwidth]{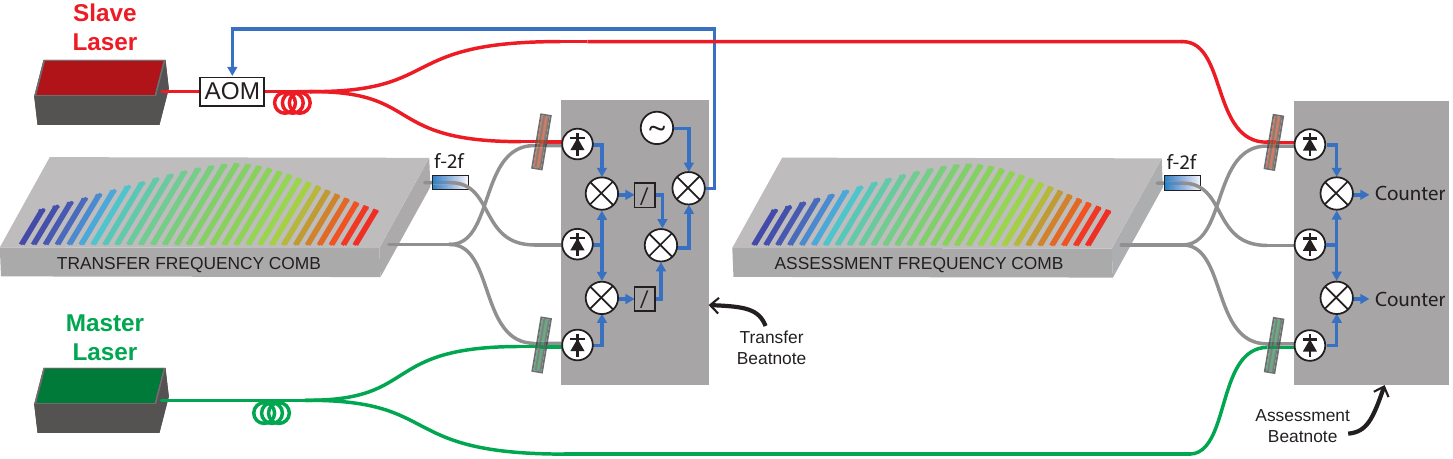}
    \caption{Out-of-loop characterisation of the transfer of spectral purity. The transfer frequency comb is used to form a transfer error signal to phase lock the slave to the master laser, while the assessment frequency comb ``reads'' the result: the two optical beat notes are formed in the same way and are post-processed to eliminate the technical parameters of this second comb.}
    \label{fig:In_Out_of_loop}
\end{figure}

As a second step, the fibres supplying the CW fields are length stabilised up to reference planes defined by partial mirrors. Most of the light power is transmitted through the semi-transparent reference mirrors. Shortly behind the mirrors, the two CW fields are superposed and from there on the fields propagate on a common path, which again efficiently suppresses differential phase fluctuations due to path-length variations~\citep{Benkler2019}. At the beam combiner, the CW light fields at the two frequencies interfere with the comb lines coming from the single branch. Hence, they can subsequently be separated using a dichroic mirror, which allows detection on separate photodiodes without adding differential phase fluctuations. The detected beat signals have a linewidth of 25\;kHz (1542-nm laser vs comb) and 80\;kHz (698-nm laser vs comb); these signals are tracked in a 2-MHz band. After mixing out the CEO beat signal of the comb, a transfer beat signal between the two CW fields can be derived following the transfer-oscillator scheme. This can be implemented either computationally by post-processing the simultaneously counted beat frequencies or in real time using RF hardware (see Figure~\ref{fig:In_Out_of_loop}). Besides the excess noise, the transfer beat signal carries information about the relative fluctuations between the two CW light fields. Hence, it can be used either for analysing the stability-transfer performance or for phase locking the clock light field such that the relative fluctuations are eliminated, i.e., the stability is actually transferred.

\subsubsection{1542\;nm towards 1062\;nm}

A slightly different single-branch transfer between pairs of metrological wavelengths was performed at LNE-SYRTE: any combination of master and slave is possible between 1062\;nm (metrological probing of Hg atoms, after quadrupling), 1160\;nm (laser based on the spectral hole burning technique, see Section~\ref{sec:spectral_hole_burning}) and 1542\;nm (reference channel for the European fibre link). Contrary to the previous case, these three wavelengths are within reach of a single EDFA without any frequency conversion, which facilitates the required setup.
In this case study, the comb at LNE-SYRTE is not free-running: after elimination of $f_0$, a tooth of the comb is phase locked to a reference infrared ultrastable laser, whose fractional stability ($6\times10^{-16}$) is transferred to $f_{\text{rep}}$. The comb is said to be in the ``narrow-linewidth regime''. It is important to note that the transfer-oscillator technique eliminates $f_{\text{rep}}$ at Fourier frequencies allowed by the tracking bandwidths; the frequency of the reference is then purely arbitrary. Stabilising the comb has only practical implications, notably the linewidths of $\widetilde{f_\mathrm{M}}$ and $\widetilde{f_\mathrm{S}}$ depend only on the stability of the lasers chosen as master, slave, and reference. In the present case, all the lasers at play have a noise floor $\num{< e-14}$, the beats are less than 1\;Hz wide, and they are tracked in a band of a few kHz in practice. The transfer oscillator was applied to this setup and the analysis of the performance is described in the next section.

\subsubsection{Out-of-loop assessment}

The performance of the transfer is in practice limited by technical effects: residual uncompensated optical paths, similar to the ones described in Figure \ref{fig:Principle_TSP}, imperfect mode matching between the beams, limited signal-to-noise ratios, cycle slips that degrade the correlation and therefore the stability, and residual contributions from dispersion fluctuations. Uncorrelated electronic noise in the different branches of the setup is unlikely to perturb the transfer by more than the equivalent of $1\times10^{-18}$ in the optical domain.

In order to quantify the residual noise, the single-branch setup must be tested by transferring the stability from the master laser to the slave laser using one frequency comb, and by testing the result using a second independent comb. This out-of-loop measurement is absolutely necessary to assess the performance of the setup. The noise of the transfer itself, $\epsilon_\text{trans}$, can be modelled by
\begin{equation}
\nu_\mathrm{S}={M_\mathrm{S}} \left( \frac{\nu_\mathrm{M}}{M_\mathrm{M}} + f_\text{synth} \right) +  \epsilon_{\text{trans}},
\end{equation}
and the detection of the transfer beat note by the second comb leads to
\begin{equation}
f'_\mathrm{t} = \frac{\nu_\mathrm{S}}{M'_\mathrm{S}}-\frac{\nu_\mathrm{M}}{M'_\mathrm{M}}+\epsilon'_\text{detect} = \left( \frac{M_\mathrm{S}}{M'_\mathrm{S} M_\mathrm{M}}-\frac{1}{M'_\mathrm{M}}\right) \nu_\mathrm{M} +\frac{M_\mathrm{S}}{M'_\mathrm{S}} f_\text{synth} + \frac{\epsilon_{\text{trans}}}{M'_\mathrm{S}} +\epsilon'_{\text{detect}},
\end{equation}
where $\epsilon'_\text{detect}$ corresponds to the detection noise in the out-of-loop setup.
The prefactor of $\nu_\mathrm{M}$ and the instability of $f_\text{synth}$ are small enough to simplify the noise analysis into
\begin{equation}
\sigma \left( f'_\mathrm{t} \right)= \sqrt{ \left(\frac{\sigma \left(\epsilon_\text{trans} \right)}{M'_\mathrm{S}}\right)^2+\sigma^2 \left( \epsilon'_\text{detect}\right)}.
\end{equation}

\begin{figure}[t]
    \centering
    \includegraphics[width=1\columnwidth]{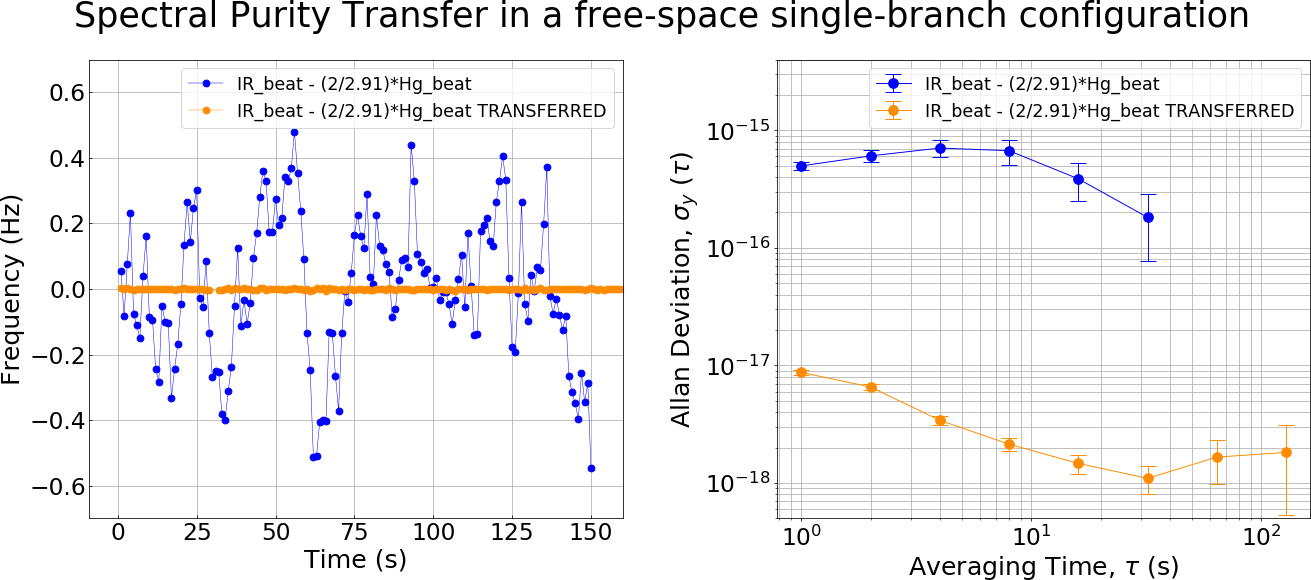}
    \caption{Results of the single-branch transfer of spectral purity from 1542\;nm to 1062\;nm. The resulting stability, around $8\times10^{-18}$ at 1\;s, includes the contribution of the transfer by the first comb and the contribution of the detection by the second comb. If this limit is due to, e.g., uncompensated paths on both sides, we can assume that the noise due to the transfer itself is around $(5\ldots 6)\times 10^{-18}$ at 1\;s in the case studied here.}
    \label{fig:SPT_single_ok}
\end{figure}

From the acquisition of $f'_\mathrm{t}$, we can directly derive an instability plot showing the Allan deviation versus averaging time $\tau$, see Figure~\ref{fig:SPT_single_ok}. The stability at 1\;s, around $8\times10^{-18}$, is a bit higher than the limit set by the signal-to-noise ratio (38\;dB in 1\;kHz corresponds to a stability limit of around $1.4\times10^{-18}/\tau\,[\mathrm{s}]$ at 1062\;nm for one photodiode). The residual noise, around a few $10^{-18}$, is likely to be due to uncompensated paths.

Estimating the total noise of the slave laser can be done only by comparing to another laser that is equally or more stable, by using a cross-correlator~\citep{Nelson2012,Xie2016} with two secondary sources, or by probing atoms in a clock and inferring the stability of the laser from the Dick effect.

\subsubsection{Case of Yb combs}
An approach that is more exploratory is to use frequency combs based on Yb-doped fibre lasers. These lasers surpass Er-doped technology in several technical aspects, such as power, power efficiency, pulse energy, and minimum achievable pulse duration. The supercontinuum generated by Yb-fibre lasers typically covers the range 650--1450\;nm. Therefore, it eliminates the need for additional conversion processes for many wavelengths of metrological interest. The large power available across the range eliminates the need for dedicated amplifiers. Hence, the Yb combs show an interesting potential for all-single-branch transfer of spectral purity. In the context of OC18, T{\"U}B\.{I}TAK has developed such an Yb frequency comb with pulses as short as $33$\;fs, which is a factor ten shorter than for typical Er combs~\citep{Senel2018}. At the same time, the total dispersion of the cavity was minimised for low-noise supercontinuum generation.

\subsection{Multi-branch transfer} \label{sec:transfer_Multi}

In cases where the aim is to transfer the stability of a single highly stable master laser to several other CW slave lasers operating at different wavelengths, a multi-branch frequency comb has some practical advantages over a single-branch comb. In a multi-branch configuration, a dedicated optical amplifier and nonlinear fibre, perhaps followed by a frequency-doubling stage, are used to generate a narrow-band (typically 1--3\;nm) frequency-comb output around each CW laser frequency of interest. Each output is therefore used to form only one beat note between the comb and one of the lasers. The advantage of this approach is that it enables independent optimisation of the signal-to-noise ratio of the multiple beats. In contrast, with a single-branch comb it can be challenging to optimise the broadened spectrum for simultaneous detection of multiple beat signals.

The increased flexibility of the multi-branch approach, however, brings with it the possibility of differential phase noise between the branches, which has been reported previously to limit the stability transfer to typically the low parts in 10$^{16}$ level for an averaging time of 1\;s~\citep{Nicolodi2014,Nakajima2010,Hagemann2013}. The CEO frequency is detected independently and mixed out, like in the single-branch case, and the differential noise at two different outputs M and S of the comb can be modelled by two repetition rates $f_{\mathrm{rep}}^{\mathrm{M}}$ and $f_{\mathrm{rep}}^{\mathrm{S}}$
\begin{equation}
    f_\mathrm{t}=\frac{\nu_\mathrm{S}}{M_\mathrm{S}}-\frac{\nu_\mathrm{M}}{M_\mathrm{M}}-\frac{N_\mathrm{S}}{M_\mathrm{S}} f_{\mathrm{rep}}^{\mathrm{S}} + \frac{N_\mathrm{M}}{M_\mathrm{M}} f_{\mathrm{rep}}^{\mathrm{M}} \simeq \frac{\nu_\mathrm{S}}{M_\mathrm{S}}-\frac{\nu_\mathrm{M}}{M_\mathrm{M}}-\frac{N_\mathrm{S}}{M_\mathrm{S}} \left( f_{\mathrm{rep}}^{\mathrm{S}} - f_{\mathrm{rep}}^{\mathrm{M}}\right).
\end{equation}

\begin{figure}[t]
    \centering
    \includegraphics[width=0.85\columnwidth]{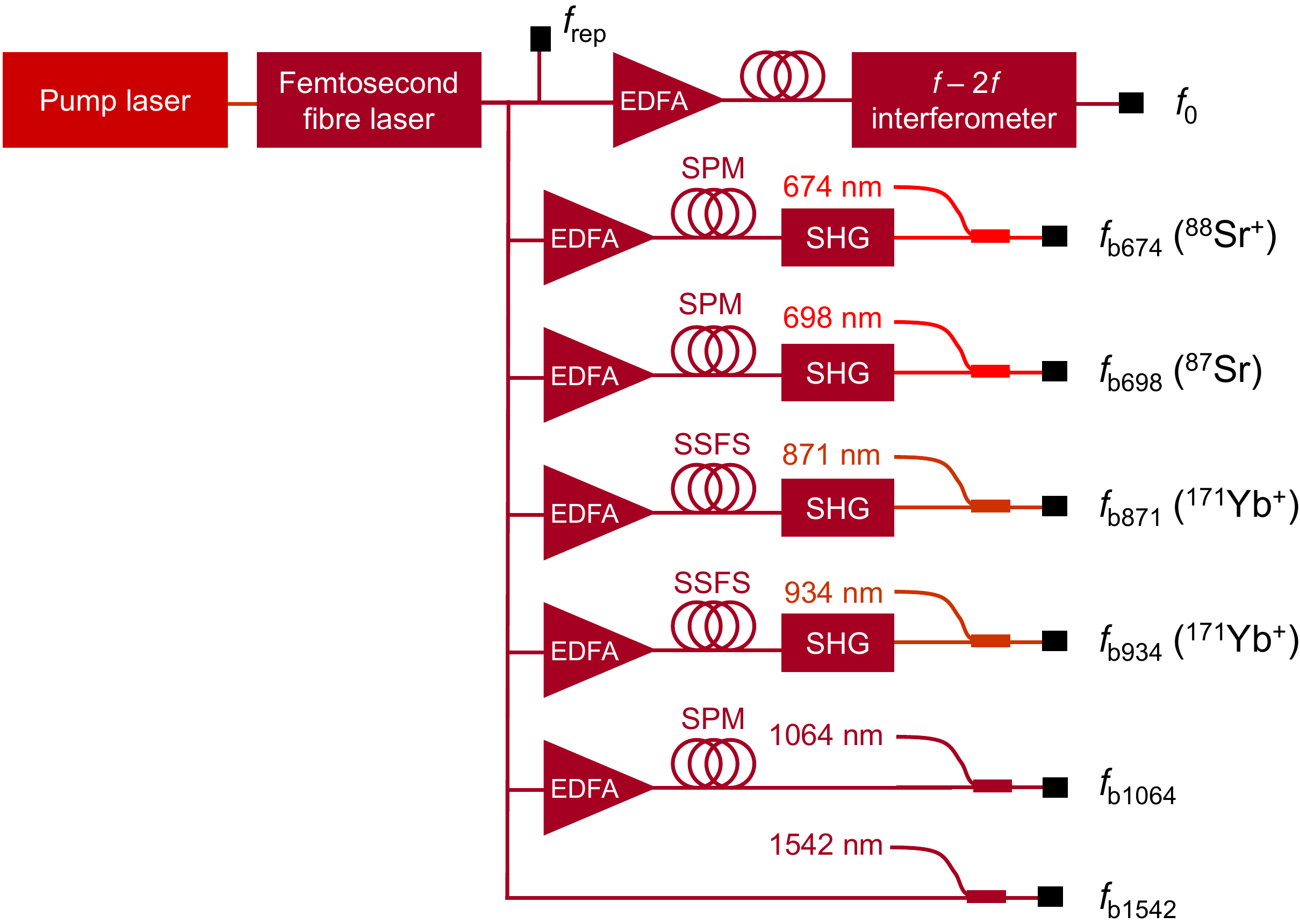}
    \caption{Schematic diagram of the NPL multi-branch erbium-doped fibre comb. The 674-nm and 698-nm branches are used to stabilise the probe lasers for the strontium ion optical clock and strontium optical lattice clock, respectively, whilst the 871-nm and 934-nm branches are used to stabilise the probe lasers for the two optical clock transitions in $^{171}$Yb$^+$. The 1542-nm branch is used to lock a transfer laser for the London--Paris optical fibre link. EDFA: erbium-doped fibre amplifier; SPM: self-phase modulation; SSFS: soliton self-frequency shift.}
    \label{fig:NPL_comb}
\end{figure}
However, experiments performed within the OC18 project have shown that with carefully designed beat-detection systems better performance can be achieved.
These experiments used the multi-branch erbium-doped fibre comb shown in Figure~\ref{fig:NPL_comb}, which is designed to transfer the stability of a 1064-nm master oscillator to five other wavelengths simultaneously. These are the wavelengths of interest for probing the optical clocks under development at NPL (strontium lattice clock, ytterbium and strontium single-ion clocks) and a 1542-nm laser required for optical-clock comparisons using optical fibre links. The femtosecond fibre laser has a repetition rate of 250\;MHz and all fibres after the output of this laser are polarisation-maintaining. The multi-branch configuration makes it straightforward to obtain signal-to-noise ratios of at least 30\;dB in 300-kHz bandwidth for all optical beat notes between the CW lasers and the comb.

A further design choice that must be made when using a frequency comb to transfer the stability of an ultrastable laser from one frequency to another is the method for stabilisation of the frequency comb. Here two degrees of freedom must be controlled, or at least measured, which determine both the mode spacing (repetition rate) $f_{\rm rep}$ and carrier-envelope-offset frequency $f_{0}$. One possibility is to tightly lock one mode of the frequency comb to the master oscillator, and then to phase-lock $f_{0}$ to either an RF synthesiser or a sub-harmonic of the repetition rate. To achieve sufficiently tight phase locking requires fast actuators and typically means that the femtosecond laser must include an intracavity electro-optic modulator. An alternative to tight phase locking is to weakly stabilise the comb to an RF oscillator such as a hydrogen maser, and to compensate for the residual noise of the comb modes using the transfer-oscillator scheme~\citep{Telle2002}. This can be more robust for long-term operation and is therefore the approach used in the NPL experiments.

\begin{figure}[t]
    \centering
    \includegraphics[width=0.9\columnwidth]{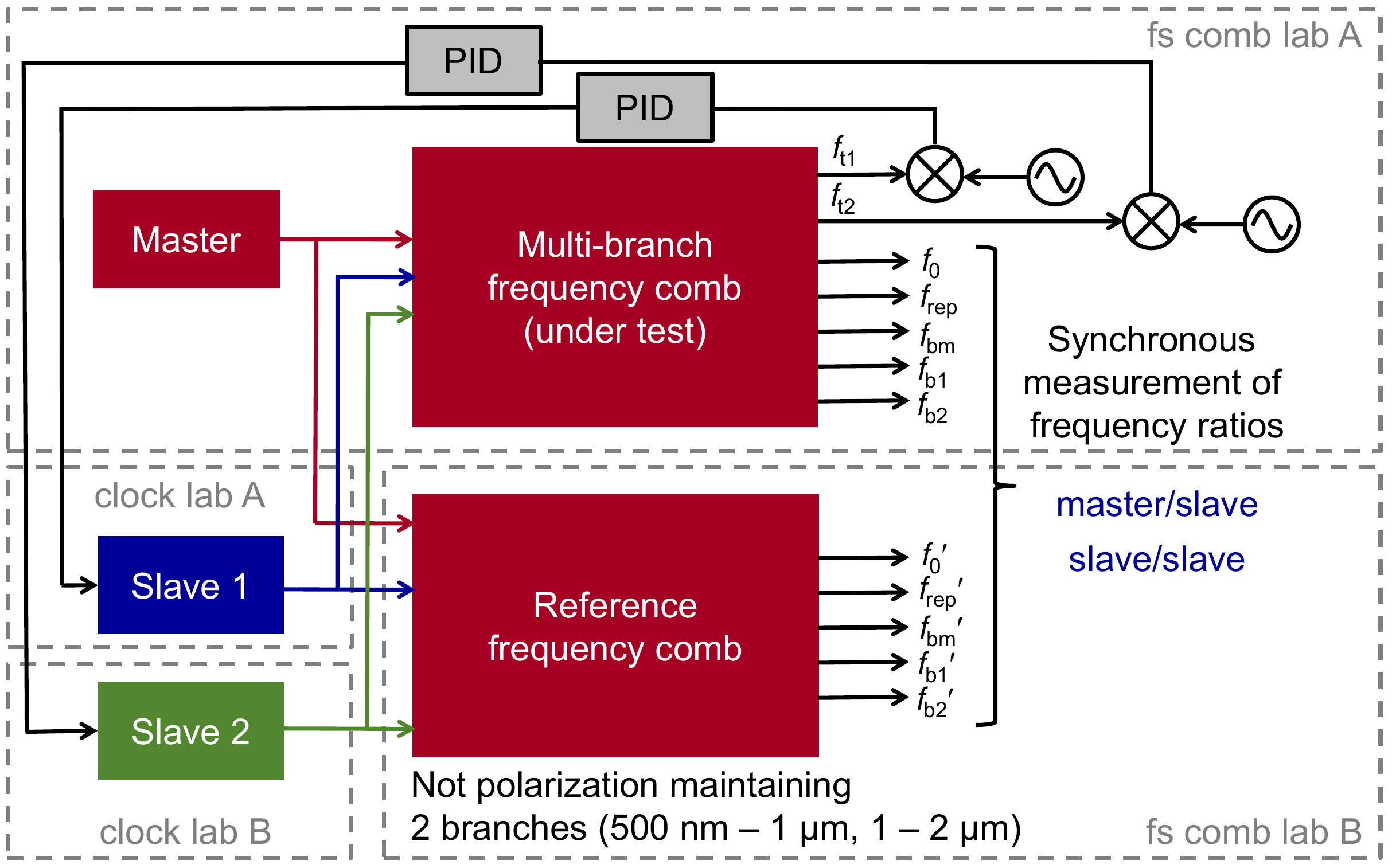}
    \caption{Out-of-loop measurement setup used to assess the performance of the stability transfer using the multi-branch frequency comb. The beat frequencies necessary to determine the various optical frequencies are counted synchronously using FXE frequency counters from K+K Messtechnik operating in $\Lambda$ mode.}
    \label{fig:out_of_loop_tests}
\end{figure}

The effectiveness of the multi-branch approach for stability was tested by stabilising the various clock lasers to the multi-branch comb using the transfer-oscillator technique, and then measuring the optical frequency ratios between pairs of lasers using a second, independent optical frequency comb (Figure~\ref{fig:out_of_loop_tests}). This reference comb was of a different design to the first, not using polarisation-maintaining fibre, and having two broadband outputs, one around 500--1000\;nm, and the other around 1--2\;\si{\micro\meter}. All clock-laser outputs were heterodyned with the first of these two broadband outputs, meaning that the reference comb is essentially single-branch for pairs of these wavelengths. However, the 1064-nm laser output was heterodyned with the longer-wavelength branch, so for optical-frequency-ratio measurements involving this wavelength, the reference comb is also multi-branch. The reference comb and the multi-branch comb under test are located in separate laboratories with different temperature-control systems, and so the correlated noise between the two systems is expected to be negligible.

\begin{figure}[p]
    \centering
    \includegraphics[width=0.66\columnwidth]{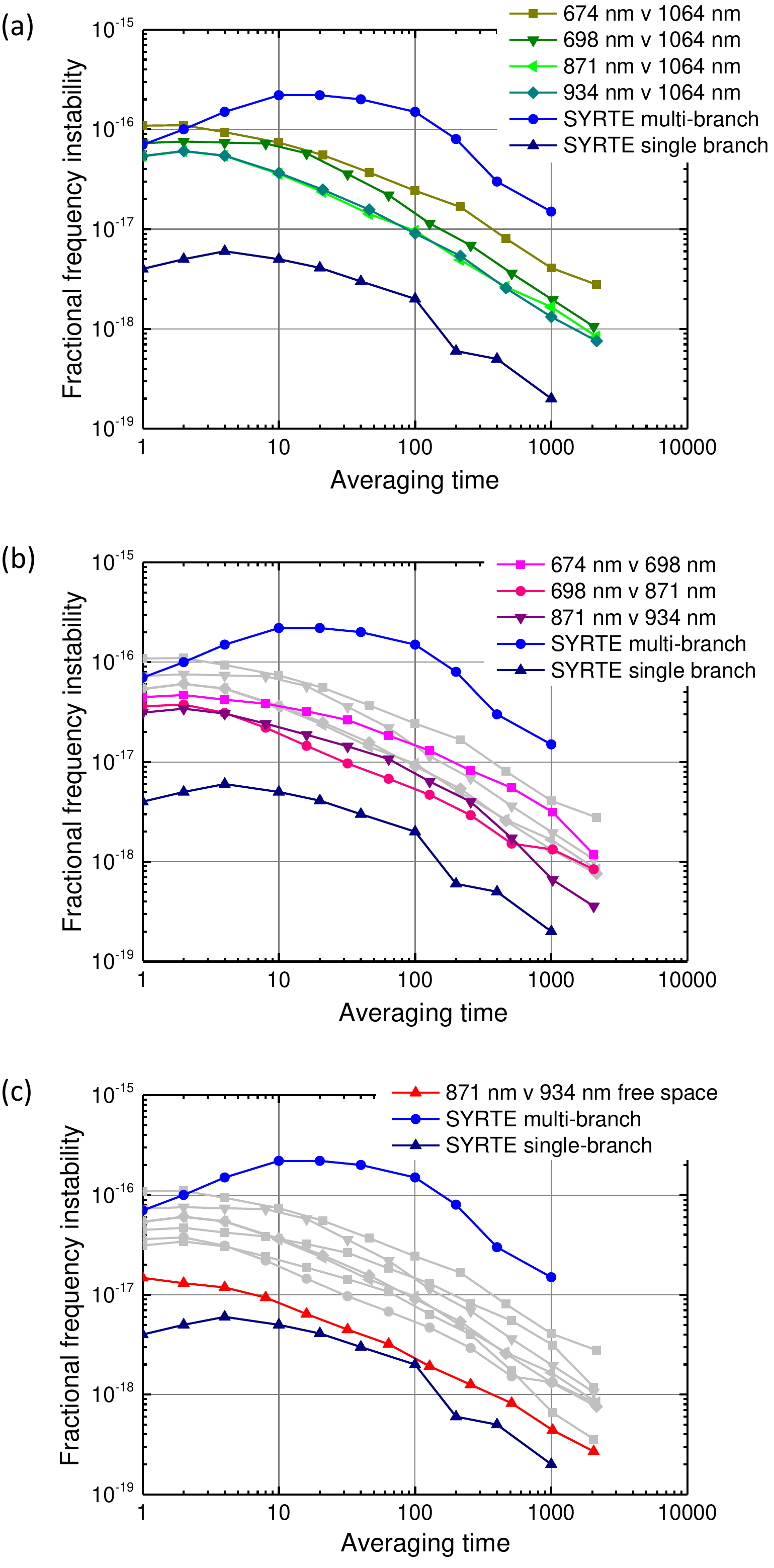}
    \caption{NPL multi-branch stability results. (a) Instability of the ratios between the optical-clock lasers and the 1064-nm master oscillator, using fibre-coupled beat-detection systems. (b) Instability of the ratios between pairs of optical-clock lasers, using fibre-coupled beat-detection systems. (c) Instability of the ratio between the 871-nm and 934-nm clock lasers, in the case where the fibre-coupled beat-detection systems were replaced by free-space beat-detection systems in which the retroreflectors used for phase-noise cancellation were moved much closer to the beat-frequency detectors. In case (a), the reference comb is multi-branch, whereas in cases (b) and (c) it is effectively single-branch. In all cases, the SYRTE results shown are taken from~\citet{Nicolodi2014}.}
    \label{fig:NPL_stability}
\end{figure}

The results obtained are shown in Figure~\ref{fig:NPL_stability} and indicate the importance that should be placed on careful design on the beat-detection systems. For initial measurements, fibre-coupled beat-detection systems were used, with the retroreflectors used for phase-noise cancellation of the incoming clock-laser light located several metres away and 2-m uncompensated (and unshielded) fibres linking these phase-noise-cancellation setups with the beat-detection systems. In later experiments, the beat-detection systems for 871\;nm and 934\;nm were replaced by systems based on free-space optics and with the retroreflectors for the phase-noise compensation located much closer to the beat-frequency detector. This enabled much better performance to be achieved, approaching the $10^{-17}$ level at 1-s averaging time, close to the single-branch performance reported previously by~\citet{Nicolodi2014}.

\section{Delivering the stability to the atoms} \label{sec:delivering}

\subsection{Outline of the problem}\label{sec:delivering_Description}

Delivering an ultrastable optical carrier to the atoms (neutrals or ions) is a problem that includes several aspects. By design, the ultrastable reference point of the clock laser is, for instance, the input mirror of an ultrastable cavity or the frequency actuator (e.g., AOM) used for transferring the spectral purity. In order to ensure that the propagation of the wave from this point to the atoms does not degrade the initial stability, it is necessary to detect the phase noise accumulated in the course of the propagation and to compensate it.

The stabilisation is equivalent to controlling the optical path length, i.e., controlling the integrated index of refraction of the different elements traversed by the wave along its path. This includes passive elements (e.g., free-space or in-fibre propagation, nonlinear crystals for frequency conversion) and active elements (optical amplifiers). The end point chosen to transfer the stability to is also a critical parameter: the atoms must be rigidly referenced to this point in order to avoid uncontrolled Doppler effects that could degrade the performance of the clock.

\subsection{Fibre stabilisation}\label{sec:Fiber_Stab}

In the simplest case, there is only a short path and an AOM for switching and frequency steering the beam between the clock-laser reference point and the atoms. Often, optical fibres are employed to transfer the light even over short distances, inducing additional phase noise that needs to be cancelled in an optical-path-length stabilisation setup. The spectral broadening induced by fibre-length fluctuations depends on the length, type and environment of the fibre used. A 100-m single-mode fibre in a typical environment causes a homogeneous spectral broadening on the order of a few~kHz. Fibre-length fluctuations can be detected in a double-pass configuration and compensated with a phase-locked loop acting on an AOM~\citep{Ma1994,Newbury2007,Foreman2007}.

\begin{figure}[p]
	\centering
    \includegraphics[width=\columnwidth]{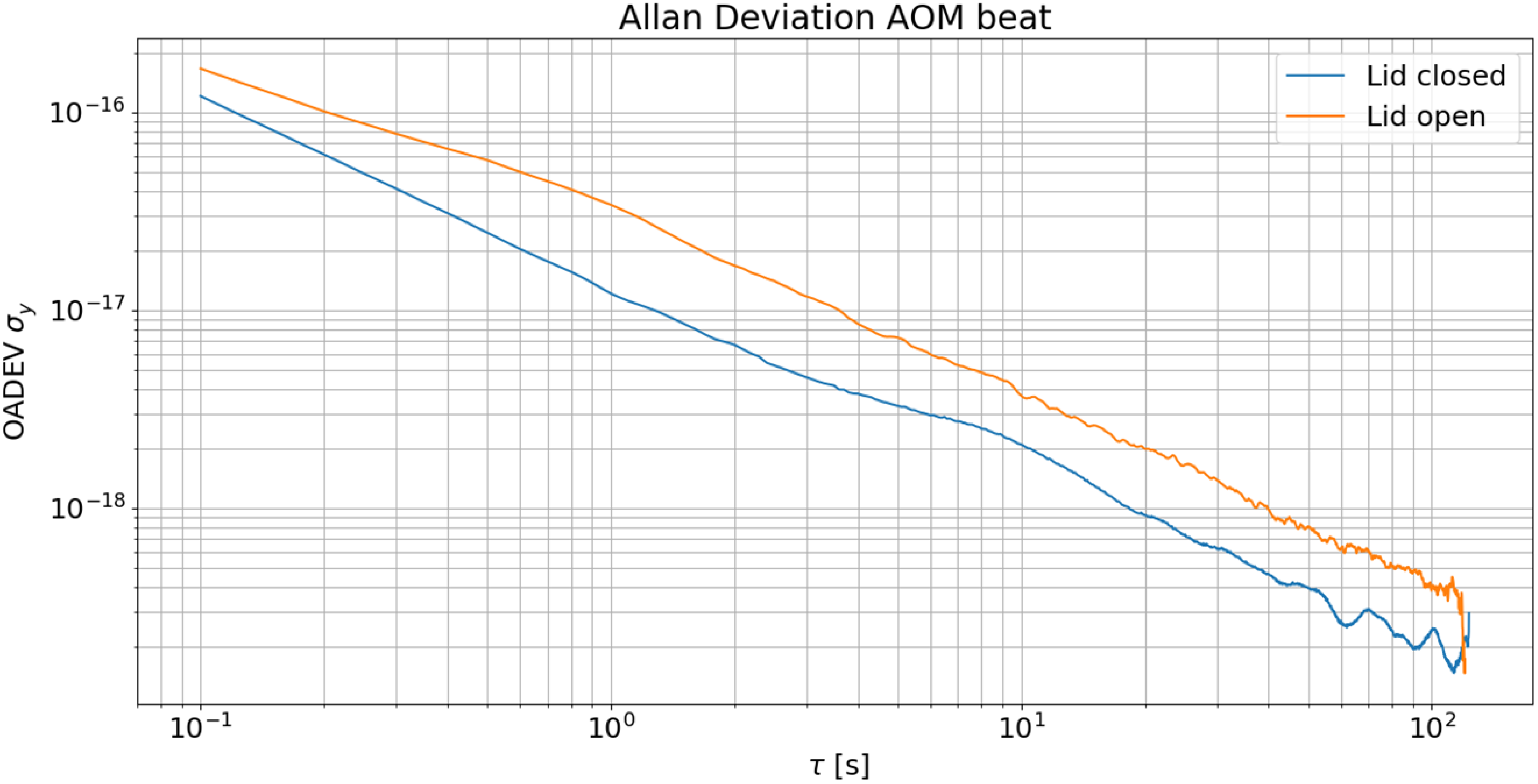}
	\caption{Comparison of the Allan deviation of residual phase noise in a short interferometer with and without enclosure. The test setup was a Mach--Zehnder interferometer with an AOM in one of its 20-cm-long arms. It was set up in a typical laboratory environment for testing the influence of pressure fluctuations with and without an enclosure for damping of the air flow along the beam path.}
	\label{fig:Short_interferometer_ADEV}
\end{figure}

However, even without using fibres, pressure and temperature fluctuations in a typical laboratory environment can cause significant phase noise on the order of $1\;\mathrm{rad/s}$. A slow drift of the optical path length by even $3\;\mathrm{nm/s}$ results in a systematic bias of $10^{-17}$. Figure~\ref{fig:Short_interferometer_ADEV} shows a comparison of phase stability for an enclosed and open free-space interferometer in a typical laboratory environment. We see that even short unstabilised beam paths should be avoided, kept as short as possible, and/or protected against pressure and temperature fluctuations~\citep{Nicolodi2014}.

\begin{figure}[p]
	\centering
    \includegraphics[width=\columnwidth]{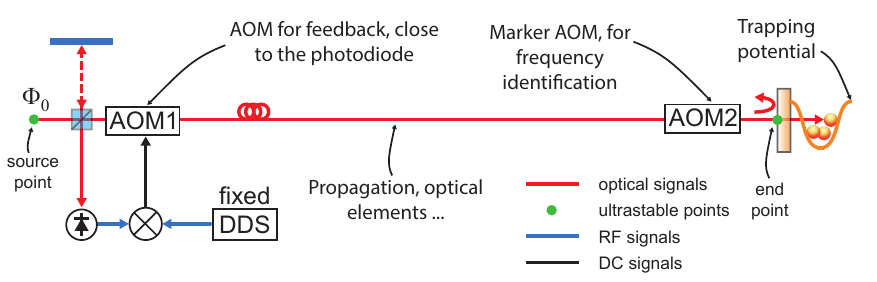}
	\caption{Stabilisation of an optical path with defined end point rigidly linked to the atoms. This configuration is the simplest one; no element (isolator, amplifier, etc.) prevents the light from being retroreflected all the way back. A Michelson interferometer compares the propagation phase between a short local-oscillator arm and the path between the ultrastable laser and the atoms. The input of the interferometer is as close as possible to the reference the laser is locked to (typically the input mirror of an ultrastable cavity), while the back end is a plane tightly linked to the atoms. This approach relies on the fact that identical phase fluctuations are experienced by the light on the way to the atoms and on the way back. The photodiode detects the phase difference that is locked to, e.g., a DDS via feedback to AOM1. AOM2 is used as a frequency discriminator in order to identify without ambiguity the part of the light effectively reflected by the end point. For typical laboratory distances (10\;m), the bandwidth is not limited by the propagation time of the light, but rather by the input bandwidth of AOM1 (a few~MHz at most).}
	\label{fig:Stabilisation_principle}
\end{figure}

For typical path-length-stabilisation approaches, a bidirectional invariant beam path with defined end point is needed.  For more complex optical elements with competing reflecting surfaces, this condition is not always fulfilled and a double-pass marker AOM is required as close as possible to the back-reflecting surface to identify the reference signal by its frequency without ambiguity. A generic setup is shown in Figure~\ref{fig:Stabilisation_principle}. It is based on interference between a part of the light taken as close as possible to the initial ultrastable point (phase $\Phi_0$) and light that has made a round trip to the plane we want to transfer the stability to (phase $\Phi_0+2\Phi_{\text{AO1}}+2\Phi_{\text{propagation}}+2\Phi_{\text{AO2}}$). The interferometric detection yields a beat note whose phase $2\Phi_{\text{AO1}}+2\Phi_{\text{propagation}}+2\Phi_{\text{AO2}}$ is phase locked to a DDS ($\Phi_{\text{DDS}}$) by inducing feedback to an AOM physically close to the detection photodiode. Therefore, the light experienced by the atoms has a phase $\Phi_0+\Phi_{\text{DDS}}/2$, which is an ultrastable quantity. A critical point is the reference arm: since it is the local oscillator used for the reference, its path must be kept as short and isolated as possible. Typically, this arm must be at most a few cm long, well shielded, and possibly put under vacuum. In the case of multiple fibres, all phase stabilised, a good practice to avoid crosstalk is to design the setup in such a way that all the beat notes to be locked are at different frequencies.

\subsection{Methods to control the noise of optical amplifiers}\label{sec:delivering_Amplifier}

\begin{figure}[t]
	\centerline{\includegraphics[width=0.75\columnwidth]{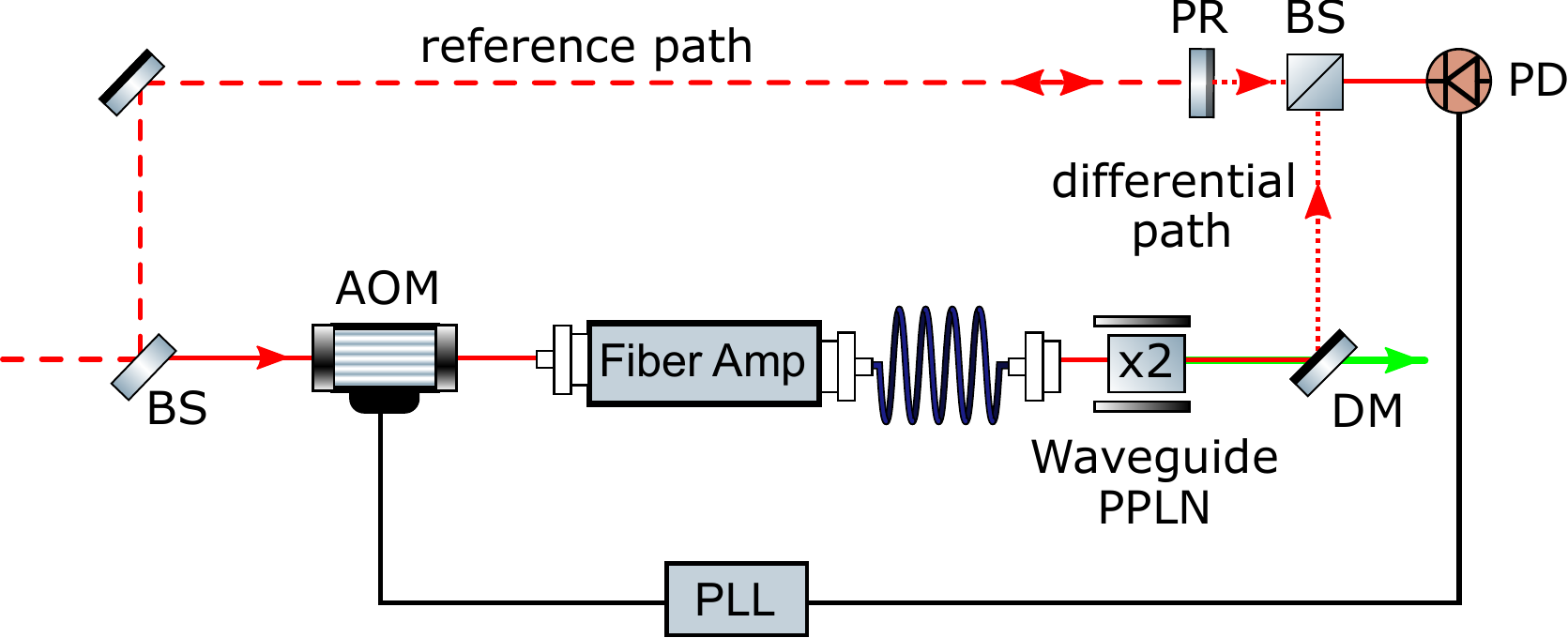}}
	\caption{Optical-path-length stabilisation with active element. The loop consists of a length-stabilised reference path (dashed line) with the reference point given by the partial reflector (PR), bypassing a fibre amplifier, which does not support bidirectional signals. The fibre amplifier pumps a PPLN waveguide doubler. Residual fundamental light is split up with a dichroic mirror (DM) and overlapped with the reference path on a photodiode (PD). The AOM is used to correct phase fluctuations between the arms. The differential path (dotted line) is unstabilised and should be kept as short as possible.}
	\label{fig:setup}
\end{figure}

\begin{figure}[t]
	\centerline{\includegraphics[trim={20mm 15mm 33mm 30mm},clip,width=0.95\columnwidth]{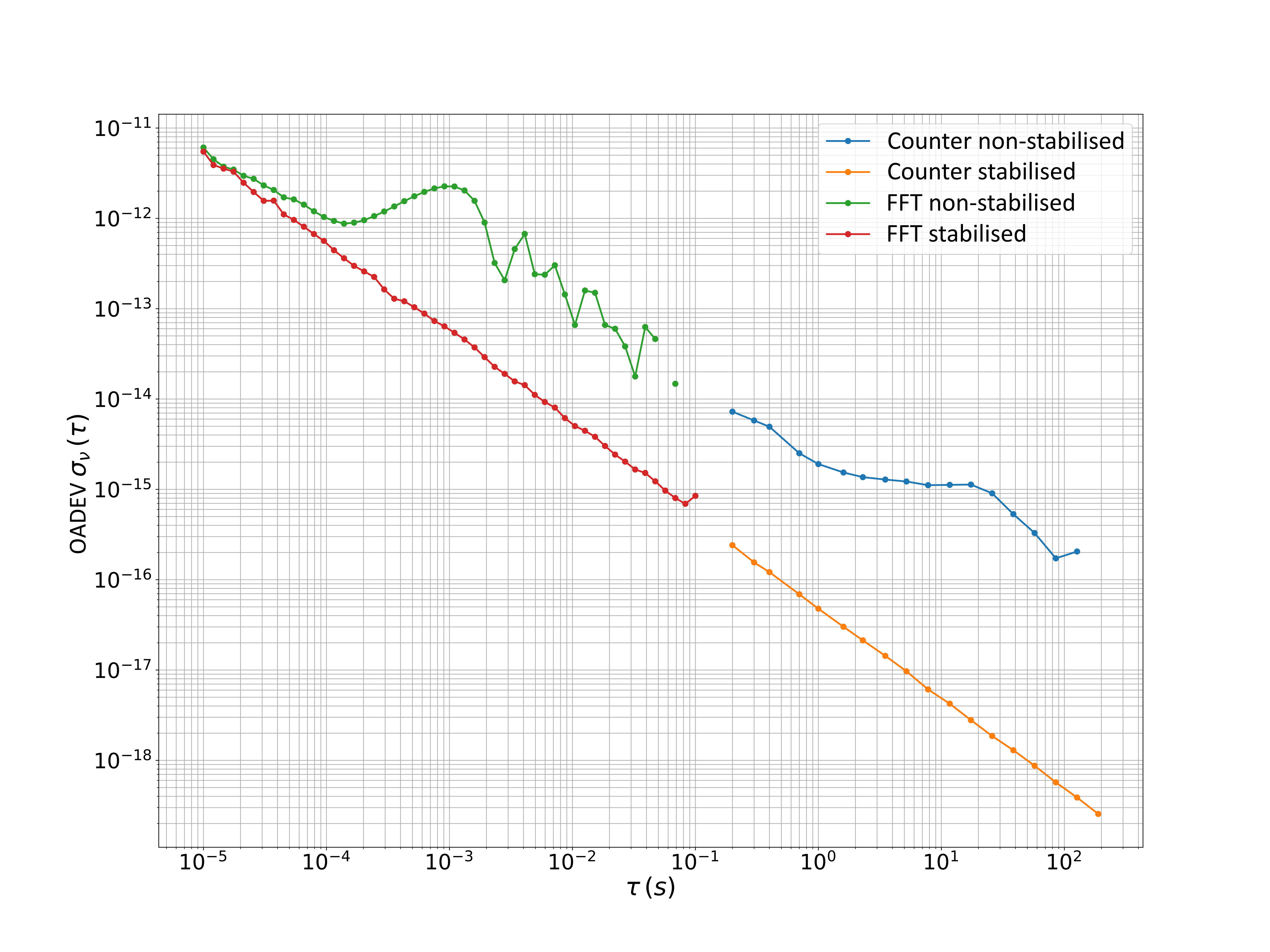}} 
	\caption{Stability assessment of optical path through fibre amplifier and waveguide doubler. Overlapping Allan deviation of main- and reference-path interference (see Figure~\ref{fig:setup}) with and without active length stabilisation. For times $\tau < 0.1$\;s, the signal was mixed down and recorded using a fast-Fourier-transform (FFT) analyser. For longer times, the signal was recorded with a frequency counter.}
	\label{fig:AmplifierNoise}
\end{figure}

With more complex beam paths including active components that allow bi-directional operation, such as bi-directional EDFAs, the previous strategy still applies. However, some components may not allow retro-reflected light propagation. In these cases the length stabilisation is realised with a Mach--Zehnder interferometer, where the phase of the main light path is locked to a stabilised reference arm bypassing the optical element as shown in Figure~\ref{fig:setup}. The feedback is applied through an AOM and again it is important to keep the differential, unstabilised path between the main and reference path as short and shielded as possible to avoid unwanted propagation noise. Typical in-loop stabilities are shown in Figure~\ref{fig:AmplifierNoise}, with residual instabilities smaller than $10^{-16}$ after 1\;s and averaging down as $1/\tau$.

\subsection{Methods to investigate the noise of nonlinear processes}\label{sec:delivering_NonLinear}

If the clock transition frequency is not directly accessible by suitable laser sources, a more sophisticated beam path is often required.
Typically, clock transitions are in the visible to ultraviolet region, e.g., Al$^+$ at 267\;nm~\citep{Chou2010} and Yb$^+$ E3 at 467\;nm~\citep{Huntemann2016}, while high-finesse mirrors with small absorption and suitable lasers are only available in the visible or infrared spectral range.
In cases where frequency conversion such as doubling is required, phase stability needs to be maintained through the conversion process. An upper bound for the phase noise added through the doubling process has been investigated in \cite{Delehaye2017}. It is of the same order of magnitude as that of a laser with $4\times 10^{-17}$ flicker floor \citep{Matei2017} for Fourier frequencies between 1 and 10\;Hz, and smaller everywhere else \citep{Herbers2019}. In the case of bidirectional waveguide doublers, phase stabilisation can in principle include the doubler. If further frequency-conversion steps are required or an extended path length is required afterwards, time-dependent dispersion can introduce additional phase noise. One option to maintain phase stability under this condition is to use residual fundamental light after the doubler for stabilisation and begin a new stabilisation loop with the reference point for the doubled frequency as close as possible to the reference point of the fundamental light. This way only the phase-noise contribution from refractive-index differences in the nonlinear medium remains, and these are small~\citep{Delehaye2017}.

\begin{figure}
	\centerline{\includegraphics[width=1\columnwidth]{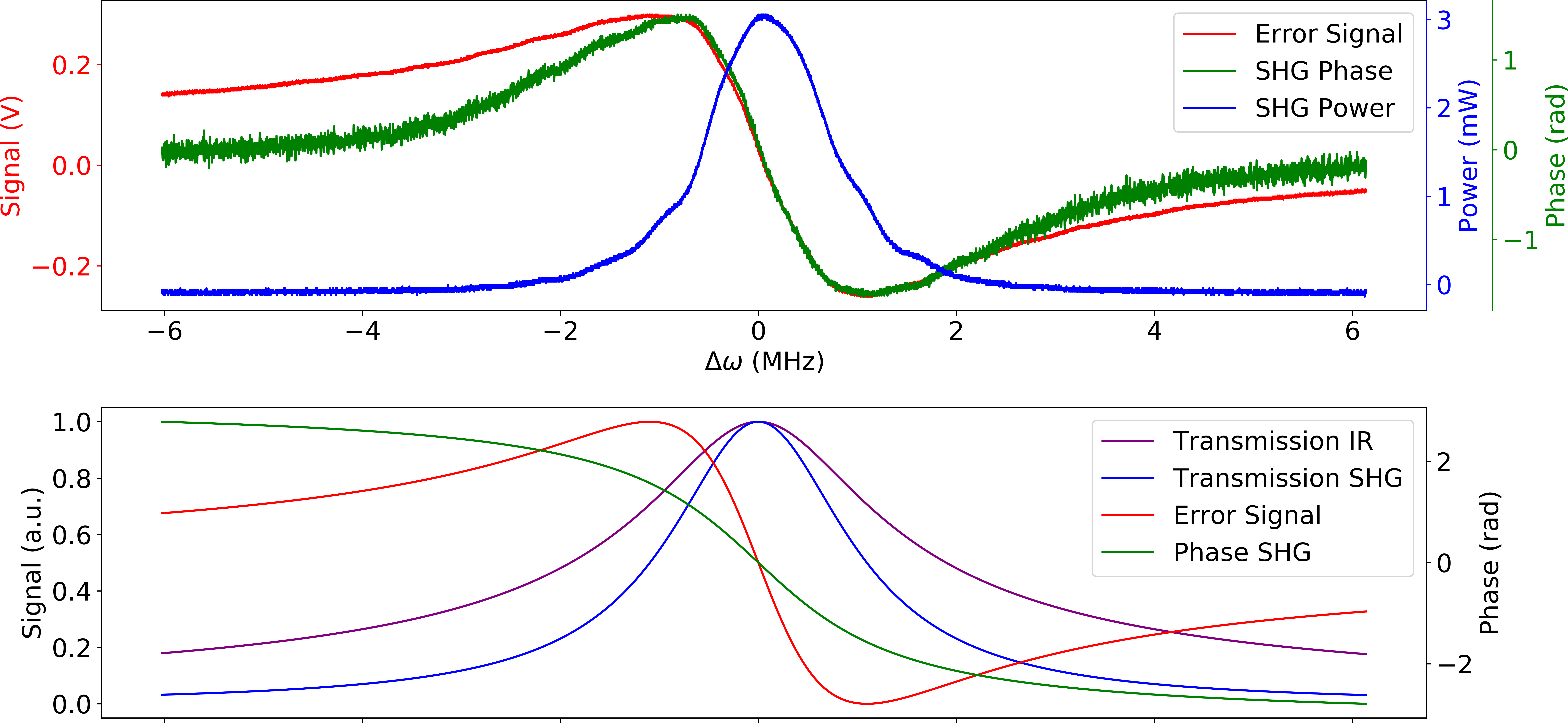}}
	\caption{Phase noise from doubling-cavity length changes. Comparison of measured (top) and calculated (bottom) transmitted field, phase, and H\"{a}nsch-Couillaud error signal for piezo-induced length changes.}
	\label{fig:piezoscan}
\end{figure}

Special care has to be taken when pump-beam resonance cavities are employed for frequency conversion.
The cavity acts as a bandpass filter for the resonant light, so it induces a frequency-dependent phase shift $\Delta \phi = \arg{( \left|E_t(\omega) \right|^2)}$ to the resonant transmitted field
\begin{equation}
  E_t(\omega)= E_0\frac{ t^2 e^{-i\omega L/c}-r^2e^{i\omega L/c}}{1 + r^4 - 2r^2\cos{(2\omega L/c)}}.
\label{eq:E_t}
\end{equation}
The residual length fluctuations of the cavity lock with respect to the resonance condition translates into phase fluctuations of the near-resonant light.
Therefore, the phase shift due to uncorrected residual length fluctuations can be monitored via the error-signal slope (see Figure~\ref{fig:piezoscan}).
For example, for a doubling cavity locked to within 1\% of its error signal slope, the added phase noise is $10^{-7}\;\mathrm{rad^2/Hz}$ (white phase noise in 10\;kHz range), limiting the stability at a level of 3$\times 10^{-17}$ at~1\;s for radiation at 1070\;nm.

\subsection{Methods to reference the clock-laser phase to the atomic trapping potential}\label{sec:delivering_reference}

Optical clocks are based on trapped atoms, which allows extended spectroscopy times, resulting in a better Fourier resolution of the resonance. Atoms are typically confined in a ``magic'' optical lattice (neutrals) or in a Paul trap (ions). Therefore, they are rigidly fixed to this external potential. It is crucial that this potential and the end point of the clock-laser phase stabilisation are tightly connected.
Relative motion of the trapping potential with respect to the reference point can result in:
\begin{itemize}
    \item A loss of stability, if the motion is not synchronous with the cycle frequency of the clock. The stability would be degraded at a time corresponding to the period of the motion.
    \item A bias if the motion is synchronous, because it then systematically induces a Doppler shift at each interrogation.
\end{itemize}

\noindent
In the case of optical lattice clocks, several strategies can be implemented:
\begin{itemize}
    \item If the standing wave (lattice) trapping the atoms is formed by a single retroreflection from a mirror, then this mirror can also be used as the end point for the clock-laser phase stabilisation. The coating must simply be adjusted to reflect \num{>99}\% at normal incidence at the magic wavelength \citep{Falke2012}, and to send around \SI{1}{\micro\watt} of clock light backwards in order to close the interferometer. Considering that a standing wave on the clock-laser light would result in a modulation of the Rabi frequency, and thus in a broadening of the transition, the lattice laser and the clock laser must come from opposite sides of the mirror.
    \item If the lattice is formed inside an enhancement cavity, a similar approach can be implemented; the cavity input mirror can be coated to reflect part of the clock light before it enters the cavity. Again, it is important that the second (output) mirror is sufficiently AR coated at the clock wavelength to avoid a standing wave.

\end{itemize}

\begin{figure}
	\centerline{\includegraphics[width=0.55\columnwidth]{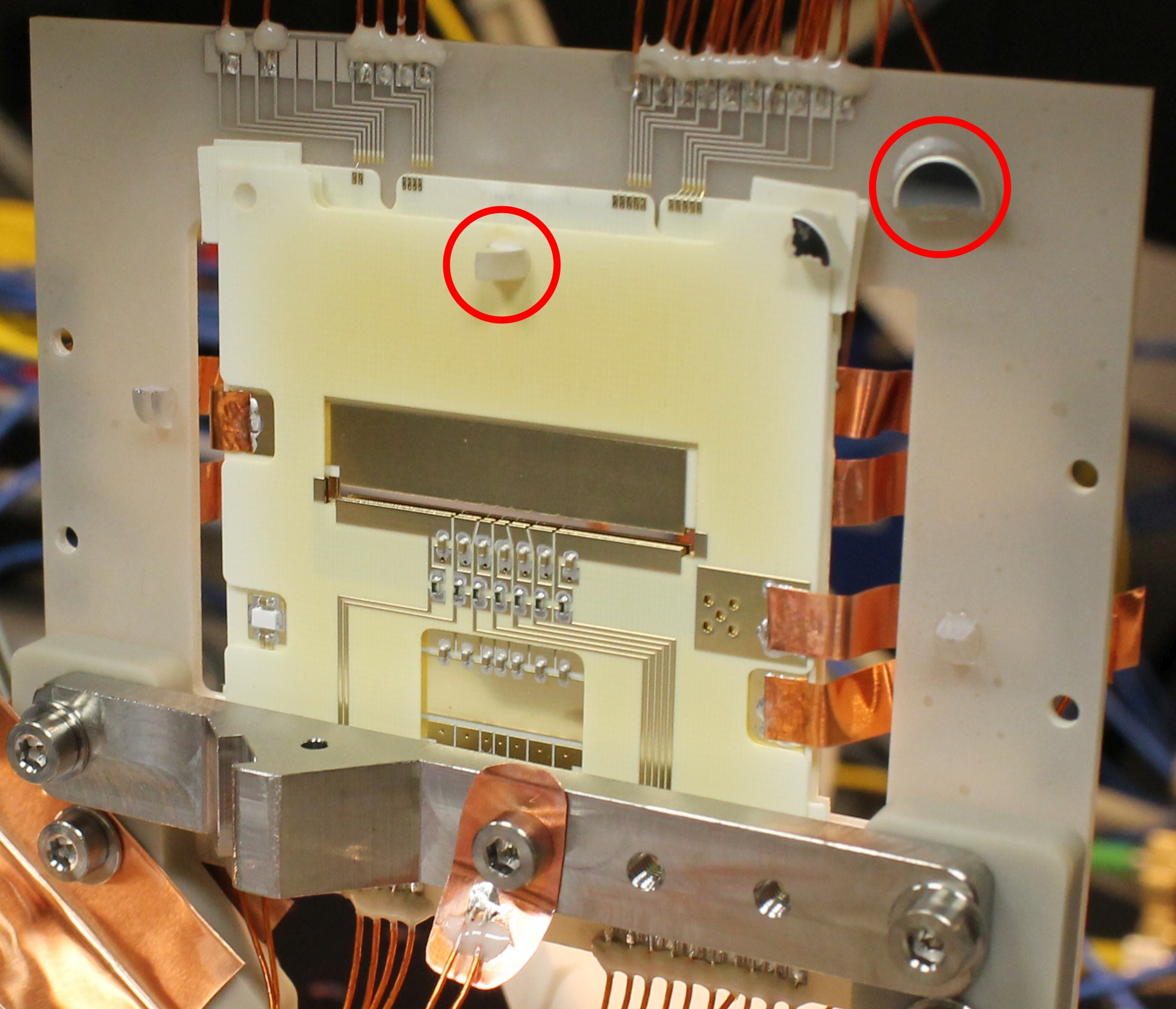}}
	\caption{Segmented Paul trap of the design presented in \cite{Pyka2014} with attached mirrors (indicated by red circles) for probing trap motion \citep{Hannig2019}.}
	\label{fig:iqloc2_trap}
\end{figure}

In the context of ion clocks, the approach studied in OC18 is to fix a mirror directly on the ion trap, see Figure~\ref{fig:iqloc2_trap} \citep{Hannig2019}. This way, all length fluctuations relative to the trap are cancelled and only the motion of the ion relative to the trap remains. While the effort of implementing phase stability all the way to the ion trap can be significant, its necessity can be tested independently of clock operation in an interferometric setup.

It is also important to activate the phase stabilisation of the clock laser as soon as the spectroscopy pulse begins. This can be achieved by using an AOM to pulse the probe laser and having the feedback loop, as described in Figure~\ref{fig:Stabilisation_principle}, catch up rapidly to avoid any systematic phase oscillations while the loop settles down \citep{Falke2012}. Another possibility is to have the clock laser on all the time, but to detune it sufficiently far away from resonance (typically ${\sim} 1$\;MHz) when no spectroscopy of the narrow resonance is ongoing.

\section{Summary}

In conclusion, this chapter has presented the typical problems that are encountered when transferring the frequency stability of an ultrastable laser source from one wavelength to another, and also when propagating the laser light towards the atoms.
In each case, practical suggestions have been given on how to overcome the difficulties, and case studies have been presented from several of the laboratories in Europe.  The techniques employed have allowed frequency stability to be transferred from the laser source to the atoms, whilst adding no more than $1 \times 10^{-17}$ to the laser noise at~1\;s.

\clearpage

\graphicspath{{D3_iontrap/}}

\chapter{Guidelines for optimised ion traps\label{chap:iontrap}} 

\authorlist{Christian Tamm$^{1,\dag}$,
Moustafa Abdel-Hafiz$^1$,
Petr Balling$^2$,
Miroslav Dole\v{z}al$^2$,
Rachel M.~Godun$^3$,
Nils Huntemann$^1$ and
Thomas Lindvall$^4$
}

\affil{1}{\PTBaff}
\affil{2}{\CMIaff}
\affil{3}{\NPLaff}
\affil{4}{\VTTaff}
\corr{christian.tamm@ptb.de}

\chapstart
This chapter summarises requirements and practical recommendations for the design of ion traps for single-ion optical clocks. Here we consider in particular the case of alkaline-earth-metal ions (like Ca$^+$ and Sr$^+$) and ions with similar level systems (like Hg$^+$ and Yb$^+$), characterised by a $^2$S$_{1/2}$ ground state and a strong electric-dipole-allowed resonance transition that is used for Doppler cooling. With a suitable ion trap setup, confinement-induced frequency shifts of the clock reference transition can be reduced to the low $10^{-17}$ range or less for these ion species~\citep{Ludlow2015}. In present single-ion clocks, the coherent interaction time between the probe laser and the trapped ion is typically limited to less than 1\;s. This appears to be the main obstacle to further improvements in stability and accuracy. Therefore, particular attention will be paid to design features that support long probe times by minimising motional heating of the trapped ion in the absence of laser cooling. Another point of interest is to eliminate sources of excessive thermal radiation in the trap setup so that the AC Stark shift induced by blackbody radiation can be evaluated more accurately~\citep{Dolezal2015}.

Section~\ref{sec:iontrap_theory} describes the basic characteristics of ion storage in radio-frequency (RF) traps, with emphasis on aspects relevant to the design of traps for optical clocks that enable long probe times. Section~\ref{sec:iontrap_design} considers advantageous options for trap electrode geometry, materials and electrical circuitry connected to the trap electrodes. In Section~\ref{sec:iontrap_OC18traps}, we describe two of the ion traps that have been newly developed in the frame of the OC18 project with consideration of the guidelines noted here.

\section{Traps for single-ion clocks: theory and design requirements
} \label{sec:iontrap_theory}

The traps employed in optical clocks serve the purpose of confining the active atoms or ions in a small volume where they are excited by cooling and probe lasers. \emph{Ideally} the trap provides a harmonic conservative potential of the form
\begin{equation}
\label{eq:iontrap_potential}
\Phi(\mathbf{r}) = \frac{1}{2}\left(\xi_x x^2 + \xi_y y^2 + \xi_z z^2 \right),
\end{equation}
where $\mathbf{r} = (x, y, z)$ denotes the particle position in a coordinate system that is aligned with the principal axes of the trap, and the steepness of the potential is described by the parameters $\xi_i$ ($i=x,y,z$). A particle with mass $m$ and non-zero kinetic energy oscillates  along the principal axes with frequencies $\omega_i = (\xi_i/m)^{1/2}$. The one-dimensional time-averaged kinetic energy  $\langle E_i \rangle$ of the particle is related to the oscillation amplitude $a_{i0}$ through $\langle E_i \rangle = m (a_{i0} \omega_i)^2/4$.

In a quantum description, a useful measure for $\langle E_i \rangle$ is the average harmonic-oscillator quantum number $\bar{n}_i = \langle E_i \rangle /(\hbar\omega_i) - \nicefrac{1}{2}$, where $\hbar$ is the reduced Planck constant. At $n_i=0$, the kinetic energy of the trapped particle is the zero-point energy $\hbar \omega_i/2$ and its localisation is described by a Gaussian wave function whose spread is given by $a_i^2(n_i=0) = \hbar/(2m\omega_i)$.

\subsection{Radio-frequency Paul trap
} \label{sec:iontrap_Paul_trap}

For ions, a trap potential that is much larger than the typical kinetic energy at room temperature can be generated through the Coulomb force. The electric field of an ion trap must have an oscillating component because Earnshaw's theorem states that static electric fields cannot provide three-dimensional confinement. In the Paul trap and related RF trap electrode geometries, the trap field can be purely oscillatory. For optical clocks based on ions with alkaline-earth-metal-type level systems, this is advantageous because the substantial linear quadrupole shift of the reference transition frequency that is caused by the gradient of the trap field is averaged out.

\begin{figure}[tb]
    \centering
    \includegraphics[width=.48\columnwidth]{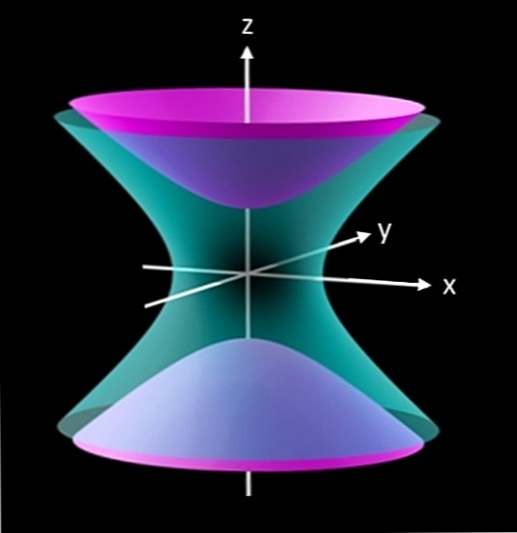}
    \caption{\label{fig:iontrap_hyperbolic} Truncated sketch of the equipotential surfaces defined by Equation~\eqref{eq:iontrap_rf} for the case $\alpha_x = \alpha_y = - \alpha_z/2 $. The purple and green surfaces correspond to the conditions $\Phi_\mathrm{rf}(t=0) < 0$ and $\Phi_\mathrm{rf}(t=0) > 0$, respectively.}
\end{figure}

Radio-frequency Paul traps use an arrangement of electrodes that supports an oscillating electric quadrupole potential of the form
\begin{equation}
\label{eq:iontrap_rf}
\Phi_\mathrm{rf}(\mathbf{r},t) = V_\mathrm{rf} \cos{\Omega t} \frac{(\alpha_x x^2 + \alpha_y y^2 + \alpha_z z^2)}{2d^2},
\end{equation}
with the condition $\alpha_x + \alpha_y + \alpha_z = 0 $ imposed by the Laplace equation. Here $V_\mathrm{rf}$ and $\Omega$ are the amplitude and frequency of the drive voltage applied between the trap electrodes, and $d$ is a geometrical scaling parameter. For an electrode system that is symmetric around the $z$ axis, $\alpha_x = \alpha_y = - \alpha_z/2 $.  With $\alpha_z=1$, Equation~\eqref{eq:iontrap_rf} describes the ideal case that $\Phi_\mathrm{rf}(\mathbf{r}, t)$ is generated by electrodes that have a distance $2d$ on the $z$ axis and whose shapes are determined by the corresponding hyperbolic equipotential surfaces of Equation~\eqref{eq:iontrap_rf}, as shown in Figure~\ref{fig:iontrap_hyperbolic}.

Equation~\eqref{eq:iontrap_rf} implies that the oscillating trap field vanishes at the symmetry centre. Such a field-free nodal point also exists if the electrodes have simplified shapes and are not precisely symmetric, provided that the phase of the RF potential is strictly spatially uniform as described by the $V_\mathrm{rf}\cos{\Omega t}$ prefactor in Equation~\eqref{eq:iontrap_rf}. In this case and near the nodal point, also simplified electrode geometries are well described by Equation~\eqref{eq:iontrap_rf} by setting $\alpha_z = \eta < 1$. Here $\eta$ can be considered as an efficiency factor that lies in the range $0.2 \lesssim \eta \lesssim 0.9 $ for commonly used trap geometries.

The equation of motion of an ion with charge-to-mass ratio $e/m$ in an axisymmetric quadrupole RF field has bounded solutions if the axial stability parameter,
\begin{equation}
\label{eq:iontrap_qz}
q_z = \frac{2 \eta V_\mathrm{rf} e}{m d^2 \Omega^2},
\end{equation}
lies in the range $0 \leq q_z \lesssim 0.92$~\citep{Paul1990}. For $q_z \geq 0.55$, imperfections of the potential can lead to unstable operation at certain values of $q_z$~\citep{Alheit1995}. The motion in the RF potential can be approximately described as a free ``secular'' oscillation in a time-independent harmonic potential like Equation~\eqref{eq:iontrap_potential}. Excursions from the potential minimum are accompanied by motion at a smaller amplitude that is driven by the inhomogeneous RF field. The average kinetic energy of this intrinsic micromotion is equal to that of the secular oscillation.

Neglecting terms of higher order in $q_z$, the secular oscillation frequencies are given by $\omega_z \approx q_z \Omega/(2\sqrt{2})$ and $\omega_x = \omega_y = \omega_z/2$. The secular oscillation frequencies can be modified by superposing a DC quadrupole field on the RF trap field. For example, for a positively charged ion the addition of a negative (positive) DC voltage to $V_\mathrm{rf} \cos{\Omega t}$, see Equation~\eqref{eq:iontrap_rf}, can decrease $\omega_z$ ($\omega_x$ and $\omega_y$) to the point where the axial (radial) trap potential vanishes. In a similar way, additional DC-biased electrodes can be used to lift the degeneracy of $\omega_x$ and $\omega_y$.

\subsubsection{Doppler cooling and Lamb--Dicke condition
} \label{sec:iontrap_cooling}

The secular oscillation of a trapped ion along the propagation direction of the clock-laser beam leads to an oscillating linear Doppler shift that can prevent coherent excitation of the ion. To suppress effects related to the linear Doppler shift, the secular oscillation amplitude must be reduced to $a_{i0} < \lambda/2\pi$, where $\lambda$ is the laser wavelength. In a quantum description, this corresponds to the Lamb--Dicke condition~\citep{Leibfried2003}
\begin{equation}
\label{eq:iontrap_Dicke}
\left[k_i a_i(n_i=0)\right]^2(2\bar{n}_i+1)\ll 1,
\end{equation}
where $k_i$ denotes the projection of the clock-laser $k$ vector on the principal trap axis $i$. For a given residual motional energy of the trapped ion, Equation~\eqref{eq:iontrap_Dicke} yields a condition on the secular frequencies because $a_i(n_i=0)$ and $\bar{n}_i$ depend on $\omega_i$.

If the secular frequencies $\omega_i$ are non-degenerate, three-dimensional Doppler cooling of a trapped ion is possible with a single cooling laser beam whose $k$ vector has non-zero projections on all principal axes of the trap potential. The lowest achievable vibrational quantum number is $\bar{n}_i \sim C\Gamma / (2\omega_i)$ where $\Gamma$ denotes the natural linewidth of the cooling transition and $C$ is a constant of order unity that depends on the cooling geometry. With $\bar{n}_i \gg 1$ and $\theta_i$ denoting the angle between the clock-laser $k$ vector and the principal axis $i$, Equation~\eqref{eq:iontrap_Dicke} leads to
\begin{equation}
\label{eq:iontrap_secular_Dicke}
\omega_i^2 \gg C(2\pi \cos{\theta_i/\lambda})^2\hbar \Gamma/(2m)
\end{equation}
as the Lamb--Dicke condition on the secular frequencies of an ion with clock transition wavelength~$\lambda$. For the alkaline-earth-metal-like ions considered here, Equation~\eqref{eq:iontrap_secular_Dicke} indicates minimum axial secular frequencies in the range $0.5\;\mathrm{MHz} \lesssim \omega_z/2\pi \lesssim 5$\;MHz, with the highest demands set by Hg$^+$ due to the comparatively short wavelength (282\;nm) and the large natural linewidth of the cooling transition (120\;MHz).

Equation~\eqref{eq:iontrap_secular_Dicke} has some direct implications on the design and operating parameters of traps for single-ion clocks. Assuming a target value $\omega_z/2\pi = 2$\;MHz together with a stability parameter $q_z = 0.5$ leads to a trap drive frequency of $\Omega/2\pi \approx 11$\;MHz. For an atomic mass of 100\;amu, the drive-voltage amplitude of a trap with efficiency factor $\eta = 1$ and axial electrode separation $2d = 1$\,mm is then calculated as $V_\mathrm{rf} \approx 330$\;V. The resulting axial potential depth is $\Phi(0, 0, d) = e^2 V_\mathrm{rf}^2/(4 m \Omega^2 d^2) \approx 20$\;eV.

\subsubsection{Excess micromotion
} \label{sec:iontrap_micromotion}

For an RF trap potential of the form of Equation~\eqref{eq:iontrap_rf} and $\Omega \gg \omega_z$, the amplitude of intrinsic micromotion is much smaller than that of the secular oscillation. However, in experiments involving laser-cooled ions in RF traps one commonly observes excess micromotion (EMM) that persists in the limit of vanishing secular oscillation amplitude. EMM can be caused by static electric stray fields that displace a trapped ion from the RF node. Another source of EMM is a nonuniform phase of the RF trap potential due to asymmetric phase delays of the RF voltages present on the trap electrodes. The two types of EMM can be distinguished through the phase shift of the motion relative to trap drive voltage~\citep{Berkeland1998, Keller2015}.

To suppress stray-field-induced EMM, adjustable DC voltages are usually applied between the trap electrodes and suitably placed compensation electrodes. The suppression of EMM caused by phase delays is more difficult because it requires adjustments in the layout of RF-carrying conductors or the application of RF voltages with well-defined phase shifts relative to the trap drive voltage.

\subsubsection{Trap-induced shifts of the clock transition frequency} \label{sec:iontrap_shifts}

\emph{Shifts related to ion motion}: For the ion species considered here, Doppler cooling can reduce the second-order Doppler (time-dilation) shift and the quadratic Stark shift due to residual interaction with the trap field to maximum relative magnitudes in the low $10^{-17}$ range \citep{Ludlow2015}. However, this requires sensitive three-dimensional detection of EMM and a stray-field compensation accuracy of  ${\lesssim} 10$ V/m \citep{Berkeland1998, Keller2015}. A much lower EMM sensitivity can be achieved if EMM-induced Doppler and Stark shifts cancel each other. This is possible if the scalar quadratic Stark shift of the atomic reference transition frequency is positive and the trap drive frequency $\Omega$ is set to a particular value. For the $^2$S$_{1/2} \rightarrow {}^2$D$_{5/2}$ reference transitions of Sr$^+$ and Ca$^+$, the required values of $\Omega$ are in an experimentally convenient range \citep{Dube2014,Cao2017}.

\vspace{0.3\baselineskip} \noindent
\emph{Quadrupole shift}: Atomic states with $J > 1/2$ can have a non-zero quadrupole moment that couples to the electric field gradient present at the trap centre.
According to Equation~\eqref{eq:iontrap_rf}, the RF trap generates electric field gradients that would cause a quadrupole shift with typical relative magnitude $10^{-13}$.  However, the shift oscillates in sign at the trap drive frequency and quickly averages to zero.  It is therefore only necessary to take account of \emph{static} electric field gradients at the ion, such as from stray electric fields and from the applied EMM compensation field.  Electric field gradients from these two effects typically lead to a static quadrupole shift of relative magnitude $10^{-15}$.
This shift can be cancelled through spectroscopic averaging schemes that also cancel the tensorial Stark shift \citep{Itano2000, Dube2005}. Like stray-field-induced EMM, also the stray-field-induced quadrupole shift can show significant variations, for example, after ion loading or illumination of the trap electrodes~\citep{Tamm2009}.

\vspace{0.3\baselineskip} \noindent
\emph{Trap-drive-induced Zeeman shift}: The capacitance between the trap electrodes gives rise to conduction and displacement currents and thus to an RF magnetic field whose amplitude is proportional to~$\Omega V_\mathrm{rf}$. Its spatial variation near the trap centre is determined by the electrode geometry and the current-density pattern in the electrodes, which is strongly influenced by the skin effect. Generally one expects that many types of symmetry imperfections can lead to a non-vanishing RF magnetic field at the nodal point of the electric trap field. For a trapped ion, the resulting oscillating Zeeman shift of the reference transition sublevels can introduce errors in EMM detection and in quadratic-Zeeman-shift measurements~\citep{Gan2018}.

\vspace{0.3\baselineskip} \noindent
\emph{Blackbody-radiation shift}: The electric field associated with the blackbody radiation (BBR) emitted by the trap structure and the inner surface of the vacuum chamber gives rise to a quadratic Stark shift of the reference transition frequency. If the thermal radiation is isotropic, the tensor contribution to the shift averages to zero. For the ion species considered here, the relative magnitude of the BBR shift is in the range of $10^{-16}$ to $10^{-15}$ at temperature $T = 300$ K and scales as~$T^4$. The components of the trap structure are subject to heating through resistive loss in RF conductors and dielectric loss in insulators exposed to the RF trap drive voltage. The BBR intensity at trap centre critically depends on the spatial arrangement of hot surfaces, their infrared emissivity and on the thermal conduction between components inside and outside the vacuum chamber. A thermal analysis of various trap designs indicates that a careful selection of materials and components can limit the effective temperature increase of the BBR incident on the trapped ion  to less than 1\;K relative to an external heat sink \citep{Dolezal2015}. See Section~\ref{sec:BBR_ions} for more details on characterising the BBR shift in ion clocks.

\subsection{Motional heating due to electrical field noise
} \label{sec:iontrap_fieldnoise}

A trapped ion is held in close proximity to the electrodes that create the trap potential. In the absence of laser cooling, the secular oscillation is not damped so that a resonant electric field originating from the electrodes can continuously raise the secular oscillation amplitude. The gain of secular oscillation energy is usually quantified through the motional heating rate $\dot{\bar{n}}_i$ that denotes the rate of increase of the corresponding harmonic-oscillator quantum number. For an electric field oriented parallel to the secular mode $i$, the heating rate is given by~\citep{Brownnutt2015}
\begin{equation}
\label{eq:iontrap_nbar}
    \dot{\bar{n}}_i \approx \frac{e^2}{4m\hbar \omega_i}S_E(\omega_i),
\end{equation}
where $S_E$ is the spectral density of the electric field noise at frequency $\omega_i$ (unit $(\mathrm{V/m})^2\,\mathrm{Hz}^{-1}$) .

Typically the largest contribution to $\dot{\bar{n}}_i$ originates from the field produced by the electrodes that are nearest to the trap centre. Noise superimposed on the trap drive voltage cannot efficiently excite the secular oscillation of an ion located close to the nodal point of the trap field. Parametric excitation through sidebands at $\Omega \pm 2 \omega_i$ is supported by the quadrupolar trap-electrode geometry but is usually well suppressed by the strong spectral filtering of the trap drive voltage (see below).

The kinetic energy of a trapped ion that is Doppler cooled to satisfy the Lamb--Dicke condition typically corresponds to a vibrational quantum number $\bar{n}_i \approx 10$. Unless limited by other effects, the maximum coherent interaction time between probe laser and ion is determined by the time when the acquired secular heating leads to a violation of the Lamb--Dicke condition. The increase in motional energy due to secular heating also leads to increased motion-induced shifts of the clock transition frequency. Therefore, it is usually desirable to reduce the secular heating rates along all trap axes in order to satisfy $\dot{\bar{n}}_i \tau \leq \bar{n}_i$, where $\tau$ is the probe pulse length.

\subsubsection{Sources of electrical field noise
} \label{sec:iontrap_sources_fieldnoise}

\emph{Thermal (Johnson) noise}: The thermal noise voltage $V_\mathrm{n}$ produced by an ideal resistance $R$ has a spectral density per unit bandwidth of $\langle V_\mathrm{n}^2 \rangle = 4 \kB TR$, where $\kB$ is the Boltzmann constant. The heating rate of an ion that is confined between two trap electrodes connected through a resistance~$R$ can be estimated using  Equation~\eqref{eq:iontrap_nbar} by setting $S_E = 4 \kB TR/d'^2$, where $d' \sim 2d$ is the effective distance between the electrodes. Using $d' = 1$\;mm, $T = 300$\;K, $m = 100$\;amu, $\omega_z/2\pi = 2$\;MHz, and $R = 1$\;k$\Omega$ as a resistance value commonly used in electronic setups, we obtain $S_E = 1.7 \times 10^{-11}\; (\mathrm{V/m})^2\,\mathrm{Hz}^{-1}$ and $\dot{\bar{n}}_z = 480\;\mathrm{s}^{-1}$. For optical clocks, such a heating rate is a serious impediment. A heating rate of this magnitude is also expected if resistors in the k$\Omega$ range are used in the input circuit of a broadband low-noise amplifier whose output is directly connected to a trap electrode. Low-pass filter networks can efficiently suppress the thermal noise of DC voltages applied to the trap electrodes. In this way, the equivalent noise resistance between the electrodes in the secular-frequency range can be reduced to well below $1\;\Omega$ so that it is comparable to the expected resistance of the electrodes and attached conductors.

\vspace{0.3\baselineskip} \noindent
\emph{Electromagnetic interference (EMI) and pickup}: Many electronic devices such as switched-mode power supplies internally generate large AC currents with harmonics extending to the RF range. Consequently, non-negligible electric and magnetic fields in the secular-frequency range can be emitted, for example, through connections to control inputs or the mains supply. The frequency spectrum of this interference typically contains multiple components at fluctuating frequencies. EMI-induced voltage differences between trap electrodes are an efficient source of motional heating even if there is only a transient resonance at a secular frequency. EMI transmitted through stray capacitances can be suppressed by conductive shielding or by providing a low AC impedance between the points where the EMI voltage appears. The magnetic part of EMI can be suppressed by minimizing the area of exposed conductor loops. This also applies to loops formed by ground connections between devices in the setup.

\vspace{0.3\baselineskip} \noindent
\emph{Anomalous heating by surface noise}: In most experiments involving single ions or Coulomb crystals in ion traps, one observes secular heating rates many orders of magnitude larger than that expected from external fields and Johnson noise. This so-called anomalous heating is attributed to impurities that are adsorbed to the trap-electrode surfaces. In fact, monomolecular layers containing oxygen and carbon are known to form on metal surfaces exposed to ambient air in less than 1\;h \citep{Taborelli2017}. The chemical structure of the adsorbates that cause anomalous heating is still not fully known.

Secular heating due to anomalous heating can be characterised through power laws that describe how the spectral density of the associated electric-field noise $S_E$ depends on frequency $\omega$ and distance $d$ of the ion from the surface
\begin{equation}
\label{eq:iontrap_fieldnoise}
S_E = S_0 \left( \frac{d_0}{d} \right) ^k \left( \frac{\omega_0}{\omega} \right) ^l ,
\end{equation}
where  $d_0$ and $\omega_0$ are normalisation constants. Observed strength factors $S_0$ vary by several orders of magnitude between apparently similar experimental setups and large changes can also occur, for example, after breaking the vacuum or cleaning procedures. Typically $S_0$ decreases strongly if a trap is operated at cryogenic temperature. In agreement with several theoretical models, most experimental observations indicate a distance scaling according to $3.5 \lesssim k \lesssim 4$ and frequency dependence according to $0.5 \lesssim l \lesssim 1.5$ in the usual secular-frequency range \citep{Brownnutt2015,Kumph2016,Boldin2018}.

In a number of investigations where high anomalous heating rates were observed and contamination of the trap electrodes with carbon compounds was detected, $S_E$ could be strongly reduced by \emph{in situ} cleaning with a rare-gas ion beam. Most of these experiments were conducted with electroplated gold electrodes in miniaturised planar traps~\citep{Hite2012}. For nonplanar traps, it seems more attractive to minimise $S_E$ by maximising the trap size and by identifying favourable electrode materials and chemical and mechanical surface-conditioning methods.

Low electric-field-noise levels corresponding to anomalous heating rates $\dot{\bar{n}}_z \lesssim 10\;\mathrm{s}^{-1}$ in mm-sized traps have been reported for gold-plated \citep{Keller2019, Poulsen2012} and molybdenum trap electrodes \citep{Diedrich1989, Turchette2000, Roos1999}. These electrode materials will be considered in more detail in Section~\ref{sec:iontrap_materials}. It may be noted, however, that comparably low heating rates have also been obtained with other electrode materials, including stainless steel \citep{Rohde2001} and evaporated aluminium and copper layers \citep{Daniilidis2014}.

\section{Trap design
} \label{sec:iontrap_design}

\subsection{Trap geometry and size
} \label{sec:iontrap_geometry}
Trap electrodes whose shape closely approximates the equipotential surfaces shown in Figure~\ref{fig:iontrap_hyperbolic} are not useful for single-ion clocks because several cooling and probe laser beams must pass through the trap centre without large geometrical constraints. Moreover, fluorescence detection typically requires a free aperture angle of at least \ang{20}.

\begin{figure}[tb]
    \centering
    \includegraphics[width=1.0 \columnwidth]{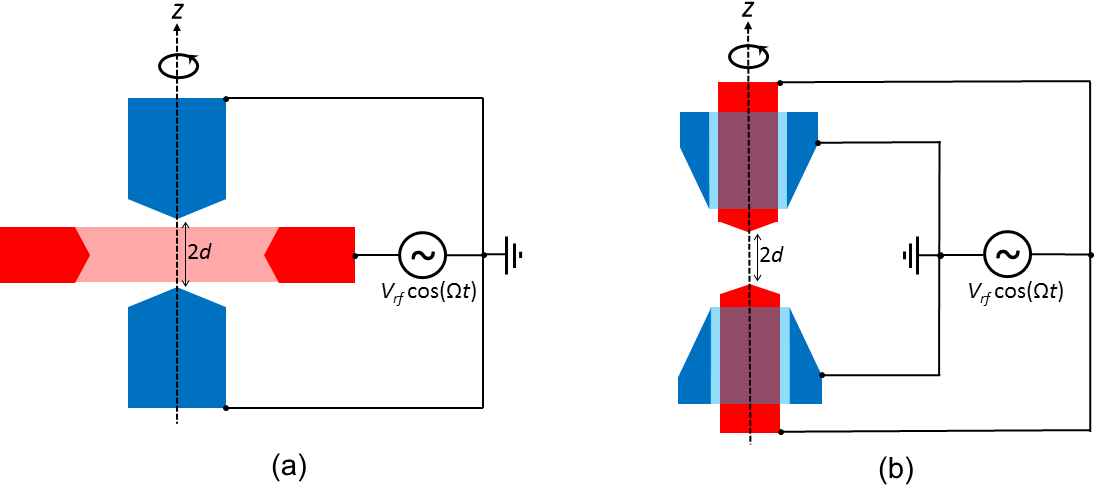}
    \caption{\label{fig:iontrap_electrodes} Simplified electrode geometries and driving schemes of axially symmetric RF quadrupole traps: (a) conical ring and cap electrodes, (b) endcap trap with concentric outer electrodes. The electrode shapes shown here are not optimised. }
\end{figure}

Two simplified RF quadrupole electrode geometries that offer sufficient optical access are shown in Figure~\ref{fig:iontrap_electrodes}. The geometrical reduction leads to reduced efficiency factors $\eta < 1$ and to anharmonic contributions to the trap potential. The geometries shown in Figure~\ref{fig:iontrap_electrodes} have enough degrees of freedom to minimise the anharmonicity by suppressing the third- and fourth-order terms in the spherical-harmonics expansion of the potential. Ring electrode geometries, see Figure~\ref{fig:iontrap_electrodes}(a), optimised with this constraint can achieve efficiency factors $\eta \approx 0.9$~\citep{Beaty1987}.  For an endcap trap with grounded outer electrodes, Figure~\ref{fig:iontrap_electrodes}(b), optimised in the same way, an efficiency factor $\eta \approx 0.4$ was found~\citep{Stein2010}. A distinctive feature of endcap trap geometries is that the trap drive voltage is applied to two spatially separate electrodes. In order to avoid EMM due to unbalanced phase shifts, any asymmetry in the connections of the electrodes to the drive voltage must be avoided. The direct connection of both electrodes to $V_\mathrm{rf}$ ensures that the equivalent noise resistance between them is negligible. Voltage differences that appear between the outer electrodes have a much smaller effect on the ion because the endcap electrodes act as a shield.

\begin{figure}[tb]
    \centering
    \includegraphics[width=1\columnwidth]{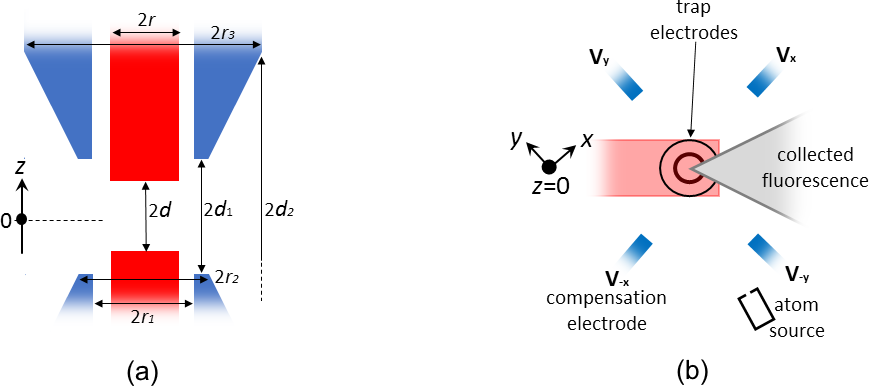}
    \caption{\label{fig:iontrap_dimensions} Endcap trap electrodes with planar endfaces (a), with dimensioning parameters discussed in the text. The shown electrode shapes are not optimised. (b): suggested arrangement of radial compensation electrodes and atom source for the trap geometry shown in (a).}
\end{figure}

A further simplified electrode geometry with planar endcap faces is shown in Figure~\ref{fig:iontrap_dimensions}(a). This design permits a higher machining accuracy and facilitates mechanical polishing of the endcap faces. The introduction of a small plane segment $r_2-r_1$ at the end of the outer electrodes supports high machining accuracy for the critical distance $d_1-d$. With this electrode geometry, only the lowest-order anharmonic term in the trap potential can be suppressed by suitable choice of $r/d$. This is sufficient to reduce the fractional anharmonicity of the trap potential to a few per cent in the region $-0.5 d \leq z \leq  0.5 d$. The resulting efficiency factor is around $\eta \approx 0.6$.

Due to the strong dependence of the anomalous secular heating rate on the distance of the ion from the electrode surfaces, it is desirable to scale traps for single-ion clocks to the largest possible size $d$. Equation~\eqref{eq:iontrap_qz} implies that fixing the parameters $\omega_z$, $q_z$, $\Omega$ and $\eta$ also determines the ratio~$V_\mathrm{rf}/d^2$. Consequently, scaling to a larger trap size requires a quadratic increase of $V_\mathrm{rf}$ .

In current single-ion clock setups, trap drive voltages are typically limited to $V_\mathrm{rf} \lesssim 1000$\;V, which limits the trap size to $d \approx 0.5$ mm. In many cases the $V_\mathrm{rf}$ limit arises either from the limited power-handling capability of the helical-resonator circuit that generates the trap drive voltage or from dielectric loss in the insulators exposed to it. Some of the construction features and components described in the following and in Section~\ref{sec:iontrap_OC18traps}  can be instrumental for upscaled trap designs that require drive voltages above 1000\;V.

\subsection{Electrode and insulator materials
} \label{sec:iontrap_materials}

\emph{Molybdenum electrodes}: As noted above, molybdenum (Mo) trap electrodes appear to be an advan\-ta\-geous choice for single-ion clocks because very low anomalous heating rates were observed in a number of experiments. This feature might be related to the low chemical reactivity of Mo at room temperature which is comparable to noble metals. Another useful property of Mo is the high thermal and electrical conductivity that is only a factor of three smaller than that of copper and silver. The hardness of Mo is comparable to iron so that similar machining and tooling recommendations apply.

It is useful to polish the trap electrode surfaces nearest to the ion because surface irregularities caused by machining can complicate cleaning and promote the adsorption of contaminants. The hardness of Mo facilitates efficient mechanical polishing. For Mo electrodes, a microroughness of approximately 20\;nm can be achieved~\citep{Nisbet-Jones2016}.

\vspace{0.3\baselineskip} \noindent
\emph{Gold-coated electrodes}: Stainless steel or Mo electrodes coated by a gold (Au) layer are another favourable option for ion traps with low anomalous heating rates. The thickness of the deposited Au layer should be not much smaller than the skin depth at the trap drive frequency (\SI{20}{\micro\meter} at $\Omega/2\pi = 15$\;MHz) so that thickness variations do not strongly affect the surface current distribution. Layers of this thickness are usually deposited by electroplating. For trap electrodes, this poses a risk because impurities transferred from the electrolytic bath can contribute to anomalous heating. Therefore, deposition by evaporation or sputtering is preferable even if much longer processing times are required.

\vspace{0.3\baselineskip} \noindent
\emph{Insulators with low dielectric loss}: Typically, the heating of insulators due to intrinsic dielectric loss is the main source of excessive blackbody radiation in RF ion traps~\citep{Dolezal2015}. Insulating elements made from fused silica (quartz glass), alumina ceramic, sapphire, or CVD (chemical-vapor-deposition) diamond have favourably low dielectric losses. With these materials, one expects that the dissipated thermal power can be reduced to negligible levels provided that the electric field strength in the material is restricted by the geometrical design of the insulator and attached conductors. It should be noted, however, that the dielectric losses of fused silica and alumina can vary widely depending on the purity of the material. Fused silica has the advantage of low relative permittivity ($\epsilon_\mathrm{r} = 3.8$) and can be readily obtained in the form of optical substrates with specified spectral transmittance, which is indicative of the impurity level.

\subsection{Compensation electrodes and atom source
} \label{sec:iontrap_atom_source}

To cancel stray-field-induced EMM, an electric field must be generated that compensates the three spatial components of the stray field at the trap centre. In an endcap trap geometry as shown in Figure~\ref{fig:iontrap_electrodes}(b), the axial stray-field component can be easily compensated by applying antisymmetric DC voltages $V_z$ and $V_{-z} = - V_z$ to the outer trap electrodes.

In order to cancel the radial stray-field components, it is useful to arrange four compensation electrodes in the symmetry plane of the trap as shown in Figure~\ref{fig:iontrap_dimensions}(b). In practice, the shape of the compensation electrodes is not critical and they can be made from wires bent towards the centre in the symmetry plane of the trap. If the distance of the wire ends from the trap centre is around $5d$, the voltages required for stray-field compensation are expected to be around $\pm 50$\;V.

For appropriate biasing of the radial compensation electrodes, it is advantageous to provide separate adjustment modes for symmetric and antisymmetric voltage variations. Antisymmetric biasing of the electrodes according to $V_{-x} = -V_x$ and $V_{-y} = - V_y$ compensates the radial stray field in the $x$ and $y$ directions. Since the field produced by the compensation electrodes is inhomogeneous, also the static field gradient at the trap centre is altered by the applied voltages. Symmetric biasing according to $V_x = V_{-x}$ and $V_y = V_{-y}$ makes it possible to rotate the principal axes of the trap potential in the radial plane without first-order effect on the EMM cancellation. This adjustment may be useful in order to ensure that the $k$ vector of the cooling laser beam has non-vanishing projections on both radial principal axes of the trap potential, as required for three-dimensional Doppler cooling.

Ions are generated at the trap centre by photoionisation of atoms produced by a source (for example, a small atomic oven) that is remote from the trap. As indicated in Figure~\ref{fig:iontrap_dimensions}(b), it is advantageous to also place the orifice of the source in the symmetry plane of the trap because this minimises the exposure of the trap-electrode end faces to the emitted atomic beam. During trap operation, the source should be either grounded or biased together with the nearest compensation electrode.

\subsection{Biasing and filter circuitry
} \label{sec:iontrap_circuits}

\begin{figure}[tb]
    \centering
    \includegraphics[width=0.9 \columnwidth]{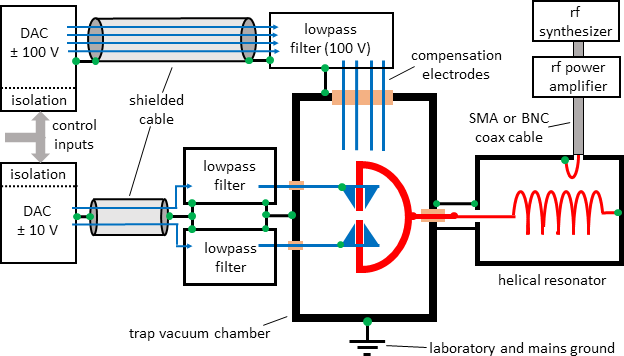}
    \caption{\label{fig:iontrap_drivevoltage} Sketch of the electrical connections of trap and compensation electrodes to external components and devices. DAC: digital-to-analog converters determining the DC voltages applied to the electrodes. The green dots denote ground connections between the components that should be realised with negligible resistance and inductance. The grounding scheme shown here has no loops if additional connections between devices (for example by shared power supplies or the mains connection) are avoided. }
\end{figure}

Figure~\ref{fig:iontrap_drivevoltage} schematically shows the essential peripheral components and devices that are required for trap operation and their connection to the trap electrodes. The trap drive voltage applied to the endcap electrodes is provided by a helical resonator that acts as a step-up transformer for the RF-amplifier output at frequency $\Omega$. The resonance quality factor $Q = \Omega/\delta \Omega$, where $\delta \Omega$ is the width between the half-power points of the resonance, of a helical resonator is mainly determined by the RF resistance of the copper wire or rod of the resonator coil, which is subject to the skin effect. At $\Omega/2\pi \approx 25$\;MHz and with impedance-matched coupling, a quality factor $Q = 640$ can be achieved using a rod diameter of 6\;mm \citep{Pyka2013}. The $Q$ value is not significantly reduced if insulators with low dielectric loss (see Section~\ref{sec:iontrap_materials}) are used in the vacuum feedthrough of the trap drive voltage and between the trap electrodes. On the other hand, transmission of the helical-resonator output even through a short length of cable strongly reduces the $Q$ value and the resonance frequency. Consequently, the helical resonator must be located close to the trap with a low-capacitance shielding of its output that connects resonator enclosure and vacuum chamber (see Figure~\ref{fig:iontrap_drivevoltage}). The short conductor length between vacuum feedthrough and resonator coil also permits efficient thermal coupling to the trap which is advantageous if its temperature is to be stabilised.

Due to stray capacitances, non-negligible RF currents at the trap drive frequency are picked up by the compensation and outer trap electrodes. Since these electrodes are to be kept at pure DC potentials, the RF currents must be absorbed through capacitors that provide a negligible impedance between the electrodes and the vacuum chamber. Another requirement that applies in particular to the outer trap electrodes is that EMI-induced and Johnson-noise voltages in the range of the secular frequencies must be suppressed to a high degree. This can be achieved by low-pass filter circuits that are mounted close to the feedthroughs of the vacuum chamber as indicated in Figure~\ref{fig:iontrap_drivevoltage}.

\begin{figure}[b]
    \centering
    \includegraphics[width=0.7 \columnwidth]{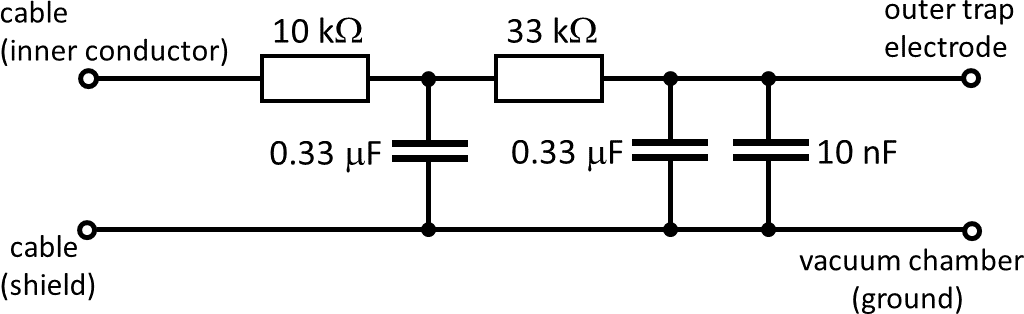}
    \caption{\label{fig:iontrap_filter} Schematic of a low-pass filter that provides low RF impedance and equivalent noise resistance for the outer trap electrodes and strongly suppresses noise superimposed on the control voltages. Best performance is achieved with the use of SMD multilayer ceramic capacitors (MLCC) because their equivalent series resistance is typically well below \SI{0.1}{\ohm}.}
\end{figure}

The schematic of a low-pass filter suitable for the outer trap electrodes is shown in Figure~\ref{fig:iontrap_filter}. The cascading of two RC circuits yields an AC-voltage attenuation factor \num{\geq 3e8} (corresponding to 170-dB power attenuation) at frequencies \SI{\geq 0.5}{MHz}. The parallel connection of two capacitors at the output improves the attenuation at frequencies \SI{\geq 10}{MHz} where it is dominated by the effective series resistance of the capacitors. The 3-dB corner frequency of the filter shown in Figure~\ref{fig:iontrap_filter} is 15\;Hz, and above 50\;Hz the power attenuation increases by 40\;dB per decade. The scheme shown in Figure~\ref{fig:iontrap_filter} is also suitable for the compensation electrodes. Here a reduced capacitance value around \SI{0.1}{\micro\farad} may be sufficient, which leads to a proportional increase in corner frequency.  As an alternative to the filter shown in Figure~\ref{fig:iontrap_filter}, LCR Pi filters can be employed to obtain a steep resonant increase in attenuation at a defined frequency \citep{Allcock2011}.

\section{Traps developed in OC18
} \label{sec:iontrap_OC18traps}

As part of the OC18 project, single-ion traps were developed in four separate institutes: CMI, NPL, PTB and VTT MIKES.  In each case, the endcap electrodes were made from molybdenum and their geometry was based on the design shown in Figure~\ref{fig:iontrap_dimensions}(a).  The way the electrodes were mounted, however, showed a greater variety.  At NPL and VTT MIKES, the electrode mounting closely followed that described in~\citet{Nisbet-Jones2016}.  At PTB, the design was modified to enable more robust alignment of the electrodes, while at CMI greater consideration was given to minimising the heating in the trap structure. More details of these modified designs are given in the following Sections~\ref{sec:iontrap_PTBtrap} and \ref{sec:iontrap_CMItrap}.

\subsection{Endcap trap with self-aligned electrodes (PTB)
} \label{sec:iontrap_PTBtrap}

For investigations that involve cleaning or coating of the trap electrodes, a trap design is desirable that permits reproducible repositioning of the electrodes without manual alignment. To achieve this, the positions of the electrodes must be uniquely determined by their mechanical contact with the trap mount and insulation elements, and a machining accuracy around 0.01\;mm must be maintained for all critical dimensions. The mechanical design must be compatible with vacuum baking at a temperature up to \SI{250}{\degreeCelsius}.

\begin{figure}[tb]
    \centering
    \includegraphics[width=1.0 \columnwidth]{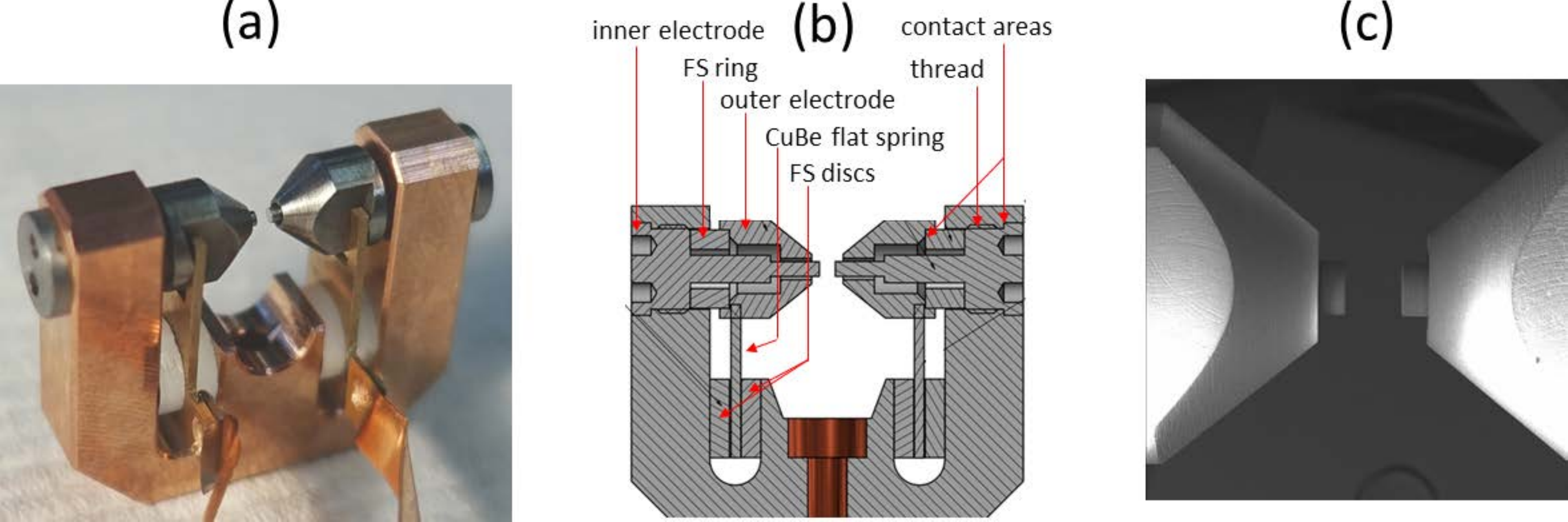}
    \caption{\label{fig:iontrap_PTBtrap} (a) Endcap trap with molybdenum electrodes developed at PTB. The size (width) of the trap mount is approximately 20\;mm. (b) Cross-section drawing showing the arrangement of electrodes and insulators. FS: fused silica. (c) SEM image of the assembled trap electrodes. The trap is mounted to a vacuum flange that contains the feedthroughs for trap drive voltage and outer trap electrodes. Radial compensation electrodes and Yb source (not shown) are attached to a separate flange.}
\end{figure}

The trap shown in Figure~\ref{fig:iontrap_PTBtrap} satisfies these conditions. Its electrodes have the shape shown in Figure~\ref{fig:iontrap_dimensions}(a) and are machined from Mo with mechanically polished inner-electrode end faces. With the dimensions $2r = 0.8$ mm, $2r_1 = 1.2$ mm, $2r_2 = 1.6$ mm, $2r_3 = 3.6$ mm and $2d = 0.86$ mm, $2d_1 = 1.26$ mm, $2d_2 = 3.22$ mm, the lowest-order anharmonic component of the trap potential is suppressed and the efficiency factor is $\eta =0.66$. The trap electrodes are held in an arc-shaped mount made from a CuCrZr alloy that is considerably harder than pure copper (Cu) and can be machined to a high precision. The mount provides a symmetric connection of the inner electrodes to the trap drive voltage. The trap is held in the vacuum chamber by attaching the mount to the massive Cu conductor (diameter 6\;mm) of a high-power vacuum feedthrough.

As shown in Figure~\ref{fig:iontrap_PTBtrap}(b), the position of the inner trap electrodes is defined by a thread and a well-defined contact area with the trap mount. The position of the outer electrodes relative to the inner electrodes is defined by insulating spacer rings cut from UV-grade-fused-silica optical substrates (Edmund Optics \#{}47-833). The low parallelism error (5\;arcsec) of the original substrates is essential for reaching the required centring accuracy of the end faces of the inner and outer trap electrodes. The outer trap electrodes are clamped down on the spacer rings by flat springs made from a CuBe alloy with high yield strength. The end of each spring is held between two unmodified fused-silica substrates. Copper strips spot-welded to the springs provide electrical connection of the outer trap electrodes to vacuum feedthroughs. Figure~\ref{fig:iontrap_PTBtrap}(c) shows an SEM (scanning-electron-microscope) image of the assembled trap electrodes.

\begin{figure}[tb]
    \centering
    \includegraphics[width=0.88 \columnwidth]{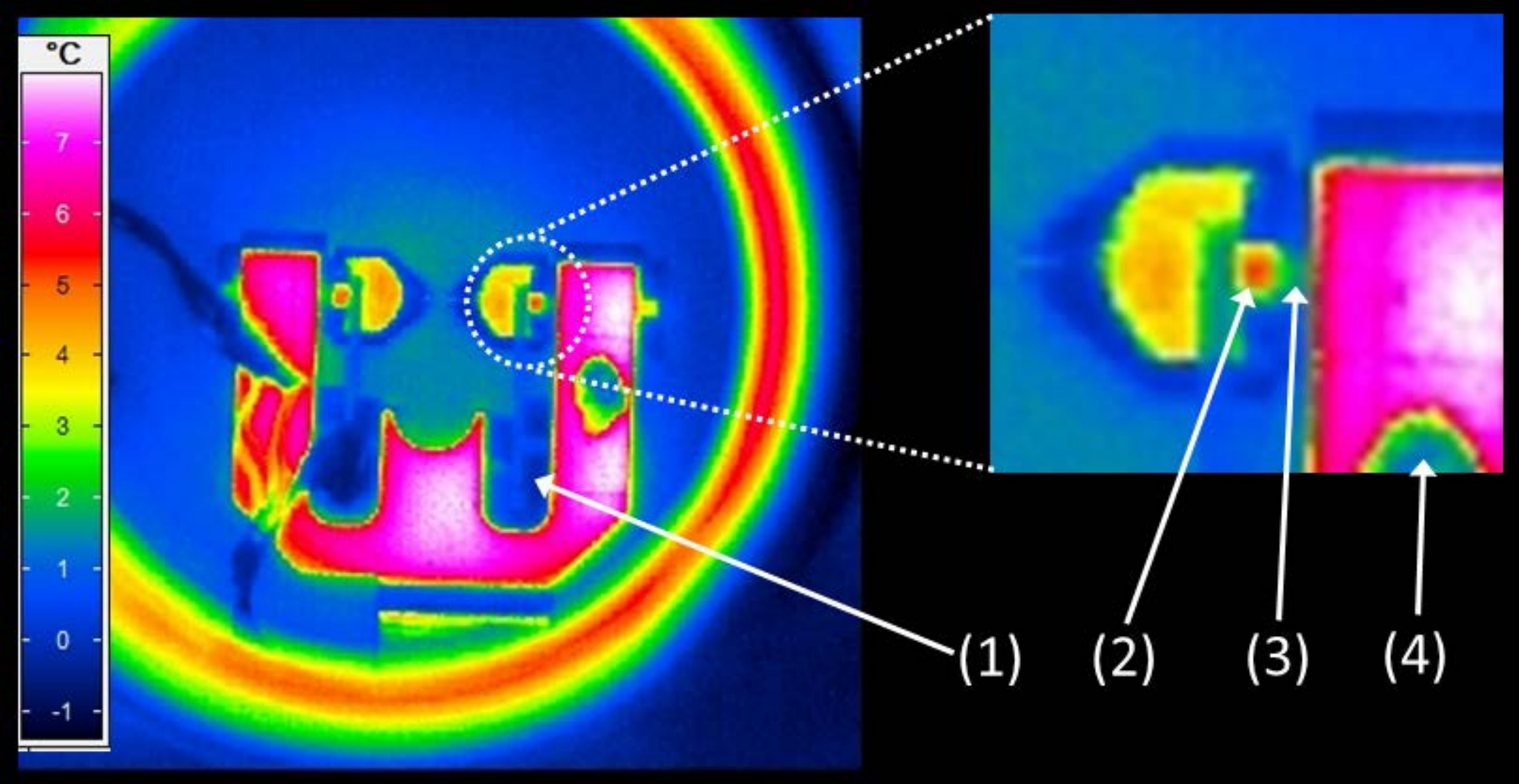}
    \caption{\label{fig:iontrap_PTBthermal} Thermal image (see text) of the trap mount and electrodes shown in Figure~\ref{fig:iontrap_PTBtrap}. The encircled section is shown enlarged on the right. (1): location of the fused-silica insulators holding the AC-grounded flat spring in the trap mount. (2): ambient radiation reflected from the Mo electrode. (3): location of fused-silica spacer ring. (4): spot of high-emissivity paint indicating the trap mount temperature.}
\end{figure}

The heating of the trap electrodes due to dielectric loss in the fused-silica spacers was investigated by operating the trap at $V_\mathrm{rf} \approx 1000$\;V in a vacuum of $10^{-3}$\;mbar. Thermal radiation in the wavelength range near the spectral maximum of room-temperature blackbody radiation ($\lambda \approx \SI{10}{\micro\meter}$) was observed with a bolometric camera through a ZnSe window. A typical camera image is shown in Figure~\ref{fig:iontrap_PTBthermal}. The thermal radiation that is registered from the largest part of the trap mount and from the outer trap electrodes is not related to the blackbody emission of these parts but originates from partly diffuse reflection of ambient heat sources. Despite this, Figure~\ref{fig:iontrap_PTBthermal} indicates a negligible temperature rise of the fused-silica spacers that are used to fix the outer trap electrodes. The temperature of the spacers between outer and inner trap electrodes is difficult to determine because their emission is partly masked by reflections from the outer trap electrodes. It seems likely that their temperature rise relative to the trap mount is smaller than \SI{0.5}{\degreeCelsius}.

\begin{figure}[tb]
    \centering
    \includegraphics[width=0.6 \columnwidth]{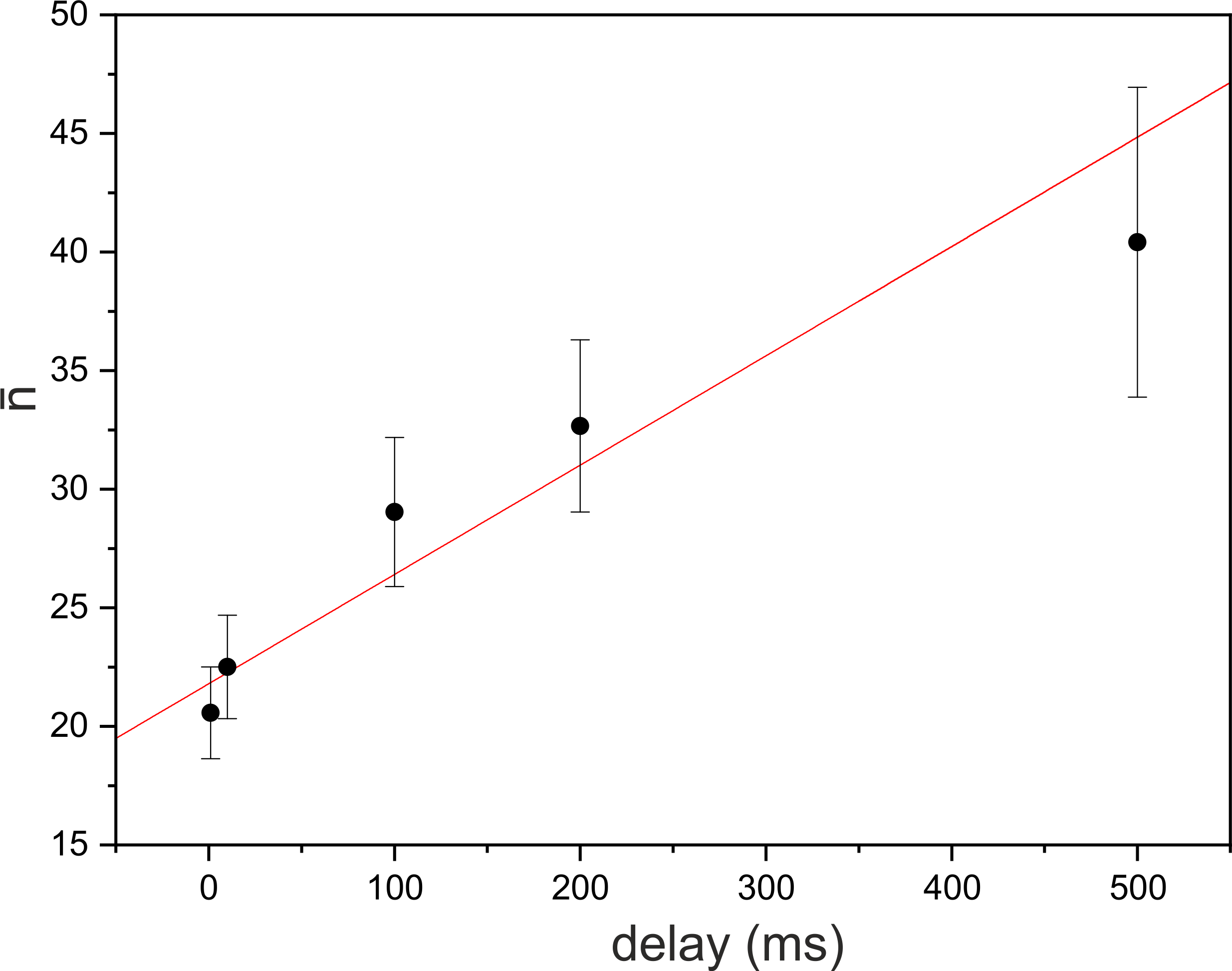}
    \caption{\label{fig:iontrap_heatingrate} The data points show the measured effective secular-oscillation quantum number for various delay times between the end of laser cooling and the start of the excitation of the $\lambda = 436$ nm reference transition of a trapped $^{171}$Yb$^+$ ion (see text) for the trap shown in Figure~\ref{fig:iontrap_PTBtrap}. The error bars denote statistical measurement uncertainty. The red line is a linear fit to the data points.}
\end{figure}

The anomalous heating rate of the secular motion of a trapped ion was determined by observation of Rabi oscillations on the $\lambda = 436$\;nm reference transition of a Doppler-cooled $^{171}$Yb$^+$ ion. The delay time between the end of the cooling and the start of the Rabi oscillations was varied, and the increase in the damping rate of the Rabi oscillations resulting from heating-induced dephasing was evaluated~\citep{Letchumanan2004}. The measurement was performed under conditions where the axial secular freqency was $\omega_z/2\pi \approx 1.2$\;MHz and radial and axial secular modes contributed approximately equally to the
 observed heating rate. The result of this measurement is shown in Figure~\ref{fig:iontrap_heatingrate} and indicates an effective heating rate $\dot{\bar{n}}= 46(11)\;\mathrm{s}^{-1}$. If the motional energy gain is distributed equally between the three secular oscillation modes, $\dot{\bar{n}} \approx \dot{\bar{n}}_x/\sqrt{2} \approx \dot{\bar{n}}_y/\sqrt{2} \approx 2 \dot{\bar{n}}_z/\sqrt{2}$. In a separate investigation of a similarly sized trap with Mo electrodes at NPL, a comparable heating rate magnitude was observed \citep{Nisbet-Jones2016}.

\subsection{Endcap trap with minimised blackbody radiation (CMI)
} \label{sec:iontrap_CMItrap}

The main objective in the design of this trap was to minimise additional sources of blackbody radiation.  Additional design considerations were (i) to obtain a very wide optical access for laser beams and fluorescence detection, (ii) to use a single vacuum feedthrough for all electrical connections and (iii) to minimise distortions of the trap field imperfections caused by the connections of the trap electrodes to the trap drive voltage.

\begin{figure}[b]
    \centering
    \includegraphics[width=1.0 \columnwidth]{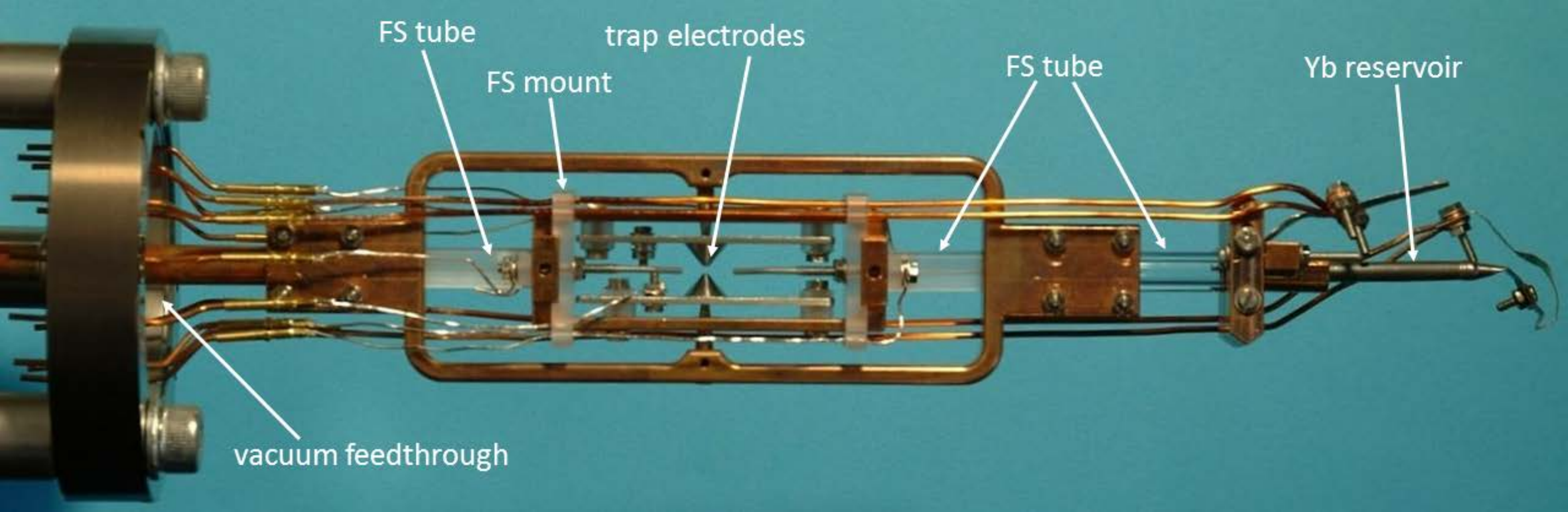}
    \caption{\label{fig:iontrap_CMItrap} Endcap trap with molybdenum electrodes developed at CMI.}
\end{figure}
The trap and its mounting on a custom-made vacuum feedthrough are shown in Figure~\ref{fig:iontrap_CMItrap}. The setup can be placed between two vacuum-chamber windows that are separated by only 20\;mm, which is favourable for obtaining a large optical access angle. The trap drive voltage is applied to the massive central copper conductor (6 mm diameter) of the vacuum feedtrough. On the vacuum side, this conductor is attached to a rectangular frame that is machined from Cu. The frame holds the inner trap electrodes. The outer trap electrodes are supported by an interior frame (also Cu) that is kept at ground potential. In this frame, the outer trap and compensation electrodes are mounted on fused-silica spacers. The inner frame is held in the outer frame by fused-silica tubes. Only these insulators are exposed to the high electric field caused by the trap drive voltage, and their geometry minimises dielectric loss and heating. In addition, there is no direct line of sight between the fused-silica tubes and trap centre so that a trapped ion is well shielded from their blackbody radiation.

\begin{figure}[tb]
    \centering
    \includegraphics[width=0.85 \columnwidth]{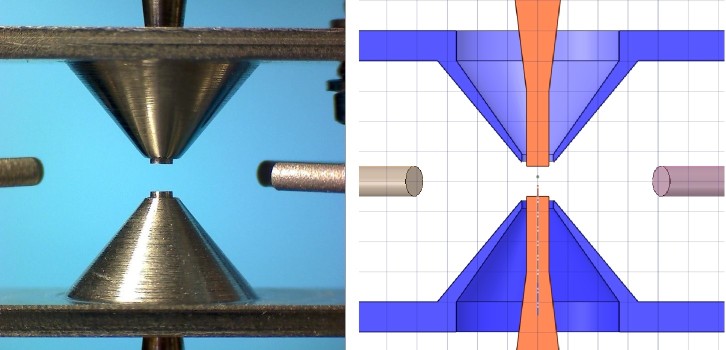}
    \caption{\label{fig:iontrap_electrodesurface} Electrode configuration of the trap shown in Figure~\ref{fig:iontrap_CMItrap}. The cross-section drawing on the right shows the shape of the inner electrodes (red) inside the outer electrodes. This results in a low capacitance between inner and outer electrodes. }
\end{figure}

The trap electrode configuration is shown in Figure~\ref{fig:iontrap_electrodesurface}. The trap electrodes are machined from Mo and their geometry corresponds to the design described in \citet{Nisbet-Jones2016}, which yields an efficiency factor $\eta \approx 0.88$. The Yb atoms used for loading the trap are evaporated from a commercially available Yb dispenser (AlfaSource by \citet{AlfaVacuo}, on the far right in Figure~\ref{fig:iontrap_CMItrap}) and are passed to the trap through the attached fused-silica tube.

\begin{figure}[t]
    \centering
    \includegraphics[width=0.6 \columnwidth]{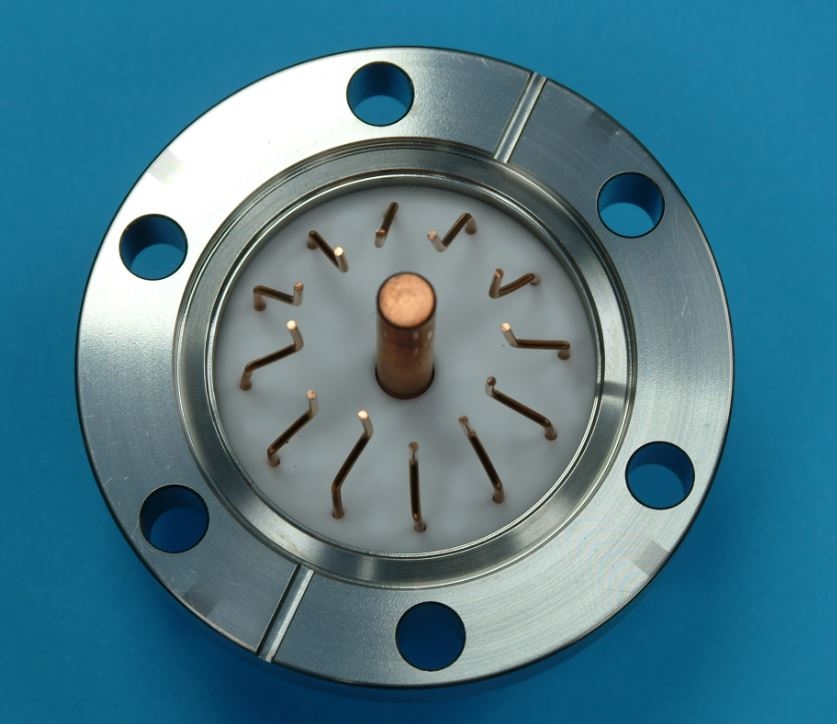}
    \caption{\label{fig:iontrap_feedthrough} Custom-made vacuum feedthrough used in the trap setup of Figure~\ref{fig:iontrap_CMItrap}. The central 6-mm conductor is the RF feedthrough; the surrounding pins are for the DC voltages and dispenser current. The insulator material (white) is alumina ceramic.}
\end{figure}

The employed custom-made vacuum feedthrough is shown in Figure~\ref{fig:iontrap_feedthrough}. Since it is based on a large alumina ceramic disc, the capacitance between the central conductor and the other conductors and the mounting flange is much smaller than that of standard vacuum feedthroughs. With this design also the dielectric load of the insulator material is strongly reduced so that heat dissipation is expected to be negligible for $V_\mathrm{rf} \lesssim 1000$\;V.

\begin{figure}[b!]
    \centering
    \includegraphics[width=0.6 \columnwidth]{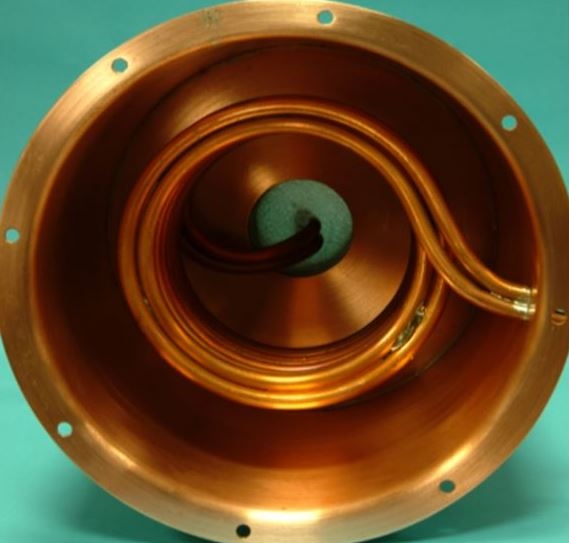}
    \caption{\label{fig:iontrap_CMIhelical} Helical resonator for the trap setup of Figure~\ref{fig:iontrap_CMItrap} with the shielding cap at the ground end of the bifilar resonator coil removed. The coil is made from copper tubes that are joined at the high-voltage end of the coil.  The cooling-water inlet and outlet are at the ground end of the coil. }
\end{figure}

The helical resonator that is attached to the central conductor of the vacuum feedthrough is shown in Figure~\ref{fig:iontrap_CMIhelical}. The design of the resonator addresses the problem that typically a power of up to several watts is dissipated in the resonator coil, which can lead to a substantial increase of the temperature of the trap electrodes and vacuum chamber. The double-helix (bifilar) coil of the resonator shown in Figure~\ref{fig:iontrap_CMIhelical} is made of Cu tubes so that cooling water can be circulated without resistive or dielectric loss. With active stabilisation of the cooling water temperature, it is possible to keep the trap electrodes at a well-defined temperature.

\section{Summary}
When designing single-ion traps for use in optical clocks, there are many issues that must be considered.  These include factors that affect how well the ion can be cooled and how the heating rate can be kept at a minimum.  Effects that lead to clock frequency shifts must also be minimised, including compensating stray electric fields and reducing blackbody radiation from the trap structure itself.  This chapter has gathered together different considerations and offered practical solutions that can be employed when designing and constructing single-ion traps for high-performance optical clocks.

\clearpage

\graphicspath{{D4_coherence/}}

\chapter{Understanding processes that limit coherence times in optical lattice clocks \label{chap:coherence}} 

\authorlist{%
Michal Zawada$^{1,\dag}$,
Ali Al-Masoudi$^2$,
William Bowden$^3$,
Roman Ciury\l{}o$^1$,
Hubert Cybulski$^1$,
S\"oren D\"orscher$^2$,
Stephan Falke$^{2,\ddag}$,
Ian R.\ Hill$^3$,
Richard Hobson$^3$,
Christian Lisdat$^2$,
Roman Schwarz$^2$,
Uwe Sterr$^2$ and
Alvise Vianello$^3$
}

\affil{1}{\UMKaff}
\affil{2}{\PTBaff}
\affil{3}{\NPLaff}
\corr{zawada@fizyka.umk.pl}
\presaddr{TOPTICA Photonics AG, Lochhamer Schlag 19, 82166 Gr\"afelfing, Germany}

\chapstart
This chapter summarises the effects of background-gas collisions, lattice-photon scattering and parametric heating in neutral-atom optical lattice clocks.  All three processes can have a detrimental effect on the clock frequency stability and must be mitigated through careful control of the vacuum pressure and the trap depth and by avoiding fluctuations in the trapping potential.

\section{Background collisions in optical lattice clocks
} \label{sec:coherence_back}

An established way of evaluating the effect of thermal background-gas collisions in optical lattice clocks is to measure the trap lifetime~\citep{Gibble2013,McGrew2018} and take advantage of theoretically determined coefficients. Therefore, the uncertainty related to the background-gas collisions heavily depends on the quality of the theoretical calculations.
As shown by \citet{Gibble2013}, clock-atom losses as well as the collisional shift of the clock transition are governed by long-range interactions which can be approximated by a van der Waals potential.
Therefore, it is crucial to determine the $C_6$ coefficients for the interaction between the clock atoms and the colliding atoms or molecules.
While literature data is available for Sr--Sr and Yb--Yb interactions, such data did not exist for Sr--H${}_2$ or Yb--H${}_2$ interactions, in particular not for the ${}^3$P${}_0$ excited clock states in Sr and Yb.

\begin{figure}[p]
    \centering
    \includegraphics[width=.92\columnwidth]{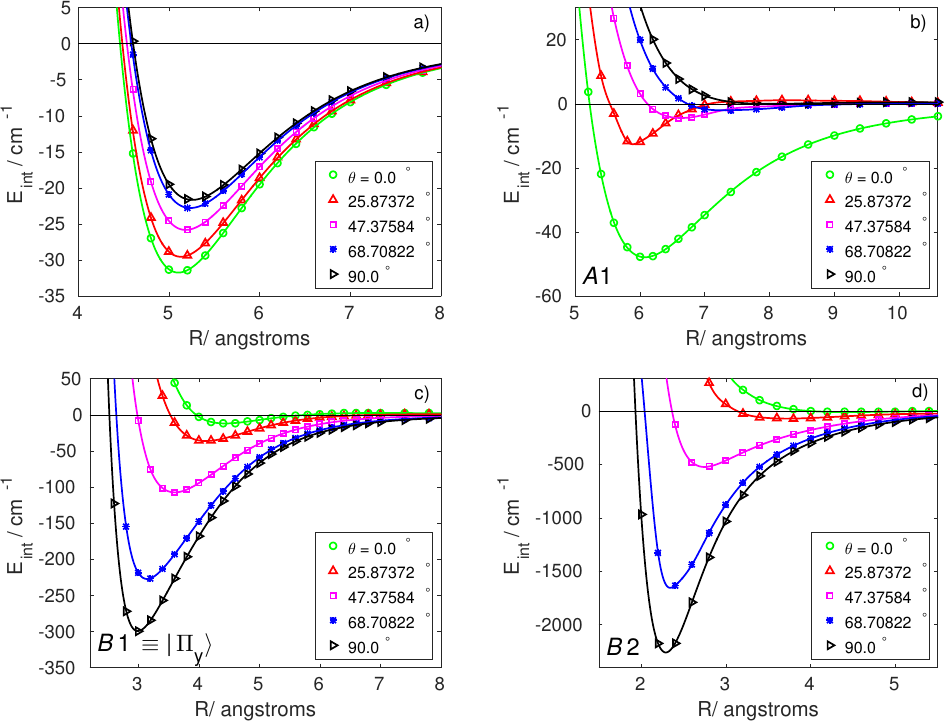}
    \caption{\label{fig:YtH2} The calculated intermolecular potential-energy curves for the ground (a) and first excited (b-d) states of the Yb--H${}_2$ complex. From \citet{Cybulski2018}.
}
\end{figure}

\begin{figure}[p]
    \centering
    \includegraphics[width=.92\columnwidth]{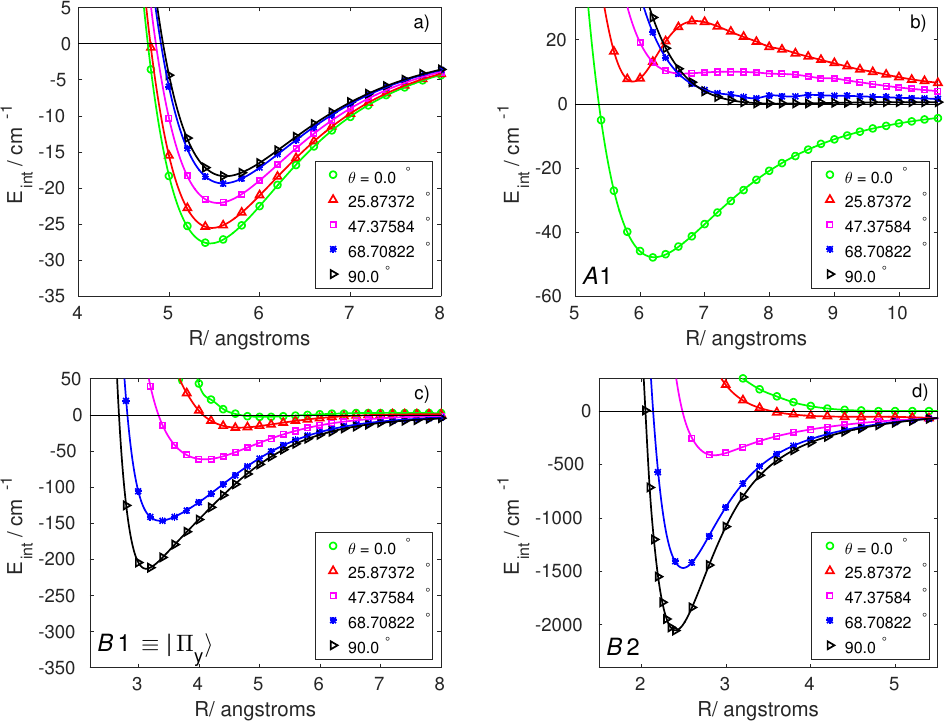}
    \caption{\label{fig:SrH2} The calculated intermolecular potential-energy curves for the ground (a) and first excited (b-d) states of the Sr--H${}_2$ complex. From \citet{Cybulski2018}.
}
\end{figure}

The missing $C_6$ coefficients can be estimated from interaction potential energy surfaces (PESs)  for Sr--H${}_2$ and Yb--H${}_2$ interactions in the electronic ground state as well as in the excited $^3$P~state.
These PESs were determined for the first time by \citet{Cybulski2018}. A series expansion in terms of spherical harmonics depending on the angle between the clock-atom--H${}_2$ axis and the H--H axis of the H${}_2$ molecule is commonly used for representing the PES. The calculated intermolecular potential-energy curves for the ground and the first excited states of the Yb--H${}_2$ and Sr--H${}_2$ complexes are shown in Figures~\ref{fig:YtH2} and \ref{fig:SrH2}, respectively.

\begin{figure}[t]
    \centering
    \includegraphics[width=0.81\columnwidth]{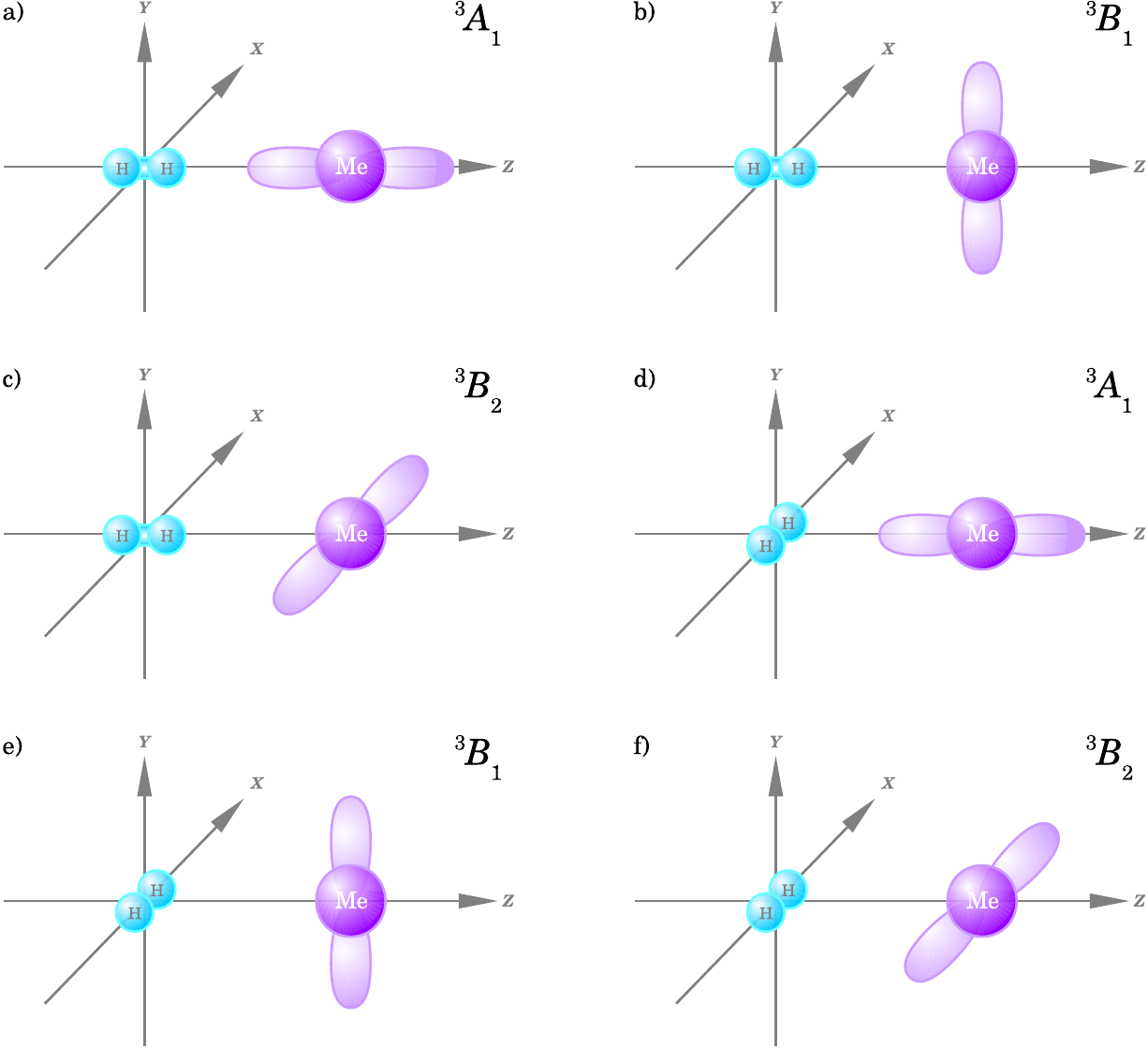}
    \caption{\label{fig:Tshape}Schematics of the singly occupied p-orbital orientation in the first excited state of the Me--H$_2$ ($\mathrm{Me} = \mathrm{Sr,Yb}$) complexes for the
collinear (a-c) and T-shape (d-f) configurations. From \citet{Cybulski2018}.}
\end{figure}

Only the isotropic potentials were used to estimate  the $C_6$ coefficients.
To determine the particular $C_6$ coefficient the isotropic potential is multiplied by $r^6$ and  a plateau between the short-range and  long-range parts of the potential is found. In the short-range part the interaction does not have a van der Waals character and in the long-range part the errors of the {\em ab initio} calculations  are large in comparison to the value of the potential. A similar approach was used to determine the $C_6$ coefficient near the asymptotes where Sr and  Yb atoms are excited to the ${}^3$P states. The $C_6$~coefficients were calculated for the isotropic parts of the PES in the $A1$, $B1$ and $B2$ electronic states.
The $A1$, $B1$ and $B2$ nomenclature is used for the T-shape geometry.  In the case of the $A1$ surface, the p-orbital is directed towards the hydrogen molecule, Figures \ref{fig:Tshape}(a) and (d), whilst in the case of the $B1$ surface the p-orbital is perpendicular to the complex symmetry plane, Figures \ref{fig:Tshape}(b) and (e). The $B2$ surface corresponds to the third situation, where the p-orbital is neither perpendicular to the complex symmetry plane nor directed towards the H${}_2$ dimer, Figures~\ref{fig:Tshape}(c) and (f). To estimate the $C_6$ coefficients  near the Sr(${}^3$P${}_0$)--H${}_2$ and Yb(${}^3$P${}_0$)--H${}_2$ asymptotes that involve the excited ${}^3$P${}_0$ clock state, it is necessary to use an averaged coefficient $\left[C_6(A1)+C_6(B1)+C_6(B2)\right]/3$. The uncertainty of the $C_6$ coefficients determined in this way was estimated to be about 10\%.

The $C_6$ coefficients necessary for the evaluation of the background-collision effects in Yb and Sr optical lattice clocks are summarised in Table~\ref{tab:C6}. These coefficients allow the ratio of the relative collisional shift $\Delta \nu_\mathrm{bg}/\nu_0$ and the collisional loss rate of clock atoms $\Gamma_\mathrm{bg}$ to be calculated. These ratios are presented in Section~\ref{sec:latticeshift_bgcoll}.
The collisional loss rate due to collisions with H${}_2$ can be extracted from the trap lifetime at a given vacuum pressure.
As an example, for Sr atoms the dependence of the loss rate (defined as the inverse of the $1/e$ trap lifetime) on the vacuum pressure is equal to
\begin{equation}
\Gamma_\mathrm{bg} = \SI{4.31(98)e8}{\second^{-1}} \, P\,[\text{mbar}].
\end{equation}

\begin{table}[b]
\centering
\caption{
    $C_6$ coefficients necessary for the evaluation of background-collision effects in Yb and Sr optical lattice clocks.
   } \label{tab:C6}
\begin{tabular}{l S l}
\toprule
\multicolumn{1}{l}{System} & {$C_6$ (a.u.)} & \multicolumn{1}{l}{Reference}   \\
\midrule
 Sr(${}^1$S${}_0$)--Sr & -3164(10) & \citet{Stein2010a}\\
Sr(${}^3$P${}_0$)--Sr & -4013(50) & \citet{Borkowski2014}\\
Sr(${}^1$S${}_0$)--H${}_2$ & -166(17) & \citet{Cybulski2018}\\
Sr(${}^3$P${}_0$)--H${}_2$ & -212(22) &\citet{Cybulski2018}\\
 Yb(${}^1$S${}_0$)--Yb & -1937.27(57) & \citet{Borkowski2017}\\
Yb(${}^3$P${}_0$)--Yb & -2561(95) & \citet{Porsev2014,Borkowski2018}\\
Yb(${}^1$S${}_0$)--H${}_2$ & -146(15) & \citet{Cybulski2018}\\
Yb(${}^3$P${}_0$)--H${}_2$ & -217(22) &\citet{Cybulski2018}\\
\bottomrule
\end{tabular}
\end{table}

\section{Lattice-photon scattering} \label{sec:coherence_scat}

Here, the main results of the investigation of scattering of lattice photons by the atoms in a $^{87}$Sr lattice clock are summarised and the consequences for the operation of the optical clock are discussed. The results are presented in more detail in the article by \citet{Doerscher2018}.

Off-resonant scattering of lattice-laser radiation by  $\Sr$ atoms in the two states $\state{1}{S}{0}$ and $\state{3}{P}{0}$ causes decoherence of the atomic superposition state during interrogation and population redistribution between electronic, fine- and hyperfine-structure, and Zeeman states. This will lead to a degradation of the signal-to-noise ratio and thus the frequency stability of the clock, and potentially to systematic frequency shifts, e.g., by line pulling.
Observation of a resulting loss of contrast at long interrogation times has recently been reported for Ramsey spectroscopy \citep{Schioppo2017,Marti2018}.

\begin{figure}[tb]
	\centering
    \includegraphics[width=0.66\columnwidth]{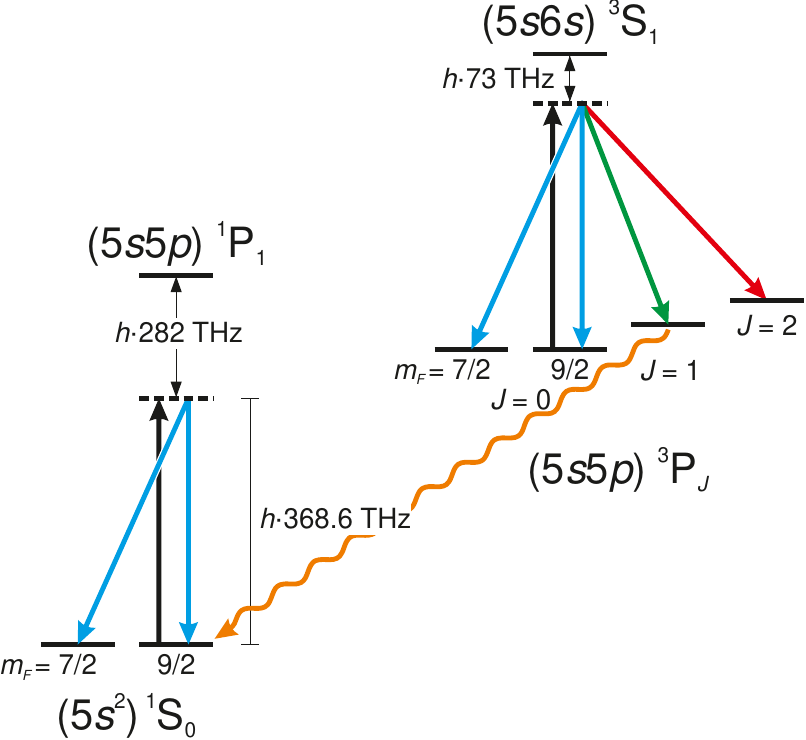}
	\caption{Schematic summary of photon scattering near the magic wavelength $\lambda \approx \SI{813}{\nano\meter}$ (i.e., a frequency of $\SI{368.6}{\tera\hertz}$) in the magnetic substates $m_F=9/2$ of the ground and excited states of a $\Sr$ optical lattice clock.
		Only the dominant intermediate states are shown. From \citet{Doerscher2018}.}
	\label{fig:scattering_summary}
\end{figure}

Figure~\ref{fig:scattering_summary} schematically illustrates the different types of photon-scattering processes at the lattice wavelength, all of which are off-resonant.
Inelastic scattering events changing the internal state of the atom are referred to as Raman scattering; elastic ones leaving the atom's internal state unchanged as Rayleigh scattering.

Raman scattering $\state{3}{P}{0} \rightarrow \state{3}{P}{1}$ is followed by radiative decay to the ground state, since the state $\state{3}{P}{1}$ has a lifetime of about \SI{21}{\micro\second} \citep{Nicholson2015}.
The atom is typically not lost from the optical lattice after such a decay process.
In contrast, Raman scattering $\state{3}{P}{0} \rightarrow \state{3}{P}{2}$ leaves atoms shelved in the metastable state $\state{3}{P}{2}$, which has an effective lifetime on the order of \SI{100}{\second} at room temperature due to blackbody radiation \citep{Yasuda2004}.

The decay rate $\state{3}{P}{0} \rightarrow \state{1}{S}{0}$ is experimentally investigated by measuring the populations in each of the two clock states as a function of hold time in the trap.
Lattice-induced scattering to the state $\state{3}{P}{1}$ is determined by varying the depth of the optical lattice, since its rate is proportional to the intensity of the lattice laser field.
The atoms are prepared in the substate $m_F = 9/2$ of the excited state and in the lowest two axial vibrational states of the lattice.
Population in any other state is removed from the trap by the preparation sequence.
The sample is held in the optical lattice for a variable amount of time $\thold$.
Finally, the populations in the ground state $\state{1}{S}{0}$ and in the metastable states $\state{3}{P}{0}$ and $\state{3}{P}{2}$ are destructively detected.
For each hold time $\thold$ and lattice potential depth $\Upeak$, the populations $\Ngnd$ in $\state{1}{S}{0}$ and $\Nexc$ in $\state{3}{P}{0,2}$ are averaged over typically thirty to forty samples.
In order to reject long-term fluctuations of the initial number of atoms, the populations are also divided by the total population found in reference measurements with $\thold = \SI{1}{\second}$, performed before and after each set of measurements, for normalisation.

\begin{figure*}[tb]
	\centering{\includegraphics[width=0.49\textwidth]{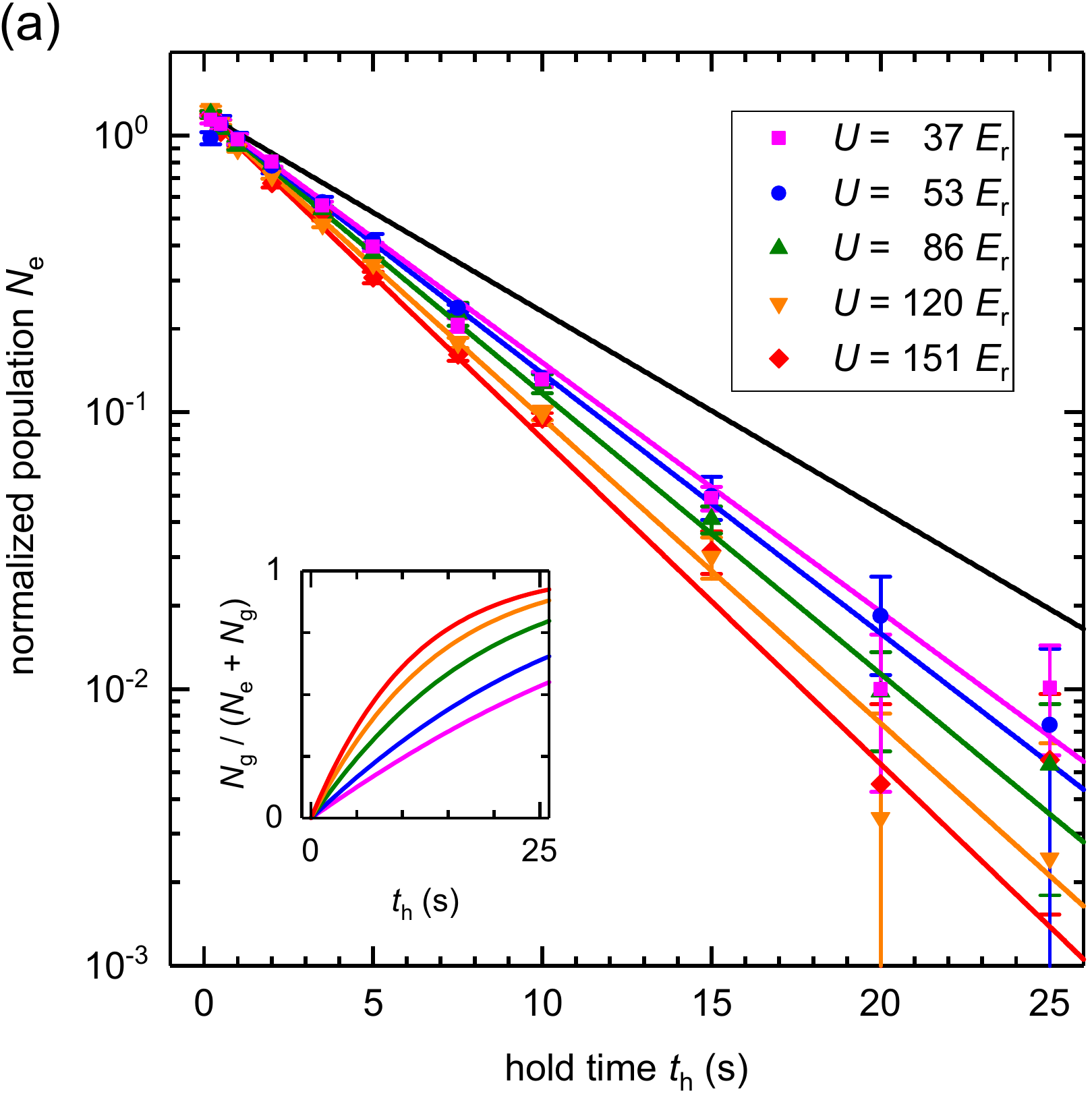} \hfill
		\includegraphics[width=0.49\textwidth]{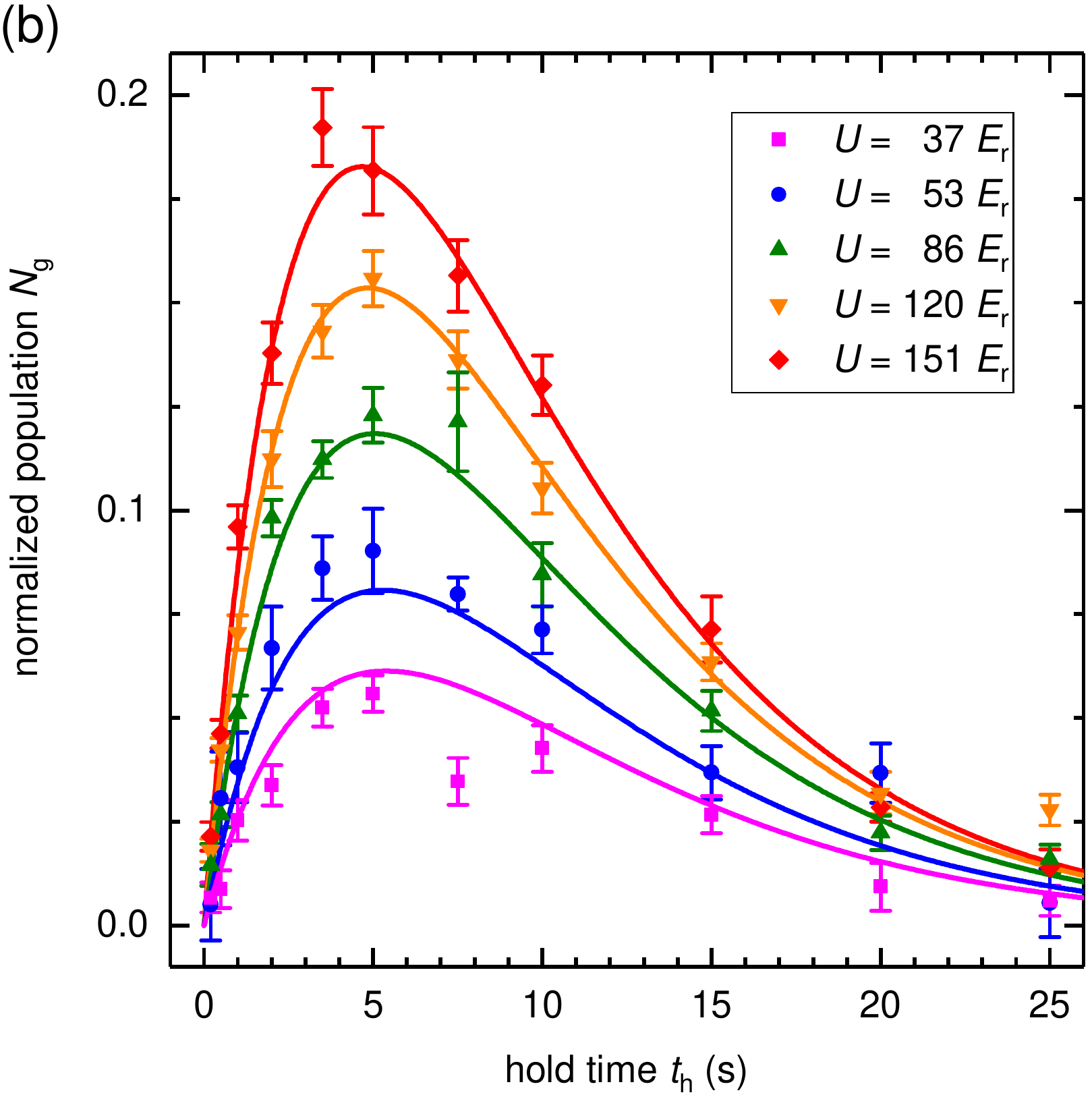}}
	\caption{	Measured populations (a) in the metastable states $\state{3}{P}{0,2}$ and (b) in the ground state $\state{1}{S}{0}$ for samples prepared in the excited state at different effective lattice depths $\Utherm$ as a function of hold time $\thold$.
		Results of a combined fit of a rate-equation model are shown as coloured lines.
		The inset in (a) shows the fraction of atoms in the ground state as derived from this model.
		For reference, the solid black line shows the decay of ground-state samples. From \citet{Doerscher2018}.}
	\label{fig:measurement_es}
\end{figure*}

This detection scheme can neither resolve any magnetic substates nor distinguish the metastable states $\state{3}{P}{0}$ and $\state{3}{P}{2}$, because the atoms are repumped via the intermediate state $\state{3}{P}{1}$ by driving the transitions $\state{3}{P}{0,2} \rightarrow \state{3}{S}{1}$.
Therefore, only lattice-induced decay to the ground state (see Figure~\ref{fig:scattering_summary}) can be studied experimentally in this scheme, whereas the lattice-induced population redistribution over the Zeeman and hyperfine states of the metastable $\state{3}{P}{}$ states cannot be resolved.

Decay $\state{3}{P}{0} \rightarrow \state{1}{S}{0}$ is clearly observed, as, in addition to enhanced losses in the excited state, atoms emerge in the ground state (Figure~\ref{fig:measurement_es}).
The population in the ground state becomes substantial after several seconds, especially for deep lattice potentials.

A quantitative analysis of these measurements is provided by the time evolution of the two populations by a pair of coupled rate equations
\begin{subequations} \label{eq:rate_equation}
	\begin{eqnarray}
	\dot{\Nexc} & = & - \traplossES \Nexc - \left( \decayrateOTHER + \decayratecoeffLAT \Utherm \right) \Nexc, \label{eq:rate_equation_es} \\
	\dot{\Ngnd} & = & - \traplossGS \Ngnd + \left( \decayrateOTHER + \decayratecoeffLAT \Utherm \right) \Nexc, \label{eq:rate_equation_gs}
	\end{eqnarray}
\end{subequations}
which describe lattice-independent losses from the trap at different rates, $\traplossGS$ and $\traplossES$, as well as decay from the excited state to the ground state.
The rate of the latter is the sum of a lattice-independent contribution $\decayrateOTHER$ and a lattice-induced contribution $\decayratecoeffLAT \Utherm$.
The analytic solution to these rate equations reads
\begin{subequations} \label{eq:model_solution}
	\begin{eqnarray}
	\Nexc(t) & = & \Nexc(0)
	               \exp\left( -\left[ \traplossES + \decayrateOTHER + \decayratecoeffLAT \Utherm \right] t \right), \\
	\Ngnd(t) & = & \left( \Nexc(0)
	               \frac{ 1 - \exp\left( - \left[ \traplossDIFF + \decayrateOTHER + \decayratecoeffLAT \Utherm \right] t \right) }
	               { 1 + \traplossDIFF / \left( \decayrateOTHER + \decayratecoeffLAT \Utherm \right) }
	                 + \Ngnd(0) \vphantom{\frac{\traplossDIFF}{\traplossDIFF}} \right) \exp\left( -\traplossGS t \right), 
	\end{eqnarray}
\end{subequations}
where $\traplossDIFF = \traplossES - \traplossGS$.
A nonlinear least-squares fitting of Equations~\eqref{eq:model_solution} to the data shown in Figure~\ref{fig:measurement_es} yields the rate coefficient of decay $\state{3}{P}{0} \rightarrow \state{1}{S}{0}$ due to inelastic scattering of lattice laser radiation
\begin{equation}
	\decayratecoeffLAT  =  \SI{556(15)d-6}{\per\second}\,\Erec^{-1}. \label{eq:res_scatterrate}
\end{equation}
The off-resonant scattering rates of lattice-laser radiation can also be calculated from atomic data in order to investigate the other processes shown in Figure~\ref{fig:scattering_summary} to which this experiment is not sensitive.

The rate $\Gamma_{i \rightarrow f}$, at which an atom in an initial state $i$ off-resonantly scatters linearly polarised incoming radiation with intensity $I$ and is transferred to a final state $f$, is given by the Kramers--Heisenberg formula \citep{Brandsen2003}
\begin{equation}
	\Gamma_{i \rightarrow f} = \frac{I {\omega^{\prime}}^3}{\left(4\pi \epsilon_{0} \right)^2 c^4 \hbar^3} \frac{8\pi}{3} \sum_{q=-1}^{1} \left| D_q^{(i \rightarrow f)} \right|^2,
	\label{eq:sc_rate}
\end{equation}
with
\begin{equation}
	\begin{split}
		D_q^{(i \rightarrow f)} = \sum_{k} &\left( \matrixelement{f}{d_{q}}{k} \frac{\matrixelement{k}{d_{0}}{i}} {\omega_{ki} - \omega} +
		                                   \matrixelement{f}{d_{0}}{k} \frac{\matrixelement{k}{d_{q}}{i}} {\omega_{ki} + \omega^{\prime}}  \right),
	\end{split}
	\label{eq:sc_amplitude}
\end{equation}
where $d_q = -e r_q$ are the elements of the electric dipole operator in spherical tensor notation, $\omega$ and $\omega^{\prime}$ are the angular frequencies of the incoming and scattered radiation, respectively, and $q$ is the polarisation state of the scattered radiation in spherical tensor notation, where the polarisation axis of the incoming light is used as quantation axis.
The sum is over all intermediate states $k$, and $\omega_{ki}$ is the frequency of the transition $i \rightarrow k$.

The matrix elements of the dipole operator can be related to Einstein coefficients using tensor-algebraic expressions \citep{Doerscher2018}. The scattering rates are calculated according to Equation~\eqref{eq:sc_rate} using line strengths as discussed in the supplement to \cite{Middelmann2012} and including those reported by \cite{Nicholson2015} for the $(4d5s)\,\state{3}{D}{}$ states and by \cite{Sansonetti2010} for the state $(5s6s)\,\state{3}{S}{1}$.

The calculated rates of all Rayleigh- and Raman-scattering processes of laser radiation at the magic wavelength near \SI{813}{\nano\meter} that occur for $\Sr$ atoms are summarised in Table~\ref{tab:all_rates}.
Interestingly, the elastic scattering amplitudes of the two states $\state{1}{S}{0}$ and $\state{3}{P}{0}$ used in lattice clocks are necessarily equal at any magic wavelength. As damping of coherences by Rayleigh scattering depends on the difference of the elastic scattering amplitudes \citep{Uys2010}, no loss of coherence is expected from this effect.

The two Raman scattering processes $\state{3}{P}{0}\rightarrow\state{3}{P}{1,2}$ result in decoherence of the atomic superposition state as well as depopulation of the excited state $\state{3}{P}{0}$.
They do not cause systematic frequency shifts directly, as they are not sensitive to the phase of the superposition state.
However, the maximum slope of the spectroscopic signal is reduced.
In Ramsey spectroscopy, these effects manifest themselves as a reduction of fringe contrast if the duration of an excitation pulse is small compared to the inverse scattering rate.
For Rabi spectroscopy using a single, long pulse, the line shape is modified more intricately; it is broadened, but remains symmetric with respect to the detuning of the interrogation laser from resonance.

\begin{table}[b!]
	\caption{Calculated off-resonant scattering rates of lattice-laser radiation by an atom in the magnetic substate $m_F = 9/2$ of the state $i$.
		The final state of the the atom is denoted by $f$ with hyperfine and magnetic quantum numbers $F^{\prime}$ and $m_F^{\prime}$, respectively.
		Rates are given for an intensity corresponding to an optical dipole potential of $\Utherm = 1\,\Erec$. From \citet{Doerscher2018}.}
	\label{tab:all_rates}
	\centering
	\begin{tabular}{crrS[table-format=3e+2, table-align-exponent = false]}
		$i \rightarrow f$							& \multicolumn{1}{c}{$F^\prime$}	& \multicolumn{1}{c}{$m_F^\prime$}	& {$\Gamma/(\SI{d-4}{\per\second})$}	\\
        \midrule \\
		$\state{1}{S}{0}\rightarrow\state{1}{S}{0}$	& $ 9/2$		& $ 7/2$		& 3e-16							\\
													&				& $ 9/2$		& 5.57							\\\\
		$\state{3}{P}{0}\rightarrow\state{3}{P}{0}$	& $ 9/2$		& $ 7/2$		& 5e-10							\\
													&				& $ 9/2$		& 5.57							\\\\
		$\state{3}{P}{0}\rightarrow\state{3}{P}{1}$	& $ 7/2$		& $ 7/2$		& 1.99							\\
													& $ 9/2$		& $ 7/2$		& 0.45							\\
													&				& $ 9/2$		& 1e-10							\\
													& $11/2$		& $ 7/2$		& 0.05							\\
													&				& $ 9/2$		& 6e-10							\\
													&				& $11/2$		& 2.49							\\
	\end{tabular}
	\hfill
	\vrule
	\hfill
	\begin{tabular}{crrc}
		$i \rightarrow f$							& \multicolumn{1}{c}{$F^\prime$}	& \multicolumn{1}{c}{$m_F^\prime$}	& $ \Gamma / (\SI{d-4}{\per\second})$	\\
        \midrule \\
		$\state{3}{P}{0}\rightarrow\state{3}{P}{2}$	& $ 7/2$		& $ 7/2$		& \SI{0.37}{}							\\
													& $ 9/2$		& $ 7/2$		& \SI{0.26}{}							\\
													&				& $ 9/2$		& \SI{0.76}{}							\\
													& $11/2$		& $ 7/2$		& \SI{0.08}{}							\\
													&				& $ 9/2$		& \SI{0.53}{}							\\
													&				& $11/2$		& \SI{0.50}{}							\\
													& $13/2$		& $ 7/2$		& \SI{0.01}{}							\\
													&				& $ 9/2$		& \SI{0.11}{}							\\
													&				& $11/2$		& \SI{0.22}{}							\\
		\\ \\ \\
	\end{tabular}
\end{table}

The resulting populations in the states $\state{3}{P}{2}$ and $\state{1}{S}{0}$ themselves may also disturb an optical lattice clock.
In particular, atoms having decayed to the ground state are highly susceptible to the interrogation laser.
Our calculations show that lattice-induced decay to the ground state most likely returns an atom to the magnetic sublevel $m_F = \pm9/2$ (Table~\ref{tab:all_rates}).
The consequences on spectroscopy are quite similar to those of decoherence and depopulation of the superposition state.
For Ramsey spectroscopy, the atoms having decayed during the free evolution time are resonantly excited by the final $\pi/2$-pulse, resulting in reduced fringe contrast.
For the case of Rabi spectroscopy with long excitation pulses, coherently driving the reference transition in those atoms after decay modifies the line shape further.
Likewise, this degrades the slope of the error signal and, in some cases, the observable linewidth, but does not give rise to systematic frequency shifts.
In contrast, population of the other magnetic substates ($m_F\neq\pm9/2$) in the ground state can cause line pulling if the laser detuning is varied to derive an error signal.

Last but not least, atoms that are incoherently transferred to different states by Raman scattering may cause systematic shifts of the transition frequency by disturbing the remaining atoms.
For instance, interactions with atoms in the original superposition state at the same lattice site are generally no longer suppressed by the Pauli principle and lead to collision-induced systematic frequency shifts at high atom density.

\section{Design of a lattice trap with low parametric heating rate} \label{sec:coherence_heating}

In any accurate atomic clock it is essential to have precise control over the motion of the atomic reference in order to avoid large uncertainties in the Doppler and recoil shifts. In the case of an optical lattice clock, the motion is controlled firstly by laser cooling the atomic sample to a temperature of a few \si{\micro\kelvin}, and secondly by trapping the atoms in a magic-wavelength optical lattice during the clock interrogation \citep{Ye2008}. The role of the optical lattice trap is to ``freeze out'' the Doppler and recoil shifts in the Lamb--Dicke regime, where each atom is tightly confined to a single lattice site to within a region much smaller than the wavelength of the probe laser \citep{Wineland1979}. However, for this technique to be effective, the atoms must remain cold and trapped in the lattice for long enough to carry out clock interrogation.

In this section some of the key design considerations for building a lattice trap with a sufficiently long lifetime and low heating rate for an optical lattice clock are described. Some of the important features of the cavity-enhanced lattice trap built at NPL, which lead to parametric heating rates below $0.1\;\text{quanta}/\mathrm{s}$,  are described.


\subsection{Scaling laws for parametric heating}

Parametric heating is caused by instability in shape and depth of the lattice potential, which can resonantly excite the atoms into higher-energy motional states. Generally the heating rate scales with the square of the motional frequency, so in a 1D lattice the rate is typically much faster in the axial direction ($\nu_z \approx 5\times10^4$\;Hz) than in the radial direction ($\nu_\rho \approx 10^2$\;Hz). In this section, results from \cite{Savard1997} are applied to calculate rough targets for realising a trap with low heating rates.

The first cause of heating is from vibrations in the position of the trap. In the case of a 1D optical lattice, the position of each lattice site can move either due to vibrations of the mirror used to reflect the lattice beam, or due to frequency noise on the lattice laser which has the effect of wobbling the relative phase of the incident and retroreflected beams. In the approximation of each lattice site as a harmonic potential with motional frequency $\nu_z$ along the lattice axis, the rate of change of motional energy $\left< E \right>$ from site position instability is given by
\begin{equation} \label{eq:parametric_heating_position}
\frac{\left< \dot{E} \right>}{\left< E \right>} = \pi^2\nu_z^2\frac{S_z(\nu_z)}{\left<z^2\right>} \approx \frac{\pi^2\nu_z^2}{\left<z^2\right>}\left(S_\mathrm{mirror}(\nu_z) + \frac{4 d^2}{\nu_\mathrm{L}^2}S_{\nu_\mathrm{L}}(\nu_z)\right),
\end{equation}
where $\left<z^2\right>$ is the mean-square spread of atoms in the axial direction on a single lattice site and $S_z(\nu_z)$ is the power spectral density (PSD) of lattice-site position fluctuations at frequency $\nu_z$.

To derive the second identity in Equation (\ref{eq:parametric_heating_position}), two separate contributions are considered --- one from physical mirror vibrations and the other from laser frequency noise --- and for this purpose the PSD of mirror position fluctuations is given by $S_\mathrm{mirror}(\nu_z)$ and the PSD of lattice laser frequency fluctuations by $S_{\nu_\mathrm{L}}(\nu_z)$. In Equation \eqref{eq:parametric_heating_position}, $d$ is the distance of the atoms from the retroreflecting mirror and $\nu_\mathrm{L}$ is the lattice-laser optical frequency. The physical mirror vibrations are usually negligible under laboratory conditions since the relevant Fourier frequencies $\nu_z \approx 5 \times 10^4$\;Hz tend not to be easily excited by the environment and are well damped by the inertia of the mirror. However, heating from laser frequency noise can be a problem if combining a relatively broad laser source (e.g., a MHz-linewidth DFB) with long lattice propagation distances of several 100\;mm: to avoid this effect, it is helpful if the laser linewidth is narrow (e.g., from a titanium-sapphire laser or extended-cavity diode laser) or if the mirror distance from atoms is kept to a few 10\;mm.

The second heating effect, and often the dominant one, is from fluctuations in the depth of the lattice potential caused by intensity noise on the trapping light. These intensity fluctuations can excite (or de-excite) the atoms in units of two motional quanta, with a transition rate $R_{n\pm2 \leftarrow n}$ given by
\begin{equation} \label{eq:parametric_heating_intensity}
R_{n\pm2 \leftarrow n} = \frac{\pi^3\nu_z^2}{4} S_\varepsilon(2\nu_z)(n+1\pm 1)(n\pm 1),
\end{equation}
where $S_\varepsilon(2\nu_z)$ is the PSD of fractional intensity fluctuations, where the peak trap intensity is $I(t) = I_0 \left[1 + \varepsilon(t)\right]$. Under the assumption that the atoms are prepared in the ground state $n = 0$ with a motional frequency of 65 kHz, and targeting a transition rate of less than $0.1\;\text{quanta}/\mathrm{s}$, Equation~\eqref{eq:parametric_heating_intensity} implies a required fractional intensity noise of $\sqrt{S_\varepsilon(2\nu_z)} < 10^{-6} \ \mathrm{Hz}^{-{1}/{2}}$.


\subsection{Design of an in-vacuum lattice enhancement cavity with low heating rates}

In order to achieve a systematic uncertainty below $10^{-17}$ in an optical lattice clock, there are several reasons to use an enhancement cavity to generate the lattice trap:
\begin{itemize}
\item The cavity cleans up the radial mode profile of the lattice beam into a near-ideal Gaussian, giving a more reliable basis to apply detailed theoretical models of the atomic motion \citep{Blatt2009} and high-order lattice shifts \citep{Brown2017, Katori2015}.
\item The running-wave component of the lattice beam is eliminated by the cavity, avoiding first-order contributions from electric-quadrupole and magnetic-dipole lattice shifts, and therefore eliminating retroreflector beam-pointing instability as a possible source of systematic error.
\item The power build-up factor, related to the cavity finesse as $\mathcal{F}/\pi$, enables the use of larger lattice beam waists --- and therefore larger trap volume --- while still achieving the same peak intensity. This reduces the density shift for a fixed trap depth and atom number, and also enables more efficient capture of atoms from the final stage of cooling.
\item The cavity reduces the amount of laser light needed to generate the same depth of trap, therefore reducing the (often unstable) Stark shift from amplified spontaneous emission (ASE) incident on the atoms --- an especially important consideration when using semiconductor laser sources.
\end{itemize}
However, the cavity presents a significant technical challenge: it has to be possible to couple a stable amount of lattice power into the TEM00 cavity mode, otherwise intensity fluctuations will parametrically heat the atoms out of the lattice trap. In practice the main difficulty is in locking the laser to the cavity to within much less than the cavity linewidth, in order to avoid conversion of lattice-laser frequency noise into fluctuations of cavity-coupled power --- i.e., to avoid FM to AM conversion. To make this task easier, it helps (i) to make both the cavity and the lattice laser as intrinsically stable as possible, and (ii) to use a cavity with a relatively broad linewidth $\Delta\nu_\mathrm{FWHM} = \nu_\mathrm{FSR}/\mathcal{F}$, ideally at least a few MHz. In general, a shorter enhancement cavity helps because the linewidth is larger for a given finesse, and because more compact spacers tend to have reduced sensitivity to vibrations.

\begin{figure}[tb]
    \centering
    \includegraphics[width=1\columnwidth]{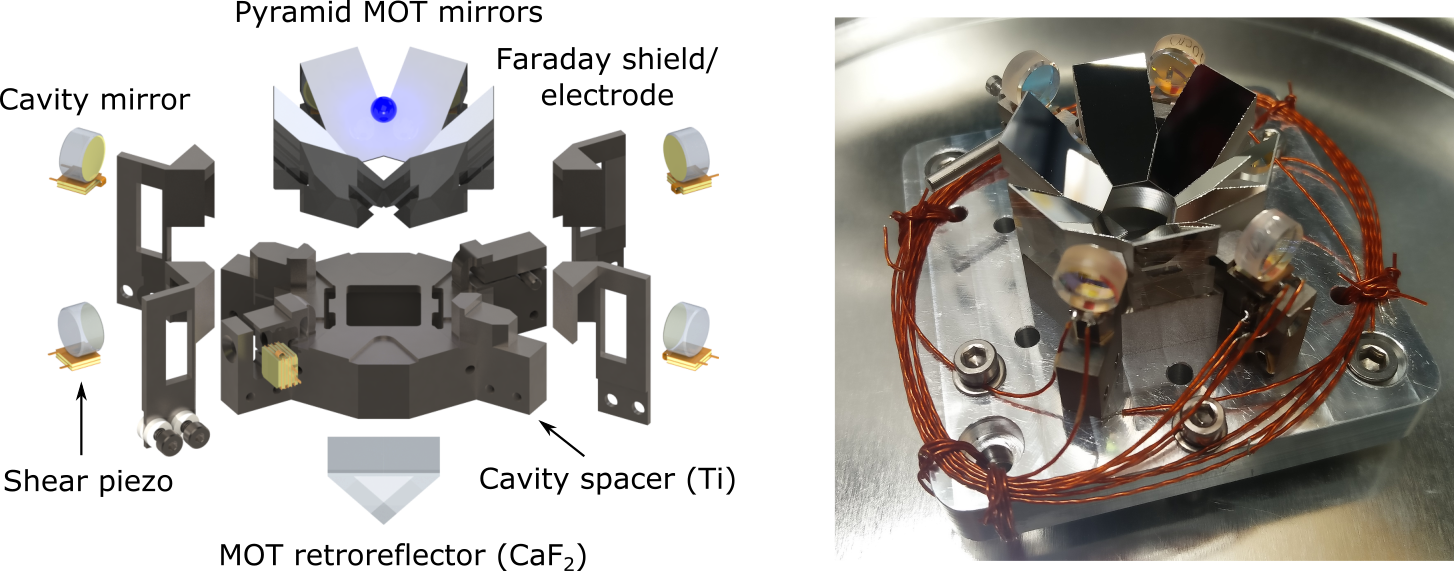}
    \caption{\label{fig:heating_design}\textit{Left:} Exploded rendering of the NPL cavity setup, which is also integrated with in-vacuum mirrors for a pyramid MOT. \textit{Right:} Photo of the partially completed assembly, before the installation of electrodes or the CaF$_2$ prism. Note: the two pairs of cavity mirrors form two separate cavities, but only one of those cavities is actually used for the optical lattice clock.}
\end{figure}

To minimise the cavity length and maximise the mechanical stability, a monolithic in-vacuum cavity like in Figure \ref{fig:heating_design} is an effective solution.  The cavity has a finesse of 6000, a length of 35.9\;mm, and a mirror radius of curvature of 10\;cm, resulting in a lattice waist of \SI{100}{\micro\meter} at the centre of the cavity where atoms are trapped. The design has been incorporated into the NPL Sr lattice clock and supports a lattice lifetime of 27\;s with negligible parametric heating of less than~$70\;\mathrm{nK/s}$.

These results show that it is possible to achieve good trap lifetime and low heating rates with a moderately high-finesse in-vacuum cavity, but there are a handful of further considerations to bear in mind for an accurate optical lattice clock. Firstly, with in-vacuum optics placed near the atoms there is a danger that stray electric fields from patch charges on the mirrors will cause DC Stark shifts at the $10^{-14}$ level. To mitigate this possibility, two precautions have been taken in the NPL design: (i) the shear piezos for cavity length tuning are actually formed of a two-layer stack, with the high-voltage electrode sandwiched in the middle between grounded electrodes top and bottom. This shields the electric field generated by the high-voltage electrode, and also ensures that the surrounding spacer and optics are all contacted to ground; (ii) in-vacuum cages are installed around the cavity mirrors and connected electrically to a vacuum feedthrough, both providing shielding for any patch charges on the mirrors, and also allowing the application of large electric fields for DC-Stark-shift evaluation. With these precautions in place, a small residual background electric field from stray charges is still observed, but only enough to cause a relatively benign DC Stark shift of $2\times 10^{-18}$.

A second design consideration is that all materials for the cavity and spacer should, if possible, be non-magnetic to avoid perturbation when switching the magnetic coils needed for the MOT cooling stages. In our case the cavity spacer, screws and most of the vacuum chamber are made from titanium, the piezo electrodes are a copper-beryllium alloy, and any electrically isolating spacers or shims are ceramic.

In future lattice clock work, there are a handful of minor modifications to the NPL design which could be recommended: (i) the inclusion of a thin damping Viton spacer in the chamber mount in order to avoid transmission of kHz acoustic vibrations onto the cavity; (ii) a reduction in the finesse of the cavity to realise a roughly 2-MHz linewidth for reduced FM to AM conversion; (iii) the use of finite-element modelling to optimise the spacer and its in-vacuum mount for reduced sensitivity to vibrations; (iv) the omission of the second pair of cavity mirrors to simplify assembly; (v) the replacement of the silver-coated glass pyramid-MOT mirrors with an aluminium substrate, which would be easier to machine, would have higher thermal conductivity for improved homogeneity of the blackbody-radiation environment, and would be electrically conductive for guaranteeing low DC Stark shifts.

\clearpage

\graphicspath{{D5_latticeshifts/}}
\chapter[Understanding and controlling light and collisional shifts in optical lattice clocks]{Understanding and controlling light and collisional \\ shifts in optical lattice clocks} \label{chap:latticeshifts}

\authorlist{%
J\'er\^ome Lodewyck$^{1,\dag}$,
Mateusz Borkowski$^{2}$,
Roman Ciury\l{}o$^{2}$,
Marco Pizzocaro$^{3}$ and
Micha\l{} Zawada$^{2}$
}

\affil{1}{\OPaff}
\affil{2}{\UMKaff}
\affil{3}{\INRIMaff}
\corr{jerome.lodewyck@obspm.fr}

\chapstart
This chapter considers the optical-lattice-induced light shifts and the collisional shifts relevant for optical lattice clocks. Section~\ref{sec:latticeshifts_Title1} starts with an overview of the lattice light shift and then summarises the evaluation of relevant parameters for different atom species, mainly strontium and ytterbium. The discussion of the collisional shifts is divided into cold collisions between clock atoms (Section~\ref{sec:cold_coll}) and collisions with hot background-gas atoms (Section~\ref{sec:latticeshift_bgcoll}).

\section{Lattice light shifts \label{sec:latticeshifts_Title1}}

The lattice light shift is induced by the lattice light used to confine the atoms in the Lamb--Dicke regime. At the so-called magic wavelength \citep{Katori2003} for the lattice light, the electric dipole shift is the same, to leading order, for the excited and ground clock states, thus effectively suppressing the light shift.
However, this suppression may only be partial due to several effects:
\begin{itemize}
		\item A possible deviation of the lattice frequency from  the magic wavelength.
		\item Effects that depend on the polarisation of the trapping light for species with a non-zero spin.
		\item Higher-order terms in the expansion of the interaction between the lattice light and the atoms, namely magnetic dipole and electric quadrupole interactions.
		\item Multi-photon processes, among which the dominant contribution is the hyper-polarisability.
		\item The finite expansion of the atomic wave function, determined by the atomic temperature, that results in a modification of the light shift.
\end{itemize}

All the effects mentioned above can be summarised in the following general expression of the frequency shift $\delta\nu$ that takes into account the dominant electric dipole (E1) interaction as well as the electric quadrupole (E2) and the magnetic dipole (M1) interactions, via the atomic polarisabilities $\aEO$, $\alpha_\text{M1}$ and $\alpha_\text{E2}$, and also hyper-polarisability effects via the coefficient $\beta$
\begin{equation}
		\label{eq:deltanu}
		h\delta \nu = -\frac{\Delta \aEO}{\aEO}\langle U \rangle - \frac{\Delta\alpha_\text{M1} + \Delta\alpha_\text{E2}}{\aEO}(n_z+ 1/2)\sqrt{\langle U\rangle_r E_\mathrm{r}} - \frac{\Delta \beta}{\aEO^2} \langle U \rangle^2.
\end{equation}
Here $\Delta x$ indicates the difference in quantity $x$ between the ground and excited clock states, $z$ ($r$) the longitudinal (transverse) coordinate, $\langle U \rangle$ the effective trap depth (see below), $E_\mathrm{r} = \hbar^2|\mathbf{k}|^2/2m$ is the recoil energy ($m$ is the atomic mass and $\mathbf{k}$ the lattice wave vector), and $n_z$ the longitudinal vibrational quantum number. The electric dipole term can be further expanded in an irreducible-tensor representation to exhibit its dependence on the lattice polarisation $\bm{\epsilon}$ and the magnetic sublevel $m_F$
	\begin{equation}
		\label{eq:tensor}
		\Delta \aEO = \frac{d \Delta \alpha_\mathrm{s}}{d\nu_\mathrm{l}}(\nu_\mathrm{l} - \nu_\mathrm{l}^\text{magic}) + \frac{m_F}{2F}\xi\mathbf{k}\cdot \mathbf{e}_z \Delta\alpha_\mathrm{v} + \frac{3m_F^2 - F(F+1)}{2F(2F-1)}(3|\bm{\epsilon}\cdot\mathbf{e}_z|^2 - 1)\Delta\alpha_\mathrm{t},
	\end{equation}
where $\alpha_\mathrm{s}$, $\alpha_\mathrm{v}$ and $\alpha_\mathrm{t}$ are the scalar, vector and tensor polarisabilities, respectively, $\nu_\mathrm{l}$ is the lattice light frequency, $\nu_\mathrm{l}^\text{magic}$ is the true magic frequency cancelling the differential scalar polarisability, $F$ is the total spin of the clock states, $\xi = ||i\bm{\epsilon} \wedge \bm{\epsilon}^*||$ is the degree of polarisation of the lattice light, and $\mathbf{e}_z$ is the quantisation axis, effectively realised by a strong static bias magnetic field. \citet{Shi2015} details the suitable treatment of the vector and tensor shifts in the practical case where the strength of the bias magnetic field is not considered infinite. It results in a quadratic dependence of the E1 light shift on the trap depth that does not impair the clock accuracy.

In Equation~(\ref{eq:deltanu}), $U$ denotes the depth of the trapping potential, which can be expressed as a function of the local intensity $I(\mathbf r)$ of the trapping light as $U(\mathbf r) = \aEO I(\mathbf r)$. It is a quantity that varies with space according to the intensity profile of the laser light and reaches its maximum $U_0$ at the position of constructive interference in the lattice. When evaluating this potential, several critical aspects have to be included:
	\begin{itemize}
		\item The lattice light shift does not directly depend on the maximum trap depth $U_0$ at the maximum intensity of the optical lattice, but rather on the effective trap depth $\langle U \rangle$ averaged over the spatial profile of the atomic wave function, because this wave function extends over regions of space for which the lattice intensity is lower than the maximum intensity. This average depends on the temperature of the atoms in the optical lattice or, in an equivalent way, on the average vibrational quantum numbers $n$ in the three dimensions of space. To correctly handle this effect, one has to express the lattice light shift of Equation~(\ref{eq:deltanu}) for given longitudinal and transverse vibrational quantum numbers, and then average the atomic distribution over these levels. This treatment is developed in \citep{Katori2015,Ovsiannikov2016,Brown2017}. Alternatively, and equivalently, one can simply use Equation~(\ref{eq:deltanu}) with the average trap depth given by the maximum trap depth $U_0$ reduced by the thermal energy
		\begin{equation}
			\label{eq:Uavg}
			\langle U \rangle = U_0 - \kB T_r - \frac 1 2 \kB T_z^*,
			\quad \text{with} \; \frac{1}{2} \kB T^*_z \equiv \left\langle n_z + \frac 1 2 \right\rangle \sqrt{U_0 E_\mathrm{r}}.
		\end{equation}
		Here $T_r$ is the temperature of the atoms in the transverse direction. Note that the E2/M1 term in Equation~(\ref{eq:deltanu}) only involves the transverse average trap depth $\langle U \rangle_r = U_0 - \kB T_r$. The intuitive reason for this is that in the longitudinal direction, the E2/M1 effects are stronger as the atoms explore large regions.
		\item The parameters $U_0$, $T_r$ and $T_z^*$ that appear in Equation~(\ref{eq:Uavg}) can be accurately measured by motional-sideband spectroscopy of the clock transition of the atoms in the optical lattice. The frequency of the longitudinal sidebands gives the maximum trap depth $U_0$, the ratio between the intensity of the blue and red sidebands yields $T_z^*$, and the width of the sidebands relates to $T_r$ through the anharmonicity of the optical lattice \citep{LeTargat2007}. Alternatively, the temperature of the atoms can be directly deduced from a measurement of the lattice light shift \citep{McDonald2015}.
		\item In practice, because of the way the atoms are cooled in the optical lattice, which may include adiabatic cooling by ramping the lattice depth, the temperature of the atoms depends on the maximum lattice trap depth $U_0$. Because of this, the various terms in Equation~(\ref{eq:deltanu}), giving $\delta\nu$ as a power series of the trapping potential, are effectively intermixed when one fits the clock frequency as a function of the maximum trap depth $U_0$~\citep{Brown2017}. In other words, the difference between the average trap depth $\langle U \rangle$ and the maximum trap depth $U_0$ depends on $U_0$.
		\item The motional quantum numbers used in Equation~(\ref{eq:deltanu}) hold for a harmonic trap. In practice, the optical lattice is an anharmonic sine wave in the longitudinal direction and a Gaussian profile in the transverse direction. This induces anharmonicity in both the $z$ and $r$ directions, with terms $z^4$ and $r^4$, but also a coupling  $z^2r^2$. These anharmonic terms have to be included in the treatment of the lattice light shift as their impact is not negligible at the level of the best accuracy of optical lattice clocks. Evaluating the light shift from Equations~(\ref{eq:deltanu}) and~(\ref{eq:Uavg}) effectively takes into account the critical anharmonicities.
	\end{itemize}

Evaluating the lattice light shift requires knowledge of the various polarisabilities involved in Equations~(\ref{eq:deltanu}) and~(\ref{eq:tensor}). They can be found in the literature (see the next sections for specific values), or evaluated locally by measuring  the clock frequency as a function of the trap depth. A proper combination of the various polarisabilities shifting the clock transition can be used to operate lattice clocks at a point for which the light shift is null and insensitive to the lattice power over a large range. This scheme was proposed in \citet{Katori2015}.

Equation~(\ref{eq:tensor}) shows the dependence of the lattice light shift on the lattice polarisation. The vector term is proportional to $m_F$ such that it is cancelled by averaging the clock frequency over resonances with opposite $m_F$, as is done to cancel the first-order Zeeman effect. The tensor component vanishes for atoms with $F < 1$, such as \ce{^{171}Yb}, \ce{^{199}Hg}, and \ce{^{88}Sr}. For \ce{^{87}Sr}, a proper evaluation of this effect must be considered (see the next section).

In the optical lattice, tunnelling between the sites may induce a frequency shift due to the band structure associated with the period structure of the lattice. For a trap depth $U_0 = 50\;E_\mathrm{r}$, this shift is as large as $6 \times 10^{-16}$. It can, however, be effectively suppressed by using a vertical lattice because in this configuration, the gravitational energy splits the degeneracy between the energy levels of adjacent sites~\citep{Lemonde2005}.

Another source of light shift may arise from the incoherent background of the laser source used to generate the optical lattice. It has been observed with semiconductor sources (laser diodes and tapered amplifiers), leading to frequency shifts as large as $10^{-15}$~\citep{LeTargat2013}. While proper filtering of the laser source using narrow interference filters or Bragg gratings can reduce the effect to $10^{-17}$~\citep{Bilicki2017}, the best suppressions to below $10^{-18}$ have been reported with titanium-sapphire lasers~\citep{Lodewyck2016,Brown2017}.

\subsection{Evaluation for strontium }

For \ce{^{87}Sr} the magic wavelength, cancelling the scalar part of the E1 differential polarisability, has been experimentally evaluated to be $\nu_\mathrm{l}^\text{magic} = \SI{368554725(3)}{MHz}$ \citep{Shi2015, Campbell2017}. The difference between the magic wavelengths of \ce{^{88}Sr} and \ce{^{87}Sr} is published in \citet{Takano2017}. Unlike \ce{^{88}Sr}, \ce{^{87}Sr} has a spin $F = 9/2 \geq 1$ such that a tensor shift adds to the scalar shift, effectively displacing the wavelength at which the E1 contribution vanishes. The tensor shift has been experimentally measured in~\cite{Westergaard2011} to be
\begin{equation}
-\frac{\Delta\alpha_\mathrm{t}}{h\aEO}\frac{1}{2F(2F-1)} = \num{-57.7(23)}\;\frac{\si{\micro\hertz}}{E_\mathrm{r}},
\end{equation}
confirmed by a theoretical evaluation~\citep{Shi2015}. If the polarisation of the lattice light is fluctuating, the lattice light shift will fluctuate accordingly. This fluctuation is extremal for a lattice polarisation parallel or orthogonal to the quantisation axis. In the case where the angle $\theta$ between these two axes is small and fluctuates by $\delta\theta$, the clock accuracy is better than $10^{-18}$ if $\theta\,\delta\theta < 5\times 10^{-4}\;\text{rad}^2$, which can be achieved in practice. The tensor shift prevents a simple 3D optical lattice clock from being realised, as the polarisation cannot be identical for three orthogonal beams due to the transverse nature of electromagnetic fields. This can be circumvented by detuning the frequency of one of the lattice beams \citep{Westergaard2011,Campbell2017}.

The vector light shift for \ce{^{87}Sr} has been experimentally and theoretically evaluated in~\cite{Westergaard2011,Shi2015}.
The sensitivity of the linear light shift to the lattice-light frequency has been experimentally measured to be \citep{Shi2015, Middelmann2012, Norcia2018}
\begin{equation}
-\frac{1}{h\aEO}\frac{d\aEO}{d\nu_\mathrm{l}} = -15\;\frac{\si{\micro\hertz}}{\mathrm{MHz}\,E_\mathrm{r}}.
\end{equation}
The hyper-polarisability coefficient has been experimentally evaluated in \citet{LeTargat2013,LeTargat2012} to be $-\Delta\beta/h\alpha_\mathrm{s}^2 = \SI{0.45(10)}{\micro Hz}/E_\mathrm{r}^2$ and in \citet{Ushijima2018} to be $-\Delta\beta/h\alpha_\mathrm{s}^2 = \SI{0.461(14)}{\micro Hz}/E_\mathrm{r}^2$. Looser experimental upper bounds were also published in \citet{Westergaard2011} and \citet{Nicholson2015}. These values agree with a recent theoretical evaluation~\citep{Porsev2018}. For a typical trap depth of $70\;E_\mathrm{r}$, the effect amounts to $5.3(2)\times 10^{-18}$.

An upper bound for the E2/M1 effect was experimentally evaluated in \cite{Westergaard2011} as $|\Delta\alpha_\text{M1} + \Delta\alpha_\text{E2}|/\aEO < 10^{-7}$, and a recent theoretical evaluation also falls within this bound \citep{Porsev2018}. In a separate experiment \citep{Ushijima2018}, the E2/M1 effect in Sr was also evaluated and found to be a little larger and with opposite sign from \citet{Porsev2018}, $(\Delta\alpha_{\text{M1}}+\Delta\alpha_{\text{E2}})/\alpha_\text{E1} = \num{-2.77(12)e-7}$.
For $U_0 = 70\;E_\mathrm{r}$, these results suggest a maximum shift at the level of mid to high parts in $10^{18}$, but this
can be further reduced by using lower trap depths, or with a refined evaluation of the coefficient.

\subsection{Evaluation for ytterbium}

The magic wavelength for \ce{^{171}Yb} is approximatively at \SI{759.4}{nm}, a wavelength easily accessible by titanium-sapphire lasers. The most accurate determinations of the magic frequency are $\nu_\mathrm{l}^\text{magic} = \SI{394798265(9)}{MHz}$ by \citet{Nemitz2016} and $\nu_\mathrm{l}^\text{magic} = \SI{394798267(1)}{MHz}$ by \citet{Brown2017}.

The linear dependence of the scalar shift around the magic wavelength is approximatively $21\;\mathrm{mHz}/(\mathrm{GHz}\, E_\mathrm{r})$ \citep{Nemitz2016, Pizzocaro2017, Brown2017}. For example, for a trap depth of $100 E_\mathrm{r}$, this implies that the lattice frequency has to be locked at \SI{<0.3}{MHz} to make the changes in the light shift \num{<1e-18}.

Unlike \ce{^{87}Sr}, \ce{^{171}Yb} has nuclear spin $1/2$ so that the tensor shift is identically zero \citep{Angel1968}. The clock transition in \ce{^{171}Yb} is separated into two $m_F = \pm 1/2$ Zeeman resonances. With the external magnetic field perpendicular to the lattice direction, the Zeeman and vector shifts of the two resonances are summed in quadrature. The vector light shift can be suppressed to \num{<1e-18} by a combination of three methods \citep{Lemke2009}: using a 1D lattice, using linear polarisation ($\xi \approx 0$) and averaging over the two $m_F = \pm 1/2$ resonances. \cite{Lemke2009} measured the vector shift by measuring the splitting between the $m_F = \pm 1/2$ resonances as a function of the lattice polarisation ellipticity, determining a maximum shift of \SI{220(30)}{Hz} at $U_0 = 500\;E_\mathrm{r}$. Assuming $\langle U \rangle/U_0 = \num{0.7(1)}$ for their measurement, this translates to $\Delta\alpha_\mathrm{v}/(h \aEO) = -0.6(1)\;\mathrm{Hz}/E_\mathrm{r}$.

\begin{figure}[t]
\centering
\includegraphics[width=0.9\textwidth]{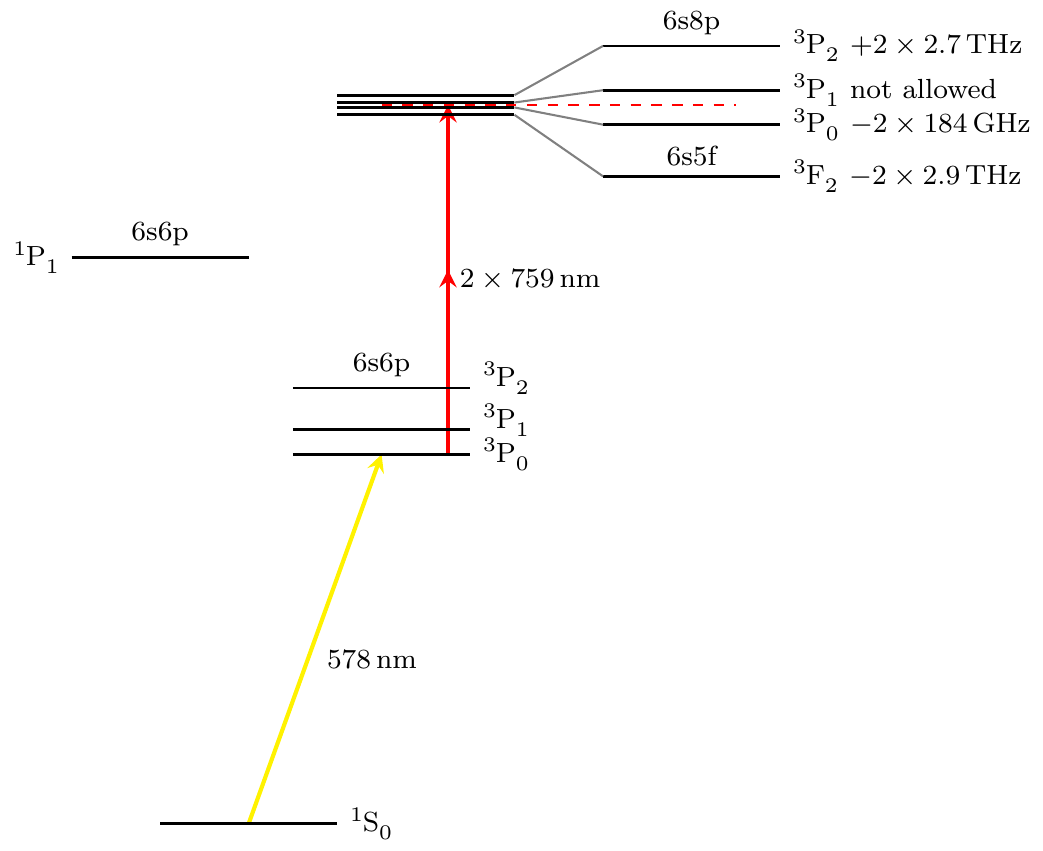}
\caption{Relevant energy levels (not drawn to scale) for the hyperpolarisability of Yb. Adapted from \cite{Barber2008}.}
\label{fig:barber_levels}
\end{figure}

The hyperpolarisability in Yb arises from the presence of three two-photon resonances near the magic wavelength \citep{Barber2008}, see Figure~\ref{fig:barber_levels}. Two of these show hyperfine splitting in the case of \ce{^{171}Yb}.
The one closest to the magic wavelength is the $\mathrm{6s6p\,{}^3P_0} \rightarrow \mathrm{6s8p\,{}^3P_0}$ resonance at \SI{394615300}{MHz}, about $\SI{-184}{GHz}$ from the magic wavelength (detuning given as single-photon). Then there are two hyperfine components of the  $\mathrm{6s6p\,{}^3P_0} \rightarrow \mathrm{6s5f\,{}^3F_2}$ resonance at \SI{391910770}{MHz} and \SI{391908650}{MHz} (wavelength \SI{765}{nm}, detuning $\SI{-2.9}{THz}$). The hyperfine splitting is \SI{4240(20)}{MHz}. Finally, there are two components of the $\mathrm{6s6p\,{}^3P_0} \rightarrow \mathrm{6s8p\,{}^3P_2}$ resonance at
\SI{397480707}{MHz} and \SI{397484603}{MHz} (wavelength \SI{754.2}{nm}, detuning \SI{+2.7}{THz}).

A recent, direct measurement \citep{Nemitz2019} of the Yb hyperpolarisability coefficient yielded $\Delta \beta/(h \aEO^2) = \SI{-1.194(89)}{\micro Hz/E_\mathrm{r}^2}$, which is in good agreement with other values, see Table~\ref{tab:Yb_hyperpol}.
Using the value from \citet{Nemitz2019} with a trap depth $100 E_\mathrm{r}$ gives an uncertainty from the hyperpolarisability coefficient of \num{2e-18}. The thermal-average method of \citet{Brown2017} does not provide a direct value of the hyperbolarisability coefficient $\Delta \beta$, but demonstrated that it is possible to reach an uncertainty of \num{1e-18}.

The E2/M1 coefficient was measured by \cite{Nemitz2016} to be $(\Delta\alpha_\text{M1} + \Delta\alpha_\text{E2})/\aEO = -\num{3.4(36)e-7}$ and again more recently by \citet{Nemitz2019} to be $-\num{5.1(19)e-7}$.  \cite{Brown2017} reported a theoretical calculation of the same coefficient to be $\num{4(4)e-8}$.

\begin{table}[b]
\centering
\caption{Summary of ytterbium hyperpolarisability measurements. \label{tab:Yb_hyperpol}}
\begin{tabular}{llS}
\toprule
{Source}	& {Method}		& {$\Delta\beta/(h\alpha^2_\text{E1})/(\si{\micro Hz}/E\ped{r}^2)$}\\
\midrule
\citet{Barber2008,Nemitz2016}		& Two-photon resonances 	& -1.9(8)\\
\citet{Brown2017}		& Direct (thermal average)	& -0.5\\	
\citet{Kobayashi2018}		& Two-photon resonances	& -1.1(4)\\
\citet{Nemitz2019}		& Direct		& -1.194(89)\\
INRIM, unpublished		& Two-photon resonances	& -1.04(9)\\
\bottomrule
\end{tabular}
\end{table}

\subsection{Other atomic species}

The evaluation of the lattice light shift for other atomic species suitable for lattice clocks is, so far, not as advanced as for Sr and Yb. For mercury, the magic wavelength and its derivative with respect to the lattice frequency have been experimentally measured in \citet{Yamanaka2015,Tyumenev2016}, and theoretical estimates of the higher-order coefficients can be found in \citet{Katori2015,Ovsiannikov2016}. The magic wavelength for magnesium can be found in \citet{Kulosa2015}.

\section{Collisions} \label{sec:collisions}

\enlargethispage{-2\baselineskip}

\subsection{Cold collisions} \label{sec:cold_coll}

Large atomic samples are desirable in the operation of an optical atomic clock because they lower the quantum projection noise, which in turn yields the fundamental limit on the instability of the clock. On the other hand, at large atomic densities, cold collisions may lead to significant undesirable clock line shifts.

In 3D optical lattice clocks the problem of collisional shifts can be eliminated altogether. One way is to make sure that there can be only one atom per lattice site by blowing away atoms in multiply occupied sites using a photoassociation laser beam, see, e.g., \citet{Akatsuka2010, Origlia2018}. Another approach is to load a quantum-degenerate, rather than thermal, gas into the lattice \citep{Campbell2017}. In this case the clock lines originating from different site occupancies are well resolved, resulting in the reduction of density shifts by many orders of magnitude.

In certain applications, such as transportable clocks, the use of a 3D lattice may be impractical. In a 1D lattice multiply occupied sites are difficult to avoid and the density shifts, even for fermions \citep{Lemke2011, Martin2013}, can be as high as several $10^{-16}$ if uncontrolled. The density of the sample can be lowered by distributing the atomic sample over many lattice sites in a cavity-enhanced lattice \citep{LeTargat2013, Nicholson2015}. This way the atomic density, and hence the density shifts, can be substantially reduced.

To evaluate the possible impact of the collisional shift, the density of the sample must first be evaluated. Atoms trapped in a harmonic potential have a Gaussian spatial density distribution and the figure of merit is a ``density-weighted density'' felt by the trapped atoms \citep{Swallows2012}
\begin{equation}
	\rho = \frac{1}{N_{\rm site}} \int n(\mathbf r)^2 d\mathbf r = \frac{N_{\rm site}}{(2\pi)^{3/2}L_x L_y L_z}\,,
\end{equation}
where $N_{\rm site}$ is the number of atoms per lattice site, while $L_{x,y,z}$ are the $1/e$ distribution widths in the $x$, $y$ and $z$ directions that are given by
\begin{equation}
	L_{x,y,z} = \sqrt{\left(2\left<v_{x,y,z}\right>+1\right) \frac{\hbar}{m \omega_{x,y,z}}}.
\end{equation}
Here $\left<v_{x,y,z}\right> = \left[\exp(-\hbar \omega_{x,y,z}/\kB T) -1\right]^{-1}$  is the average vibrational quantum number. This expression correctly describes the width of the Gaussian density distribution expected for both the lowest vibrational mode of the quantum harmonic oscillator (for $T\to 0$) and a thermal sample with finite~$T$. The density can be controlled by changing the average number of atoms per site $N_{\rm site}$, by changing the trapping frequencies using the trapping laser intensity and, in the case of the radial frequencies $\omega_{x,y}$, also by changing the beam waist.

For example, the JILA SrI clock \citep{Swallows2012} had $N_{\rm site} \approx 23$, $\omega_z = 2\pi \times 80$\;kHz, $\omega_x=\omega_y=2\pi \times 450$\;Hz and $T_{x,y,z} \approx \SI{3}{\micro\kelvin}$, leading to an estimated average density of $\rho = 3.9\times 10^{11}\;\mathrm{atoms/cm}^3$. On the other hand, a setup at LNE-SYRTE \citep{LeTargat2013} had atoms loaded into the optical lattice directly from a blue MOT using a deep lattice and drain lasers overlapped with the lattice to pump the atoms to metastable states. This reduces the density because the blue MOT has a larger cloud than a red MOT.
The parameters were $N_{\rm site}\approx 1.9$, $\omega_z = 2\pi \times 120$~kHz, $\omega_x=\omega_y=2\pi \times 250$~Hz and  temperatures $T_{x,y} \approx \SI{15}{\micro\kelvin}$ (after Doppler cooling in the transverse direction) and $T_z \approx \SI{4}{\micro\kelvin}$ (after sideband  cooling in the longitudinal direction).
The corresponding density is $\rho = 2.7\times10^9\;\mathrm{atoms/cm}^3$. It should be noted that at sufficiently low temperatures, the density in a 1D lattice scales as $\rho \propto T^{-1}$ (rather than $\rho \propto T^{-3/2}$ as expected for a trapped gas) due to the reduced dimensionality.

Depending on whether the atomic species used is a boson or a fermion, there are fundamental differences in the collisionally induced density shifts the atoms experience. Due to symmetry considerations, bosonic atoms in the same electronic and nuclear spin states may only collide in even partial waves --- corresponding to the rotational quantum number $l=0,2,4,\ldots\ $ (s-, d-, g-,\ldots\ waves, respectively). On the other hand, spin-polarised fermions in the same electronic states may only collide in odd partial waves p, f,\ldots\ (for $l=1,3,\ldots$, respectively). Collisions in all partial waves except for the s-wave are associated with a centrifugal barrier that grows as $\propto l(l+1)$; therefore, in the ultracold regime, only the lowest partial waves s and p are expected to be relevant because the collisional kinetic energy is too low to penetrate the barrier. In these two partial waves, the elastic collisions are completely described by the respective s- and p-wave scattering lengths $a$ and $b$. As an example we list these parameters for $^{87}$Sr in Table~\ref{tab:scatlen_sr}.

\begin{table}[b]
\centering
\caption{Scattering lengths in s-wave and p-wave cold collisions of ground and excited state $^{87}$Sr atoms.} \label{tab:scatlen_sr}
\begin{tabular}{l c r@{}l l}
  \toprule
	Channel & Partial wave & \multicolumn{2}{c}{Scattering length $a_0$} & \multicolumn{1}{l}{Reference} \\
    \midrule
	\hspace{10pt}$gg$    & s			  & 96&.2(1) & \cite{Martinez2008}; \\
    		& 			  &		&		& \cite{Stein2010a} \\
            & p			  & 74&.6(4) & \cite{Zhang2014a} \\
    \hspace{10pt}$ge^+$  & s 		  & 160&.0(5)$_{\rm stat}$(23)$_{\rm sys}$  & \cite{Goban2018}\\
            & p           & $-169$&(23)  & \cite{Zhang2014a} \\
    \hspace{10pt}$ge^-$  & s           & 69&.1(2)$_{\rm stat}$(9)$_{\rm sys}$ & \cite{Goban2018}\\
            & p 		  & $-42$&$^{+103}_{-22}$ & \cite{Zhang2014a}\\
    \hspace{10pt}$ee$	& s           & 176&(11)   & \cite{Zhang2014a}\\
            & p			  & $-119$&(18)  & \cite{Martin2013}\\
    \bottomrule
\end{tabular}
\end{table}

Most of the work on optical clocks thus far has concentrated on fermionic species because of their substantially reduced s-wave interactions and the expected ``freezing out'' of the p-wave interactions. It turns out, however, that there are at least three cold-collisional mechanisms that can contribute to density shifts in optical clocks that employ fermionic atoms in 1D optical lattices:
\begin{itemize}
	\item imperfect spin polarisation of the sample (s-wave),
    \item inhomogenous excitation due to beam misalignment (s-wave), and finally
    \item p-wave interactions.
\end{itemize}

In a mixture of atoms in ground $\ket{g}$ and excited $\ket{e}$ states there are four possible collision channels: a collision of two ground-state atoms $\ket{gg} = \ket{g_1} \otimes \ket{g_2}$, of two excited atoms $\ket{ee} = \ket{e_1} \otimes \ket{e_2}$, and two channels for collisions between ground-state and excited atoms in ``triplet'' $\ket{ge^+}$ and ``singlet'' $\ket{ge^-}$ states, where
\begin{equation}
\ket{ge^{\pm}} = \frac{1}{\sqrt{2}} \left( \ket{g_1}\otimes\ket{e_2} \pm \ket{e_1}\otimes\ket{g_2} \right) .
\end{equation}
Due to symmetry considerations for spin-polarised fermions, p-wave collisions are only possible in the $\ket{gg}$, $\ket{ee}$ and $\ket{ge^+}$ states. On the other hand, the ``singlet'' $\ket{ge^-}$ states allow s-wave collisions \citep{Campbell2009}. However, if the atomic sample is homogeneously driven by the laser field, the $\ket{ge^-}$ state remains unpopulated.

For distinguishable atoms, all partial waves contribute in all of these four collision channels. Therefore, spin-polarisation impurities in fermionic species can lead to s-wave density shifts. For a sample of distinguishable atoms, the shift of the atomic-clock transition frequency $\nu_{ge}$ is \citep{Campbell2009}
\begin{equation}
	\Delta \nu_{ge} = \frac{2\hbar}{m}\rho\left[a_{ge^+}(1-2f)+a_{ee}f-a_{gg}(1-f)\right] ,
\end{equation}
where $m$ is the mass of the colliding atoms, $\rho$ is the density of the atomic sample, $f$ is the excited fraction and $a_{xy}$ is the s-wave scattering length for the collision channel $\ket{xy}$. If a small fraction $\eta$ of the atomic sample is in a different spin state from the otherwise purely polarised sample, this could lead to a frequency shift of $\eta \Delta \nu_{ge}$. A simple estimate of an upper bound for this effect can be calculated if the respective scattering lengths and the density $\rho$ are known. The largest absolute value of this shift can occur at $f=0$ or $f=1$. For $^{87}$Sr this occurs at $f=0$, where $|a_{gg}-a_{ge+}|\approx 64 a_0$. For a completely unpolarised sample of density $\rho = 10^{11}\;{\rm cm^{-3}}$, this corresponds to a shift of about~0.5\;Hz, or a relative shift of about $1.2 \times 10^{-15}$. By reducing the density to below $\rho = 10^{10}\;{\rm cm^{-3}}$ and the spin impurities to below $\eta=1$\%, this shift can be reduced to below $10^{-18}$. As a real-world example, \citet{McGrew2018} report a spin purity of ${\geq} 99.5$\% in their $^{171}$Yb clock.

Even in a perfectly spin-polarised sample there is a possibility of s-wave density shifts originating from the $\ket{ge^-}$ collision channel if an excitation inhomogeneity is present \citep{Campbell2009, Swallows2011}. The inhomogeneity may be a result of slightly different Rabi oscillation frequencies for different atoms, for example due to a residual Doppler effect caused by angular misalignment between the probe and lattice laser beams. It should be noted that even if the laser beams were \emph{perfectly} aligned, there is a residual angular spread because a Gaussian laser beam always has some variation in its radius as it propagates. This effect can be eliminated by carefully aligning the probe laser beam and using large beam waists.

The p-wave density shifts were initially expected to be marginal at ultracold temperatures because the p-wave interaction energy scales as $V\propto k_T^2b^3$, where $k_T=\sqrt{2\pi m \kB T}/\hbar$ is the wavenumber corresponding to the thermal de Broglie wavelength \citep{Lemke2011}. On the other hand, in a 1D optical lattice $\rho \propto T^{-1}$ due to the reduced dimensionality. Therefore, for a fixed number of atoms, p-wave collisional shifts cannot be ``frozen out'' by lowering the sample temperature \citep{Zhang2014a}, but they can be controlled by lowering the density by other means. Also, the p-wave density shift can be suppressed by up to about two orders of magnitude by carefully controlling the excitation fraction \citep{Lemke2011, Martin2013}. Within the mean field approximation, the density shift $\Delta\omega$ is a linear function of the excitation fraction $f$ \citep{Martin2013},
\begin{equation}
	\Delta\omega = N \left[ 2\chi (f - 1/2) + C\right] .
\end{equation}
Here $N$ is the total number of atoms, whereas $\chi = V_{gg}+V_{ee}-2V_{ge^+}$ and $C = V_{ee}-V_{gg}$ are p-wave interaction parameters. The values $V_{gg}$, $V_{ge^+}$, and $V_{ee}$ are proportional to the respective p-wave scattering lengths cubed. The proportionality factor depends, among other things, on the trap geometry and the spatial distribution of atoms and could, in principle, be calculated theoretically using the theory in \citet{Martin2013, Rey2014}. In practice, however, $\chi$ and $C$ can be measured by varying the number of atoms in the trap and using Ramsey spectroscopy to measure the density shift using the local oscillator as a short-term frequency reference \citep{Lemke2011, Martin2013}. For a fixed atom number $N$, both $\chi$ and $C$ are independent of the sample temperature \citep{Zhang2014a}.

In principle, if the density shift in a given setup can be constrained to be on the order of $10^{-16}$ and the excitation fraction $f$ can be controlled to within 1\%, then the density shift could be further suppressed to 10$^{-18}$. This requires the use of Ramsey spectroscopy. The initial Ramsey pulse has to be tailored so that the excitation fraction $f$ is at a value
\begin{equation}
	f = \frac{1}{2}\left(1-\frac{C}{\chi}\right) = \frac{1}{2}\left(1-\frac{b^3_{ee}-b^3_{gg}}{b^3_{gg}+b^3_{ee}-2b^3_{ge^+}}\right) .
\end{equation}
This result is universal: it does not depend on the trap geometry, but only on the p-wave scattering lengths. For example, using the scattering data in Table~\ref{tab:scatlen_sr} one can calculate that the collisional shift is suppressed in $^{87}$Sr for $f\approx 0.625$. For an $^{171}$Yb clock, this technique enabled \citet{Ludlow2011} to reach density shifts at the 10$^{-18}$ level despite a relatively high density of $\rho\approx3\times10^{10}\;\mathrm{cm}^{-3}$.
Finally, we note that a suppression of density shifts in Rabi spectroscopy is also possible by care\-fully tailoring the length of the probe pulse. For details, see \citet{Lee2016}.

\subsection{Background-gas collisions \label{sec:latticeshift_bgcoll}}

The collisional shift of a clock transition due to the residual background gas is governed by the long-range interactions, which can be approximated by the van der Waals potential \citep{Gibble2013,Gibble2013a}.
In optical atomic clocks the background gas consists mostly of hot clock atoms that were not trapped and molecular hydrogen outgassing from the walls of the vacuum system.
In Section~\ref{sec:coherence_back}, we determined all the $C{}_6$ coefficients needed to estimate the ratio of the relative collisional shift, $\Delta \nu_\mathrm{bg}/\nu_0$, to the collisional loss rate of clock atoms, $\Gamma_\mathrm{bg}$.
This ratio can be calculated with the expression given by \citet{Gibble2013,Gibble2013a}
\begin{equation}
\frac{\Delta \nu_\mathrm{bg}}{\nu_0 \Gamma_\mathrm{bg}} = \frac{\Delta C_6}{13.8 \pi C_6 \nu_0},
\end{equation}
where $\Delta C_6$ is the difference between the $C_6$ coefficients of the ground and exited states of the clock transition.

Only the collisions that do not lead to clock-atom loss from the optical lattice will contribute  to the collisional shift. Moreover, the optical transition requires the involvement of atoms that are in both the ground and excited states. To estimate the loss of atoms in a spectroscopic clock experiment, the interactions that have longer range and lead to greater losses should be used. For the collisions considered here, this corresponds to the $C_6$ coefficients of interactions between background gases and clock atoms in the excited states.

Experimentally, the background-gas collisional shift can be measured by artificially increasing the residual H${}_2$ pressure, for instance by heating a non-evaporable getter (NEG) pump to induce outgassing. With residual gas analysis, it has been verified at NIST  that the gas released by the NEG pump  is ${>}95$\% H${}_2$. The ratios for Yb--H${}_2$ and Sr--H${}_2$ were measured with this method at NIST \citep{McGrew2018} and LNE-SYRTE, respectively.

The theoretical and experimental values of the ratio $\Delta \nu_\mathrm{bg}/(\nu_0\Gamma_\mathrm{bg})$ in Yb and Sr optical lattice clocks are summarised in Table \ref{tab:backratio}.
The uncertainties of the theoretically calculated values provided in Table \ref{tab:backratio} reflect only the influence of the uncertainties of the $C_6$ coefficients.

\begin{table}[t]
\begin{center}
\caption{Theoretical and experimental values of the ratio $\Delta \nu_\mathrm{bg}/(\nu_0\Gamma_\mathrm{bg})$ in Yb and Sr optical lattice clocks.  Uncertainties of the theoretically estimated values reflect the influence of the uncertainties of the $C_6$ coefficients only.
   } \label{tab:backratio}
\begin{tabular}{l S[table-format=+4.4] S[table-format=+1.6e+2] S[table-format=+1.5e+2] l}
\toprule
System & {$\max{C_6}/\text{a.u.}$} & {$[\Delta \nu_\mathrm{bg}/(\nu_0\Gamma_\mathrm{bg})]/\mathrm{s}$} & {$[\Delta \nu_\mathrm{bg}/(\nu_0\Gamma_\mathrm{bg})]/\mathrm{s}$} & \multicolumn{1}{l}{Reference}   \\
 & & {theory} & {experiment} & \\
\midrule
Yb--Yb & -2561(95)  &  -1.08(13)e-17 & & \\
Yb--H${}_2$ & -217(22) &  -1.46(44)e-17 & -1.64(12)e-17 & \cite{McGrew2018} \\
Sr--Sr & -4013(50) &  -1.137(55)e-17 & & \\
Sr--H${}_2$ &-212(22) &  -1.17(62)e-17 & -3.0(3)e-17 & LNE-SYRTE \\
\bottomrule
 \end{tabular}
\end{center}
\end{table}

\clearpage

\graphicspath{{D6_BBR/}}
\chapter[Controlling the thermal environment for optical clocks]{Controlling the thermal environment for optical \\ clocks} \label{chap:BBR}

\authorlist{Marco Pizzocaro$^{1,\dag}$,
Piotr Ablewski$^{2}$,
Ali Al-Masoudi$^3$,
Petr Balling$^4$,
Geoffrey Barwood$^5$,
Marcin Bober$^2$,
Miroslav Dole\v{z}al$^4$,
S\"oren D\"orscher$^3$,
Thomas Lindvall$^6$,
Christian Lisdat$^3$,
Roman Schwarz$^3$ and
Micha\l{} Zawada$^{2}$
}

\affil{1}{\INRIMaff}
\affil{2}{\UMKaff}
\affil{3}{\PTBaff}
\affil{4}{\CMIaff}
\affil{5}{\NPLaff}
\affil{6}{\VTTaff}
\corr{m.pizzocaro@inrim.it}

\chapstart
The blackbody radiation (BBR) emitted by the environment surrounding the atoms in optical clocks causes an important shift of their frequency. This chapter explains how to control the thermal environment in both optical lattice clocks and ion clocks and the techniques necessary to evaluate the BBR shift uncertainty to $\num{1e-18}$. In Section \ref{sec:BBR_intro} a general introduction to blackbody radiation is given. Section \ref{sec:BBR_atomicresponse} presents the theory of the atomic response to the BBR necessary to calculate the BBR shift. The role of emissivities and thermal inhomogeneity will be discussed in Section~\ref{sec:BBR_emissivity}.
Finally, practical techniques for controlling the thermal environment of ion clocks will be presented in Section \ref{sec:BBR_ions} while Section \ref{sec:BBR_lattice} presents practical considerations for optical lattice clocks.

\section{Blackbody radiation} \label{sec:BBR_intro}

The electromagnetic energy per unit volume at angular frequency $\omega$ in equilibrium at temperature $T$ is given by Planck's law \citep{Cohen-Tannoudji1977}
\begin{equation}
    u(\omega)\dd\omega = \frac{\hbar}{\pi^2 c^3} \frac{\omega^3\dd\omega}{e^{\hbar\omega/k\ped{B} T} - 1 },
\end{equation}
where  $u(\omega)$ is the spectral energy density, $c$ the speed of light, $\hbar = h/2\pi$ the reduced Planck constant and $k\ped{B}$ the Boltzmann constant.
Since the electric-field spectral density $E(\omega)$ is related to the spectral energy density by
\begin{equation}
    u(\omega)\dd \omega = \epsilon_0E^2(\omega)\dd\omega,
\end{equation}
where $\epsilon_0$ is the electric constant, we can write
\begin{equation}\label{eq:planck}
    E^2(\omega)\dd\omega = \frac{\hbar}{\pi^2\epsilon_0c^3} \frac{\omega^3\dd\omega}{e^{\hbar\omega/k\ped{B} T} - 1 }.
\end{equation}
The mean square electric field can then be obtained by integration
\begin{equation} \label{eq:BBR_S-B1}
    \langle E^2(t) \rangle = \int_0^{\infty} E^2(\omega)\dd\omega = \frac{4\sigma T^4}{\epsilon_0 c},
\end{equation}
where $\sigma = \pi^2 k\ped{B}^4/(60\hbar^3c^2)$ is the Stefan--Boltzmann constant.
Expanding the numerical values for the constants \citep{Mohr2012}, the last relation can be written
\begin{equation}\label{eq:bbr-rms-field}
    \langle E^2(t) \rangle = \left[\SI{8.319430(15)}{V/cm}\right]^2 \left(\frac{T}{\SI{300}{K}}\right)^4.
\end{equation}
The maximum of $E^2(\omega)$ is given by Wien's displacement law
\begin{equation}\label{eq:wien}
    \omega\ped{max} \approx 2.281 \frac{k\ped{B}}{\hbar} T.
\end{equation}
At $T = \SI{300}{K}$, Equation (\ref{eq:wien}) gives $\omega\ped{max} =2\pi\times\SI{17.6}{THz}$.

While the electric-field spectral density $E(\omega)$ is useful for calculating the atomic response to the BBR field (Section \ref{sec:BBR_atomicresponse}), other spectral densities are more useful when dealing with heat transfer (Section \ref{sec:BBR_emissivity}) or thermal imaging (Section \ref{sec:BBR_thermal_imaging}). Moreover, many applications prefer to use spectral densities in units of wavelength $\lambda = 2\pi c/\omega$. Because equations like \eqref{eq:planck} include differential quantities, these should also be considered while doing the substitution. For example, Planck's law for spectral radiance of a blackbody at temperature $T$, i.e., the radiated power per unit area of the body, per unit solid angle, per unit wavelength, is \citep{Rybicki2008}
\begin{equation} \label{eq:BBR_radiance}
B(\lambda,T) = \frac{2 h c^2}{\lambda^5} \frac{1}{e^{hc/(\lambda \kB T)}-1}.
\end{equation}
Its integral is proportional to $T^4$, $B_\text{tot}(T) = \int_0^\infty B(\lambda,T) \dd\lambda = \sigma T^4 / \pi$.
The total power emitted by a blackbody with area $A$ at temperature $T$ can instead be obtained by integrating (Stefan--Boltzmann's law)
\begin{equation}\label{eq:sb}
P = A \int_0^\infty B(\lambda,T) \dd\lambda \int \cos\theta\dd\Omega = A \sigma T^4,
\end{equation}
where $\Omega$ is the solid angle. The cosine of $\theta$, the angle between the light propagation and the surface normal, appears because blackbody radiation is diffuse (Lambert's cosine law).

\section{Atomic response to the BBR field} \label{sec:BBR_atomicresponse}

The BBR electric field causes a temperature-dependent shift in the transition frequency of atoms and ions by the frequency-dependent Stark shift \citep{Itano1982}.
The frequency shift between the states $e$ and $g$ of an atom is given by
\begin{equation}\label{eq:bbr-full}
    \Delta\nu_{eg} = -\frac{1}{2h} \int_0^{\infty} \Delta\alpha_{eg}(\omega) E^2(\omega)\dd \omega,
\end{equation}
where $\Delta\alpha_{eg}(\omega) = \alpha_e(\omega) - \alpha_g(\omega)$ is the differential frequency-dependent dipole polarisability for the transition.
Theoretically the polarisability for the state $i$ can be expressed as
\begin{equation}\label{eq:polarizab}
    \alpha_i(\omega) = \frac{2}{\hbar}\sum_{j\neq i} |d_{ij}|^2 \frac{\omega_{ij}}{\omega_{ij}^2 - \omega^2},
\end{equation}
where $d_{ij}$ and $\omega_{ij}$ are the dipole-moment matrix element and transition frequency for the transition between the states $i$ and $j$. The summation runs over every state with dipole coupling to the state $i$.

\begin{figure}[tb]
    \centering
    \includegraphics[width=0.8\columnwidth]{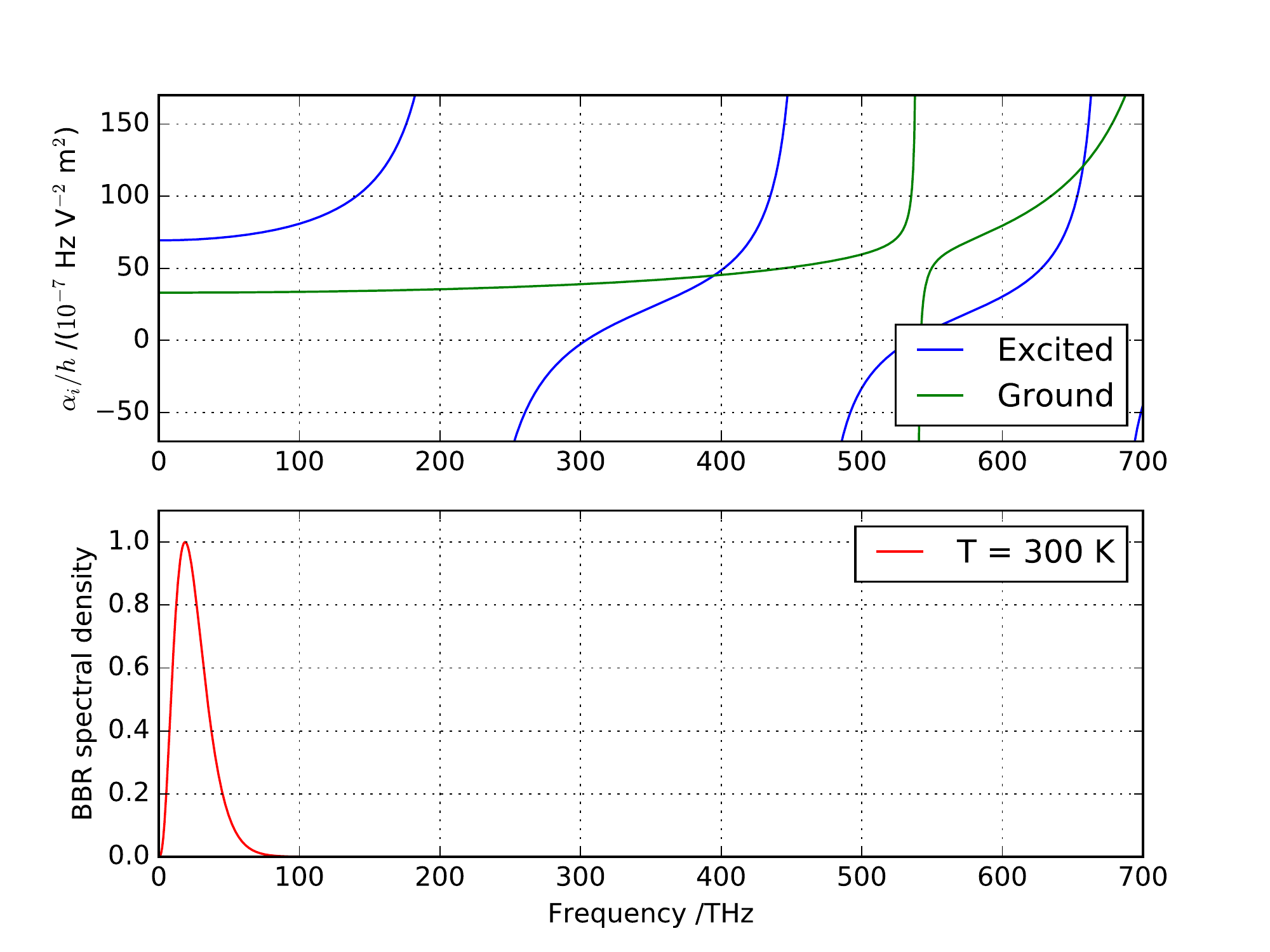}
    \caption{Polarisabilities of the ground and excited clock states for Yb compared with the normalised BBR spectral density at $T = \SI{300}{K}$.}
    \label{fig:yb-bbr}
\end{figure}

For most optical clocks the peak frequency of the BBR radiation $\omega\ped{max}$ at room temperature, Equation~(\ref{eq:wien}), is small compared to the transition frequencies that appear in Equation~(\ref{eq:polarizab}). For example, Figure~\ref{fig:yb-bbr} shows the polarisability of the ground and excited clock states for Yb atoms up to $\omega = 2\pi \times\SI{700}{THz}$ compared to the normalised spectral density for a blackbody at $T=\SI{300}{K}$. The dispersive shapes corresponding to the low-frequency transitions in Equation~(\ref{eq:polarizab}) can be recognised.  Also note the crossing between the two polarisabilities at $\SI{\approx 400}{THz}$ that is the magic frequency for Yb lattice clocks \citep{Derevianko2011}.
Assuming $\omega\ped{max} \ll \omega_{ij}$, Equation~(\ref{eq:bbr-full}) can be simplified as \citep{Porsev2006}
\begin{equation}\label{eq:bbr-short}
     \Delta\nu_{eg} = -\frac{\Delta\alpha_{eg}(0)}{2h}   \langle E^2(t)\rangle [1+\eta(T)],
\end{equation}
where $\Delta\alpha_{eg}(0)$ is the \textit{static} differential polarisability between the $e$ and $g$ states and a small dynamic correction factor $\eta(T)$ accounts for the residual frequency dependence of $\Delta\alpha_{eg}(\omega)$.

Given the importance of these parameters in the evaluation of the BBR shift of Equation~(\ref{eq:bbr-short}), considerable efforts have been made to determine their values either theoretically \citep{Porsev2006,Hachisu2008,Lea2006} or experimentally \citep{Middelmann2012,Sherman2012,Beloy2014,Dube2014,Huntemann2016}. Table~\ref{tab:atomic} summarises the differential static dipole polarisabilities  and dynamic corrections for the clock transitions of some atomic species used in optical clocks.
The table reports the polarisabilities divided by the Planck constant, $\Delta\alpha_{eg}(0)/h$, using the units of the International System (SI).
Conversion factors to other units, including the popular atomic units (a.u.), can be found in \citet{Mitroy2010}.

\begin{table}[bh]
    \centering
    \caption{Atomic response to BBR for the clock transitions of some atomic species used in optical clocks.}
    
\resizebox{\textwidth}{!}{%
\begin{tabular}{ 
l
S[table-format=+2.7, tight-spacing=true]
S[table-format=+1.8, tight-spacing=true, table-comparator=true, table-align-comparator = false]
l 
}
\toprule
Atom    & {$\Delta\alpha_{eg}(0)/h$}    & {$\eta(\SI{300}{K})$}   & References \\
        &   {$/(\SI{e-7}{Hz V^{-2} m^2})$}           & &\\
\midrule
Hg      &   5.2(5)          & < 0.02                     & \cite{Hachisu2008, Tyumenev2016}\\
Sr      &   61.5558(17)     & 0.06980(33)           & \cite{Middelmann2012, Nicholson2015}\\
Yb      &   36.2612(7)      & 0.01804(38)           & \cite{Sherman2012,Beloy2014}\\
Sr$^+$  &   -7.235(11)       & -0.00951(15)          & \cite{Dube2014,JiangD2009}\\
Yb$^+$(E2)  &  7.85(24)    &              & \cite{Lea2006}\\
Yb$^+$(E3)  & 1.340(24)     & -0.0015(7)             & \cite{Huntemann2016}\\
\bottomrule
\end{tabular}
}

    \label{tab:atomic}
\end{table}

The static contributions for neutral atoms have been determined experimentally by measuring the differential polarisability in a static electric field for the $(5\mathrm{s}^2)\, {}^1\mathrm{S}_0 \rightarrow (\mathrm{5s5p)\, {}^3P}_0$ transition in strontium \citep{Middelmann2012} and the corresponding transition in ytterbium \citep{Sherman2012}. For ion clocks the static contributions have been measured by using infrared lasers \citep{Huntemann2016} or by comparing the micromotion-induced scalar Stark shift (if positive) to the time-dilation shift \citep{Dube2014}.

\begin{figure}[tbp]
	\centering
	\includegraphics[width=0.8\columnwidth]{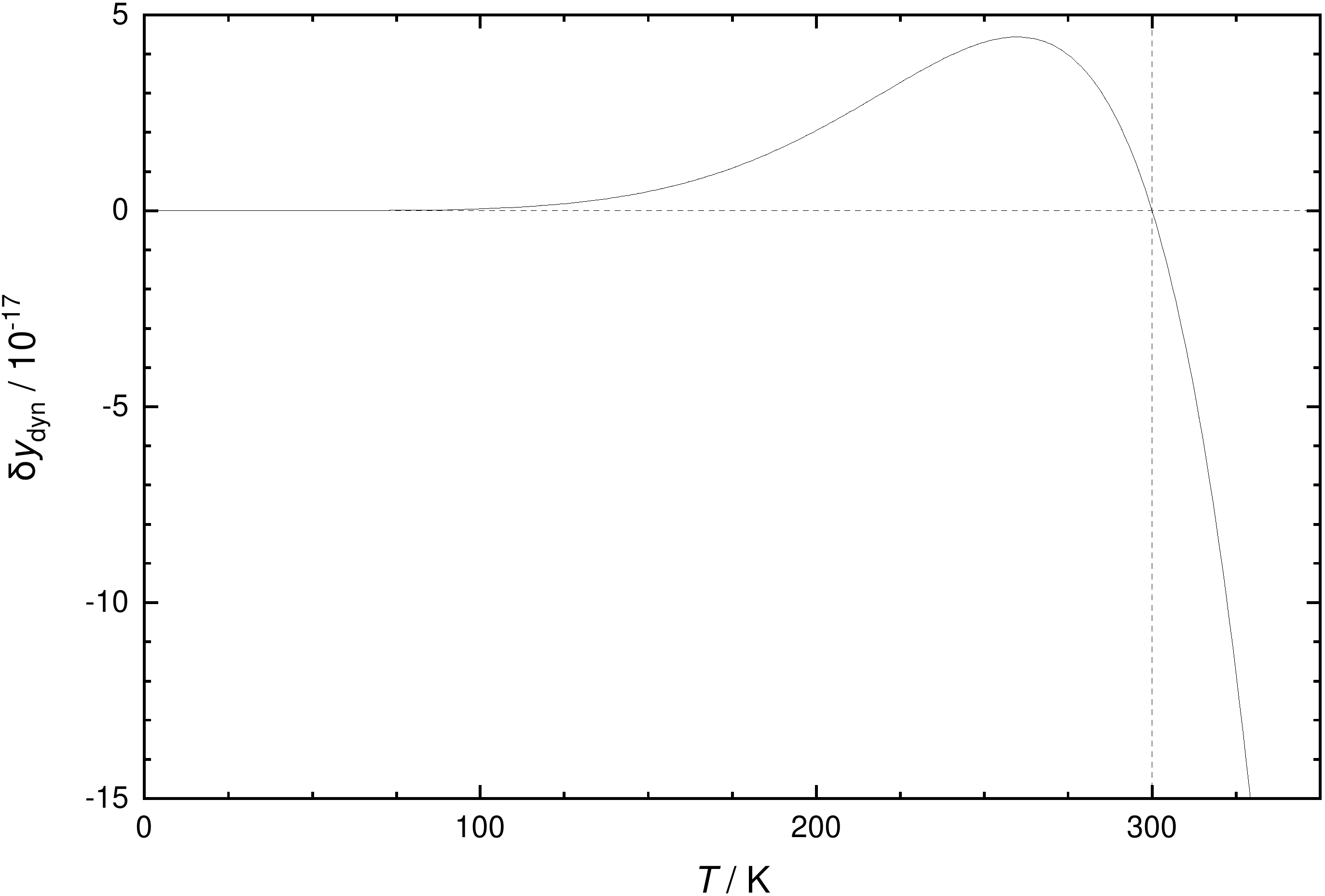}
	\caption{\label{fig:DynamicCorrection_LeadingOrderError} Fractional frequency difference $\delta y_{\mathrm{dyn}}$ between using only the leading-order $(T/T_0)^2$ term and including the $(T/T_0)^4$ term in Equation (\ref{eq:dyn}) to estimate the dynamic contribution to the BBR-induced frequency shift in strontium for $T_0 = \SI{300}{\kelvin}$. Approximation by the leading-order term underestimates the magnitude of the shift for $T>T_0$. Based on the results reported by \cite{Middelmann2013}.}
\end{figure}

The dynamic factor $\eta(T)$ can be expanded in a power series around a reference temperature $T_0$ \citep{Beloy2014,Middelmann2012,Middelmann2013}
\begin{equation}\label{eq:dyn}
   \eta(T) = \eta_2 \left(\frac{T}{T_0}\right)^2 + \eta_4 \left(\frac{T}{T_0}\right)^4 + \ldots
\end{equation}
Further note that approximating the dynamic contribution by the leading $T^2$ term is only appropriate within a small window around $T_0$. The $T^4$ term in Equation~\eqref{eq:dyn} is suppressed by a factor of about 30 for ytterbium lattice clocks \citep{Beloy2014}, but by less than an order of magnitude for strontium lattice clocks \citep{Middelmann2013}. Figure~\ref{fig:DynamicCorrection_LeadingOrderError} shows the difference between using only the leading  term and including the $T^4$ term to estimate the dynamic contribution for the case of strontium. The allowed deviation from $T_0 = \SI{300}{\kelvin}$ for a fractional error below $1\times10^{-18}$ is less than \SI{4}{\kelvin}; higher-order terms in Equation (\ref{eq:dyn}) must thus  in many cases be taken into account in lattice clocks operated around room temperature.

\begin{figure}[tbp]
	\centering
	\includegraphics[width=0.8\columnwidth]{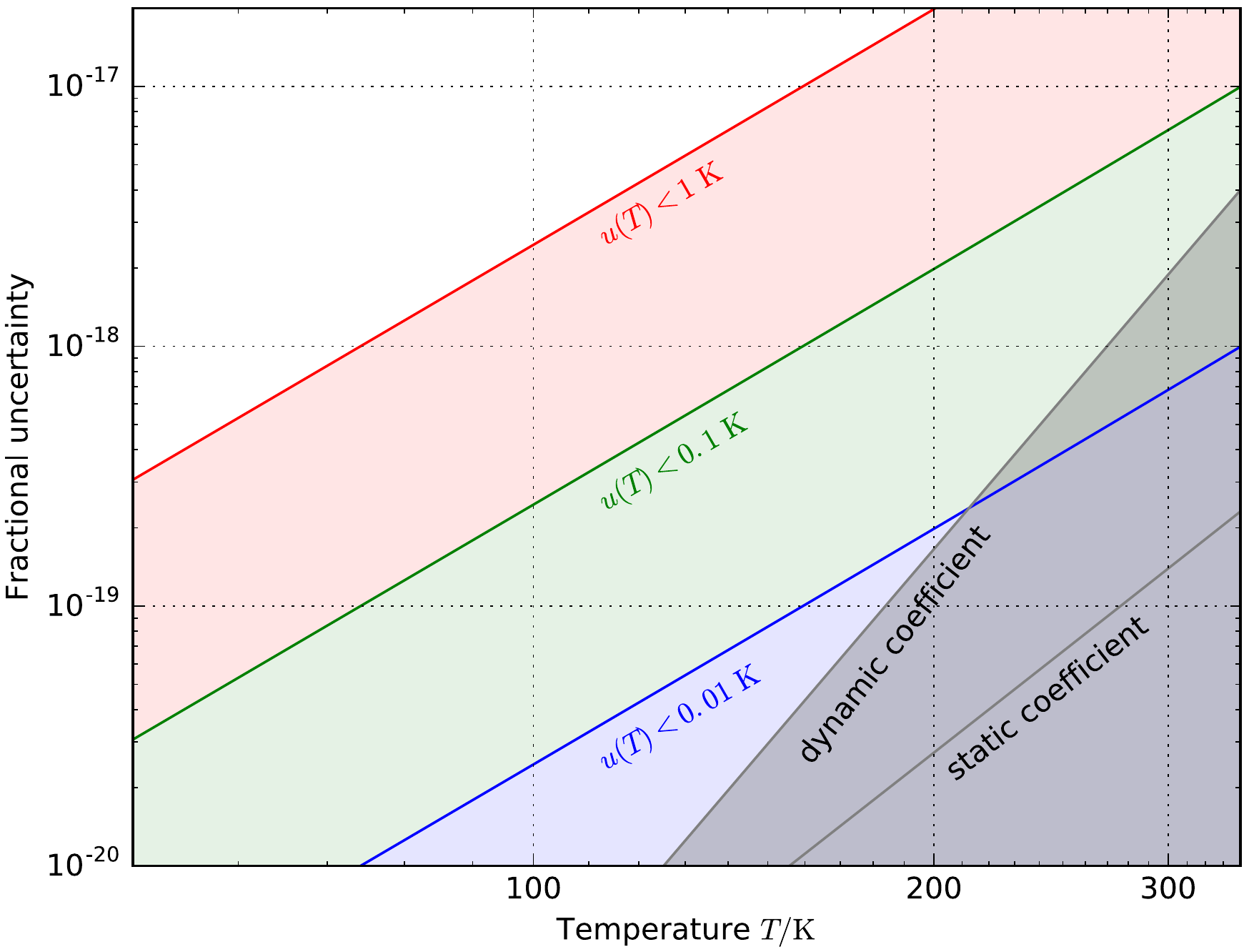}
	\caption{\label{fig:BBRTemperatureUncertainty} Fractional uncertainty of the BBR frequency shift of Sr due to the uncertainty of the representative temperature $T$ and the atomic response coefficients as a function of $T$.}
\end{figure}

Figure~\ref{fig:BBRTemperatureUncertainty} reports the uncertainty of the BBR shift in Sr clocks  due to the uncertainty of the representative temperature $T$ and the atomic response coefficients as a function of $T$.
In a similar fashion, Table \ref{tab:uncs} reports the uncertainty of the BBR shift for different optical clocks at different temperatures using Equation~\eqref{eq:bbr-short} and the data in Table \ref{tab:atomic}. The uncertainty is calculated at room temperature with different temperature uncertainties and at liquid nitrogen temperature with an uncertainty of \SI{1}{K}. For the Yb and Sr lattice clocks, a fractional uncertainty of $\num{1e-18}$ requires temperature uncertainties of about \SI{10}{mK} at room temperature or cryogenic environments, where the requirements on temperature uncertainty are relaxed. Owing to the lower value of the static differential polarisability,  Yb$^+$(E3) clocks can achieve BBR uncertainty at the level of $\num{1e-18}$ with 1-K uncertainties at room temperature. The Sr$^+$ clock falls between the lattice clocks and Yb$^+$(E3); a 100-mK temperature uncertainty at room temperature corresponds to $\num{1e-18}$ fractional uncertainty with approximately equal contributions from the temperature and the static differential polarisability. Even for the ion clocks, the uncertainty on the temperature plays a critical role and this is mostly coming from the inhomogeneity of the thermal radiation in the presence of thermal gradients, as explained in the next section.

\begin{table}[b]
\centering
\sisetup{table-format=3.1e2,table-number-alignment = center,table-column-width=2.2cm}
    \caption{Uncertainty from the BBR shift for Yb, Sr, Sr$^+$ and Yb$^+$(E3) clocks assuming different temperatures and temperature uncertainties, considering the uncertainty contributions from temperature, static and dynamic coefficients.}
	\begin{tabular}{SSSSS}
	\toprule
	{$T /\si{K}$}	&	{Sr}	&	{Yb}	&	{Sr$^+$}   & {Yb$^+$(E3)}\\
	\midrule
	297(1)	    & 7.1e-17  & 3.2e-17	& 7.2e-18   &   1.6e-18\\
    297.0(1)    & 7.3e-18  & 3.3e-18    & 1.1e-18   &   1.2e-18\\
	297.00(1)	& 1.6e-18  & 1.0e-18	& 0.8e-18   &   1.2e-18\\
	77(1)	    & 1.1e-18  & 0.6e-18	& 0.1e-18   &   1.7e-20\\
	\bottomrule
	\end{tabular}
	\label{tab:uncs}
\end{table}

\section{Emissivity and temperature inhomogeneity}\label{sec:BBR_emissivity}

For a body that is not perfectly black, a so-called graybody, the Stefan--Boltzmann law of Equation~(\ref{eq:sb}) is modified as
\begin{equation}\label{eq:BBR_graybody}
P = A \epsilon \sigma T^4,
\end{equation}
where we have introduced the emissivity $0 < \epsilon < 1$ of the body to quantify the difference with respect to a perfect blackbody. For a perfect graybody the emissivity $\epsilon$ is assumed  wavelength independent and reflections are assumed to be diffuse. For a general material it is possible to introduce a wavelength- and temperature-dependent emissivity $\epsilon(\lambda, T)$ in Equation~\eqref{eq:sb}, but in most cases the approximation of Equation (\ref{eq:BBR_graybody}) is sufficient.

\begin{figure}[tb]
    \centering
    \includegraphics[width = 0.52\textwidth]{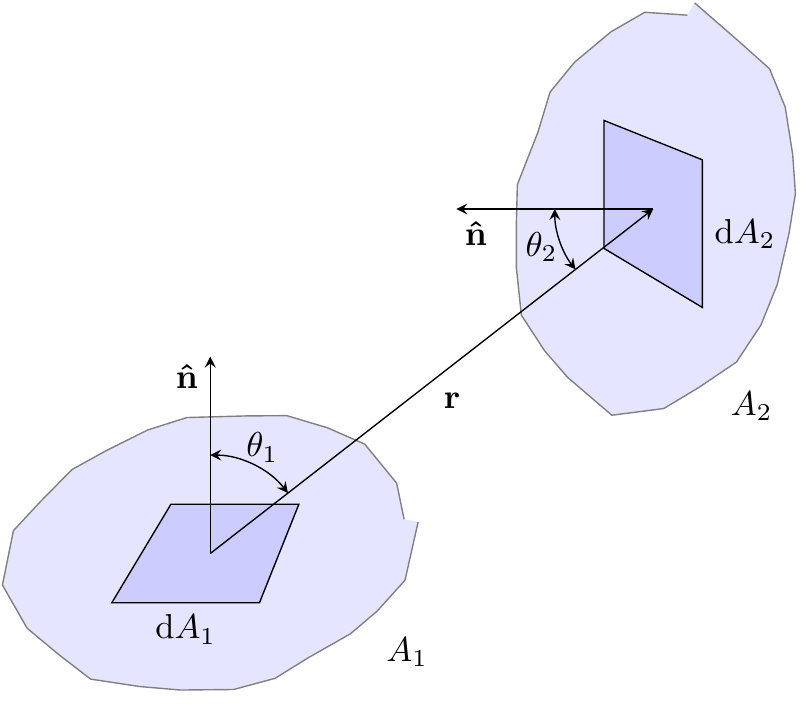}
    \caption{Geometry for the determination of the view factor between two surfaces.}
    \label{fig:view-factor}
\end{figure}

In general, the power radiating from a blackbody with temperature $T_1$ and area $A_1$ towards a second body with  temperature $T_2$ and area $A_2$ is
\begin{equation}
P_{1\rightarrow2} = A_1 F_{1\rightarrow2} \sigma T_1^4,
\end{equation}
where we have introduced the view factor $F_{1\rightarrow2}$ as the fraction of radiation leaving body 1 that reaches body 2 \citep{Cengel2015}.
With the geometry illustrated in Figure~\ref{fig:view-factor}, the view factor for infinitesimal areas can be expressed as
\begin{equation}
\dd F_{\dd A1\rightarrow \dd A2} = \frac{\cos\theta_1 \cos\theta_2}{\pi r^2} \dd A_2,
\end{equation}
where $r$ is the distance and $\theta_1$ and $\theta_2$ the angles between the normal to the surfaces and the ray between the two surfaces.
The total view factor $F_{1\rightarrow2}$ can be found by integration
\begin{equation}
F_{1\rightarrow 2} = \frac{1}{A_1}\int_{A_1}\int_{A_2}\frac{\cos\theta_1 \cos\theta_2}{\pi r^2} \dd A_1 \dd A_2.
\end{equation}
We note that the following reciprocity relation holds true
\begin{equation}
A_1 F_{1\rightarrow2} = A_2 F_{2\rightarrow1},
\end{equation}
so that the total net heat transfer from one body to the other can be written
\begin{equation}\label{eq:BBR_NetHeatBlack}
P = P_{1\rightarrow2} - P_{2\rightarrow1} = A_1 F_{1\rightarrow2} \sigma (T_1^4- T_2^4),
\end{equation}
where we have chosen the sign so that $P>0$ for $T_1>T_2$.
Similar equations can be derived for graybodies by introducing the proper emissivities, but in this case care has to be taken to also consider reflections from the surfaces. For two graybody surfaces  with different emissivity coefficients $\epsilon_1$ and $\epsilon_2$, Equation~\eqref{eq:BBR_NetHeatBlack} can be written as
\begin{equation}\label{eq:NetHeatBlack}
P  = \frac{\sigma \left( T_1^4 - T_2^4 \right)}{  \frac{ 1 - \epsilon_1}{ \epsilon_1 A_1 } + \frac{1}{A_1 F_{1 \rightarrow 2}} + \frac{1 - \epsilon_2}{\epsilon_2 A_2} }.
\end{equation}
These equations are the basis of radiative heat transfer. In the next section we will illustrate their use with a simple geometry,   while complex systems like optical clocks often need to be modelled by finite-element analysis \citep{Dolezal2015}.

\begin{figure}[t]
    \centering
    \includegraphics{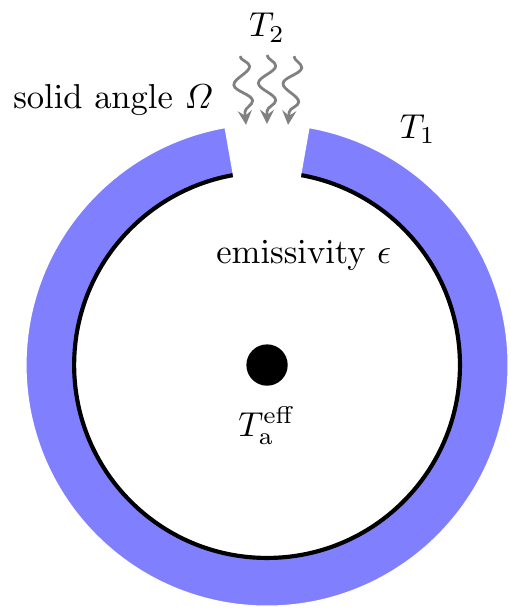}
    \caption{Simple model of a spherical vacuum chamber with temperature $T_1$ and emissivity $\epsilon$. A hole with solid angle $\Omega$ leaks BBR radiation at temperature $T_2$ inside the chamber. Atoms at the centre of the chamber are modelled as a small spherical blackbody with effective temperature $T\ped{a,eff}$.}
    \label{fig:toy-model}
\end{figure}

\subsection{Basic model of temperature inhomogeneity\label{sec:BBR_temperature_inhomogeneity}}

The role of temperature inhomogeneity can be better understood by analyzing the simple model depicted in Figure \ref{fig:toy-model}. A similar model is presented by \citet{Middelmann2011}. Atoms are located at the centre of a spherical vacuum chamber which is held at temperature $T_1$. The chamber wall has an emissivity $\epsilon$. BBR radiation at temperature $T_2$ leaks into the chamber through a hole which spans a solid angle $\Omega$ with respect to the position of the atoms. We want to calculate the effective temperature $T\ped{a,eff}$ seen by the atoms. To do so we model the atoms as a small, perfect blackbody with temperature $T\ped{a,eff}$ and we impose steady state. The power balance on the atoms should consider the following:
\begin{itemize}
    \item The BBR at $T\ped{a,eff}$ lost by the atoms.
    \item The BBR at $T_2$ leaking through the hole. The simple geometry allow us to calculate the view-factor between the atoms and the hole as $F_{a\rightarrow2} = \Omega/4\pi$.
    \item The BBR at $T_1$ from the chamber wall. The view-factor between the atoms and the wall is  $F_{a\rightarrow1} = (1 - \Omega/4\pi)$.
    \item The BBR at $T\ped{a,eff}$ lost by the atoms but reflected back from the wall with reflectivity $(1 - \epsilon)$ and reabsorbed.
    \item The BBR at $T_2$ leaking from the hole, reflecting from the wall and reaching the atoms. This contribution goes to zero as the size of the atom blackbody goes to zero for the simple geometry considered.
\end{itemize}

\begin{figure}[t]
    \centering
    \includegraphics[width=0.8\columnwidth]{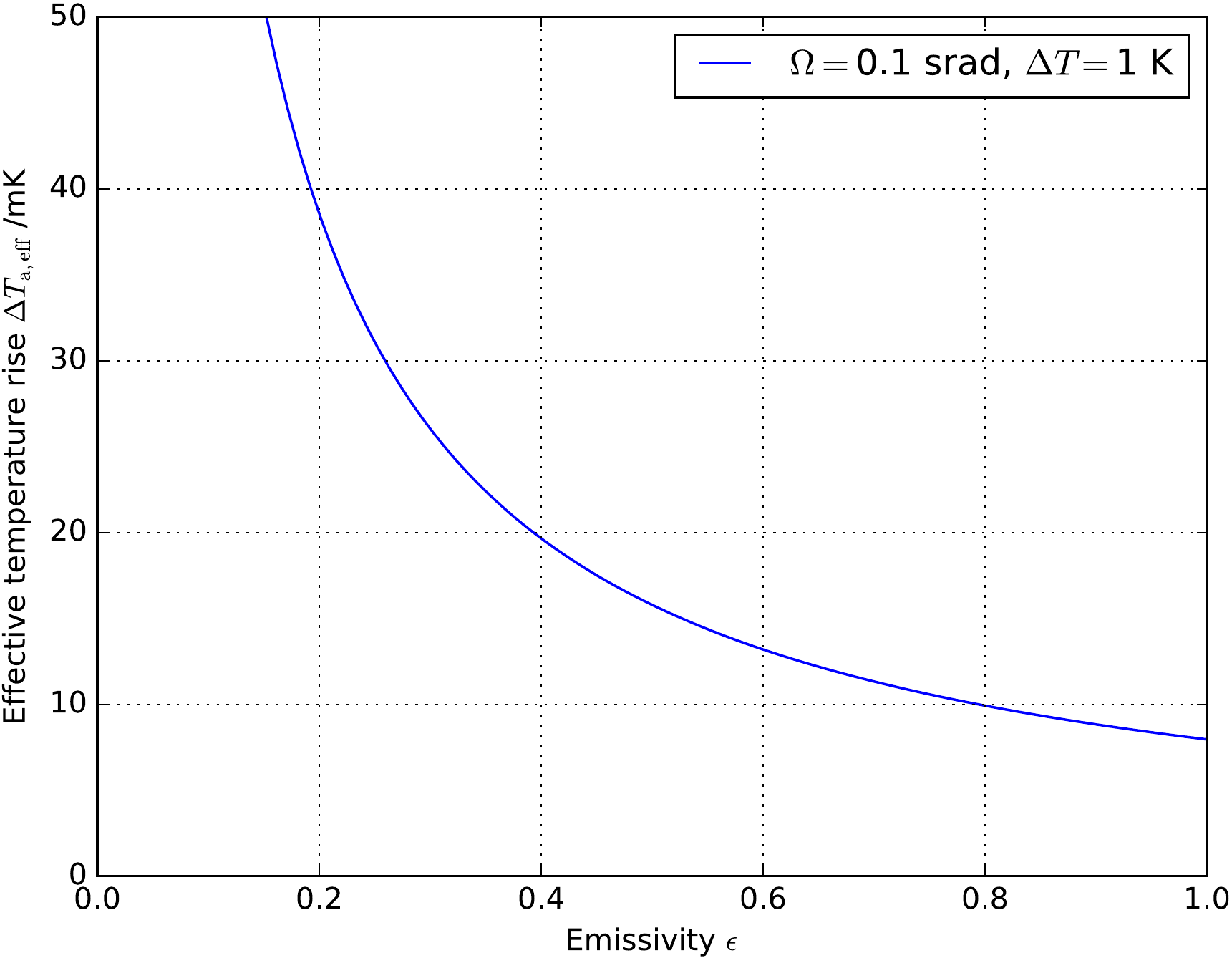}
    \caption{Temperature increase seen by the atoms as a function of the wall emissivity $\epsilon$ in the simple model of Figure \ref{fig:toy-model}, for $\Omega = \SI{0.1}{srad}$ and  $\Delta T = \SI{1}{K}$.}
    \label{fig:rise}
\end{figure}

Assuming the surface area of the atom blackbody is $A\ped{a}$ and making use of the reciprocity formula for the view factors, the steady-state condition implies
\begin{equation}
    -\sigma A\ped{a} T\ped{a,eff}^4 + \sigma A\ped{a}(\Omega/4\pi) T_2^4 + \sigma A\ped{a} (1-\Omega/4\pi) \epsilon T_1^4 + \sigma A\ped{a} (1-\Omega/4\pi) (1-\epsilon) T\ped{a,eff}^4 = 0.
\end{equation}
All prefactors to temperatures should add up to zero. Dropping $\sigma A\ped{a}$ from the equation and solving for $T\ped{a,eff}$ we obtain
\begin{equation}\label{eq:toy-model}
    T\ped{a,eff}^4 = \frac{\Omega T_2^4 + (4\pi - \Omega)\epsilon T_1^4}{\Omega + (4\pi - \Omega)\epsilon}.
\end{equation}
This equation can be interpreted as the weighted average of $T_1$ and $T_2$, with the effective solid angle for the hole at $T_2$ given by
\begin{equation}
   \Omega\ped{eff} = \frac{4\pi\Omega}{\Omega + (4\pi - \Omega)\epsilon}.
\end{equation}
Thus $\Omega\ped{eff} > \Omega$ for $\epsilon < 1$.
For a small temperature difference $\Delta T = T_2 - T_1$, Equation~\eqref{eq:toy-model} can be linearised to calculate the temperature increase seen by the atoms $\Delta T\ped{a,eff} = T\ped{a,eff} - T_1$
\begin{equation}
    \Delta T\ped{a,eff} \approx \frac{\Omega}{\Omega + (4\pi - \Omega)\epsilon}\Delta T.
\end{equation}
Figure~\ref{fig:rise} shows this temperature increase for $\Omega=\SI{0.1}{srad}$ and  $\Delta T = \SI{1}{K}$ as a function of $\epsilon$.

While the model of Figure \ref{fig:toy-model} is simplified, it represents well the problems encountered in real-life optical clocks, such as the leaking of BBR from hot sources (e.g., atomic ovens) or the apertures found in cryogenic fingers. The model suggests that to reduce the uncertainty coming from temperature inhomogeneity, an optical clock should be designed with as little as possible temperature gradients with line of sight to the atoms, with reduced solid angles to hot spots and with high-emissivity surfaces to reduce reflections. The knowledge of the surface emissivities is also important to predict the effective temperature seen by the atoms. Some of the techniques used in practice will be discussed in Sections \ref{sec:BBR_ions} and \ref{sec:BBR_lattice}.

\subsection{Maximum-entropy approach to temperature inhomogeneity}\label{sec:maximum-entropy}

A simple but powerful approach based on the principle of maximum entropy can be used if the maximum and minimum temperatures seen by the atoms are known, even when a complete thermal model of the system is unavailable.
This approach follows the ``Guide to the expression of uncertainty in measurement'' (GUM, \cite{BIPM2008a}) and has been used for several optical clocks, e.g., \cite{Falke2014,Pizzocaro2017}.

If the hottest and coldest points of the experiment within line of sight of the atoms are known, we can assume that the effective BBR temperature seen by the atoms lies between the two extremes $T\ped{max}$ and $T\ped{min}$. The principle of maximum entropy then prescribes that a rectangular probability density function between $T\ped{max}$ and $T\ped{min}$ should be used \citep{BIPM2008}, as this is the least informative choice one can make given the initial assumptions. The mean and standard deviation of the rectangular distribution are then given by
\begin{subequations}
\begin{eqnarray}
    T\ped{eff} &=& \frac{T\ped{min} + T\ped{max}}{2}, \\
    u(T\ped{eff}) &=& \frac{T\ped{max} - T\ped{min}}{\sqrt{12}}.
\end{eqnarray}
\end{subequations}
These can be used to evaluate the BBR shift. Note that this approach does not rely on the temperature distribution seen by the atoms (for example measured by different thermal sensors) but requires the hottest and coldest points to be identified (for example by thermal imaging) and their temperature to be measured.

For example, applying this method to the model of Figure \ref{fig:rise} with $T\ped{max} - T\ped{min} = \SI{1}{K}$ results in $\Delta T\ped{a,eff} = \SI{0.5(3)}{K}$.
The uncertainty is high compared to the calculation of the previous section but this result can be used without any knowledge of the emissivity and geometry of the system.

\subsection{Measuring material emissivities}\label{sec:BBR_emissivity-measures}

The emissivities of materials may be found tabulated, for example in \cite{Cengel2015}. In general, however, the emissivities of materials may depend on temperature, roughness and surface finish \citep{Wen2006} and on the state of oxidation for the case of metals. When the emissivity needs to be precisely known for FEM simulations or BBR-shift evaluation, it has to be measured directly.

In this section we present a method to measure emissivity coefficients and give values for the most popular vacuum-compatible materials used for vacuum chambers and thermal shields in optical atomic clocks. This method evaluates the total emissivity of the surface. While thermal imaging can be used to deduce emissivities, these are weighted by the spectral response of the camera (see Section \ref{sec:BBR_thermal_imaging}).

\begin{figure}[tb]
\centering
    \includegraphics[width=8.5 cm]{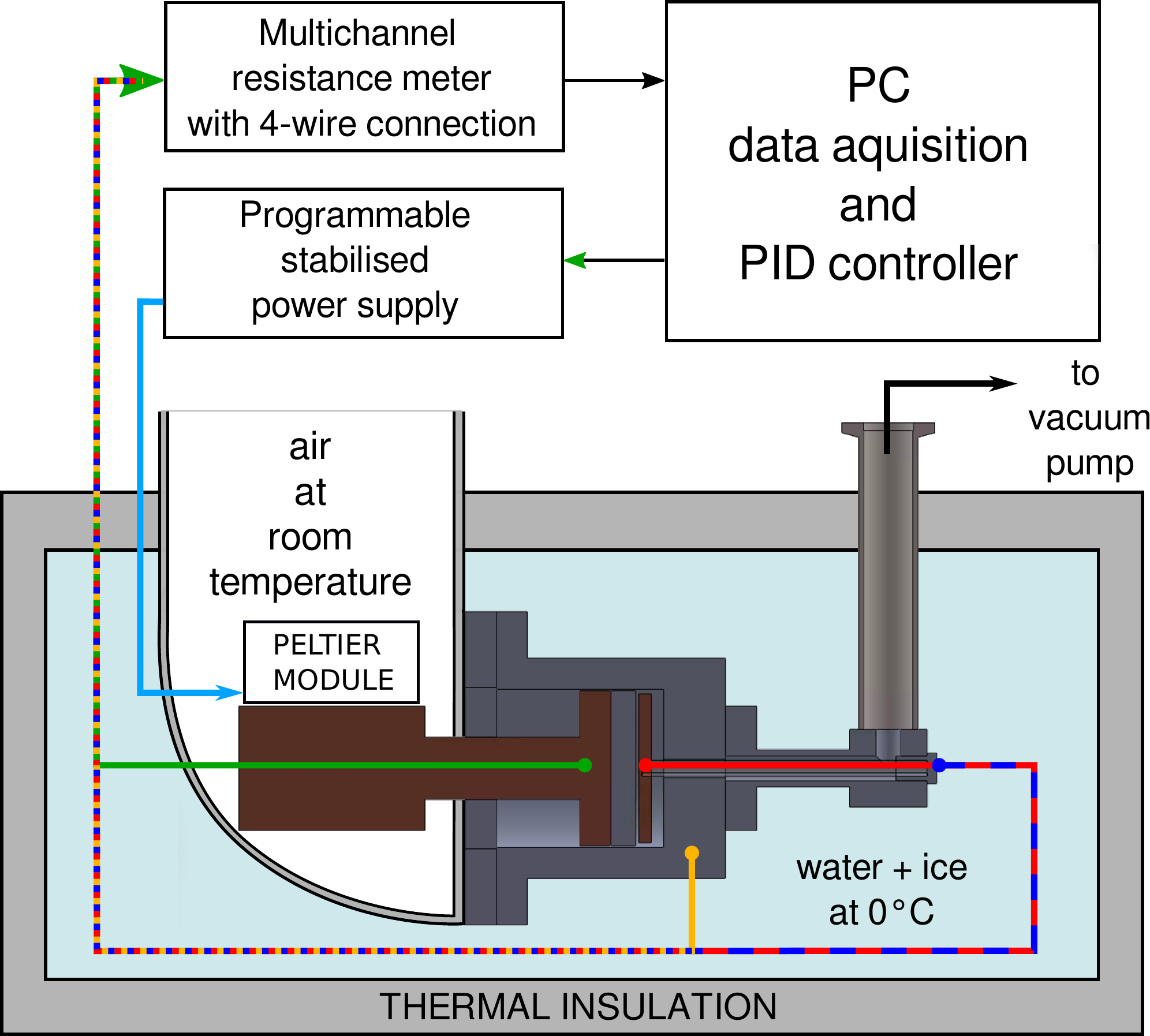}
    \caption{Schematics of the emissivity measurement setup. Pt100 sensors inside the heater near the sample (green wire), chamber (orange), thermometer plate (red) and glass tube (blue) are read by the resistance meter.
\label{fig:emiss_scheme}}
\end{figure}

To measure emissivity coefficients, a dedicated experimental setup has been designed and built at UMK, see Figure~\ref{fig:emiss_scheme}.
The temperatures of the internal surfaces were calculated with a FEM method for conductive heat transfer, while the radiative heat transfer was calculated analytically.
The setup was designed to maximise the view factor (see Figure~\ref{fig:view-factor}) between a measured sample and a calibrated thermometer plate. The measured samples are of cylindrical shape with a diameter of 50\;mm and a thickness of 8\;mm. The thermometer plate, a 4-mm-thick copper cylinder of diameter 48\;mm, is  placed opposite to the sample. It is mounted co-axially and in parallel with the sample, with a separation of 1\;mm. Thus, the diameter-to-separation ratio of the plates is around 50. The view factor between the plates is equal to 0.94, which is a good approximation of the ideal case of two infinite plates, which has a view factor of one.

The sample is mounted with an indium gasket to a copper heater, the temperature of which is controlled by a Peltier module. The copper element is made of Cu-ETP (electrolytic tough pitch) and has a 20-mm diameter at its smallest cross-section. Two thermometers for temperature stabilisation are installed inside the copper heater, close to the surface that stays in thermal contact with the sample.

The thermometer plate absorbs the thermal radiation from the nearby surface of the sample.
The heat conductivity of the copper thermometer plate dominates over radiation losses so that its temperature is homogeneous to within 1\;mK  at thermal equilibrium.
The sample and thermometer cylinders are enclosed in a vacuum chamber (pressure below 10$^{-4}$\;mbar) to ensure that conductive and convective heat transfer via air is negligible \citep{Hablanian1990}.

The thermometer plate is installed on a 90-mm-long glass tube with an inside diameter of 3\;mm and an outside diameter of 5\;mm. The glass tube and the copper heater are attached to the surrounding vacuum chamber by a Torr Seal compound gasket. Torr Seal epoxy has a low thermal conductivity of $0.435\;\mathrm{W/(K\,m)}$ and prevents conductive heat exchange between elements at different temperatures.
The surrounding chamber is  made of PA38/AW-6060 aluminium alloy with a high thermal conductivity of $200\ldots 220\;\mathrm{W/(K\,m)}$.

All surfaces inside the chamber, except the radiating surface of the sample plate, are covered with high-emissivity Krylon Ultra-Flat Black paint to eliminate reflections. This paint is vacuum compatible at the $10^{-6}$\;mbar level. The whole setup is submerged in a thermally insulated container filled with a mixture of distilled water and ice to ensure stable thermal conditions and to prevent heating of the chamber by thermal radiation from the measured sample. The only part that stays in contact with the room-temperature environment is a 20-cm-long stainless steel pipe used for connection to a turbomolecular pump.

Four Pt100 class 1/3B resistance temperature detectors (RTD) with phosphor-bronze four-wire connections are installed in the setup: one in the chamber, one  inside the thermometer plate, one inside the sample heater and one inside the glass tube that holds the thermometer plate at the point where the tube is mounted to the vacuum chamber, see Figure~\ref{fig:emiss_scheme}. The hysteresis of the Pt100 RTDs upon heating and cooling is below 1\;mK.
The thermometers were calibrated against a master thermometer\footnote{Fluke 1595A Super-Thermometer with Fluke 5640 high-accuracy thermistor.}, which was calibrated in the Polish Central Office of Measures with $3.5$-mK uncertainty. The total Pt100 calibration uncertainty is less than $10$~mK in the temperature range from $0.1\;^{\circ}$C to $35\;^{\circ}$C.

\begin{figure}[tbp]
\centering
    \includegraphics[width=0.45\columnwidth]{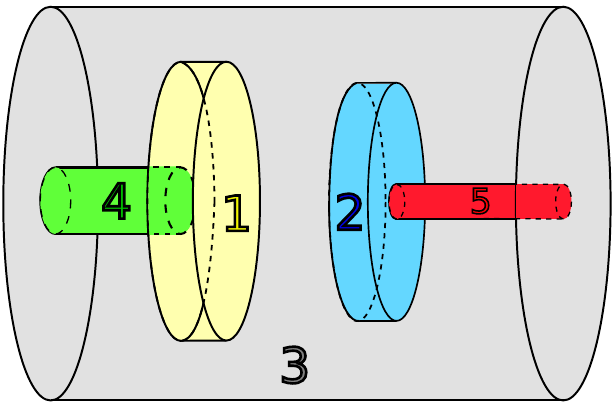}
        \caption{Schematics of the bodies exchanging heat in the measurement setup. 1: sample, 2: thermometer plate, 3: internal surfaces of the vacuum chamber, 4: sample heater, 5: thermometer holder. }
        \label{fig:SetupSurfaces}
\end{figure}

The analysis of the measurements that were carried out requires calculations of the view factors between each pair of surfaces of the setup shown in Figure~\ref{fig:emiss_scheme}. Due to the simple geometry of the experimental system (co-axial cylinders and parallel flat surfaces, see Figure~\ref{fig:SetupSurfaces}), the relevant view factors can be  calculated analytically.

The process of modelling the heat transfer in the experimental setup must be performed for the three main steps in the measurement procedure:
\begin{enumerate}
\item calibration of the conductive heat transfer via thermometer holder to the chamber,
\item calibration of the emissivity coefficient of the chamber's internal coating, and
\item measurement of the sample emissivities.
\end{enumerate}
The two calibration steps are necessary to calculate the conductive heat transfer via the thermometer holder and the emissivity coefficient of the chamber’s internal coating and their uncertainties, which are required for the measurement data analysis.

To determine the effective heat-transfer coefficient of the glass tube acting as the thermometer-plate holder, an analytical heat-transfer model has been constructed.
The holder (element 5 in Figure \ref{fig:SetupSurfaces}) is a glass tube filled with air. The combination of two parallel, independent heat transfers with different conductive heat-transfer coefficients can be substituted with the total heat transfer with an effective heat-transfer coefficient $k_\mathrm{eff}$ and the holder can then be treated as a solid rod. The total heat transfer needs to include heat added to the system via the resistive wire mounted in the thermometer plate (2 in Figure \ref{fig:SetupSurfaces}) and all possible heat losses: conductive, convective and radiative.

\begin{figure}[tb]
\centering
    \includegraphics[width=0.5\columnwidth]{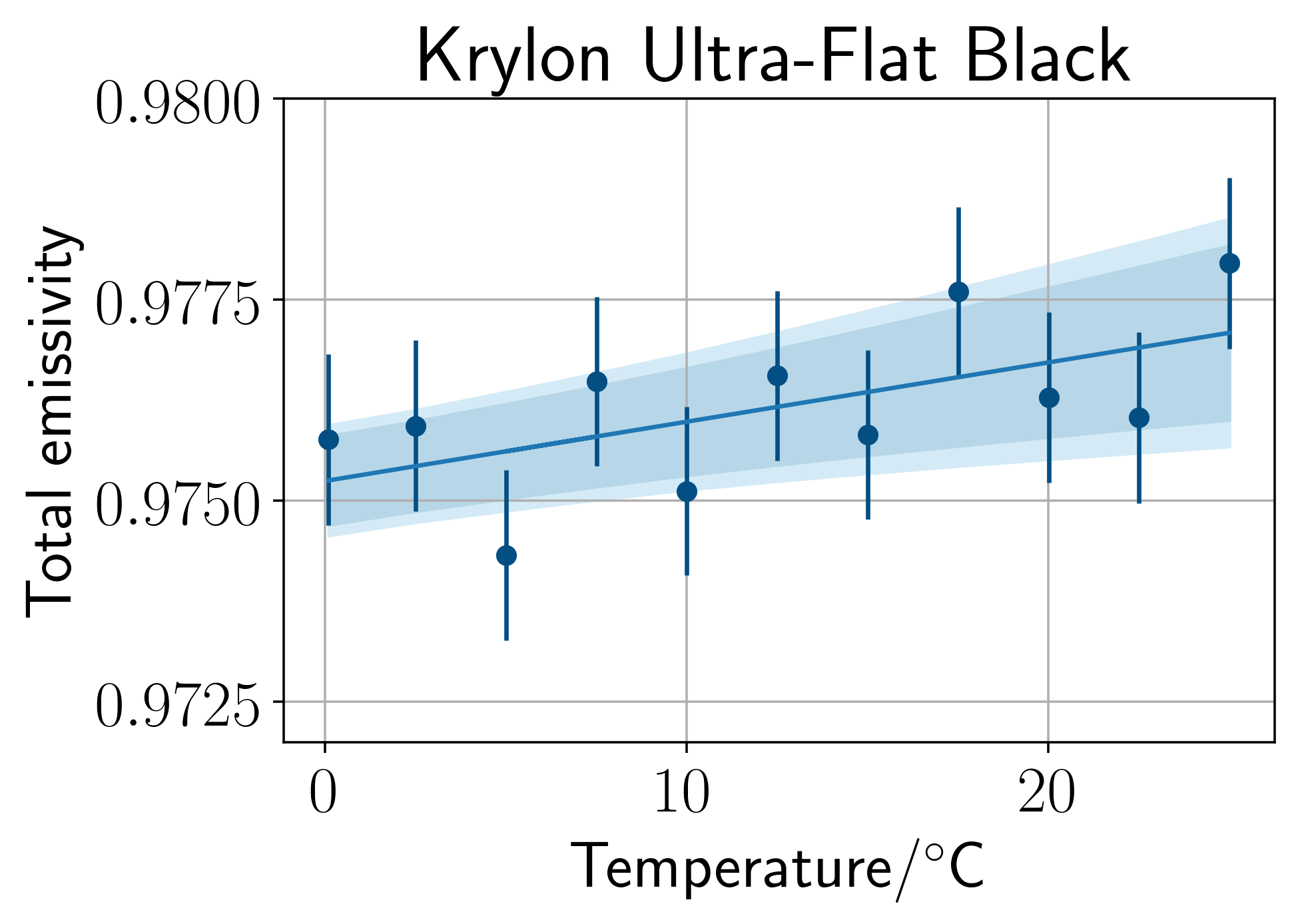}
    \caption{\label{krylon_emissivity} Measured emissivity coefficient of Krylon Ultra-Flat Black paint as a function of temperature. The error bars show the total measurement uncertainty, including statistical and systematic contributions. The line is a linear fit and the shaded areas show its statistical (dark blue) and total uncertainty (light blue).}
\end{figure}

The surfaces of the internal parts of the setup, which are  kept under vacuum, are covered with high-emissivity Krylon Ultra-Flat Black paint. Its emissivity was known approximately, but the uncertainty was unknown.
To characterise the paint emissivity, a stainless-steel sample was covered with the paint. That way, all surfaces inside the vacuum chamber were covered with the same paint.
The temperatures of the vacuum parts were measured for 11 values of the stabilised temperature of the painted sample, ranging from 0.1\;$^{\circ}$C to 25.0\;$^{\circ}$C with a step of 2.5\;$^{\circ}$C. The results are shown in Figure~\ref{krylon_emissivity}. A linear function $\epsilon(T) = a T+b$  was fitted to the data, yielding the coefficients $a= 0.000\,074(37)_{\rm stat}(33)_{\rm sys}\;{}^{\circ}\mathrm{C}^{-1}$ and $b=0.975\,25(55)_{\rm stat}(41)_{\rm sys}$. These values were used to calculate the emissivities of the internal parts of the setup during further measurements.

The measured metals are known to be vacuum compatible and suitable for industrial machining and finishing.
Another selection criterion was to have high thermal conductivity to ensure low temperature gradients, which is important for optical-clock vacuum chambers and thermal shields.
Hence, we chose the materials stainless steel 1.4301/304, aluminium PA38/AW-6060, copper Cu-ETP and titanium grade~5/6A1-4V.
Apart from the choice of material, the other crucial parameter is the finishing grade of the surface.
We describe this by the roughness of the surface, which is  defined by the DIN ISO 1302 standard. This standard describes the quality of the surface with a roughness grade number which corresponds to the mean deviation of the assessed profile denoted as $R_a$. In local national standards, this value also has an equivalent in the minimum and maximum acceptable average distance between the highest peak and lowest valley in each sampling length, denoted as $R_z$.

Two types of surface finishing were selected. The first one is machine finishing which is a typical surface achieved in industry processes, also in off-the-shelf vacuum components. This type of finishing corresponds to the roughness grade number N12, where $R_a = \SI{50}{\micro\meter}$. A typical $R_a$ value for rough machining is in the range of $40$--$80\;\si{\micro\meter}$.

The second type of finishing was polishing of the materials with grade 600 polishing paste, which corresponds to the N4 roughness grade number and $R_a = \SI{0.2}{\micro\meter}$. This type of polishing is highly reflective and can be applied both manually and mechanically for most of the common metals used in vacuum systems due to the availability of the polishing media.

\begin{figure}[p]
\centering
    \includegraphics[trim={2.5mm 3mm 2.5mm 3mm},clip,width=0.83\columnwidth]{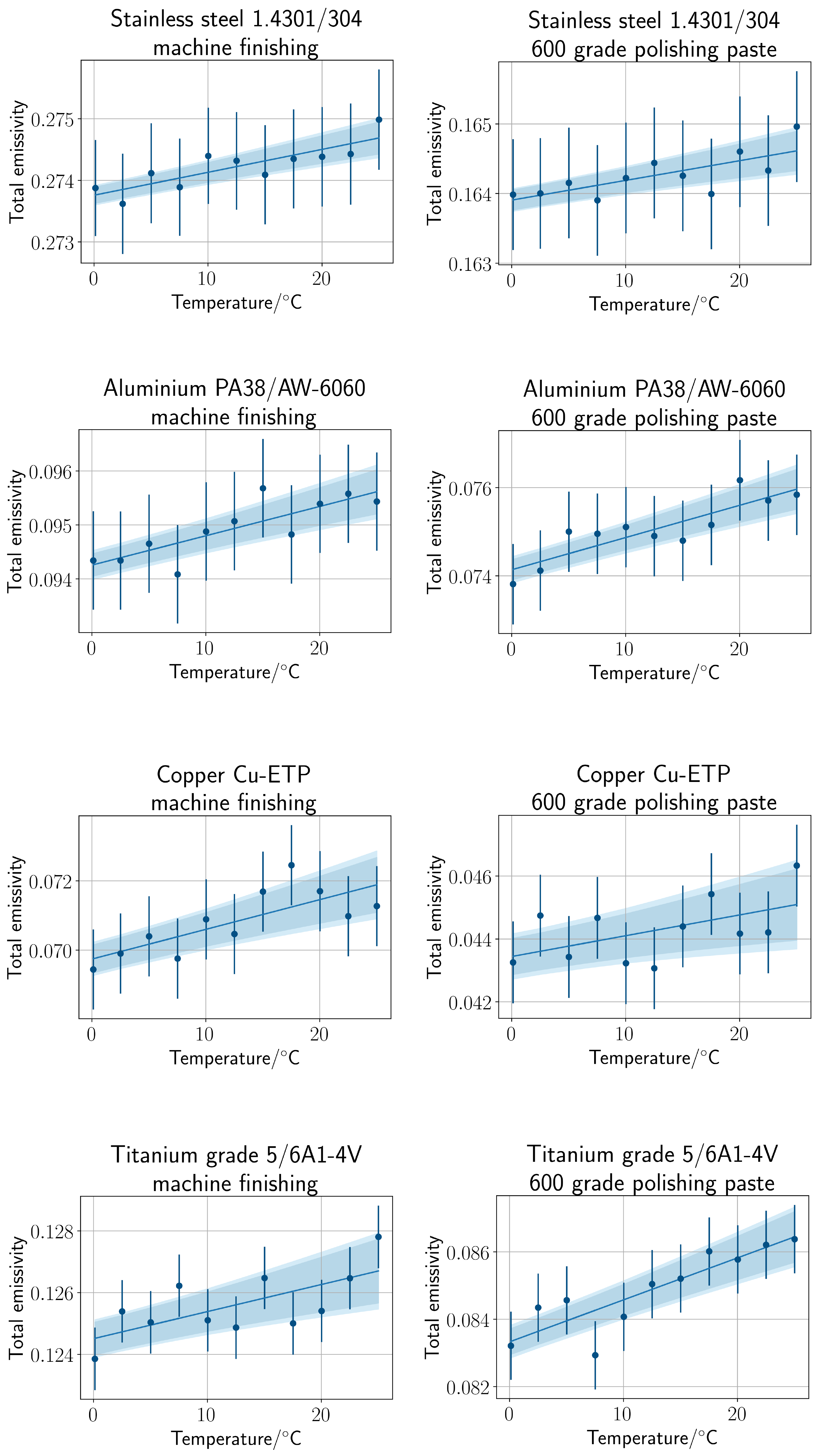}
    \caption{\label{fig:TempStab} Measured emissivity coefficients as a function of temperature with error bars showing total uncertainty. The lines are linear fits with statistical (dark blue) and total (light blue) uncertainty boundaries.}
\end{figure}

The results of the measurements are shown in Figure~\ref{fig:TempStab}. The fitted linear coefficients are gathered in Table~\ref{tab:emiss_coeff}.

\begin{table}[b!]
\centering
\caption{Coefficients for the temperature-dependent emissivity of different materials and surface finishes. \label{tab:emiss_coeff}}
\begin{tabular}{lc r@{}l r@{}l}
\toprule
\multirow{2}{*}{Material}        & \multirow{2}{*}{{Linear fit coefficients}} & \multicolumn{4}{c}{{Type of finishing}} \\
                                 &                                         & \multicolumn{2}{c}{Machined N12}  & \multicolumn{2}{c}{Polished N4}      \\
\midrule
Stainless steel  & $a/{}^{\circ}$C${}^{-1}$       &  0&.000\,037(9)$_{\rm stat}$(7)$_{\rm sys}$       &  0&.000\,028(10)$_{\rm stat}$(7)$_{\rm sys}$     \\
1.4301/304                  & $b$                 &  0&.273\,76(13)$_{\rm stat}$(9)$_{\rm sys}$       &  0&.163\,91(14)$_{\rm stat}$(8)$_{\rm sys}$      \\ \midrule
Aluminium        & $a/{}^{\circ}$C${}^{-1}$       &  0&.000\,054(14)$_{\rm stat}$(9)$_{\rm sys}$      &  0&.000\,073(16)$_{\rm stat}$(10)$_{\rm sys}$    \\
PA38/AW-6060                       & $b$          &  0&.094\,26(21)$_{\rm stat}$(17)$_{\rm sys}$      &  0&.074\,14(23)$_{\rm stat}$(19)$_{\rm sys}$     \\ \midrule
Copper           & $a/{}^{\circ}$C${}^{-1}$       &  0&.000\,086(27)$_{\rm stat}$(20)$_{\rm sys}$     &  0&.000\,066(39)$_{\rm stat}$(29)$_{\rm sys}$    \\
Cu-ETP                           & $b$            &  0&.069\,74(40)$_{\rm stat}$(26)$_{\rm sys}$      &  0&.043\,44(58)$_{\rm stat}$(43)$_{\rm sys}$     \\ \midrule
Titanium         & $a/{}^{\circ}$C${}^{-1}$       &  0&.000\,088(36)$_{\rm stat}$(24)$_{\rm sys}$     &  0&.000\,125(26)$_{\rm stat}$(14)$_{\rm sys}$    \\
grade~5/6A1-4V                        & $b$       &  0&.124\,51(53)$_{\rm stat}$(30)$_{\rm sys}$      &  0&.083\,33(38)$_{\rm stat}$(30)$_{\rm sys}$     \\                                                      \midrule
Krylon          & $a/{}^{\circ}$C${}^{-1}$        &  \multicolumn{4}{l}{\hspace{20mm}0.000\,074(37)$_{\rm stat}$(33)$_{\rm sys}$}                      \\
Ultra-Flat Black             & $b$                &  \multicolumn{4}{l}{\hspace{20mm}0.975\,25(55)$_{\rm stat}$(41)$_{\rm sys}$}                         \\
\bottomrule
\end{tabular}
\end{table}

\section{Techniques for ion clocks \label{sec:BBR_ions}}

Ion clocks generally have smaller BBR shifts than lattice clocks, see Table~\ref{tab:atomic}.  Table~\ref{tab:uncs} shows that the fractional-frequency-shift uncertainties for Sr$^+$ and Yb$^+$ are below the $10^{-17}$ level for temperature uncertainties of 1\;K or below.  The uncertainty for Yb$^+$ is dominated by the uncertainty in the differential polarisability coefficient.  For Sr$^+$, however, the uncertainty can be significantly reduced with better knowledge of the BBR field (effective temperature). Evaluation of the BBR field seen by the ion is, however, complicated by the fact that the trap structure itself, which covers a significant fraction of the solid angle visible to the ion, heats up due to joule heating in conductors and dielectric losses in insulators. One way to determine the effective temperature would be to use \emph{in situ} temperature sensors like in some lattice clocks \citep{Nicholson2015}, but this is difficult due to several reasons. The trap structures are typically small and the presence of a sensor may disturb the trap potential. Electrical sensors such as resistance temperature detectors (RTDs) and thermistors suffer from self-heating and interference from the RF field. Surface-mounted Pt100 RTDs have been successfully used on an aluminium nitride (AlN) chip trap \citep{Keller2019}, but this technique is not easily adapted to the traps used for typical single-ion clocks, see  Chapter~\ref{chap:iontrap}. 
Therefore, in Section~\ref{sec:BBR_real_time_thermom} we consider the possibility of using fibre-optic sensors for real-time thermometry of ion traps.

A more common approach is to determine the effective temperature by a combination of  thermal imaging and FEM simulations \citep{Dolezal2015,Zhang2017a}. These techniques will be described in Sections~\ref{sec:BBR_thermal_imaging} and \ref{sec:BBR_simulations}, respectively. Design recommendations on how to minimise RF heating are given in  Chapter~\ref{chap:iontrap}.

\subsection{Thermal imaging \label{sec:BBR_thermal_imaging}}

\begin{figure}[tb]
\centering
    \includegraphics[width=1\columnwidth]{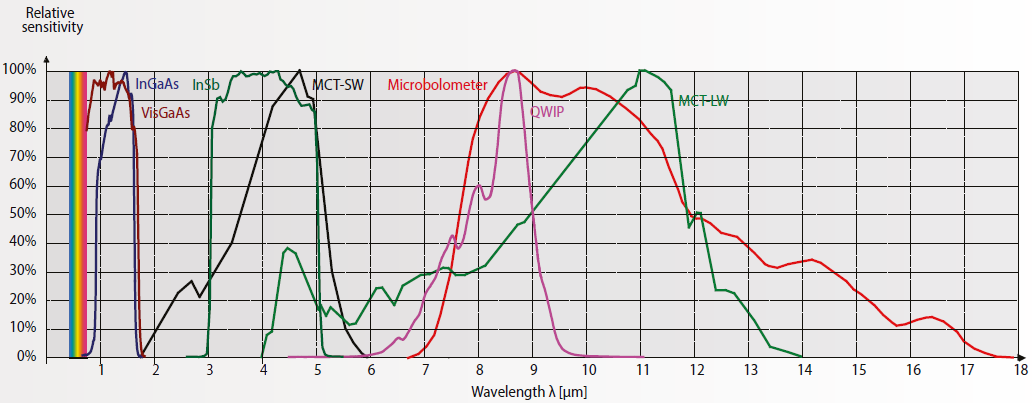}
    \caption{\label{fig:BBR_IRcam} Relative response curves for different kinds of thermal cameras. From \citet{FLIR} courtesy of FLIR Systems.}
\end{figure}

In a thermographic camera, an IR lens focuses the radiation onto a focal-plane array (FPA) of micrometer-size detectors. There are a number of detector types, see Figure~\ref{fig:BBR_IRcam}. The most common type is the thermal uncooled microbolometer detector, which has a relatively low sensitivity and slow response. The other detector types shown in Figure~\ref{fig:BBR_IRcam} are quantum (photon) detectors of different materials. These require cooling, which increases cost and complexity. Apart from detector response curves, the two distinct wavelength ranges, mid-wave (MW, ${\sim} 3\ldots 5\;\si{\micro\meter}$) and long-wave (LW, ${\sim} 7.5\ldots 13\;\si{\micro\meter}$), are governed by the so-called atmospheric windows, where the atmospheric attenuation is reduced. Thermal imaging of ion traps takes place at short distances, where this attenuation can be neglected, at least in the LW band.

For imaging ion traps, we note that LW cameras have a good overlap with the room-temperature BBR spectrum, while there is a wider range of vacuum-window materials for MW. We can distinguish between imaging of dummy traps and operational ion traps. With a dummy trap, the vacuum window can be chosen solely based on its IR transmission. For example, in \citet{Dolezal2015} and the current work on the CMI trap, a LW camera and an AR-coated ``Cleartran'' (ZnS) window (Edmund Optics \#{}64-147) were used. In addition, one can apply high-emissivity tape or paint to low-emissivity parts in order to improve the imaging contrast. With an operational trap, this is not possible and reliable temperature readings can mainly be obtained from high-emissivity parts like insulators. In principle it is possible to have a dedicated IR viewport in the ion-trap chamber, but for compactness and optical access the IR window can also be used as the exit window for the laser beams, requiring good transmission in the visible and near-infrared as well. For example, in \citet{Nisbet-Jones2016} a MW camera and a MgF$_2$ window were used, while current work on the VTT MIKES trap used a LW camera and a CaF$_2$ window. In the following, we consider using a FLIR A615 LW camera and a CaF$_2$ window as an example.

\subsubsection{Modelling camera response and window properties}

\begin{figure}[tb]
\centering
    \includegraphics[width=0.55\columnwidth]{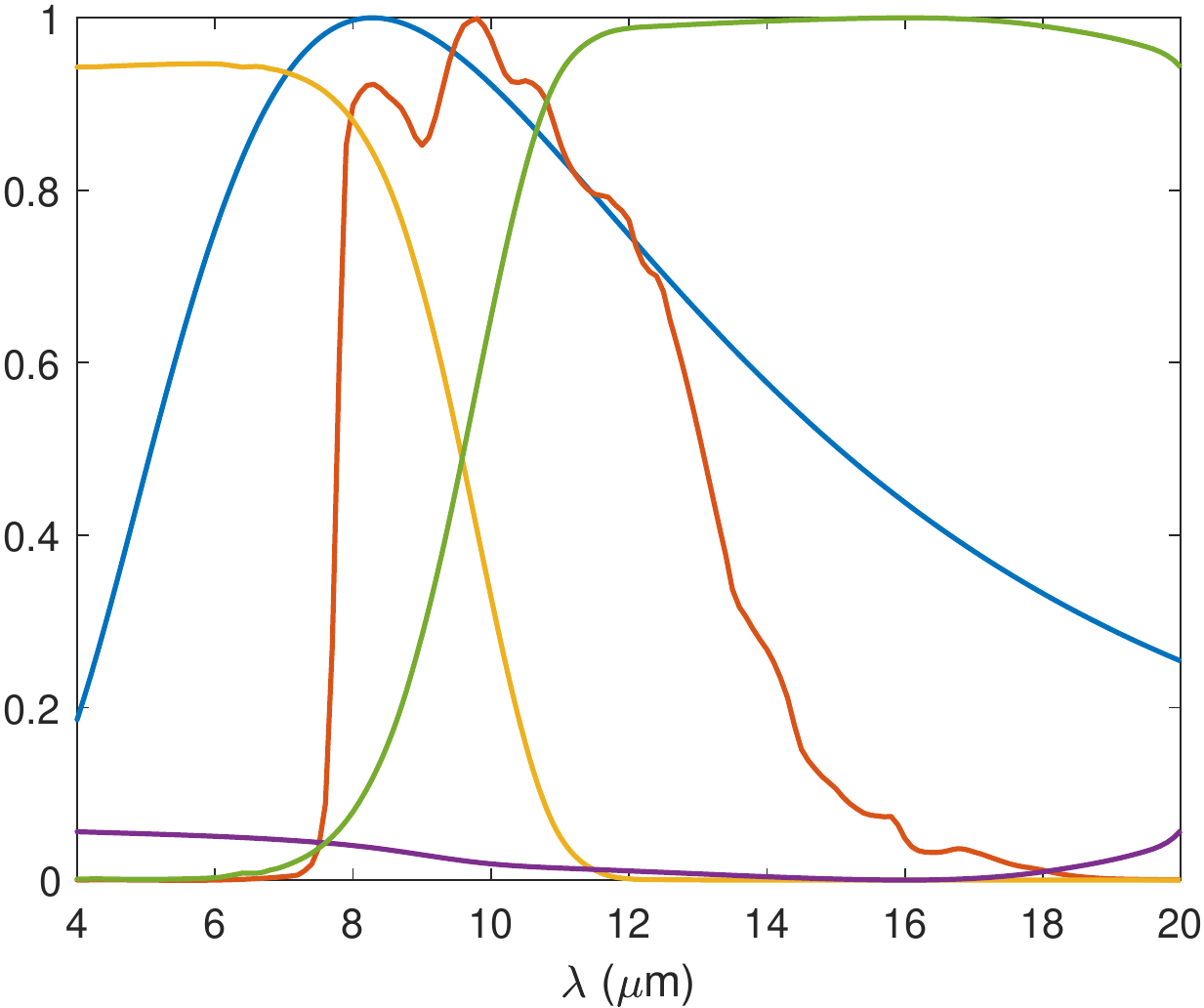}
    \caption{\label{fig:BBR_cam_win} Normalised BBR spectrum $B(\lambda,T = 300\;\mathrm{K})$ (blue), normalised camera response curve $S(\lambda)$ (red, adapted from \citet{FLIRcurve}), calculated  transmittivity $\tau(\lambda)$ (orange), reflectivity $\rho(\lambda)$ (purple), and emissivity $\epsilon(\lambda)$ (green) of a 5-mm-thick CaF$_2$ window.}
\end{figure}

Figure~\ref{fig:BBR_cam_win} shows a typical camera response curve $S(\lambda)$ compared to a normalised BBR spectrum at room temperature. As $S(\lambda)$ is not flat over the BBR spectrum, it is clear that the simple $T^4$ dependence cannot be valid. However, by numerical integration one can show that for finite temperature ranges around room temperature, the integral can be well approximated by a linear function in $T^4$, but with an added constant,
\begin{equation} \label{eq:BBR_cam_int}
B_S(T) = \int_0^\infty S(\lambda) B(\lambda,T) d\lambda \approx k_S T^4 + B_0.
\end{equation}
Figure~\ref{fig:BBR_cam_win} also shows the wavelength-dependent reflectivity, transmittivity, and emissivity of the CaF$_2$ window. In the same way as in Equation~\eqref{eq:BBR_cam_int}, we can linearise (in $T^4$) these window properties weighted by the BBR and camera curves
\begin{equation} \label{eq:BBR_win_int}
B_x(T) = \int_0^\infty x(\lambda) S(\lambda) B(\lambda,T) d\lambda \approx k_x T^4 + B_{x,0}, \quad x = \tau, \rho, \epsilon.
\end{equation}
From the closure relation $\rho + \tau + \epsilon = 1$ follows $k_\rho + k_\tau + k_\epsilon = k_S$ and $B_{\rho,0} + B_{\tau,0} + B_{\epsilon,0} = B_0$. As we will see later, the pre-factors in our measurement formulae always add up to one, which means that the $B_{x,0}$ constants will cancel. We can then define the effective reflectivity, transmittivity and emissivity as
\begin{equation}
\langle\rho\rangle = k_\rho/k_S, \quad
\langle\tau\rangle = k_\tau/k_S, \quad
\langle\epsilon\rangle = k_\epsilon/k_S.
\end{equation}
Similarly we can define ``higher-order'' coefficients like $\langle\tau\epsilon\rangle$, which is needed for radiation emitted by the window (away from the camera), reflected from the object (assumed to be a graybody) and transmitted through the window on the way to the camera. Note that $\langle\tau\epsilon\rangle$ cannot be expressed as a simple function of $\langle\tau\rangle$ and $\langle\epsilon\rangle$.

The above is valid approximately in a certain, finite temperature range. The window's $\tau$, $\rho$ and $\epsilon$ actually have a small linear dependence on the temperature, but since the measurement formulae are based on linearity in $T^4$, we cannot account for this.

\subsubsection{The measurement formulae}

We start by considering measuring a sample in air, with no window. We can then (formally) write what the camera measures as
\begin{equation}
B_S(T_\text{app}) = \int_0^\infty S(\lambda) B(\lambda,T_\text{app}) d\lambda
= \int_0^\infty S(\lambda) \left[\epsilon_\text{obj} B(\lambda,T_\text{obj}) + (1-\epsilon_\text{obj}) B(\lambda,T_\text{refl})\right] d\lambda.
\end{equation}
Here $T_\text{app}$ is the apparent temperature measured by the camera, $\epsilon_\text{obj}$ and $T_\text{obj}$ are the emissivity and temperature of the measured object (assumed to be a graybody), and $T_\text{refl}$ is the temperature reflected into the camera from the measured object. After linearisation, we obtain
\begin{equation} \label{eq:emiss_formula}
T_\text{app}^4 = \epsilon_\text{obj} T_\text{obj}^4 + (1-\epsilon_\text{obj}) T_\text{refl}^4.
\end{equation}
In principle, the emissivity can  be evaluated as the slope of Equation~\eqref{eq:emiss_formula} after measuring the sample at several known temperatures.
In practice, due to the thermal conduction and convection of the air, it is difficult to estimate the true surface temperature even with a sensor embedded in the sample. Thus, it is better to measure the emissivity values in vacuum, through an IR window identical to the one used for the actual trap-heating measurements.

Measuring through a window, the linearised measurement formula becomes
\begin{equation} \label{eq:emiss_formula_win}
T_\text{app}^4 = \epsilon_\text{obj} \langle\tau\rangle T_\text{obj}^4 + \langle\epsilon\rangle T_\text{win}^4 + \langle\rho\rangle T_\text{r,win}^4 + (1-\epsilon_\text{obj}) \langle\tau\rangle T_\text{r,obj}^4,
\end{equation}
where the terms describe radiation emitted by the object and transmitted through the window, radiation emitted by the window, radiation reflected from the window, and radiation reflected from the object and transmitted through the window. Since we are typically working at close to normal incidence, we can approximate the radiation reflected from the object as $T_\text{r,obj}^4 \approx  \langle\tau\rangle T_\text{r,win}^4 + \langle\epsilon\rangle T_\text{win}^4 + \langle\rho\rangle T_\text{obj}^4$, where the terms describe the external reflected temperature transmitted through the window, the radiation emitted towards the object from the window, and radiation emitted by the object and reflected back from the window. One might be tempted to add $\epsilon_\text{obj}$ in front of the last term, but the prefactors must always add up to one. One can also see this as a consequence of not explicitly considering multiple reflections. In practice, the reflections are not specular and there are materials with different emissivities side by side, so this term is approximate, but fortunately $\langle\rho\rangle$ is small (${\sim} 0.02$). The formula becomes
\begin{eqnarray} \label{eq:app_win}
T_\text{app}^4 &=&\left[\epsilon_\text{obj} \langle\tau\rangle + (1-\epsilon_\text{obj}) \langle\tau\rho\rangle\right]
T_\text{obj}^4 + \left[\langle\epsilon\rangle + (1-\epsilon_\text{obj}) \langle\tau\epsilon\rangle \right]
T_\text{win}^4 \nonumber \\
&& + \left[\langle\rho\rangle + (1-\epsilon_\text{obj}) \langle\tau^2\rangle\right] T_\text{r,win}^4.
\end{eqnarray}

\subsubsection{Correcting for window and sensor temperature}

Since the effective transmission of the CaF$_2$ window over the spectral range of the camera is only 34\%, the window will absorb thermal radiation from the heated sample and heat up. In order to measure the window temperature we put a
\SI{10 x 5}{mm}
sandwich of black and aluminium tape on the rear (sample) side of the window. The black tape provides a high-emissivity ($\epsilon \approx 0.93$) surface that is in direct contact with the window surface, while the aluminium tape prevents direct radiative heating of the black tape itself. The tape patch is positioned away from the measurement areas to minimise the risk that reflections from the aluminium tape disturbs the measurements. If we fit the function $T_\text{win}^4 = k_\text{win} T_\text{ref}^4 + T_{\text{win},0}^4$, where $T_\text{ref}$ is the bulk temperature given by the sensor, to the data, we can substitute this into Equation~\eqref{eq:app_win} as a correction. Without accounting for this, we obtain a window transmission and emissivities that are ${\sim} 0.01$ too large. (If care is not taken to minimise window heating, the error can be larger.)

Similarly we can monitor the temperature of the camera sensor by putting a metal mirror halfway between the camera and the object so that one can record the sensor reflection simultaneously with the object. If there is correlation with $T_\text{ref}^4$ like for the window, we can take this into account in the reflected temperature, but this effect is generally smaller than that of the window.

\subsubsection{Measurements \label{sec:BBR-th-imaging-meas}}

The first measurement is to characterise the window transmission. This is done by measuring a heated plate of known emissivity at different temperatures, first without and then with the window. At small working distances, the reflected temperature is dominated by the reflection of the heated camera microbolometer FPA from the object and thus it varies over the field of view. It can, however,  be obtained for each sample separately from the measurement at room temperature. Equation~\eqref{eq:emiss_formula} gives the true surface temperature as a function of the bulk temperature measured by a sensor. Equation~\eqref{eq:app_win} can then be used to solve $\langle\tau\rangle$ ($\langle\rho\rangle$ and the second-order coefficients can here be considered as small corrections that we can use the theoretical values for). To fit the model to the measurement results, we introduce a wavelength offset $\Delta\lambda = \SI{0.217(14)}{\micro\meter}$ in the camera response curve. This offset is reasonable considering that we are dealing with merely a typical response curve.

The next measurement is to measure material emissivities in vacuum. The samples are glued onto a temperature-controlled copper block together with reference samples of known emissivity that are used to assess the surface temperature. Again, we determine the reflected temperatures at room temperature and then we measure at different sample temperatures. Using Equation~\eqref{eq:app_win}, we can then determine the sample emissivities.

\begin{figure}[tb]
\centering
    \includegraphics[width=0.75\columnwidth]{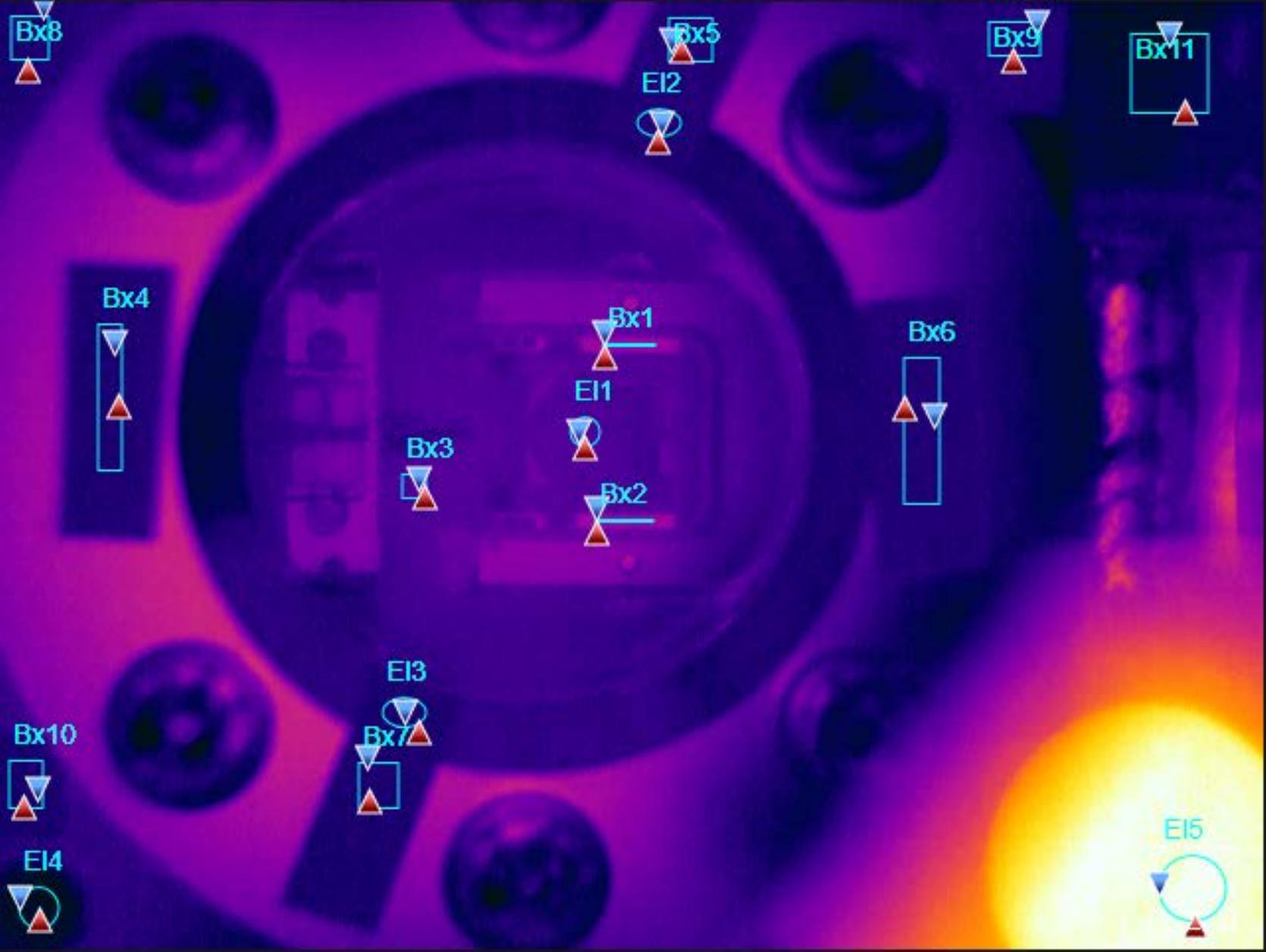}
    \caption{\label{fig:BBR_VTT_meas} Thermal image of the VTT MIKES trap. Of the trap parts, only the fuse silica insulators (Bx1 and Bx2, emissivity \num{0.721(9)}) gave reliable readings. Bx3 and El1 measure the fused-silica window on the other side of the chamber. Pieces of black tape were attached to the flange and chamber to measure the temperature of these. In the lower-right corner one can see the reflection of the camera sensor from the metal mirror used to monitor the sensor temperature. This preliminary measurement was carried out at a higher-than-normal RF voltage.}
\end{figure}

Finally, we measure the temperature of the ion trap with RF applied. Again, the analysis uses Equations~\eqref{eq:emiss_formula} and \eqref{eq:app_win} for parts outside and inside the vacuum, respectively. Figure~\ref{fig:BBR_VTT_meas} shows a thermal image of the VTT MIKES trap. It was found that the heating was dominated by the helical resonator (imaged separately, not shown).

\subsection{Simulations \label{sec:BBR_simulations}}

For the finite-element-method (FEM) simulations described here and in \citet{Dolezal2015}, the commercial ANSYS software has been used.
First the electromagnetic part of the model is solved, which gives the losses (heat generation:  joule heating in conductors and dielectric heating in insulators) in the trap parts. These are used as a load for the steady-state thermal part of the simulation, which gives a temperature map of the trap structures (according to given heat generation, boundary conditions, heat conductivity and thermal radiation).
The temperature seen by an ion is evaluated as the temperature of a small ($\SI{\approx 20}{\micro\meter}$ diameter)  ``blackbody sphere'' inserted at the expected ion position (usually the trap centre for single ion traps).
Since the temperatures of the trap parts rise approximately by the same value above ambient temperature in usual laboratory conditions (ambient temperature $\SI{\approx 20 \pm 5}{\celsius}$), we can set up the FEM model for one selected ambient temperature and then compare the predicted temperature rises with experiments. The uncertainty of the  computed temperature rise seen by the ion is evaluated by carrying out many simulations while varying the parameters used in the model in the range of their expected uncertainties and deducing the sensitivity of the ion temperature rise to the different parameters.

When developing a new trap design, the FEM simulation can be done with estimated parameters to get an order-of-magnitude estimate of the temperature rise seen by the ion. Experimental measurements of temperatures in the trap at operating conditions (applied trap voltage, frequency, ambient temperature etc.) is necessary for understanding the heat dynamics in the trap and for fine tuning the FEM model parameters. A detailed description of ion trap FEM simulations with some examples can be found in \citet{Dolezal2015}.

In the following we list a few problematic areas and tips that one should consider when developing FEM models for ion traps:

\textit{RF feedthrough is recommended to be included in the model:}
In some cases,  the observed temperature of traps designed for low heating is much higher than expected from the heat generation in the trap parts. The excessive heat can be generated in the feedthrough insulator and then conducted to the trap by wires.

\textit{Helical resonator coil is a substantial source of heat:}
Ion traps are often powered by a voltage that is amplified using a helical resonator. The high-voltage end of the helical resonator coil can have quite high temperature (several degrees above ambient temperature). In particular if the coil end is directly joined to the massive feedthrough pin that supports the trap structure, it can have considerable influence on the experimentally observed trap temperatures. From the FEM simulation perspective it is good to measure the temperature of the high-voltage feedthrough pin and use it as a boundary condition. If the feedthrough pin is not experimentally accessible, one can measure the temperature of some nearby test point and then adjust the boundary conditions to get temperature agreement at this point.

\textit{Thermal contact between parts:}
The thermal properties of the bond between two joined trap parts is represented by the thermal contact conductivity (at least in ANSYS). The values for contact conductivities even for common trap materials placed in vacuum are hard to find in the literature, but can be estimated by measuring the temperature gradient across the bond. Sometimes it is good practice to measure these gradients with the high voltage disconnected and by inserting a known amount of heat to a particular part of the trap. This avoids uncertainties in the heat generated by the RF high voltage resulting from imperfect knowledge of the loss tangent of dielectrics or of the precise voltage amplitude influencing joule heat generation in conductors.
For example, if there are Pt100 sensors mounted on the trap structures like in \citet{Keller2019}, these can be used as heaters by running a DC current through them.

\textit{Vacuum gaps between two parts that seem to be connected:}
Apart from the bond contact conductivity issue described above, there can be another problematic area for endcap trap designs that use endcap and shield electrodes in the form of concentric cylinders separated by a dielectric with high loss tangent and high relative permittivity. The gap between insulator and electrode can reduce the effective voltage applied across the dielectric and consequently the heat generation is lower than expected. It is possible to verify whether we have problems of this kind by measuring the electrical capacitance of the trap and comparing the result with the analytic formula for the capacitance of a multilayer concentric capacitor.

\textit{FEM software limitations:}
Commercial FEM software usually implements reflections as diffuse and assuming graybodies. Specular reflections are not captured.

\begin{figure}[tb]
	\centering
	\includegraphics[width=1\textwidth]{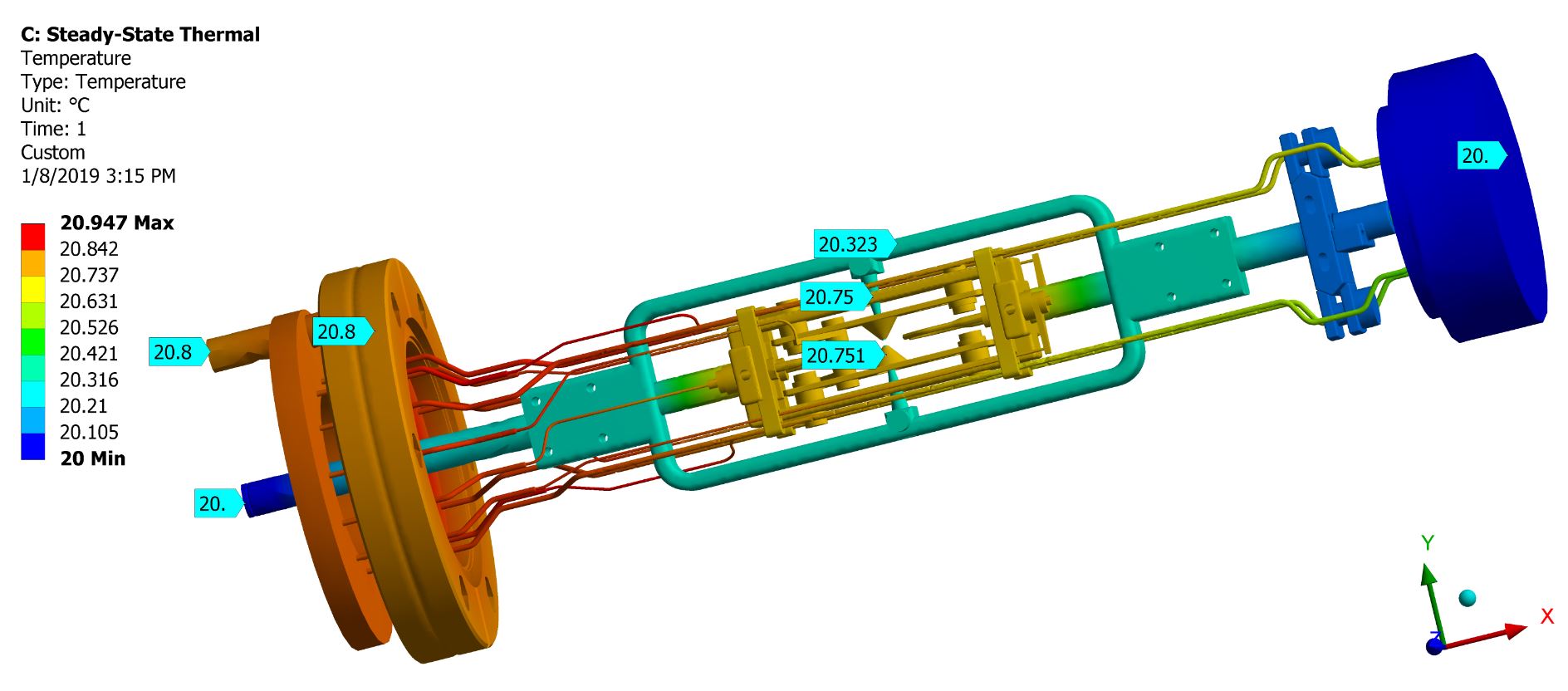}
	\caption{FEM simulation of the CMI trap. The ambient temperature is \SI{20}{\celsius} and the amplitude of the applied RF voltage is \SI{1011}{V} at a frequency of \SI{18.717}{MHz}.}
	\label{fig:CMI_WP3_CMI_FEM}
\end{figure}

An example of a FEM simulation is shown in Figure~\ref{fig:CMI_WP3_CMI_FEM} for the CMI trap design which is described in Section~\ref{sec:iontrap_CMItrap}.
The heat generated in the feedthrough insulator is reduced by lowering the electrical field (and consequently the heat generation) in the dielectric by making the diameter of the insulator disk large and by separating the high voltage from ground as much as possible. The heat generated in the helical resonator coil is removed by water cooling, which keeps the feedthrough high-voltage pin at a stable selected temperature (e.g., at the temperature of the vacuum chamber).

Model verification was done for three test points (outer frame, inner frame and shield electrode) by thermal imaging.  The experimentally observed temperature of the feedthrough flange was used as a boundary condition in the FEM model.
The agreement between the FEM model and two measurements is shown in Table~\ref{tab:CMI_Measurement}.
An ion located at the trap centre would see a temperature rise $\Delta T_\text{ion} = \SI{0.42 \pm 0.30}{\celsius}$ (for the conditions mentioned in Figure  \ref{fig:CMI_WP3_CMI_FEM}).

\begin{table}[b]
	\centering
	\caption{Temperature rise in $^\circ$C for selected parts of the CMI trap: measurement vs FEM model.\label{tab:CMI_Measurement}}
	\begin{tabular}{l
S[table-format=1.5, table-space-text-post = ~(boundary~cond.), table-align-text-post = false, table-text-alignment = left]
S[table-format=1.5]
S[table-format=1.5]
S[table-format=1.5]}
		\toprule
		 & {Feedthrough flange} & {Outer frame} & {Inner frame} & {Shield electrode}  \\
        \midrule
		Measurement 1 & 0.85\pm0.05 & 0.28\pm0.13 & 0.74\pm0.20 & 0.74\pm0.20 \\
		Measurement 2 & 0.80\pm0.05 & 0.29\pm0.08 & 0.64\pm0.12 & 0.67\pm0.13 \\
		FEM simulation & 0.8 ~(boundary~cond.) & 0.32 & 0.75 &  0.75  \\
		\bottomrule
	\end{tabular}
\end{table}

\subsection{Real-time thermometry \label{sec:BBR_real_time_thermom}}

This section discusses the feasibility of evaluating blackbody shifts in an ion clock using real-time thermometry of the trap electrodes. Thermistors are unsuitable for this application since RF voltages on the ion-clock trap electrodes induce high local RF pick-up resulting in their rapid destruction. Typically, trap voltages are a few hundred volts and at frequencies of between 10\;MHz and 20\;MHz. We therefore investigated the use of novel fibre-optic temperature sensors as an alternative to thermistors to monitor electrode temperature.

Two GaAs fibre-optic temperature sensors were purchased from OpSens, together with a multi-channel signal conditioner. The sensors were type ``OTG-F'', selected on the basis of their small size (150-\si{\micro\meter} outer diameter), fast response time (5\;ms), specified accuracy (0.3\;K) and resolution (0.05\;K). The fibre-optic sensors were mounted within the same aluminium block as Thorlabs thermistors (part TSP01) that could be read by a PC via a USB interface. These thermistors were, in turn, verified using calibrated ``Edale'' thermistors with a serial port connection. Measurements in the same Al block gave excellent agreement between calibrated Edale thermistors and allowed calibration of the Thorlabs sensors at the 0.1-K level. The fibre-optic sensors could also be calibrated to a similar level of short-term temperature uncertainty.  (For Sr$^+$ at room temperature, a 0.1-K temperature uncertainty gives rise to a relative BBR shift uncertainty of about 1 part in $10^{18}$).

We also checked the effect of electrode RF voltages on the measured temperature by switching the local RF on and off at typical 5-s intervals; there was no observable change in the reading at the 0.1-K level.  Significant problems arose, however, when the fibre-optic sensors were connected to the read-out unit via a lossy fibre link. Although simple reconnection of a fibre link resulted in a measured change in the observed temperature of less than 0.1\;K, losses introduced by a fibre-optic vacuum feedthrough increased this change to ${\sim} 0.2$\; K. Measurement variations at this level will result in a frequency uncertainty due to the blackbody shift in room-temperature Sr$^+$ of just over 1 part in $10^{18}$. Since the fibre loss affected the temperature reading in a way that was not completely reproducible in the short term, our conclusion is that fibre-optic temperature sensing is unsuitable for Sr$^+$ ion trap optical clocks with $10^{-18}$ overall uncertainty.

\section{Techniques for lattice clocks}\label{sec:BBR_lattice}

Several techniques have been developed to cope with the BBR shift in lattice clocks, both at room temperature and in cryogenic environments. As discussed below, these designs should work under ultra-high vacuum, should have sufficient optical access to manipulate atoms and should also have enough physical access for atoms. The specific requirements for characterisation of the BBR field experienced by the atoms, especially its representative temperature, depend strongly on the temperature regime, as is discussed below. Notwithstanding, any temperature measurement must take into account that the International Temperature Scale of 1990 (ITS-90) deviates from the thermodynamic temperature scale \citep{Fischer2011}.

\subsection{Cryogenic clocks}
Lattice clocks featuring interrogation in a cryogenic environment benefit from a suppression of the frequency shift induced by BBR by several orders of magnitude: the BBR-induced frequency shift in a strontium lattice clock, for instance, is reduced from $\SI{-2}{\hertz}$ at room temperature ($T \approx \SI{300}{K}$) to  $\SI{-9}{\milli\hertz}$, i.e., about two parts in $10^{17}$, near the boiling point of liquid nitrogen ($T \approx \SI{77}{K}$). This substantially relaxes the uncertainty to which both the BBR field's representative temperature and the atomic response must be determined (see Figure~\ref{fig:BBRTemperatureUncertainty}). A moderate uncertainty $u(T) \approx \SI{0.1}{\kelvin}$ of this temperature is sufficient to reduce the fractional uncertainty in the systematic frequency shift to below $10^{-19}$ in the case of strontium, where the shift is larger than in other species such as ytterbium or mercury.

There are several different approaches for realising a cryogenic lattice clock, e.g., regarding translation of the atomic sample. Two schemes will be discussed in the remainder of this section: firstly, translation of the optical lattice to transport atoms into a separate temperature-controlled cryogenic environment within the apparatus for interrogation only and, secondly, a fully cryogenic approach where the entire experimental procedure takes place in a cryogenic environment. The former has already been demonstrated \citep{Ushijima2015}. An approach similar to the latter has been demonstrated at room temperature \citep{Beloy2014}. However, there are additional concerns that must be addressed for the case of cryogenic temperatures.

The main challenge of the fully cryogenic approach is the optical access required for implementing an optical lattice clock. Atom translation trades reduced optical access for additional complexity and dead time required for transport.

\subsubsection{Translation into a cryogenic environment}

Atom translation is most easily achieved by translation of the optical lattice, particularly for two-electron atoms as used in lattice clocks. Since interrogation is performed along the lattice direction (only the case of a 1D optical lattice is considered here), optical and physical access to the cryogenic compartment can be restricted to a single axis. Laser cooling, state preparation and, typically, read-out are performed outside --- with the possible exception of techniques that only require access along the optical lattice, e.g., sideband cooling.

Pinholes provide both the physical access required to transport the atomic sample into and out of the cryogenic compartment and optical access for the lattice and interrogation lasers. Diameters of $d \lesssim \SI{0.5}{\milli\meter}$ are typically sufficient to accommodate the laser beams. As discussed in Section~\ref{sec:BBR_temperature_inhomogeneity}, the contribution of a pinhole to the BBR field experienced by the atoms depends on its diameter $d$ and distance $L$ from the atomic sample. The fractional frequency shift due to a pinhole with $d \le \SI{0.5}{\milli\meter}$ and $L \ge \SI{15}{\milli\meter}$ can be kept below $5\times10^{-19}$ \citep{Middelmann2011, AlMasoudi2016}. Windows can be used for more convenient optical access, e.g., to extract the optical lattice for retroreflection, but need to be thick enough to block out residual transmitted BBR (see below).

\begin{figure}[tbp]
	\centering
	\includegraphics[width=10cm]{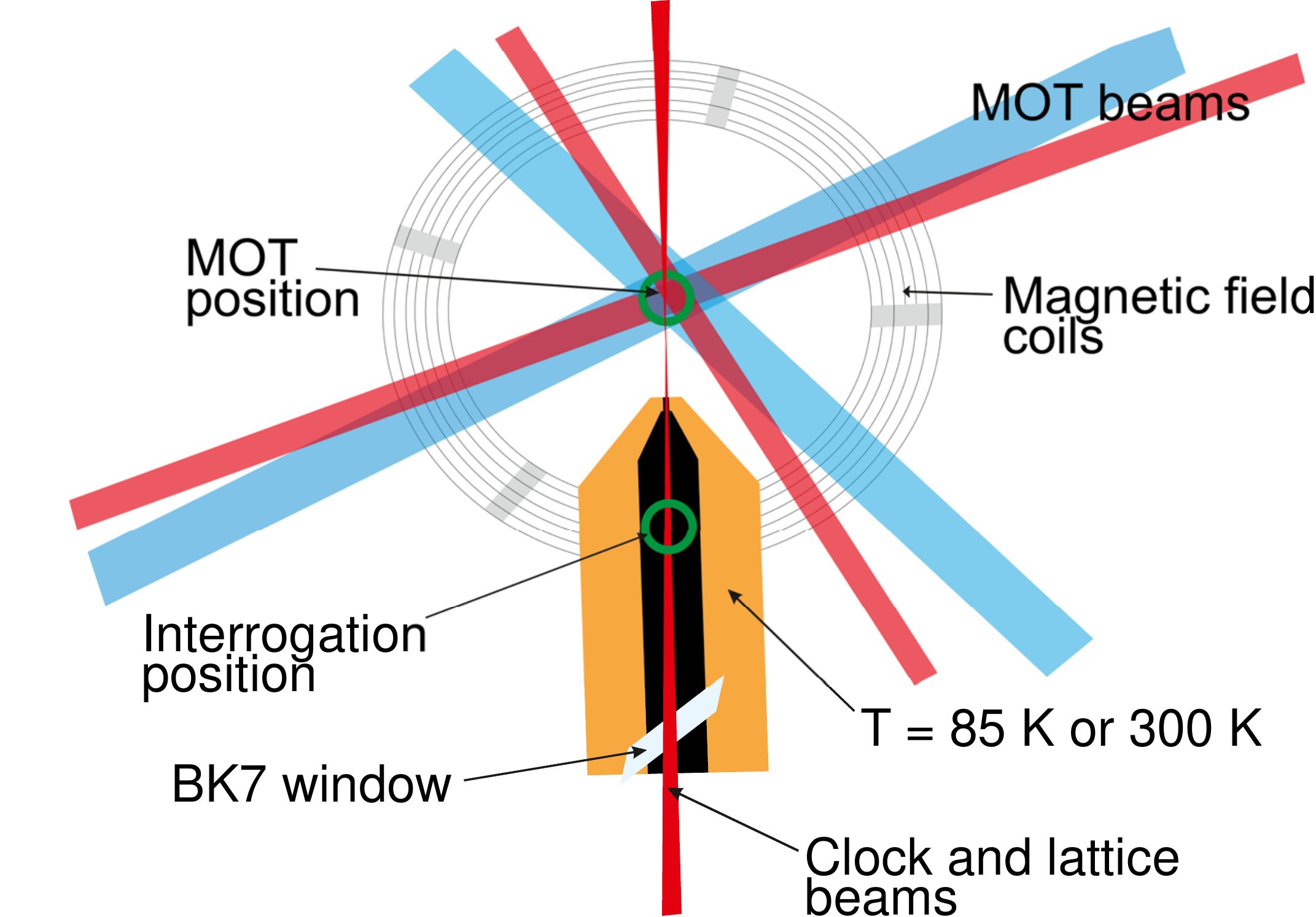}
	\caption{\label{fig:Coldfinger}Schematic illustration of the cryogenic interrogation environment using atom translation developed for the strontium lattice clock at PTB, From \cite{AlMasoudi2016}.}
\end{figure}

The mechanical design of the cryogenic compartment is usually dictated by the need to transport the atomic sample far into it and optical access to the preparation volume in front of it. Figure~\ref{fig:Coldfinger} shows the cylindrical design with a cone-shaped tip used in the strontium lattice clock at PTB as an example. The bulk of the compartment must be manufactured from a material with high thermal conductivity, typically copper, to reduce thermal inhomogeneity within the cryogenic environment. Radiative contact with the surrounding room-temperature vacuum system can lead to substantial heat load. Moreover, it is crucial to increase the emissivity of the compartment's inner surfaces as much as possible, e.g., by applying vacuum-compatible black paint or using a graphite tube, to minimise the indirect perturbation of the atomic transition frequency by BBR leaking through pinholes or windows. Objects inside the cryogenic compartment, such as the aforementioned graphite tube, require careful consideration of their thermal contact with the remainder of the system and may lead to extended thermalisation time constants.
Placement of the temperature sensors is not particularly critical, aside from general caveats such as parasitic heat conduction via the wires and self-heating in vacuum.

Due to the small distance of its inner surfaces from the atomic sample, potential patch charges must be considered when designing the cryogenic environment. Characterisation of the DC Stark shift caused by residual static electric fields within should be possible, e.g., by installing electrodes that allow applying electric fields in a controlled fashion.

Furthermore, the residual background pressure inside the cryogenic volume must be kept in mind. Pinholes act as differential pumping stages leading to an increased background pressure on the inside without pumping (e.g., by non-evaporable getter materials). Materials that are prone to outgassing should thus be avoided.

Translation of the atomic sample is achieved either by inducing a frequency detuning between counterpropagating lattice beams to create a travelling standing wave \citep{Ushijima2015} or by mechanically translating the lattice optics \citep{Middelmann2012b}. The distance across which the atoms can or need to be transported is an important factor to be considered when designing the cryogenic compartment. In the latter case, the focus position of the beams moves along with the atomic sample. Avoiding Doppler frequency shifts induced by residual atomic motion with respect to the reference system of the interrogation laser is crucial in both cases. Mechanical translation allows the retroreflection mirror of the lattice to be used as a phase reference for the interrogation laser and to cancel the dominant effect of residual longitudinal motion. However, controlling motional effects at the $10^{-18}$ level of uncertainty has proven challenging \citep{AlMasoudi2016}. In any case, the time required for translation across a given distance is ultimately limited by the acceptable perturbation of the atomic sample in either case.

\subsubsection{Fully cryogenic clock}

Providing optical access to the relevant volume that is adequate for operating an optical lattice clock while keeping the thermal environment exposed to the atomic sample well-controlled and highly homogeneous at cryogenic temperatures is crucial for a fully cryogenic lattice clock. A suitable environment for room-temperature operation has already been demonstrated \citep{Beloy2014}. For a cryogenic clock, however, several additional problems must be considered --- in particular, transmission of BBR through any windows and the substantial heat load on any surfaces with high emissivity that are exposed to the surrounding room-temperature environment.

Although the glasses typically used for laser optics are non-transmitting across most of the spectral region that is relevant for BBR, there are two distinct cases where BBR leaking through a window may cause a substantial frequency shift of the clock transition. The first is terahertz radiation. The window's absorption eventually decreases toward low frequency, where the power spectral density in the BBR field is still quite substantial. The second case is the high-frequency tail of the BBR spectral power distribution, reaching into the transmission region of the window. Here, the polarisability of the electronic states used by the clock may be substantial due to transitions in the visible or near-infrared spectral range. For strontium, the $(5s5p)\, {}^3\mathrm{P}_0 \rightarrow (5s4d)\, {}^3\mathrm{D}_1$ transition at a wavelength of about \SI{2.6}{\micro\meter} is the most relevant.

Absorption of terahertz radiation by various types of glasses has been studied by \cite{Naftaly2007}. Their investigation shows that ionic and disorded glasses like BK7 exhibit much stronger absorption in this frequency regime than more frequently used materials such as silica. For a strontium lattice clock operating in a cryogenic chamber, leakage of room-temperature BBR radiation with $\nu \le \SI{2}{\tera\hertz}$ through a 3-mm-thick window alone leads to residual fractional shifts of about $2\times10^{-18}$ for silica, but only $3\times10^{-20}$ for BK7 glass. (These values are upper limits that do not include any reduction due to the actual field of view.) Therefore, BK7 glass should be favoured as a material for windows in the cryogenic environment.

We have combined the results reported by \cite{Naftaly2007} with absorption measurements of N-BK7 glass samples at frequencies in the range from \SI{1}{\tera\hertz} through \SI{120}{\tera\hertz} carried out at PTB in Berlin\footnote{We gratefully acknowledge the support from Ch.\ Monte.} and the manufacturer's specifications \citep{Schott2014} at high frequencies to model the frequency-dependent absorption across the relevant spectral range.
Together with the spectral density of the BBR field (see Equation~(\ref{eq:planck})) at room temperature ($T_\text{env}=\SI{300}{\kelvin}$), the DC Stark polarisability and the frequency-dependent corrections due to the most relevant low-frequency resonances, we have then used this model to estimate the maximum frequency shift (excluding any reduction due to the actual field of view) due to BBR leakage through windows as a function of their thickness for the case of a strontium lattice clock. We find that a window thickness of about \SI{10}{\milli\meter} is required to reduce it to less than 1 part in $10^{18}$. However, the shift in an actual clock will be reduced further due to the field of view (see Section \ref{sec:BBR_temperature_inhomogeneity}). Note that the dominant contribution (about \SI{75}{\percent}) to the frequency shift arises from the aforementioned ${}^3P_0 \rightarrow {}^3D_1$ transition.

\begin{figure}[tbp]
	\centering
	\includegraphics[width=0.65\columnwidth]{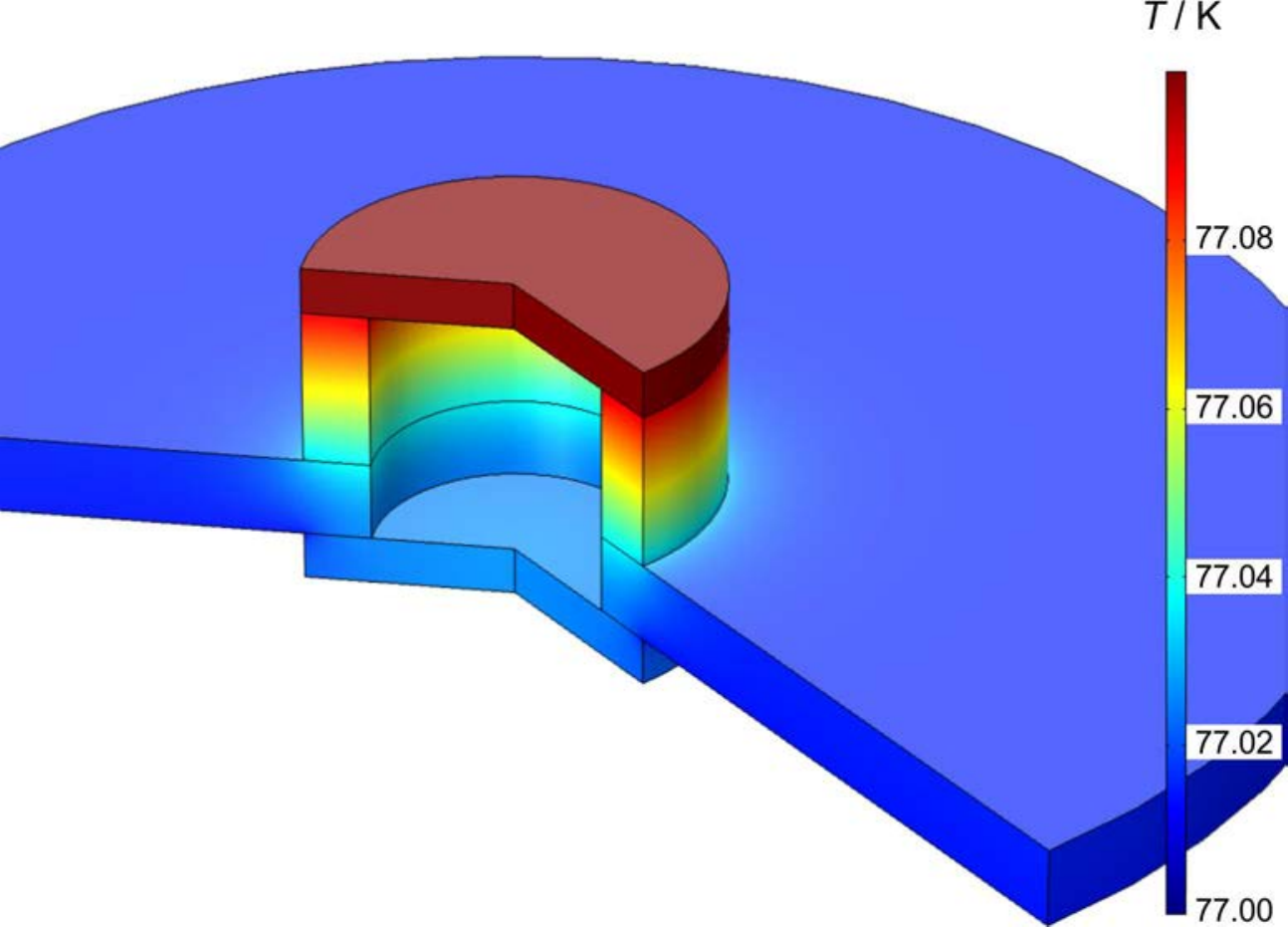}
	\caption{\label{fig:DoubleWindowFEM}FEM simulation of the temperature distribution in a dual-layer window system. The windows are connected to a copper disk, which is cooled to $\SI{77}{\kelvin}$. The upwards facing exterior surfaces are in radiative thermal contact with a room-temperature environment at a temperature of \SI{300}{\kelvin}. The temperature distribution within the exterior window has been truncated. The maximum temperature found in the window is \SI{82.7}{\kelvin}, i.e., almost \SI{6}{\kelvin} above the disk temperature. In contrast, the temperature increase of the inner window does not exceed a few $\SI{10}{\milli\kelvin}$.}
\end{figure}

Additionally, a window exposed to room-temperature BBR also suffers from substantial radiative heating due to its high emissivity, while heat dissipation is hindered by its relatively low thermal conductivity. Thus, it becomes substantially warmer than surrounding parts made from high-conductivity, low-emissivity materials such as copper. In particular, a hot spot in the atoms' thermal environment forms on the inner surface of the window. Its temperature may be elevated by several kelvin. Therefore, designs with a single layer of windows are not suitable for cryogenic lattice clocks. The problem can be mitigated by adding a secondary layer of windows. The inner windows do not experience substantial radiative heating and block out BBR emitted from the hot spot. Figure~\ref{fig:DoubleWindowFEM} shows an example of the temperature distribution in such a setup, determined by a FEM simulation. The secondary layer can be mounted to the same superstructure as the primary layer or to a separate superstructure, which facilitates removing the heat transferred to the windows without increasing temperature gradients across the environment exposed to the atomic sample.

Physical access to the cryogenic environment is still required for loading from an external atomic source and for pumping, unless a suitable pump is installed within the cryogenic environment, e.g., using non-evaporable getter materials. Due to the various sources of residual room-temperature BBR, ensuring high emissivity of the cryogenic environment's interior surfaces is highly important, e.g., by applying a suitable vacuum-compatible black paint.

Using these guidelines, a suitable temperature-controlled environment for fully cryogenic operation can be designed. Its dimensions are typically limited by the volume and optical access required for a conventional magneto-optical trap setup.

\subsection{Room-temperature clocks}

As an alternative to cryogenic clocks, room temperature solutions can be looked for.
For example, the temperature gradients across the vacuum chamber can be monitored and an uncertainty calculated as in Section \ref{sec:maximum-entropy}. BBR uncertainty at the level of $\num{1e-17}$ can be achieved in this way \citep{Falke2014,Hachisu2015,Pizzocaro2017}. Alternatively, \cite{Nicholson2015} proposed the use of radiation thermometry, where \emph{in situ} thermometers were used to measure the thermal environment close to the atom position, and achieved a BBR uncertainty of $\num{2e-18}$.

Finally, high-thermal-uniformity BBR shields can be used and operated at room temperature achieving uncertainty at $\num{1e-18}$, as first developed at NIST by \cite{Beloy2014}. The shield described by \cite{Beloy2014} has been used for an Yb lattice clock \citep{McGrew2018}. It is made of copper for high thermal conductivity and has seven windows for optical access and two apertures to allow the atomic beam through.
A high-emissivity carbon-nanotube coating is applied to all internal surfaces of the shield body. Eight calibrated thermometers are embedded in the shield for real-time measurements of the shield's absolute temperature. The shield is kept under ultra-high vacuum of $\SI{2e-7}{Pa}$, for both thermal isolation and atom trapping.

The uncertainty budget for the BBR shift in the shield considered the calibration and reading of the thermometers, and the temperature inhomogeneity caused by the windows and by the aperture.
The role of the effective solid angles of the two apertures is similar to the problem presented in Section \ref{sec:BBR_temperature_inhomogeneity}, but here the effective solid angles of the apertures were estimated by FEM simulation. The calculations are simplified given the high emissivity of the black coating and are not limited by the knowledge of the emissivity of the windows. Given that the windows are not as good a thermal conductor as copper, a thermal gradient can be established  on their surfaces, even if the effect is smaller at room temperature than at cryogenic temperatures. Given the low temperature uncertainty of the NIST system, the dominant contribution to the uncertainty is given by the dynamic coefficient of Yb.
\cite{Beloy2012} also changed the temperature of their shield between \SI{280}{K} and \SI{360}{K}, verifying Equation~\eqref{eq:bbr-short} in this temperature range.

\begin{figure}[tb]
    \centering
    \begin{subfigure}[t]{0.42\columnwidth}
        \centering
         \includegraphics[height=6.3cm]{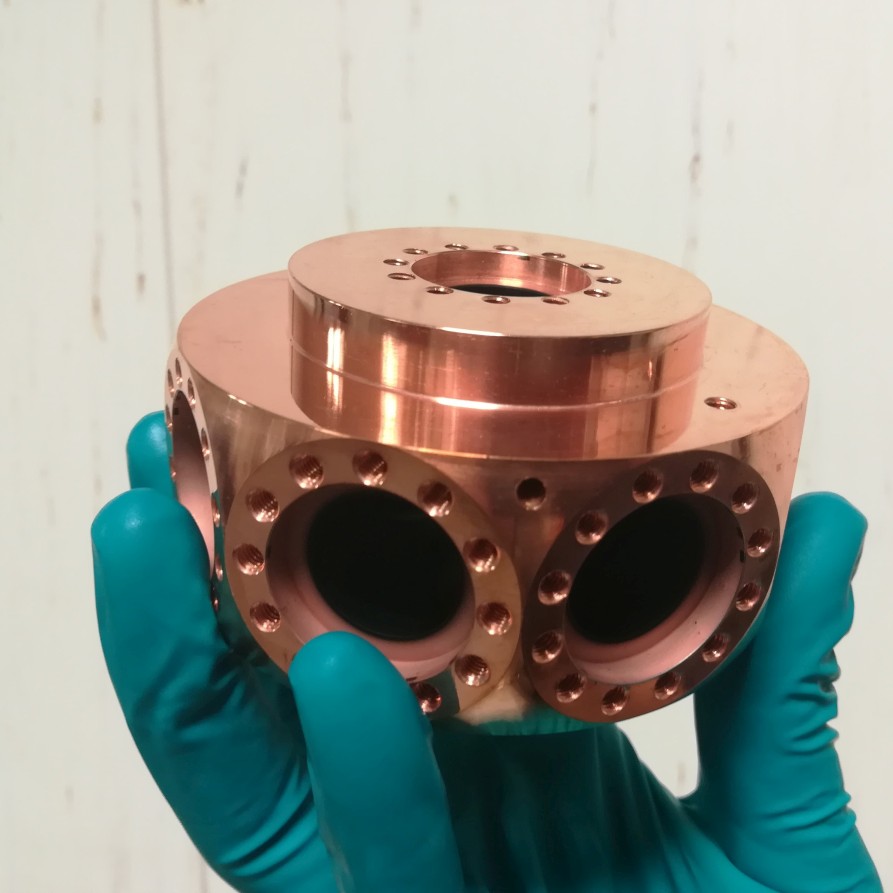}
         \caption{ }
    \end{subfigure}%
    \hfill
    \begin{subfigure}[t]{0.56\columnwidth}
        \centering
         \includegraphics[height=6.3cm]{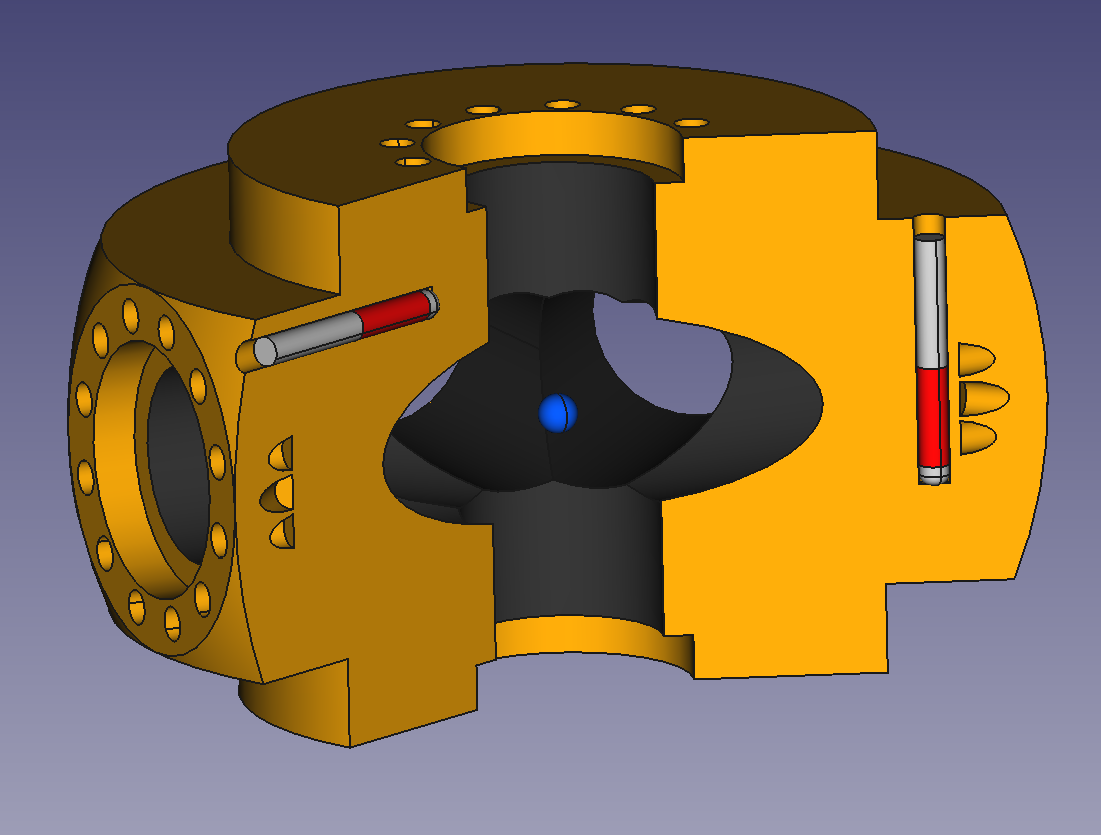}
         \caption{ }
    \end{subfigure}

    \caption{a) Picture of the BBR shield developed at INRIM. The shield is a single copper block with the inner faces coated with a high-emissivity material. b) Cutaway drawing of the shield showing the position of two of the six embedded thermometers (in red).}
    \label{fig:inrim-bbr}
\end{figure}

A picture of a similar shield developed at INRIM is shown in Figure \ref{fig:inrim-bbr}.
The shield is a single copper block and is coated with a  black coating compatible with ultra-high vacuum (Acktar Black). Six Pt100 thermometers are embedded in the shield for temperature evaluation.
A single aperture will be used for atomic physical access.
An uncertainty of \SI{10}{mK} is expected for the shield temperature.
Unlike the NIST design, this chamber will separate the inner vacuum that will be kept at ultra-high vacuum for increased atomic lifetime and the outer vacuum that will be maintained for thermal insulation only.

\cite{Beloy2018} updated the shield design so that it also works as a Faraday cage to shield atoms from external electric fields. High voltages (up to \SI{2}{kV}) can be applied to the conductive coating on the windows to directly measure the Stark shift. This application of high voltages requires the shield to be protected by insulating plastic to avoid arcing. This insulation further helps with the thermal uniformity of the shield.

\clearpage

\graphicspath{{D7_interrogation/}}

\chapter{Optimised interrogation methods \label{chap:interrogation}} 

\authorlist{Ian R.\ Hill$^{1,\dag}$,
Rachel M.\ Godun$^1$,
Nils Huntemann$^2$,
J\'er\^ome Lodewyck$^4$,
Ekkehard Peik$^2$,
Nils Scharnhorst$^{2,3}$ and
Piet O. Schmidt$^{2,3}$
}

\affil{1}{\NPLaff}
\affil{2}{\PTBaff}
\affil{3}{\LUHaff}
\affil{4}{\OPaff}
\corr{ian.hill@npl.co.uk}

\chapstart
This chapter explains how to minimise optical-clock inaccuracy and instability through careful choice of probing sequences and timings.  In Section~\ref{sec:interrogation_probe-shifts}, methods are presented to reduce inaccuracy associated with frequency shifts induced by the probe laser itself.  This is followed in Section~\ref{sec:interrogation_optimum-probe-time} by recommendations for the optimum probe times to minimise instability, given typical frequency-noise characteristics of the probe laser. Section~\ref{sec:interrogation_Dick} demonstrates how the Dick effect in optical lattice clocks can be eliminated and Section~\ref{sec:interrogation_interleaving} discusses how interleaving different measurements can be used to characterise and remove frequency shifts arising from the interaction of atoms with their environment.  Finally, the applicability of these interrogation methods is demonstrated in Section~\ref{sec:interrogation_operational-practices} with a case study based on Yb$^+$ optical clocks.

\section{Methods to reduce uncertainty from probe-induced shifts} \label{sec:interrogation_probe-shifts}

The desired output frequency of an atomic clock is that of the \emph{unperturbed} atomic transition.  It is inevitable, however, that probing the atom will disturb the very transition that is being used as the reference, leading to frequency shifts proportional to probe intensity.  For many optical clocks, the intensity of the probe beam is sufficiently weak that the resulting disturbances lead to fractional frequency shifts well below the $10^{-18}$ level and can be neglected.  Notable exceptions to this are the zero-spin bosonic optical lattice clocks based on the $^1$S$_0$ $\rightarrow$ $^3$P$_0$ transition (e.g., $^{88}$Sr, $^{174}$Yb, $^{24}$Mg)~\citep{Baillard2007, Poli2008, Akatsuka2010, Kulosa2015} and the ytterbium ion optical clock, based on an electric octupole transition~\citep{Roberts1997, Roberts2000, Godun2014, Huntemann2016}.

\subsection{Outline of the problem}

In fermionic optical lattice clocks, the nuclear spin gives rise to hyperfine structure and quenching of the $^3$P$_0$ excited clock state, giving natural linewidths on the order of a few millihertz. Sufficient Rabi frequencies for clock interrogation are obtained with very small probe powers, \SI{\ll 1}{\micro\watt}, and cause negligible probe-induced frequency shifts. In the boson there is no such favour and the clock state is very long lived ($\sim$years), decaying eventually through a weak E1M1 transition. To drive the clock transition requires a multi-photon excitation (e.g., by E1M1 or E1E1E1) using moderately large probe powers which incur large probe-induced AC Stark shifts. A popular approach is so-called magnetically induced spectroscopy (MIS) which drives the clock state using a resonant photon (E1) from an intense probe beam, $\sim$mW, and large applied static magnetic field, $\sim$mT, (M1) \citep{Taichenachev2006, Barber2006}. While attractive for its ease of implementation, it has both AC Stark and Zeeman probe-induced shifts to contend with.

In the ytterbium ion, the electric octupole transition is highly forbidden and has such a narrow linewidth that the spectral overlap with the probe laser is very low. Probe powers at the mW level are typically needed in a beam waist of ${\sim} \SI{10}{\micro\meter}$ in order to drive the clock transition on a timescale of ${\sim} 100$\;ms.  At these intensities, the AC Stark shift from the probe beam can induce fractional frequency shifts of ${\sim} 10^{-13}$.  Care must be taken to eliminate these frequency shifts from the final clock output, and the following subsections outline some different approaches.  Although these methods are demonstrated below with ytterbium ion optical clocks, they are equally applicable to bosonic optical lattice clocks or for removing smaller probe-induced shifts in other optical clocks.

\subsection{Extrapolation to zero frequency shift} \label{sec:interrogation_extrapolation}

\begin{figure}[tb]
    \centering
    \includegraphics[width=.5\columnwidth]{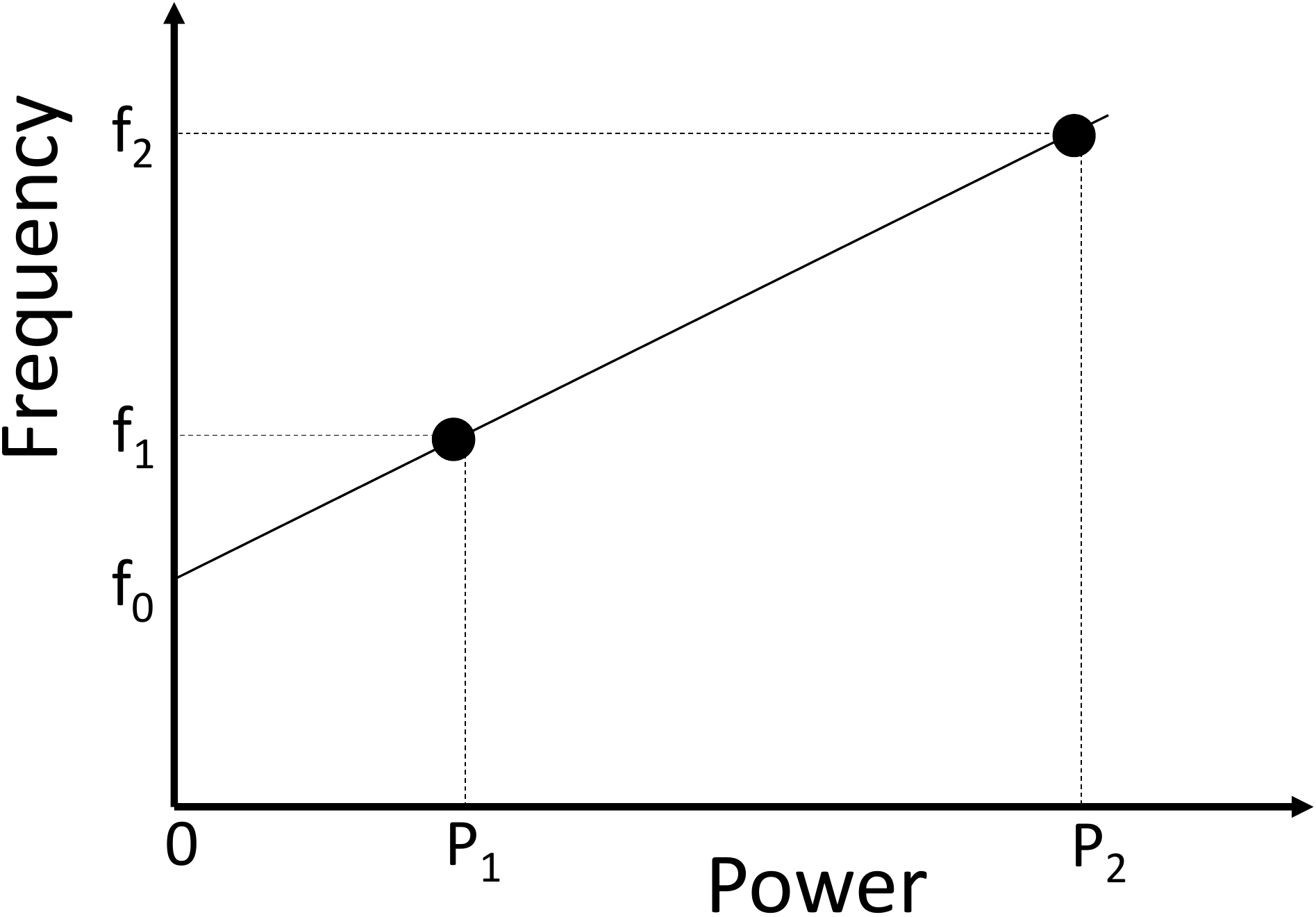}
    \caption{\label{fig:interrogation_extrapolate} The perturbed clock frequencies $f_1$ and $f_2$ are measured with probe powers $P_1$ and $P_2$ and then extrapolated to the unperturbed clock frequency $f_0$ at zero power.}
\end{figure}

A conceptually straightforward approach to eliminate the AC Stark shift induced by the probe is to measure the clock frequency with two different probe-beam intensities.  The clock frequency can then be extrapolated to the case of a probe beam with zero intensity~\citep{King2012,Huntemann2012a}.  In practice, with sufficiently stable beam pointing, the intensity remains proportional to the power and so only the power of the probe beam needs to be changed.  Assuming the powers do not exceed the range of linear scaling of the AC Stark shift, then Figure~\ref{fig:interrogation_extrapolate} shows the frequencies $f_1$ and $f_2$ that are measured with probe powers $P_1$ and $P_2$.

The unperturbed clock frequency $f_0$ can be evaluated in terms of $f_1$, $f_2$ and the ratio of the probe powers, $\kappa = P_2/P_1$,
\begin{eqnarray} \label{eq:interrogation_extrapolation}
\frac{f_1-f_0}{P_1} &=& \frac{f_2-f_0}{P_2}, \nonumber \\
f_0 &=& \frac{\kappa f_1 - f_2}{\kappa - 1}.
\end{eqnarray}
Note that $f_0$ depends only on the ratio of the powers, $\kappa$, rather than their absolute values. This is very convenient to measure experimentally as it requires only a power meter with good linearity, and does not require the power meter to be calibrated for absolute powers.

The experiment can be run in a mode that alternates between frequency readings at the two different power levels, with software carrying out on-the-fly extrapolation.  The timescale for switching between the different power levels should be dictated by how quickly other parameters begin to drift.  For example, if the beam position were to move during the extrapolation cycle, the power would no longer be proportional to the intensity at the atom and the extrapolation would be incorrect.  In general, it is best to switch power levels as often as possible.  But note that an exact alternation between high and low powers may be unwise if there is frequency drift in the probe laser that is predominantly in one direction.  The issue of how best to alternate probe pulses is dealt with in Section~\ref{sec:interrogation_interleaving}.

To account for the uncertainty introduced in the extrapolated clock frequency, $\sigma (f_0)$, a propagation of errors in Equation~(\ref{eq:interrogation_extrapolation}) leads to
\begin{equation} \label{eq:interrogation_extrapolation_error}
\sigma^2(f_0) = \left(\frac{\kappa}{\kappa-1}\right)^2 \sigma^2(f_1) + \left(\frac{1}{\kappa-1}\right)^2 \sigma^2(f_2) + \left(\frac{f_2 - f_1}{(\kappa-1)^2}\right)^2 \sigma^2(\kappa).
\end{equation}
The existence of a term depending on $\sigma(\kappa)$ shows that any error in the knowledge of the power ratio $\kappa$ would introduce a bias in the clock frequency $f_0$, but larger values of $\kappa$ lead to smaller errors.  The frequency extrapolation in the ytterbium ion clock is typically carried out with power ratios around two or  three.  With sufficient care over the power stability of the probe beam and the electronics used to monitor the power, it is possible to have knowledge of $\kappa$ to a level that introduces no more than $10^{-18}$ fractional offset to the extrapolated clock frequency.

\subsection{Tailored pulse sequences} \label{sec:interrogation_tailored-sequences}

Instead of measuring and accounting for light shifts by extrapolation, as outlined above, we can employ tailored interrogation pulse sequences based on generalised Ramsey excitation schemes, which allow for suppression of the light shift and other interrogation-related shifts.

In the Ramsey scheme \citep{Ramsey1990}, the two levels of the clock reference transition are brought into a coherent superposition by a first excitation pulse followed by a free-evolution period. After a second excitation pulse the population in one of the levels is detected, which shows the effect of the interference of the second pulse with the time-evolved superposition state.
In the case of excitation with high probe-light intensity that is required to drive a strongly forbidden transition, the spectrum obtained with Ramsey excitation indicates the presence of the light shift: the position and shape of the envelope reflects the excitation spectrum resulting from one of the pulses, whereas the Ramsey fringes result from coherent excitation with both pulses and the intermediate dark period. The fringes are less shifted than the envelope because their shift is determined by the time average of the intensity. This results in a shifted and asymmetric Ramsey pattern.

This intuitive picture suggests that the effect of the light shift $\Delta_\text{L}$ on the spectrum can be compensated by introducing a frequency step of the probe light, $\Delta_\text{S}=\Delta_\text{L}$, during the interrogation pulses as proposed in \cite{Taichenachev2010}. Additionally, the scheme can be made insensitive against small changes of the laser intensity or errors in $\Delta_\text{S}$ by inserting an additional pulse with identical intensity and frequency, but with a doubled duration, between the Ramsey pulses. The additional pulse compensates the dephasing between the atomic coherence and the probe laser field caused by the Ramsey pulses. The phase of the additional pulse is shifted by $\pi$ relative to the Ramsey pulses in order to improve the robustness against variations of the pulse area. The frequency steps are applied in a phase-coherent way so that they do not introduce additional phase changes of the probe field. These are the essential elements of the hyper-Ramsey spectroscopy (HRS) scheme as proposed in \cite{Yudin2010} and first experimentally demonstrated on the electric octupole transition $^2S_{1/2}\rightarrow {}^2F_{7/2}$ in $^{171}\text{Yb}^+$ \citep{Huntemann2012b}.

In order to stabilise the frequency of an oscillator to the line centre of the atomic resonance signal, in many clocks resonance signals are alternately recorded with a fixed positive and negative detuning around the line centre. The difference between the excitation probabilities obtained with these detunings yields a discriminator signal that varies antisymmetrically around the resonance centre. For Ramsey excitation, a discriminator signal can also be produced by alternately applying phase steps of $\phi=\pm\pi/2$ to one of the excitation pulses while the excitation frequency is kept constant \citep{Ramsey1951, Letchumanan2004}. If applied to the HRS excitation scheme, the latter technique is particularly advantageous because the immunity to light-shift fluctuations that lead to asymmetries in the line shape is further enhanced.

The light-shift-compensating step frequency $\Delta_\mathrm{S}$ is readily obtainable from Rabi spectroscopy. In order to confine $\delta_\mathrm{L}=\Delta_\mathrm{L}-\Delta_\mathrm{S}$ to the region where the HRS clock error remains small, use can be made of the significantly larger sensitivity of $\nu_\text{clock}=\nu_\text{Rabi}$ to $\delta_\mathrm{L}$ in Rabi spectroscopy. A feedback loop that employs the difference $\varepsilon=\nu_\text{HRS} -(\nu_\text{Rabi}-\Delta_\mathrm{S})$, as determined with HRS and Rabi excitations, as the discriminator signal can be used to steer $\Delta_\mathrm{S}$ so that $|\delta_\mathrm{L}|$ approaches zero.

Although the combination of interrogation schemes converts the uncertainty due to the light shift into a predominantly statistical contribution, systematic shifts can be caused by drifts of the light shift and by a difference of the shifts present during the Rabi and the HRS interrogations. In the realisation with the $^{171}\text{Yb}^+$ optical clock at PTB \citep{Huntemann2016}, slow variations of the light shift corresponding to a drift of $\Delta_\mathrm{L}$ in the range of $\SI{50}{\micro\hertz/\second}$ are typically observed. With a servo time constant of 200\;s for $\Delta_\mathrm{S}$, the resulting servo error of $\delta_\mathrm{L}$ is $10$\;mHz. This error could be avoided by using a drift-compensating second-order integrating servo algorithm \citep{Peik2005}. The main reason for differences in the frequency shifts present during the interrogation pulses are transient thermal effects of the crystal of the AOM that shapes the pulses. The resulting phase variations (AOM chirp) were found to lead to a frequency difference of less than 1\;mHz \citep{Kazda2015}. Beam pointing and focussing variations induced by different crystal temperatures were found to lead to light-shift differences between the pulses of less than 0.2\;mHz. The combination of these systematic effects and the residual sensitivity $\partial\nu_\text{HRS}/\partial\delta_\mathrm{L}=0.07$ yielded a probe-light-related fractional uncertainty of $1.1\times 10^{-18}$.

\begin{figure}[t]
\begin{center}
\includegraphics[width=0.95\textwidth]{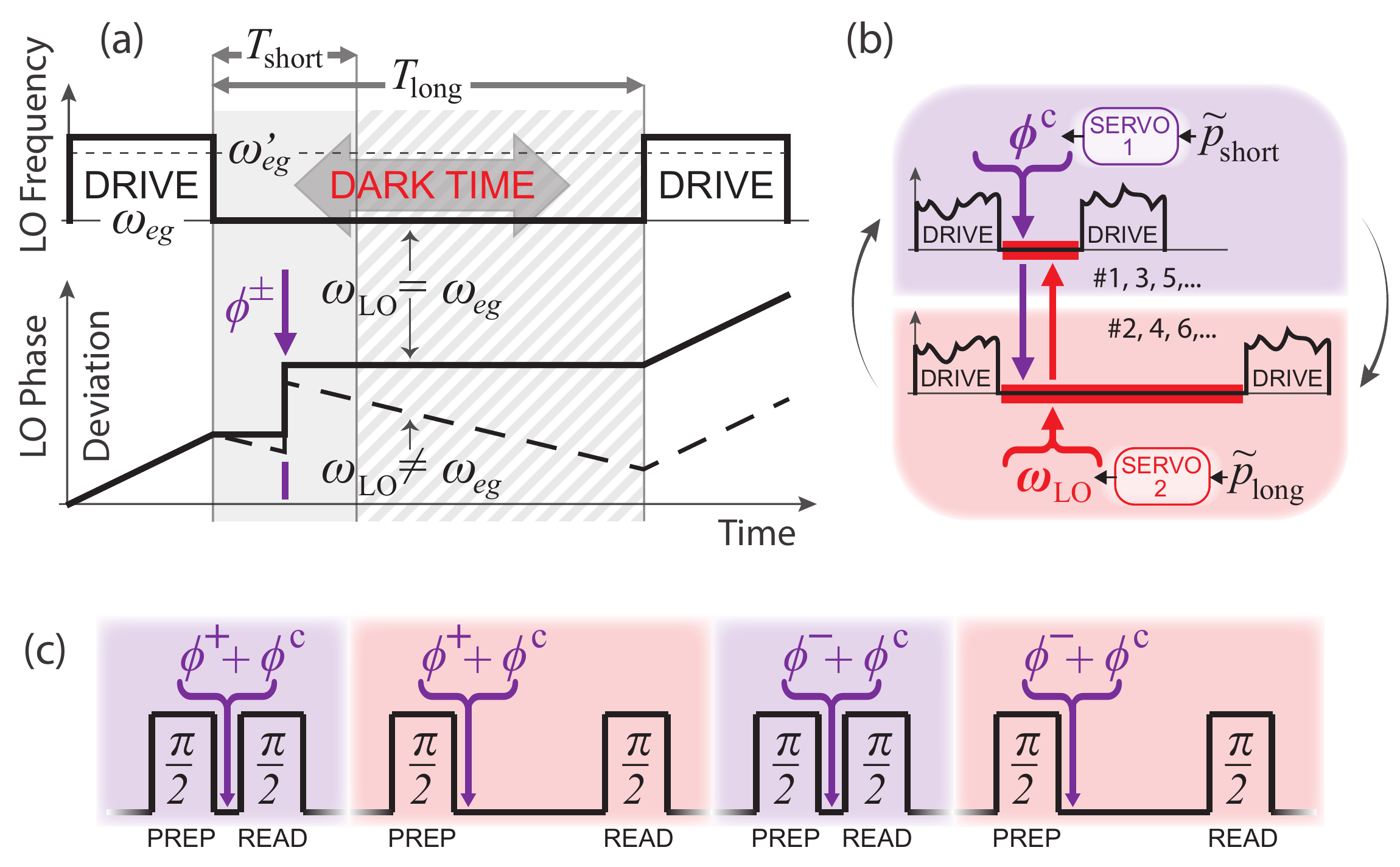}
\caption[]{
Ramsey interrogation with shift-inducing drive pulses. The drive pulse frequency $\omega_{\mathrm{LOdrive}}$ in (a) is assumed to be slightly higher than the light-shifted transition frequency $\omega'_{eg}$. This corresponds to an increasing LO phase deviation during the drive period compared to an oscillator that evolves precisely with $\omega'_{eg}$ and $\omega_{eg}$. Over the following dark Ramsey time interval of variable duration $T$, the acquired phase difference between LO field and atomic oscillator will remain unchanged (aside from the $\phi^{\pm}$ modulation) if the LO phase advances with the unperturbed transition frequency $\omega_{\mathrm{LO}} = \omega_{eg}$ (solid curve). In this case, the phase-to-population conversion performed by the second drive pulse will produce identical outcomes after short and long Ramsey times. Otherwise, for nonzero Ramsey detuning (dashed curve), a $T$-dependent relative phase is accumulated resulting in differing populations. (b) An autobalancing control scheme comprises two servo loops acting in parallel on alternately operated short and long Ramsey sequences. Servo~1 equals the phase-modulated excited-state populations $p_{\mathrm{short}}^{+}$ and $p_{\mathrm{short}}^{-}$, i.e., it nulls $\tilde{p}_{\mathrm{short}}$ by adjusting $\phi^{\mathrm{c}}$ in \textit{both} the short and long sequences. Servo 2 nulls $\tilde{p}_{\mathrm{long}}$ via $\omega_{\mathrm{LO}}$, also addressing \textit{both} sequences. The two example plots, displaying LO frequency vs time, represent distorted but isomorphic sequences which only differ in their dark times. (c) One complete probe cycle for balanced acquisition consists of four Ramsey pulse pairs combining two dark times with two phase hopping directions. From \cite{Sanner2017}. \label{fig:interrogation_pulsesandloops}}
\end{center}
\end{figure}

A more universal approach to eliminate a wide variety of interrogation-induced shifts like light shifts, phase chirps, and transient Zeeman shifts was developed within the OC18 project, based on actively balancing the spectroscopic responses from phase-congruent Ramsey probe cycles of unequal durations, see Figure~\ref{fig:interrogation_pulsesandloops} \citep{Sanner2017}.
This autobalanced Ramsey spectroscopy (ABRS) scheme has been experimentally demonstrated and it was shown that no systematic clock errors are incurred for arbitrarily detuned or otherwise defective drive pulses.

The fact that the spectroscopically relevant phase information is acquired during an interaction-free period makes Ramsey's protocol the natural choice when aiming to undisturbingly extract the transition frequency $\omega_{eg} = E_{e} / \hbar$, where $E_{e}$ is the energy of the excited state $|e\rangle$ relative to the ground state $|g\rangle$. Nonetheless, the deterministic preparation of the initial superposition state and the subsequent phase readout, which maps phase differences to observable population differences, require interactions with the oscillatory drive pulses. Hence, pulse defects (e.g., frequency deviations) potentially affect the outcome of the spectroscopic measurement leading to erroneous phase values and corresponding clock errors. If the local-oscillator phase and the atomic phase evolve at identical pace during the dark period, one can expect to find the same excited-state population when comparing the outcome of two ``isomorphic'' interrogation sequences which only differ in their dark times $T_{\mathrm{short}}$ and $T_{\mathrm{long}}$ (see also \citet{Morgenweg2014}).

Based on the simple insight that interrogation-induced errors are common mode for isomorphic Ramsey sequences it is now straightforward to combine $T_{\mathrm{short}}$ and $T_{\mathrm{long}}$ Ramsey interrogations. The respective populations $\tilde{p}_{\mathrm{short}}$ and $\tilde{p}_{\mathrm{long}}$ can be combined into a defect-immune spectroscopy scheme by using the differential population $\tilde{p}_{\mathrm{bal}} = \tilde{p}_{\mathrm{long}} - \tilde{p}_{\mathrm{short}}$ as an error signal for the local-oscillator frequency $\omega_{\mathrm{LO}}$. From this one obtains a passively balanced frequency feedback with $\tilde{p}_{\mathrm{bal}} = 0$ exclusively for $\omega_{\mathrm{LO}} = \omega_{eg}$.

This approach, however, comes with a major drawback that is, to a lesser extent, also encountered in recently proposed coherent composite-pulse schemes \citep{Hobson2016, Zanon2016}: for $\tilde{p}_{\mathrm{short}} \neq 0$ the frequency discriminant $\tilde{p}_{\mathrm{bal}} (\delta)$ is no longer an odd function around $\delta = 0$, where $\delta = \omega_\text{LO} - \omega_{eg}$ is the detuning, which leads to skewed sampling distributions and corresponding clock errors. To avoid this issue we implement an \textit{active} balancing process with two interconnected control loops as schematically displayed in Figure \ref{fig:interrogation_pulsesandloops}(b). The first feedback loop, acting on the short Ramsey sequence, ensures $\tilde{p}_{\mathrm{short}} = 0$ by injecting together with $\phi^{\pm}$ an additional phase correction $\phi^{\mathrm{c}}$. Of course, this requires that the drive detuning $\delta' = \omega_{\mathrm{LOdrive}} - \omega'_{eg}$ is smaller than the Rabi frequency $\Omega_{0}$ for sufficient fringe contrast. Depending on the dominant source of error, one could also use $\omega_{\mathrm{LOdrive}}$ as the control variable. During $T_{\mathrm{short}}$ the local oscillator evolves with $\omega_{\mathrm{LO}}$ as determined via the second feedback loop which controls the long Ramsey sequence by steering $\omega_{\mathrm{LO}}$ so that $\tilde{p}_{\mathrm{long}} = 0$. Vice versa, $\phi^{\mathrm{c}}$ (or $\omega_{\mathrm{LOdrive}}$) as obtained from the short sequence is identically applied in the long Ramsey sequence. In this way a constantly updated common-mode correction autobalances the long interrogation whose outcome is then denoted by~$\tilde{p}_{\mathrm{auto}}$.

\begin{figure}[tb]
\begin{center}
\includegraphics[trim={1mm 2mm 7mm 5mm},clip,width=0.75\textwidth]{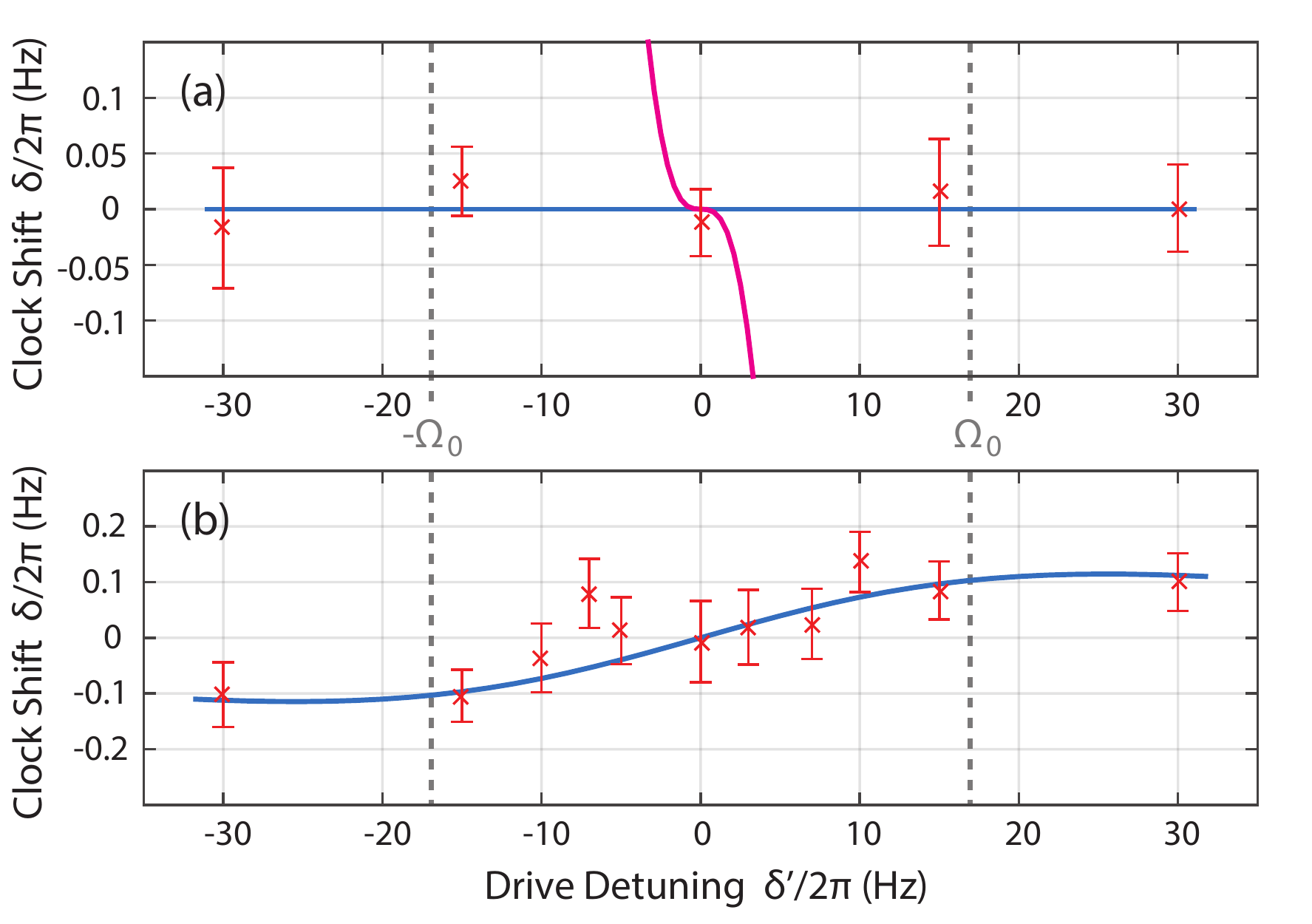} 
\caption[]{Clock shifts of the $^{171}$Yb$^{+}$ E3 transition frequency as obtained through autobalanced Ramsey spectroscopy when operated with intentionally detuned drive pulses. (a) For isomorphic short and long interrogation sequences the autobalanced clock is fully immune against drive-frequency deviations and the measured data points ($1\sigma$ error bars shown) line up on the blue zero-crosstalk axis. The magenta curve illustrates the $\delta'$-to-$\delta$ error propagation expected with the original hyper-Ramsey protocol. (b) Even without heating compensation the coupling between clock shift and drive detuning is strongly suppressed. Within statistical uncertainty the measured clock deviations reproduce the numerically simulated dependence (blue curve). From \cite{Sanner2017}. \label{fig:interrogation_transfercurves}}
\end{center}
\end{figure}

Experiments were carried out with a single $^{171}$Yb$^{+}$ ion confined in an endcap-type radio-frequency Paul trap. Figure \ref{fig:interrogation_transfercurves}(a) shows the results of autobalanced-clock runs with $\Omega_{0} = 2 \pi \times 17$\;Hz corresponding to a $\pi / 2$-pulse duration of 15\;ms, $T_{\mathrm{short}} = 6$\;ms, and $T_{\mathrm{long}} = 60$\;ms. Using about 5\;mW of drive laser light focused in a 50-\si{\micro\meter}-diameter spot causes a large light shift of $\omega'_{eg} - \omega_{eg} \approx 2 \pi \times 660$ Hz. Within the statistical uncertainty no clock error is observed for drive detunings $\delta'$ of up to $2 \Omega_{0}$. Beyond this detuning the response curve stays flat but the data points' statistical uncertainties eventually increase due to the reduced fringe amplitude. In contrast, operating a HRS interrogation at $\delta' > \Omega_{0}$ yields large clock errors that scale linearly with $\delta'$; only for $\delta' \ll \Omega_{0}$ is the error propagation suppressed to first order.  Without compensating for the motional heating of the ion (Section~\ref{sec:iontrap_fieldnoise}), one finds a residual dependence
of $\delta$ on $\delta'$ as displayed in Figure~\ref{fig:interrogation_transfercurves}(b). This dependence
is rather weak and does not introduce a clock
error as long as $\delta'$ is on average zero; yet it confirms that
heating violates the isomorphism of short and long Ramsey
sequences.

\begin{figure}[tb]
\begin{center}
\includegraphics[trim={2mm 3.5mm 3mm 2.5mm},clip,width=0.85\textwidth]{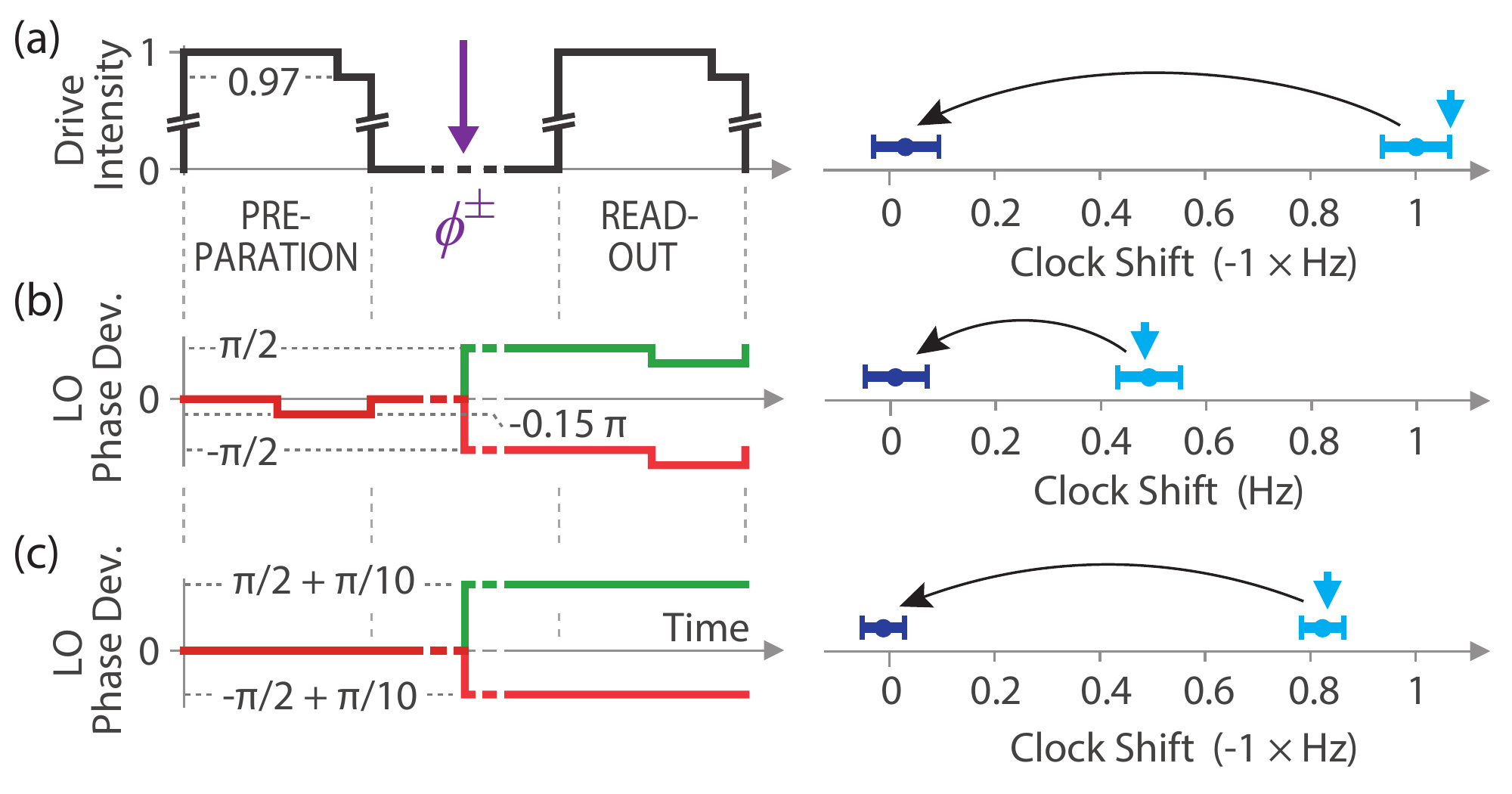}
\caption[]{Defective drive sequences and resulting clock shifts for the case of standard Ramsey spectroscopy (light blue data points with vertical arrows indicating the theoretically expected shift values) and for autobalanced Ramsey spectroscopy (dark blue data points). The applied intensity defect (a), phase excursion (b), and phase lag (c) lead to large clock offsets that are eliminated in autobalanced Ramsey mode. Both phase deviation plots display $\phi^{-}$ (red) and $\phi^{+}$ (green) traces simultaneously. From \cite{Sanner2017}. \label{fig:interrogation_pulsedefects}}
\end{center}
\end{figure}

In order to illustrate the universality of the autobalancing approach, the Yb$^{+}$ clock's response to various intentionally introduced interrogation defects was measured. Figure \ref{fig:interrogation_pulsedefects} displays three defect scenarios together with their resulting clock errors. In the first scenario the $\pi / 2$-pulses are delivered with 97\% of the nominal intensity for the last 3\;ms of their 15\;ms on-time. Considering a total light shift of 660\;Hz this intensity defect is equivalent to a temporary $\omega_{\mathrm{LOdrive}}$ deviation of more than $\Omega_{0}$ and gives rise to a clock shift of about 1\;Hz when using an unbalanced Ramsey sequence. Autobalancing the acquisition recovers the undisturbed clock frequency to within the statistical uncertainty. Similarly, defect immunity is verified for drive pulses suffering from engineered phase defects with $0.15 \pi$ phase excursions during the second half of each atom-light interaction. Finally, a third scenario assumes a phase lag $\theta = \pi / 10$ after the first $\pi / 2$-pulse, i.e., one effectively uses $\phi^{\pm} = \pm \pi / 2 + \theta$ instead of employing a symmetric phase modulation. In this case the injected servo-controlled phase correction $\phi^{\mathrm{c}}$ compensates the encountered phase lag one-to-one.

While these pulse defects are exaggerated for demonstration purposes they represent a wide variety of less pronounced and often unnoted parasitic interrogation side effects. For instance, many clocks incorporate magnetic-field switching for the atomic ground-state preparation. Certain clock transitions also require magnetic-field-induced state admixtures to enable direct optical Ramsey excitation \citep{Taichenachev2006}. Due to transients associated with the switching of the field, the dark Ramsey time gets perturbed by a transient Zeeman shift, which has in the past limited the accuracy of such clocks. Phase excursions triggered by laser beam shutters or acousto-optic modulators are another possible source of error. By choosing a proper $T_{\mathrm{short}}$ that still registers the transient perturbations, one can preventively address all such issues.

\section{\texorpdfstring{\sloppy}{}Optimum probe time and achievable instabilities for various clock candidates}
\label{sec:interrogation_optimum-probe-time}

The fractional frequency instability of an optical clock is an important performance criterion, since it determines the averaging time required to achieve a certain frequency resolution. Fundamentally, it arises from quantum projection noise (QPN) of the measured atomic populations and is typically expressed as an Allan deviation \citep{Riehle2004}, which  for Ramsey interrogation is given as
\begin{equation} \label{eq:interrogation_allanclock}
\clkadev(\tau)=\frac{1}{\omega T\sqrt{N}}\sqrt{\frac{T_\mathrm{c}}{\tau}},
\end{equation}
where $\omega$ is the angular frequency of the clock transition, $N$ the number of interrogated atoms, $T$ the Ramsey time, $T_\mathrm{c}$ the total cycle time including dead time, and $\tau$ the averaging time. This relation shows that the probe time should be as long as possible to achieve small statistical uncertainties. The maximum probe time is limited by either the excited-state lifetime of the clock transition \citep{Peik2005} or the coherence time of the laser (local oscillator, LO) \citep{Leroux2017}, which typically follows a flicker-floor-limited instability (flat Allan deviation) for the probe times of interest. In a simple picture, the maximum probe time due to laser phase noise is given by the noise accumulating to more than $\pm\pi$. Such a large phase error during interrogation results in ambiguities in the assignment of the Ramsey fringe being probed and thus in so-called ``cycle slips'' and loss of lock.
Within the OC18 project, we have analytically and numerically investigated optimum probe times for different LO noise characteristics and atom numbers \citep{Leroux2017}, complementing previous work on lifetime-limited probing \citep{Peik2005} for single ions.

In the following we summarise our findings and provide explicit values for optimal interrogation times as a function of typical LO noise levels and clock candidates. The calculation assumes a dead-time-free interrogation scheme ($T=T_\mathrm{c}$) and therefore neglects the Dick effect. Furthermore, spontaneous emission from the excited clock state is neglected. Including both effects into the analysis is subject to future work.

We define a characteristic timescale $Z = 1/(\omega \loadev)$ for which the LO Allan deviation $\loadev$ (independent of its noise characteristic) equals the instability of the clock after one probe cycle of a single atom as given by Equation~(\ref{eq:interrogation_allanclock}). The first important result is given in Table~\ref{tab:interrogation_Topt}, which lists the recommended optimal probe times $T_\mathrm{o}$ for a given noise spectrum and noise level, together with the asymptotic optimum for large $N$.

\begin{table}[tb]
\begin{center}
  \caption{
    Recommendations for the choice of Ramsey interrogation time.
    The last column gives the optimal probe time $T_\mathrm{o}$ scaled by the characteristic timescale $Z$ as a function of the number of atoms $N$ in the limit of many atoms.
    There is nothing to be gained by probing longer than this time.
    It may be necessary to use shorter probe times to avoid fringe hops;
    a suggested safe upper bound on the probe time is given in the second column. From \cite{Leroux2017}.
   } \label{tab:interrogation_Topt}
  \begin{tabular}{lcc}
    \toprule
    LO noise type & Safe $T_\mathrm{o}/\Tnought$ &  Asymptotic optimum $T_\mathrm{o}/\Tnought$\\
    \midrule
    White & -- & $\min({N^{-1/5},\,1.4\,N^{-1/3}})$ \\
    Flicker & $(0.4 \ldots 0.15)\,N^{-1/3}$ & $0.76\,N^{-1/6}$ \\
    Random walk & $(0.4\ldots 0.25)\,N^{-1/3}$ & $0.79\,N^{-1/9}$ \\
    \bottomrule
  \end{tabular}
\end{center}
\end{table}

The reduction of $T_\mathrm{o}$ with increasing atom number can be understood from the interplay between the laser phase noise and the quantum projection noise from the atoms. For pure white LO frequency noise, no optimal probe time exists, since QPN and LO noise both scale as $1/\sqrt{\tau}$ and there is no need to feed back onto the laser's frequency, since it already averages down as a clock would. For LO flicker and random walk frequency noise, the LO noise integrated over the interrogation time equals that of the atomic QPN noise level near the optimum probe time. Since the QPN reduces with increasing $N$, this optimum probe time becomes shorter.

The second noteworthy result is that the instability of a clock with $N>1$, operated with the optimal probe time for each $N$, deviates from the $1/\sqrt{N}$ scaling of Equation~(\ref{eq:interrogation_allanclock}) as shown in Table~\ref{tab:interrogation_Nscaling}. The slight reduction in instability gain as $N$ increases is again a consequence of the slightly shorter optimum probe times for increasing atom number.

\begin{table}[b]
\begin{center}
\caption{
    Asymptotic scaling of LO-limited clock instability with atom number $N$.
    The scaling differs from the conventional $N^{-1/2}$ QPN
    because the optimum probe time decreases with increasing atom number. From \cite{Leroux2017}.
   } \label{tab:interrogation_Nscaling}
  \begin{tabular}{ll}
    \toprule
    LO noise type & Asymptotic scaling of $\clkadev(\tau)\wnom\sqrt{\Tnought\tau}$\\
    \midrule
    White & \hspace{20mm} $\propto N^{-1/3}$ \\
    Flicker & \hspace{20mm} $\propto N^{-5/12}$ \\
    Random walk & \hspace{20mm} $\propto N^{-4/9}$ \\
    \bottomrule
  \end{tabular}
\end{center}
\end{table}

Table~\ref{tab:interrogation_results} lists the achievable clock uncertainty,
\begin{equation}
	\clkadev(1\;\mathrm{s}) = \frac{C}{\omega\sqrt{T_\mathrm{o}\times 1\;\mathrm{s}}},
\end{equation}
for the most common clock candidates probed with a flicker-noise-limited clock laser with a noise floor $\loadev$. Here, $C$ is a numerical factor accounting for the trade-off between LO and quantum projection noise derived from simulations. For ion (neutral atom) clock candidates with $N=1$ ($N=1000$) the correction factor is $2$ ($0.07$). Following \cite{Peik2005}, we have limited the maximum probe time to the excited-state lifetime $\tau_0$ and loss of contrast from spontaneous emission is taken into account with a scaling factor $\exp[-T_\mathrm{o}/(2\tau_0)]$ for the instability. Our results for the instability are more optimistic compared to \cite{Peik2005}.

\begin{table}[b!]
\begin{center}
\sisetup{table-format=2.2}
	\begin{threeparttable}
		\caption{\label{tab:interrogation_results} Calculated values for the optimum clock-transition probe time $T$ (in seconds) and the achievable uncertainty $\clkadev$ at \SI{1}{\second} averaging time for a laser with pure flicker noise with different noise floors $\loadev$ for various clock candidates. The numbers in square brackets denote the transition wavelength and excited-state lifetime for each species. The latter represents at the same time the maximum probe time. For ions (atoms) $N=1$ ($N=1000$) is assumed, where $N$ is the number of interrogated trapped ions (neutral atoms).}
		\begin{tabularx}{\textwidth}{Sp{2mm}SSSSSSSSSS}
			\toprule
			& & \multicolumn{2}{c}{{Ca$^+$}} & \multicolumn{2}{c}{{Sr$^+$}} & \multicolumn{2}{c}{{Yb$^+$ E2}} & \multicolumn{2}{c}{{Yb$^+$ E3}} & \multicolumn{2}{c}{{Al$^+$}} \\
			& &	\multicolumn{2}{c}{{[\SI{729}{nm}]}} & \multicolumn{2}{c}{{[\SI{674}{nm}]}} & \multicolumn{2}{c}{{[\SI{436}{nm}]}} & \multicolumn{2}{c}{{[\SI{467}{nm}]}} & 			\multicolumn{2}{c}{{[\SI{267}{nm}]}} \\
			& &	\multicolumn{2}{c}{{[\SI{1.13}{s}]}} & \multicolumn{2}{c}{{[\SI{0.4}{s}]}} & \multicolumn{2}{c}{{[\SI{51}{ms}]}} & \multicolumn{2}{c}{{[\SI{1.6e8}{s}]}} & 		\multicolumn{2}{c}{{[\SI{20.7}{s}]}} \\			\cmidrule(r){3-4}\cmidrule(r){5-6}\cmidrule(r){7-8}\cmidrule(r){9-10}\cmidrule(r){11-12}
			{$\loadev$} & & {T} & {$\clkadev$} & {T} & {$\clkadev$} & {T} & {$\clkadev$} & {T} & {$\clkadev$} & {T} & {$\clkadev$}\\
			{$10^{-16}$} & & {s} & {$10^{-16}$} & {s} & {$10^{-16}$} & {s} & {$10^{-16}$} & {s} & {$10^{-16}$} & {s} & {$10^{-16}$}\\
			\midrule 	
10.00 & & 0.10 & 11.92 & 0.09 & 10.70 & 0.05 & 6.22 & 0.06 & 9.96 & 0.04 & 7.52 \\
5.00 & & 0.19 & 8.08 & 0.18 & 6.76 & 0.05 & 6.22 & 0.12 & 7.04 & 0.07 & 5.32 \\
1.00 & & 0.97 & 2.56 & 0.40 & 3.43 & 0.05 & 6.22 & 0.62 & 3.15 & 0.35 & 2.36 \\
0.50 & & 1.13 & 2.21 & 0.40 & 3.43 & 0.05 & 6.22 & 1.24 & 2.23 & 0.71 & 1.66 \\
0.10 & & 1.13 & 2.21 & 0.40 & 3.43 & 0.05 & 6.22 & 6.20 & 1.00 & 3.54 & 0.69 \\
0.05 & & 1.13 & 2.21 & 0.40 & 3.43 & 0.05 & 6.22 & 12.40 & 0.70 & 7.09 & 0.45 \\
			\midrule
			& & \multicolumn{2}{c}{{Sr}} & \multicolumn{2}{c}{{Yb}} & \multicolumn{2}{c}{{Hg}}&&&&\\
			& & \multicolumn{2}{c}{[\SI{698}{nm}]} & \multicolumn{2}{c}{[\SI{578}{nm}]} & \multicolumn{2}{c}{[\SI{266}{nm}]}&&&&\\
			& & \multicolumn{2}{c}{[\SI{159}{s}]} & \multicolumn{2}{c}{[\SI{15.9}{s}]} & \multicolumn{2}{c}{[\SI{1.59}{s}]}&&&&\\
			\cmidrule(r){3-4}\cmidrule(r){5-6}\cmidrule(r){7-8}
			{$\loadev$} & & {T} & {$\clkadev$} & {T} & {$\clkadev$} & {T} & {$\clkadev$}&\\
			{$10^{-16}$} & & {s} & {$10^{-16}$} & {s} & {$10^{-16}$} & {s} & {$10^{-16}$} \\
			\midrule
10.00 & & 0.14 & 0.43 & 0.12 & 0.39 & 0.05 & 0.26 \\
5.00 & & 0.29 & 0.30 & 0.24 & 0.27 & 0.11 & 0.18 \\
1.00 & & 1.43 & 0.13 & 1.18 & 0.12 & 0.54 & 0.07 \\
0.50 & & 2.85 & 0.09 & 2.36 & 0.08 & 1.09 & 0.04 \\
0.10 & & 14.27 & 0.04 & 11.81 & 0.03 & 1.59 & 0.03 \\
0.05 & & 28.53 & 0.03 & 15.90 & 0.02 & 1.59 & 0.03 \\
			\bottomrule
		\end{tabularx}
	\end{threeparttable}
\end{center}
\end{table}


\section{Methods to remove instability caused by the Dick effect}
\label{sec:interrogation_Dick}

The instability of many optical clocks, especially optical lattice clocks for which the QPN level is extremely low (lower than $10^{-17}/\sqrt{\tau}$ for $10^4$ atoms, and possibly even lower for long interrogation times of several seconds), is limited by the Dick effect. This is an aliasing effect arising from the sampling of the clock-laser noise by the clock sequence. This sampling converts the high-frequency noise of the clock laser (typically from 1 to 100\;Hz) into undesirable noise at the frequency of the clock cycle with an efficiency that decreases when the fraction of the cycle spent probing the atoms with the clock laser (the ``duty cycle'') increases.

Several techniques have been developed in order to reduce the Dick effect. First, better clock lasers are designed in order to decrease the laser frequency noise. This includes several technical developments, such as long cavities, the use of crystalline coatings, cryogenic silicon cavities, or more speculative research such as spectral hole burning techniques (see Chapter~\ref{chap:ultrastable}). However, another way to reduce the Dick effect is to optimise the interrogation of the atoms with the clock laser, which is generally less demanding than building state-of-the-art clock lasers.

\begin{figure}[tb]
	\begin{center}
	\includegraphics[trim={2mm 2mm 6mm 2mm},clip,width=0.95\textwidth]{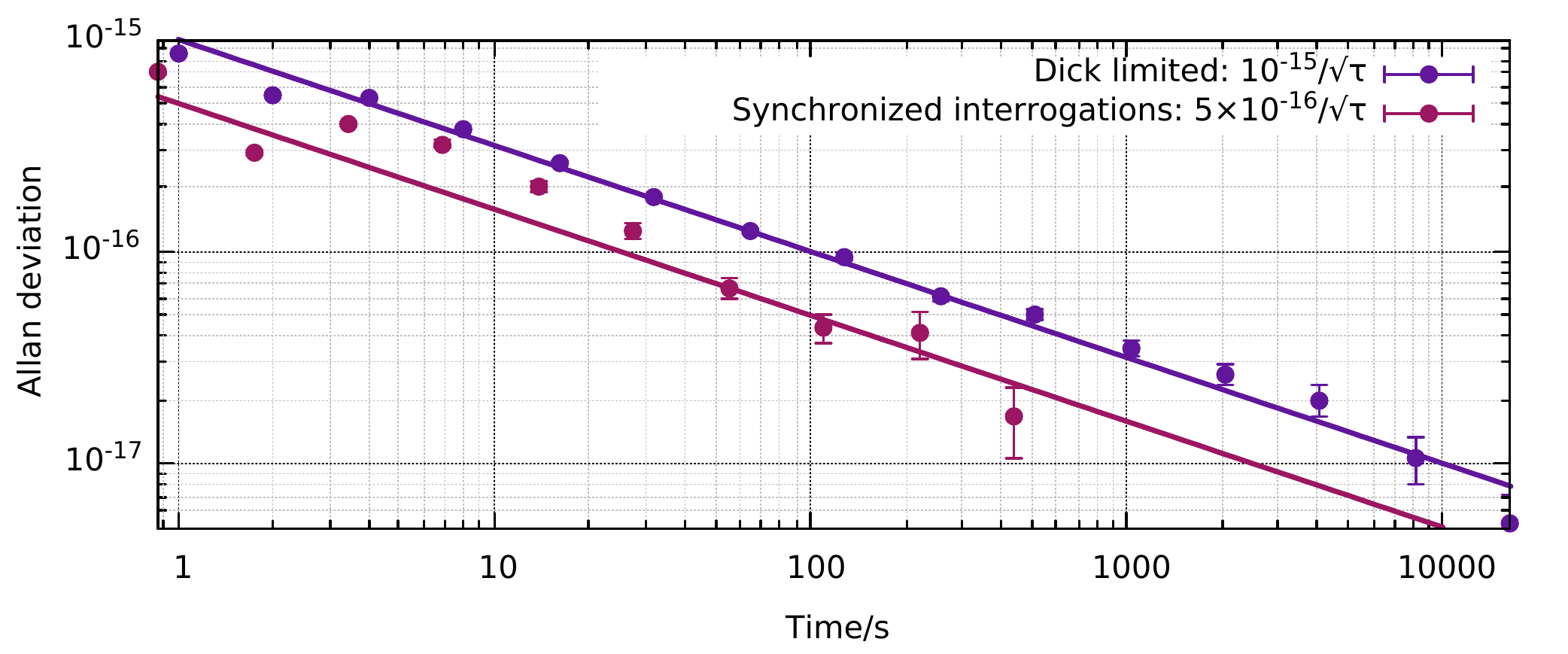}
	\caption{\label{fig:interrogation_dickfree} Allan deviation of the frequency difference between the two Sr clocks at SYRTE. When the clock interrogation windows are synchronised, the Dick effect is reduced. The reduction is effective after the servo loops have relaxed (a few tens of seconds). This leads to a reduction of the time required to probe systematic effects by a factor of four.}
	\end{center}
\end{figure}

One technique is to probe two separate clocks with the same clock laser, with synchronised interrogation windows. In this scheme, the clock-laser noise is sampled in the same way by the two atomic ensembles and is therefore eliminated when computing the frequency difference between the clocks~\citep{Lodewyck2010, Takamoto2011}. Figure~\ref{fig:interrogation_dickfree} shows an experimental realisation of such a rejection scheme. This method is suitable, e.g., to measure systematic effects with a short integration time when two clocks are available. It can even be implemented between clocks that use two different species by using a frequency comb to transfer the spectral purity of the clock laser over the optical spectrum. When implemented efficiently, the Allan deviation of the frequency difference between the two samples averages down as $1/\tau$, starting from the clock-laser frequency noise and down to the residual noise level of the clock (from detection noise, QPN or residual instabilities due to uncompensated optical paths). However, this technique has two drawbacks: first, it does not actually implement a physical oscillator with a reduced instability, and second, it is not suitable for long-range comparisons over fibre networks as this comparison method is not capable of transferring the high-frequency noise of the ultra-stable lasers.

A second method is to probe two atomic ensembles with the same clock laser, but with complementary interrogation windows using Ramsey spectroscopy. These two interleaved atomic ensembles virtually form a single clock in which the frequency of the clock laser is continuously measured by atoms, and therefore without any aliasing effect~\citep{Lodewyck2010,Schioppo2017}. For this, both clocks feed back successively to the clock-laser frequency. Anti-correlations between successive corrections enable the final output to overcome the Dick limit. This technique can be used for remote-clock comparisons. However, its realisation is experimentally involved because the dead time in the clock cycle used to prepare and detect the atoms must be as short as the duration of clock interrogation in order to reach a 50\% duty cycle. This can be achieved by extending the clock interrogation window using a clock laser with a very long coherence time~\citep{Schioppo2017}. However, this defeats the purpose of an interleaved clock whose advantage is precisely to enable very low instabilities without resorting to complicated laser-stabilisation techniques.

\begin{figure}[tb]
	\begin{center}
	\includegraphics[width=1\textwidth]{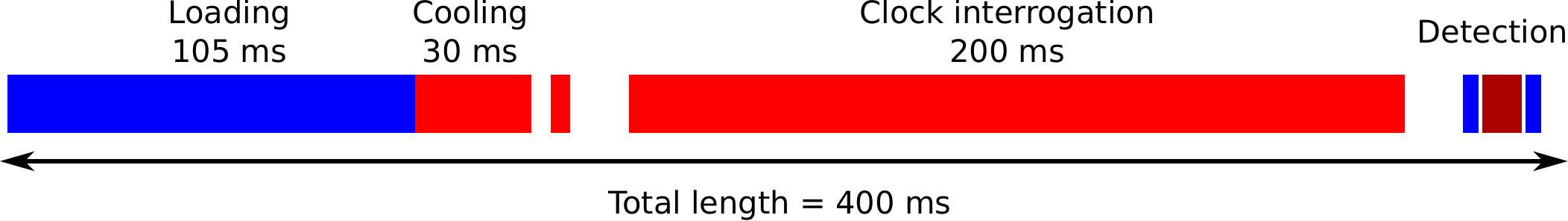}
	\caption{\label{fig:interrogation_sequence}Optimised sequence for an interleaved operation of two atomic ensembles forming a dead-time-free clock. The length of the sequence is 400~ms, with only 200~ms of dead time used to load, prepare and detect a few hundred atoms.}
	\end{center}
\end{figure}

In the framework of the OC18 project, we designed a clock sequence suitable for a dead-time-free clock by taking advantage of cavity-assisted non-destructive detection of Sr atoms in an optical lattice clock~\citep{Vallet2017}. The exquisite signal-to-noise ratio of this detection, capable of resolving just a few atoms in the optical lattice, makes it possible to operate the clock with a reduced number of atoms (a few hundred) without being limited by the detection noise. With this feature, we were able to demonstrate a workable short clock sequence with a duration of 400\;ms and a 50\% duty cycle (see Figure~\ref{fig:interrogation_sequence}) that is compatible with the frequency noise of compact ultra-stable laser systems. Furthermore, this detection technique also makes it possible to reuse the atoms from cycle to cycle, thus offering the possibility to further reduce the clock dead time. Realisation of a similar detection scheme in a second clock system is required in order to demonstrate a dead-time-free clock with enhanced stabilities.

\section{Interleaved measurement techniques to reduce uncertainty} \label{sec:interrogation_interleaving}

In order to monitor and then remove frequency shifts arising from the interaction of atoms with their environment, it is common practice to interleave independent frequency servos to the atomic transition. The separate servos can be operated with different magnitudes of perturbing fields, such as magnetic field, probe power etc., and then the frequency shifts can be extrapolated to zero, as described in Section~\ref{sec:interrogation_extrapolation}.

Care should be taken to avoid introducing any errors in the clock frequency from the ordering of the probes in the interleaved sequence.  For example, if the clock laser is drifting linearly in frequency and is used to probe servo 1 and then servo 2, always in the order of $[1, 2]$, the servos will experience different frequency offsets compared to the situation of probing the pair of servos in the reverse order $[2, 1]$.  For a cavity drift of ${\sim} 50$\;mHz/s and 1-s cycles for each servo, the bias introduced into the clock frequency would be at the level of 1 part in $10^{16}$.  Alternating the order of the servo pairs, so as to average over the sequence $[1, 2, 2, 1]$, would give immunity to linear cavity drifts.  Quadratic drift immunity could then be obtained by additionally reversing the order of these sequences, so as to average over $[1, 2, 2, 1; 2, 1, 1, 2]$.  Note that similar patterned sequences can also be used in frequency locking when alternating probes between the high and low frequency sides of a particular transition~\citep{Itano2007}.

\subsection{Ytterbium-ion E2 and E3 clock transitions}
\label{sec:interrogation_E2andE3}

\begin{figure}[tb]
\begin{center}
\includegraphics[width=.6\columnwidth]{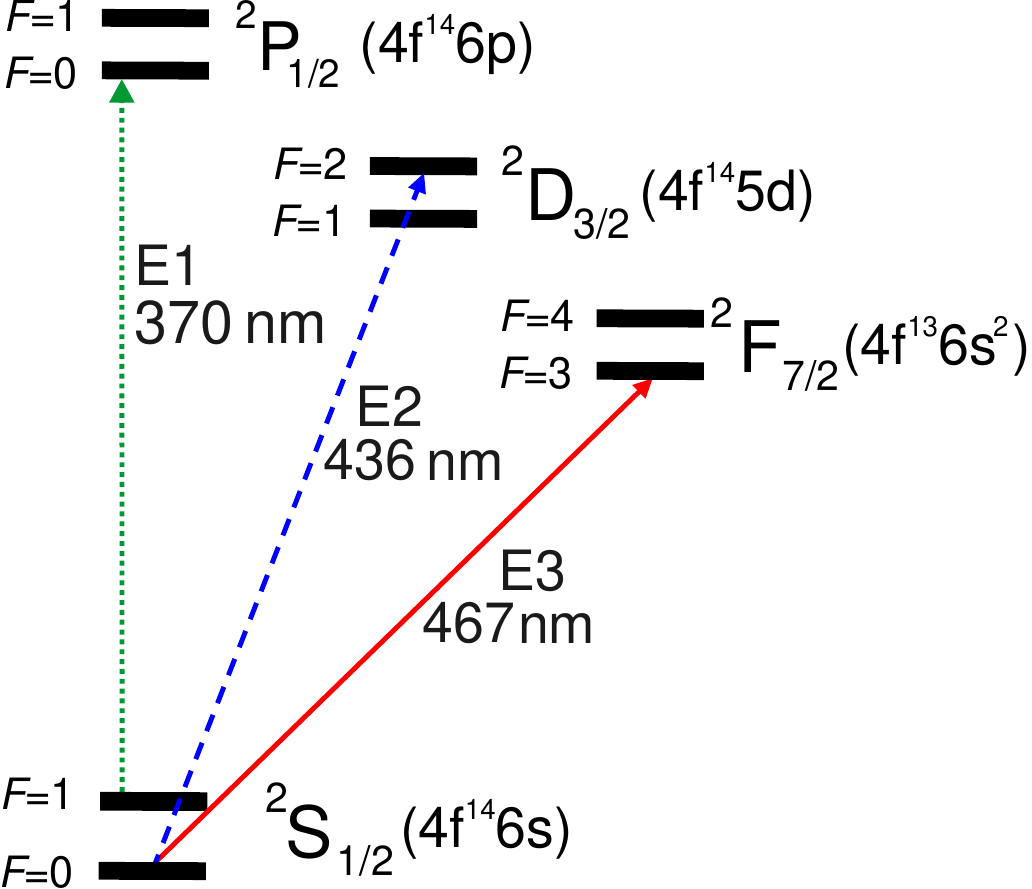}
\caption[]{Parts of the $^{171}$Yb$^{+}$ term scheme are shown. The strong $^2$S$_{1/2}\to {}^2$P$_{1/2}$ electric dipole (E1) transition at 370\;nm is used to laser cool the ion. The electric quadrupole (E2) and electric octupole (E3) transitions serve as the reference for optical frequency standards. \label{fig:interrogation_Yblevels}}
\end{center}
\end{figure}

Singly-ionised ytterbium is unique amongst optical frequency standards in that it has two different transitions that are used as optical clock frequencies: an electric quadrupole transition (E2, $^2\mathrm{S}_{1/2} \rightarrow {}^2\mathrm{D}_{3/2}$) at 436\;nm and an electric octupole transition (E3, $^2\mathrm{S}_{1/2} \rightarrow {}^2\mathrm{F}_{7/2}$) at 467\;nm, see Figure~\ref{fig:interrogation_Yblevels}.

The two transitions have different sensitivities to external perturbations and this provides an opportunity to run interleaved servos with the E2 and E3 transitions exposed to the \emph{same} external perturbations, but undergoing \emph{different} frequency shifts~\citep{Godun2014,Huntemann2016}.  Note that this is different from the previous examples of extrapolating the frequency shifts in a single transition to zero, but is particularly useful if the shifts in a transition are too small to resolve directly, such as for the E3 transition.  The significantly higher sensitivity of the E2 transition to electric and magnetic fields permits diagnosis of field-induced shifts of the E3 transition frequency on a magnified scale.  It is essential, of course, to know the relative scaling factors for the frequency shifts in the E2 and E3 transitions.  These are generally measured in auxiliary experiments and the examples given in Table~\ref{tab:interrogation_E2E3scaling} show how much more readily the frequency offsets can be resolved with the E2 transition.

\begin{table}[b]
\caption{Relative magnitudes of atomic parameters affecting frequency shifts in the E2 and E3 optical clock transitions in $^{171}$Yb$^+$.\label{tab:interrogation_E2E3scaling}}
\label{Tab:E2E3shifts}
\begin{minipage}[t]{0.71\textwidth}

\begin{tabular}{l S[table-format=-2.5, table-space-text-post = ~$^a$]
S[table-format=-2.5, table-space-text-post = ~$^a$]
S[table-format=1.6, table-space-text-post = ~$^a$]}
\toprule
Transition 	& {Quadrupole}  	& {Quadratic Zeeman}  & {Differential DC}  \\
 			& {moment} 		& {coefficient}       & {polarisability}  \\
 			& {$ea_0^2$} 		& {$\mathrm{mHz}/\si{\micro\tesla}^2$}    & {$\times 10^{-40} \mathrm{Jm^2/V}^{2}$}  \\
\midrule
E2 & 2.08(11) ~$^\mathrm{a}$  &52.13(9) ~$^\mathrm{b}$ & 6.9(14) ~$^\mathrm{a}$ \\
E3 & -0.041(5) ~$^\mathrm{c}$  &-2.08(1) ~$^\mathrm{b}$ & 0.888(16) ~$^\mathrm{d}$ \\
\midrule
Ratio\\
$\mathrm{E}2/\mathrm{E}3$ & -50.7(67)   &-25.06(13)    & 7.8(16)  \\
\bottomrule
\end{tabular}
\end{minipage}
\hfill
\begin{minipage}[t]{0.28\textwidth}
\vfill
\begin{tabular}{l}
\small $^\mathrm{a}$ \cite{Schneider2005}\\
\small $^\mathrm{b}$ \cite{Godun2014}\\
\small $^\mathrm{c}$ \cite{Huntemann2012a}\\
\small $^\mathrm{d}$ \cite{Huntemann2016}\\
\end{tabular}
\end{minipage}
\end{table}

For continuous monitoring, it would be necessary to interleave an E2 servo cycle with every E3 servo cycle.  This, however, has the undesirable effect of increasing the dead time of the E3 interactions.  In practice, it may not be necessary to probe the E2 transition so frequently and a judgment must be made about the timescale on which different systematic effects are likely to be varying when choosing how often to monitor shifts with a separate transition.

\section{Case study of operational practices with \texorpdfstring{Yb$^{+}$}{Yb+} clocks}
\label{sec:interrogation_operational-practices}

\begin{figure}[t]
\begin{center}
\includegraphics[width=.6\columnwidth]{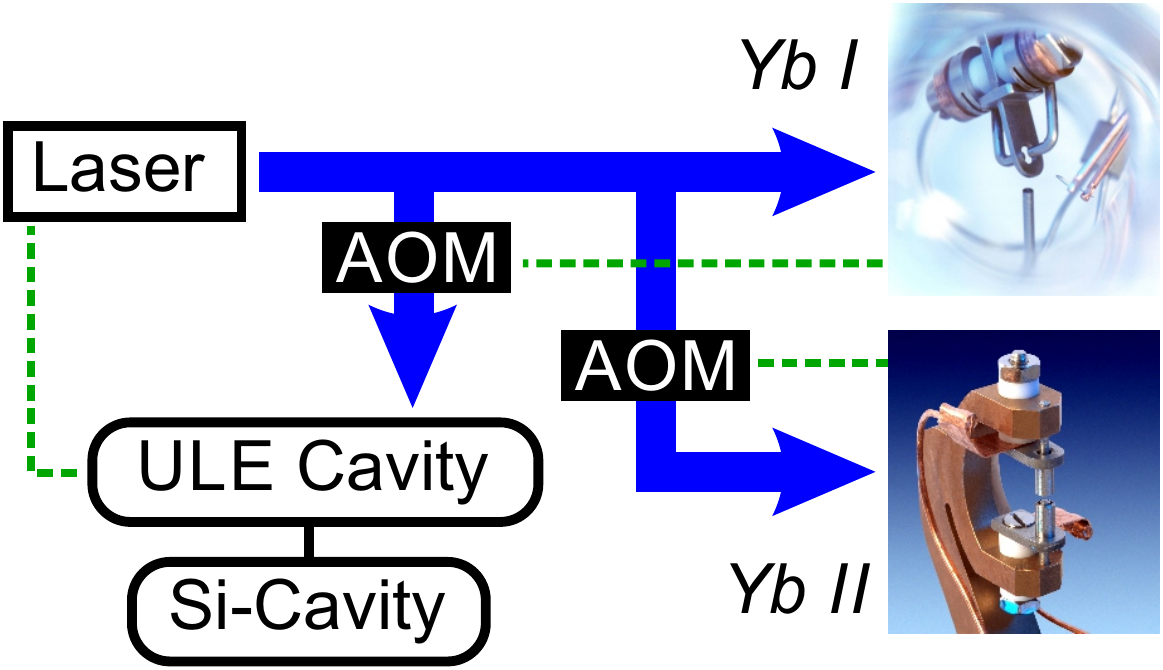}
\caption{Experimental setup for the comparison of the two Yb$^+$ single-ion clocks \textit{Yb I} and \textit{Yb II} at PTB. The blue solid arrows show the laser beam path, and the green dashed lines indicate servo loops. The clock laser is locked to an ultra-low-expansion (ULE) glass cavity with high bandwidth. The stability of the laser is further improved by referencing a single-crystal Si cavity via a frequency comb generator. With a digital loop filter that uses spectroscopic information obtained from \textit{Yb I} the laser frequency is controlled via an acoustic-optic modulator (AOM) in front of the cavity. For \textit{Yb II} a digital loop filter controls a frequency offset with an AOM before the trap. \label{fig:interrogation_Expsetup}}
\end{center}
\end{figure}

Many of the probing techniques described in this chapter have been demonstrated with $^{171}$Yb$^{+}$ optical clocks at PTB, and tested via a long-term comparison of two systems that differ significantly in trap geometry, control software and interrogation sequence (see Figure~\ref{fig:interrogation_Expsetup}).

The older clock \textit{Yb I} employs a spherically symmetric Paul trap \citep{Beaty1987}, while \textit{Yb II} uses an endcap trap design \citep{Schrama1993, Stein2010}. For the latter clock system a new control software has recently been developed that is based on MATLAB and uses National Instruments hardware for digital and analog signals. When referencing the E3 transition, the two clocks utilise different techniques, described in Section~\ref{sec:interrogation_tailored-sequences}, to cancel the probe-field-induced AC Stark shift. For \textit{Yb I}, HRS is performed with interleaved Rabi interrogations to control the step frequency $\Delta_\mathrm{S}$. For \textit{Yb II}, the autobalanced Ramsey technique is implemented. For both systems, a free evolution period between the Ramsey pulses of about 360\;ms is chosen and the contributions to the systematic uncertainty from the AC Stark shift in each clock are expected to be less than $1\times 10^{-18}$.

As described in Section~\ref{sec:interrogation_E2andE3}, the Yb$^+$ E2 transition is used as a sensitive probe of the external fields for evaluating the E3 field-induced frequency shifts. In order to decide how frequently the E2 transition should be probed, the timescale on which external fields drift should be well understood.  For \textit{Yb I}, the electric fields need to be corrected on a daily basis to minimise excess micromotion, whereas the compensation voltages in \textit{Yb II} remain constant over weeks. This is attributed to isolator material in the vicinity of the trap in the \textit{Yb I} setup that is not present in \textit{Yb II}. Both systems make use of mu-metal shielding and the magnetic fields have been found to be constant over periods of several weeks, if the mechanical setup is unchanged. Changes of the optical setup by adding, removing or changing the position of opto-mechanical components close to the trap, however, lead to a change in the magnetic field at the position of the ion. This change is compensated by adjusting the currents through an orthogonal set of coils around the ion's vacuum chamber.  With knowledge of the timescales on which the external fields change in each trap, the E2 frequency shifts can be measured at appropriate intervals. Scaling down the magnitude of the frequency shifts for the E3 transition, it is seen that electric field gradients and magnetic fields contribute no more than $5 \times 10^{-19}$ to the total frequency uncertainty of the E3 clock.

Besides the characterisation of residual electric and magnetic fields perturbing the ion, the E2 transition is also well suited for characterising the residual motion of the trapped ion. Here, measurements of the carrier-to-sideband ratio yield information on residual second-order Doppler shifts after cooling and during the interrogation. The micromotion part of this shift can also be inferred from measurements of a correlation between trap drive phase and the arrival of single photons at the photomultiplier tube \citep{Berkeland1998, Keller2015}.

In a long-term comparison of the two clocks over a period of six months, with  an uptime of up to 95\% per day, we found an agreement of the clock frequencies within the combined systematic uncertainty of $4 \times 10^{-18}$ \citep{Sanner2019}. The observed frequency instability averages as $1.4\times 10^{-15} / \sqrt{\tau\,[\mathrm{s}]}$ and reaches a statistical uncertainty below $1\times 10^{-18}$ after a total measurement time of about 25 days, thus giving only a minor contribution to the total uncertainty.  The excellent agreement seen between the two clocks with different-style traps, operating with different interrogation sequences and in different external fields, demonstrates the applicability of the techniques presented in this chapter.

\clearpage

\bibliographystyle{apsrmp4-2_OC18}

\clearpage
\phantomsection
\addcontentsline{toc}{chapter}{Bibliography}
\bibliography{OC18_bib}

\end{document}